\pdfoutput=1 
\documentclass[%
aip,pof,
amsmath,amssymb,
preprint,%
author-year,%
]{revtex4-2}
\usepackage{color}
\usepackage{hyperref}
\hypersetup{
	colorlinks=true,
	citecolor=blue,
	linkcolor=blue,
	filecolor=black,      
	urlcolor=red,
}
\usepackage{graphicx}% Include figure files
\usepackage{dcolumn}% Align table columns on decimal point
\usepackage{bm}% bold math
\usepackage[mathlines]{lineno}% Enable numbering of text and display math
\usepackage{mathrsfs}
\usepackage[utf8]{inputenc}
\usepackage[T1]{fontenc}
\usepackage{mathptmx}
\usepackage{subfigure}
\usepackage{tabularx}%控制表格宽度
\usepackage{booktabs}%控制表格线宽
\usepackage{bm}
\usepackage{CJKutf8}
\usepackage{amsthm}
\usepackage{amsfonts}
\usepackage{float}
\usepackage{extarrows}
\usepackage{pifont}
\usepackage[normalem]{ulem}
\usepackage{natbib}
\usepackage{multirow}

\usepackage{tabulary}
\usepackage{makecell}

\begin{document}
	\begin{center}
		{\large\bf Vorticity-dynamical analysis of Richtmyer-Meshkov instability based on orbital-spin vorticity decomposition\\}
	\end{center}
	\vspace{5pt}
	\noindent
	\begin {CJK*} {UTF8} {gbsn} 
	\begin{center}
		\textcolor{blue}{Xi Chen$^1$ (陈曦), Tao Chen$^{1,\dagger}$ (陈涛), Tianshu Liu$^{2}$ (刘天舒)}
	\end{center}
\end{CJK*}
\begin{center}
	$^1$ School of Physics, Nanjing University of Science and Technology, Nanjing 210094, China\\
%	$^2$ Department of Modern Mechanics, University of Science and Technology of China, Hefei 230026, China\\
	$^2$ Department of Mechanical and Aerospace Engineering, Western Michigan University, Kalamazoo, Michigan 49008, USA
\end{center}
\begin{center}
	\textcolor{black}{$^\dagger$Email address for correspondence: chentao2023@njust.edu.cn}
\end{center}

\section*{Abstract}
\vspace{5pt}
\noindent
The present study provides a detailed vorticity-dynamical analysis of shock-driven hydrodynamic physics in canonical Richtmyer–Meshkov instability (RMI) flows: shock interaction with a single-mode perturbed interface, and a cylindrical air bubble immersed in Krypton. Vorticity and vortex structures are extracted and quantified using a recently proposed streamline-based direction-dependent vorticity decomposition (DVD), $\bm{\omega}=\bm{R}+\bm{s}$, where $\bm{R}$ and $\bm{s}$ denote the orbital-rotation and spin modes, respectively [\citet{chen2025kinematic}, Physics of Fluids 37, 027187]. Both Klein-Kaden-Betz (KKB) (same sign of $\bm{R}$ and $\bm{s}$) and anti-KKB (opposite signs of $\bm{R}$ and $\bm{s}$) configurations of the vorticity modes are observed within the same flow field. For both RMI cases, the analysis reveals a three-layer structure of the primary vortex pair deposited at the interface. For each primary vortex, in the inner core region, the vorticity modes exhibit an anti-KKB configuration (KKB antagonistic effect), resembling the classical Burgers vortex solution, with $\bm{\omega}$ dominated by $\bm{R}$ and substantially attenuated by $\bm{s}$. The KKB configuration appears in an interface-attached annular vortex band that forms the periphery of the primary vortex core. This synergistic effect drives further interface rolling-up and the development of mushroom-shaped structures. External to this region, the velocity field approaches a potential-flow regime with an anti-KKB configuration satisfying $\bm{R}\approx-\bm{s}$, where the flow is induced by the interface-deposited vorticity via the Biot-Savart law. Furthermore, we propose a double decomposition of the total vorticity flux in a control volume $(\Gamma_{\omega}=\Gamma_{\rm R}+\Gamma_{\rm s})$, and a triple decomposition of the well-known $Q$-criterion (namely, the second principal invariant of the velocity gradient tensor). The transient evolution of the vorticity fluxes $(\Gamma_{\omega},\Gamma_{\rm R},\Gamma_{\rm s})$ elucidates the physical roles of and interactions among interface geometric deformation, vortex structures, and dilatational wave fields. The upper and lower bounds of these vorticity fluxes are expressed using the integrals of the characteristic vorticity modes in the invariant vorticity decompositions (IVDs). Notably, the orbital-spin coupling effect $(\bm{R}\bm{\cdot}\bm{s}/2)$ makes a primary negative contribution to $Q$ inside the inner vortex core, whereas the line-element stretching effects become significant only in the spiral arms of the annular vortex band. These findings provide insights into vortical flow physics beyond the conventional vorticity paradigm, and the DVD approach holds promise for the diagnosis of practical instability-induced flows.

%\newpage
\section{Introduction}\label{Introduction}
Rayleigh–Taylor and Richtmyer–Meshkov instabilities (RTI and RMI) play a fundamental role across a broad spectrum of engineering, geophysical, and astrophysical flows~\citep{Zabusky1999VORTEX,zhai2017review,zhou2017rayleigh,Zhou2017partII,Zhou2024}, with particular significance in high-energy-density physics~\citep{zhou2025instabilities}. 
In inertial confinement fusion (ICF), RMI-induced mixing constitutes a key hydrodynamic obstacle, degrading fusion yield by promoting atomic-scale diffusion and potential ignition failure \citep{haines2016detailed}. Extending to astrophysical scales, these instabilities are critical in driving supernova explosions \citep{wheeler1990type} and shaping the morphological evolution of nebular ejecta \citep{balick2002shapes}. Consequently, a thorough understanding of RMI evolution is essential for both fundamental physics and practical engineering applications.

The concept of RTI originates from the observation of cirrus cloud formation by Lord~\citet{Rayleigh1883}, and the modern interpretation of an underwater explosion by~\citet{Taylor1950}. It is triggered when a perturbed interface separates a heavy fluid from a light one, with the acceleration directed from light to heavy. As the implusive counterpart of  RTI (i.e, high-acceleration, short-duration RTI),  RMI arises when an incident shock wave passes through a perturbed interface separating two fluids initially at rest. The underlying mechanism was analyzed both theoretically~\citep{richtmyer1960taylor} and experimentally~\citep{Meshkov1969} (via a shock tube experiment). Notably, RMI was earlier recognized by~\citet{Markstein1957} in observations of shock-flame interaction, where the essential differences between these two instabilities were also discussed. Both instabilities typically lead to the development of the Kelvin–Helmholtz instability (KHI) characterized by fascinating rolling-up KH billows, which describes the instability that arises at the interface between two adjacent fluid layers in relative motion~\citep{Helmholtz1868,Thomson1871}. These three types of instabilities are often coupled within a single fluid system~\citep{ranjan2011shock}. 
The present study focuses on the analysis of RMI, particularly from a vorticity-dynamical perspective.  To provide the necessary background, relevant advances in the existing literature are reviewed as follows.

The first category of RMI studies considers relatively simple interface morphologies, specifically those with quasi-single-mode and multi-mode perturbations~\citep{Liang2019,Liang2021}.
As the shock wave travels across the initial interface and accelerates it along the propagation direction, the baroclinic torque, arising from the misalignment between the density and pressure gradients, induces vorticity and circulation deposition at the interface, which serves as the primary mechanism driving interfacial instability~\citep{Morgan2012,samtaney1994circulation,RobertsJacobbs2016}.
Subsequently, secondary instabilities (RTI and KHI) develop, resulting in a nonlinear, complex flow field~\citep{shankar2010numerical}. 
This induced velocity field amplifies initial perturbations and transitions the flow into a nonlinear phase, where heavy-fluid spikes penetrate the light fluid while bubbles rise~\citep{Sohn2003}. As the instability matures, the interface rolls up to form characteristic mushroom-shaped structures~\citep{Morgan2012}. In three-dimensional (3D) configurations, they evolve into intricate systems of distorted vortex rings, eventually generating a chaotic mixing layer~\citep{Chapman2006,Zhou2024}. Numerous theoretical models have been proposed to predict the growth rates of interfacial amplitude and global mixing widths during the linear and weakly nonlinear stages~\citep{richtmyer1960taylor,meyer1972numerical,Vandenboomgaerde1998}, as well as the velocity decay of bubbles and spikes in the nonlinear saturation stage~\citep{Sadot1998, Mikaelian2003, Zhang2016Universality}. Subsequent studies extended these theories to incorporate compressibility effects \citep{Fraley1986} and higher-order harmonic contributions \citep{zhang1997nonlinear, wouchuk1997asymptotic, jacobs1996experimental}.
Concurrently, advances in experimental diagnostics, particularly schlieren imaging, have enabled extended quantitative analysis in complex scenarios involving multi-mode perturbations~\citep{Yosef-Hai2003, Vandenboomgaerde2014} and non-planar shock interactions~\citep{Li2025}.

The second category of RMI studies focuses on the coupling effects arising from interface geometry configuration and the presence of multiple interfaces.
During shock interactions with smooth interfaces (e.g., cylindrical, elliptical, and spherical), hydrodynamic phenomena such as shock focusing or defocusing, anisotropic vorticity generation, and azimuthal instabilities have been observed~\citep{zhai2011evolution,Hejazialhosseini2013,Singh2021impact}.
For polygonal interfaces with flat faces and sharp corners (e.g., square, triangular, and pentagonal bubbles), the flow-geometry coupling mechanism leads to stronger baroclinic vorticity generation, more pronounced localized deformation, and the emergence of instability modes not observed in smoother configurations. In particular, secondary flows are amplified, wave patterns become more complex, and the pathways to instability vary significantly with the interface geometry~\citep{Zhai2014lp,singh2023investigation,singh2023shock,fan2012numerical}. Direct numerical simulations of RMI have also been extended to multiple-interface configurations with varying geometries, such as tandem light square/cylindrical interfaces~\citep{Singh2025ccc,Alsaeed2025} and parallel light/heavy square interfaces~\citep{Alsaeed_Singh_PhysicaD2025,Singh2024ScienceChina}, etc. Notably, the inversion of the density-gradient reverses the baroclinic vorticity deposition at the light bubble interface, resulting in fundamentally different interaction mechanisms compared to those of heavy bubbles. In these works, the authors analyzed the effects of initial interface separation distance, Atwood number, and shock Mach number on baroclinic vorticity generation, shock-interface and vortex-interface interactions, coupling-driven instability growth and turbulence mixing rates.
The practical application values of these fundamental studies were also highlighted in the contexts of ICF and propulsion engineering. Different contributions to the material evolution rate of vorticity were quantified based on the spatially integrated individual mechanisms in the vorticity transport equation~\citep{Alsaeed2025,Singh2025ccc,LiXuan2026reactive}.

Although the vorticity paradigm often demonstrates compelling advantages over primitive variables (velocity and pressure), vorticity alone could not distinguish pure shearing motions from true swirling motions, and is therefore not equivalent to vortex~\citep{Kolar2007IJHMF}. The existing studies have revealed that vorticity can be further decomposed into two elementary modes, i.e., the rigid-rotation mode (swirling) and the spin mode (shear), either by adopting characteristic algebraic approaches and invariant vorticity decompositions (IVDs)~\citep{LiZhen2014,Liu2018,chen2025kinematic}, or by employing material and field descriptions~\citep{chen2025kinematic,Chen2025arxiv}. Subsequently, a unifying principle proposed by~\citet{Chen2026operator} links these distinct vorticity modes to two communicative vorticity/vortex operators within a novel operator-form vorticity decomposition framework. These studies have proven valuable as analytical tools for characterizing the fundamental forms of vortex structures (which typically manifest as axial vortices and shear layers at high Reynolds numbers), as well as their interactions and mutual transformations during dynamic evolution.
Classical fluid mechanics has largely focused on volume-element descriptions (per unit volume or mass), leaving the rotational kinematics of directed material elements, with its extra degrees of freedom beyond vorticity, largely unexplored until recently. 
By combining field theory and differential geometry,~\citet{chen2025kinematic} examined the rotational kinematics of a pair of orthogonal line elements across all possible scenarios, and proposed a streamline-based, direction-dependent vorticity decomposition (DVD) for generic 2D flows. However, this decomposition has not yet been applied to analyze the vorticity and vortex dynamics in RMI. Therefore, it is of particular interest to perform the first study of RMI with this novel dual vorticity paradigm (i.e., the orbital-spin decomposition). 

The remainder of this article is structured as follows. Section \S\ref{NS} begins with the Navier–Stokes (NS) equations, from which the vorticity transport equation is derived and accompanied by physical interpretations. In \S\ref{DVD_IVD}, we briefly introduce the recently proposed kinematic vorticity decompositions within both algebraic and field frameworks. Building upon these results, we propose a double decomposition of the total vorticity flux within a fixed control volume and establish inequalities that provide upper and lower bounds for its constituent vorticity fluxes in \S\ref{Decomposition and Estimate of total vorticity}. The numerical model and computational approach are described in \S\ref{Numerical aspects}. For the first time, the proposed DVD approach is applied to analyze the shock-driven hydrodynamic physics of RMIs in two distinct scenarios: a planar interface subjected to a single-mode initial perturbation (\S\ref{RMI of a single-mode perturbed interface}), and a cylindrical air bubble immersed in Krypton (\S\ref{RMI in shock-accelerated cylindrical air bubble in Krypton}). The analysis places particular emphasis on the spatiotemporal evolution of and mutual interaction among material interface morphology, vorticity/vortex structures, and wave fields. Notably, the physical constituents of the well-known $Q$-criterion are clearly separated and interpreted using the streamline-based DVD. Finally, concluding remarks and discussions are documented in \S\ref{Conclusions and discussions}.

%\newpage
\section{Navier-Stokes equations and vorticity evolution}\label{NS}
For compressible viscous flow, the Navier-Stokes (NS) equations governing the evolution of the fluid system are given by
\begin{subequations}\label{NS12}
	\begin{equation}\label{NS1}
		\frac{\partial \rho}{\partial t} +\bm{\nabla}\cdot\left(\rho \bm{u}\right)=0,
	\end{equation}
	\begin{equation}\label{NS2}
		\frac{\partial\bm{u}}{\partial t} + \bm{u}\bm{\cdot}\bm{\nabla }\bm{u}= -\frac{1}{\rho}{\bm{\nabla} p} + \frac{1}{\rho}\bm{\nabla} \cdot \bm{\sigma},
	\end{equation}
\end{subequations}
where $\rho$ is the density, $\bm{u}$ is the macroscopic velocity, and $p$ is the pressure.
The continuity equation \eqref{NS1} depicts the mass conservation law. The left-hand side of \eqref{NS2}
represents the material rate of change of velocity (i.e., ${D\bm{u}}/{D t}$), driven by the pressure gradient $(-\rho^{-1}{\bm{\nabla} p})$ and the viscous force $(\rho^{-1}\bm{\nabla} \cdot \bm{\sigma})$. The symmetric–antisymmetric decomposition (SAD) of the velocity gradient tensor (VGT) $\mathbf{A}\equiv\bm{\nabla}\bm{u}$ yields 
\begin{eqnarray}
\mathbf{A}=\mathbf{D}+\bm{\Omega},~~\mathbf{D}\equiv\frac{1}{2}(\mathbf{A}+\mathbf{A}^{\rm{T}}),~~\bm{\Omega}\equiv\frac{1}{2}(\mathbf{A}-\mathbf{A}^{\rm{T}}),
\end{eqnarray}
where $\mathbf{D}$ and $\bm{\Omega}$ are the strain-rate and rotation-rate tensors, respectively.
Here, $\bm{\sigma}\equiv2\mu\mathbf{D}-(2/3)\mu{\vartheta}\mathbf{I}$ is the viscous stress tensor, where $\mathbf{I}$ is the identity tensor, $\vartheta\equiv{\rm tr}(\mathbf{A})={\rm tr}(\mathbf{D})=\bm{\nabla}\bm{\cdot}\bm{u}$ is the dilatation ($\rm{tr}$ denotes the trace operator), and $\mu$ is the dynamic visicosity. $\bm{\nabla}$ denotes the spatial gradient operator. Notably, the divergence of the viscous stress tensor can be expressed as $\bm{\nabla \cdot\sigma}=-\mu\bm{\nabla}\times\bm{\omega}+(4/3)\mu\bm{\nabla}\vartheta$, where $\bm{\omega}\equiv \bm{\nabla} \times \bm{u}$ is the vorticity as the dual vector of $\bm{\Omega}$. For convenience, the linear diffusion approximation introduced by~\citet{lighthill1956viscosity} is adopted, in which the viscous and thermal conductivity coefficients are treated as constants. This yields an alternative form of~\eqref{NS2}:
\begin{eqnarray}
		\frac{\partial\bm{u}}{\partial t} + \bm{\omega}\times\bm{u}+\bm{\nabla}\left(\frac{1}{2}u^2\right)=-\frac{1}{\rho}\bm{\nabla}\hat{p}-\nu\bm{\nabla}\times\bm{\omega},
\end{eqnarray}
where $\hat{p}\equiv{p}-(4/3)\mu\vartheta$ is the modified pressure incorporating the dilatation correction.

The vorticity paradigm offers distinct advantages in elucidating the evolutionary characteristics of the RMI~\citep{zhou2017rayleigh, zhou2021rayleigh,Zhou2024}.
Vorticity deposited at the perturbed interface drives the development of interfacial perturbations through induced velocity fields, which constitutes the fundamental mechanism underlying interfacial instability in compressible fluid systems~\citep{Morgan2012, singh2025shock}. Taking the curl of~\eqref{NS2} yields the vorticity transport equation for compressible flow, given by
\begin{equation}\label{wol}
	\frac{D \bm{\omega}}{D t}=\bm{\omega} \cdot \bm{\nabla}\bm{ u}-\vartheta\bm{\omega}+\frac{1}{\rho^{2}} \bm{\nabla} \rho \times \bm{\nabla} p+\nu\bm{\nabla}^2\bm{\omega}.
\end{equation}
The left-hand side of~\eqref{wol} represents the material evolution rate of vorticity. On the right-hand side of~\eqref{wol}, the first term is the vortex-stretching term, which accounts for the stretching, turning, and tilting of vorticity lines by the velocity gradient~\citep{Wu2015}. Vortex stretching also reflects the principle of conservation of angular momentum: stretching reduces the moment of inertia of the fluid elements composing a vortex line, thereby increasing their angular speed~\citep{ranjan2011shock}. This term vanishes in 2D flows, as the vorticity vector remains orthogonal to the plane of motion. The second term is the vorticity-dilatation coupling term, which becomes significant in highly compressible flows, particularly during the instantaneous interaction between a shock and an interface. The third term represents the baroclinic torque, which arises from the misalignment between the local pressure gradient (induced by the shock) and the density gradient (associated with the interface). This mechanism typically generates vorticity in the fluid interior and deposits it on the fluid–fluid interface.

The final term represents the diffusion of vorticity by viscosity. In the present study, this term can be neglected generally because of the low physical viscosities of the fluids considered ($\mu \sim10^{-5} \rm{Pa\cdot s}$) and the short timescales ($t\sim 10^{-6}\rm{s}$) over which the flow evolution is studied \citep{zhai2017review,ranjan2011shock}. Then, \eqref{wol} is simplified into
\begin{equation}\label{jianh}
	\frac{D \bm{\omega}}{D t}=-\vartheta\bm{\omega}+\frac{1}{\rho^{2}} \bm{\nabla} \rho \times \bm{\nabla} p. 
\end{equation}

\section{Kinematic vorticity decompositions for generic 2D flows}\label{DVD_IVD}
\subsection{Streamline-based direction-dependent vorticity decomposition}\label{Streamline-based DVD}
At each point along a regular streamline $\mathscr{C}\subset\mathbb{R}^{3}$ in a selected reference frame, one can define a right-handed orthonormal triad $(\bm{t},\bm{n},\bm{b})$, known as the Frenet-Serret triad, where $\bm{t}$ is the unit tangent vector, $\bm{n}$ is the principal normal vector (directed toward the local center of curvature), and $\bm{b}\equiv\bm{t}\times\bm{n}$ is the binormal vector. The local curvature of $\mathcal{C}$ is denoted by $\kappa$.
The velocity vector $\bm{u}=q\bm{t}$ is tangent to the streamline, with $q\equiv\lVert\bm{u}\rVert$ representing the speed (or the velocity magnitude).
For generic 2D flows in $\mathbb{R}^{2}$, a streamline-based, direction-dependent vorticity decomposition (DVD) was recently proposed~\citep{chen2025kinematic},
\begin{subequations}\label{vd}
	\begin{eqnarray}\label{vd1}
		\bm{\omega}=\bm{R}(\bm{t})+\bm{s}(\bm{t}),
	\end{eqnarray}
	\begin{eqnarray}\label{vd2}
		\bm{R}(\bm{t})\equiv2\bm{t}\times\left(\bm{t}\bm{\cdot}\mathbf{A}\right),~
		\bm{s}(\bm{t})\equiv-2\bm{t}\times\left(\bm{t}\bm{\cdot}\mathbf{D}\right),
	\end{eqnarray}
\end{subequations}
The quantities $\bm{R}(\bm{t})$ and $\bm{s}(\bm{t})$ are referred to as the streamline-based orbital-rotation (or rigid-rotation) and spin modes, respectively. For viscous flows, the term ``spin'' may be appropriately replaced by ``shear'' or ``shear strain'' without loss of physical meaning. However, in inviscid potential flows (e.g., the exterior flow of a 2D point vortex, a classic example in fluid mechanics), both the orbital-rotation and spin modes coexist, despite the absence of shear due to the lack of viscosity. Hence, we adopt ``spin'' as a generalized term that encompasses both scenarios. These two modes admit clear physical interpretations. In the Frenet-Serret frame, their $\bm{b}$-components of $\bm{R}(\bm{t})$ and $\bm{s}(\bm{t})$ are given by~\citep{chen2025kinematic,Chen2025arxiv}
\begin{eqnarray}
	{R}(\bm{t})\equiv\bm{R}(\bm{t})\bm{\cdot}\bm{b}=2\kappa{q},~{s}(\bm{t})\equiv\bm{s}(\bm{t})\bm{\cdot}\bm{b}=-\kappa{q}-\frac{\partial q}{\partial n}=\frac{1}{r_0}\frac{\partial(rq)}{\partial r},
\end{eqnarray}
where $r$ denotes the radial coordinate measured from the local center of curvature of $\mathcal{C}$. The orbital-rotation mode $\bm{R}(\bm{t})$ corresponds a local circular motion about the binormal axis $\bm{b}$ with speed $q$ and radius of curvature $r_{0}=1/\kappa$, analogous to a speed skater on a curved track. Notably, $\bm{b}$ emerges as the unique physically identifiable axis of rotation external to the streamline orbit. The spin mode $\bm{s}(\bm{t})$, in contrast, is characterized by the radial gradient of the velocity field and finds its analogy in a figure skater performing a quadruple jump. Its physical significance is underscored by the fact that the rate of change of the angle between two orthogonal material line elements instantaneously aligned with $(\bm{t},\bm{n})$ is precisely $\bm{s}(\bm{t})\bm{\cdot}\bm{b}$~\citep{chen2025kinematic}. A deeper connection has recently been revealed by~\citet{Chen2026operator}, who demonstrated that the spin mode $\bm{s}(\bm{t})$ is related to the spin vorticity operator $\bm{\Gamma}_{S}\equiv2\bm{\Gamma}$, where $\bm{\Gamma}$ (called the spin angular velocity operator) is proportional to the spin angular momentum operator for an electron in quantum mechanics. This unusual parallel between macroscopic vortex dynamics and quantum-mechanical electron spin provides compelling justification for retaining the term ``spin'' in the present study.

Under the Cartesian basis $(\bm{e}_{x},\bm{e}_{y},\bm{e}_{z})$, the velocity field is expressed as $\bm{u}=u_{x}\bm{e}_{x}+u_{y}\bm{e}_{y}$. The essential algorithm can be implemented in a compact form as follows~\citep{Chen2026operator}:

\textbf{Step1.} Calculate the unit tangent vector of $\mathcal{C}$ $(\bm{t}=\bm{e}_{x}\cos\varphi+\bm{e}_{y}\sin\varphi)$.
\begin{eqnarray}
	\cos\varphi=\frac{u_{x}}{\sqrt{u_{x}^{2}+u_{y}^{2}}},~~\sin\varphi=\frac{u_{y}}{\sqrt{u_{x}^{2}+u_{y}^{2}}}.
\end{eqnarray}

\textbf{Step2.} Calculate the velocity gradient tensor $\mathbf{A}$ and the strain-rate tensor $\mathbf{D}$.
\begin{subequations}
	\begin{eqnarray}\label{VGT_matrix}
		\mathbf{A}=\begin{bmatrix}
			A_{xx} & A_{xy} \\
			A_{yx} & A_{yy}
		\end{bmatrix}
		=\begin{bmatrix}
			\partial_{x}u_{x} & \partial_{x}u_{y}\\
			\partial_{y}u_{x} &\partial_{y}u_{y}
		\end{bmatrix};
	\end{eqnarray}
	\begin{eqnarray}
		\mathbf{D}=\begin{bmatrix}
			D_{xx} & D_{xy} \\
			D_{yx} & D_{yy}
		\end{bmatrix}
		=\begin{bmatrix}
			\partial_{x}u_{x} & \frac{1}{2}\left(\partial_{x}u_{y}+\partial_{y}u_{x}\right)\\
			\frac{1}{2}\left(\partial_{x}u_{y}+\partial_{y}u_{x}\right) &\partial_{y}u_{y}
		\end{bmatrix}.
	\end{eqnarray}
\end{subequations}
For brevity, we use the same symbol to denote both a tensor and its matrix counterpart.

\textbf{Step3.} Calculate the $\bm{e}_{z}$-component of the orbital-rotation mode $(\tilde{R}(\bm{t})\equiv\bm{R}(\bm{t})\bm{\cdot}\bm{e}_{z})$.
\begin{eqnarray}
	\tilde{R}(\bm{t})=2\begin{bmatrix}
		\cos\varphi \sin\varphi
	\end{bmatrix}
	\begin{bmatrix}
		A_{xy} & -\frac{1}{2}(A_{xx}-A_{yy}) \\
		-\frac{1}{2}(A_{xx}-A_{yy}) & -A_{yx}
	\end{bmatrix}
	\begin{bmatrix}
		\cos\varphi \\
		\sin\varphi
	\end{bmatrix}.
\end{eqnarray}

\textbf{Step4.} Calculate the $\bm{e}_{z}$-component of the spin mode $(\tilde{s}(\bm{t})\equiv\bm{s}(\bm{t})\bm{\cdot}\bm{e}_{z})$.
\begin{eqnarray}
	\tilde{s}(\bm{t})=-2\begin{bmatrix}
		\cos\varphi \sin\varphi
	\end{bmatrix}
	\begin{bmatrix}
		D_{xy} & -\frac{1}{2}(D_{xx}-D_{yy}) \\
		-\frac{1}{2}(D_{xx}-D_{yy}) & -D_{yx}
	\end{bmatrix}
	\begin{bmatrix}
		\cos \varphi \\
		\sin \varphi
	\end{bmatrix}.
\end{eqnarray}

\textbf{Step5.} Calculate the $\bm{e}_{z}$-component of the vorticity $(\omega_{z}\equiv\bm{\omega}\bm{\cdot}\bm{e}_{z})$.
\begin{eqnarray}
	\omega_{z}=A_{xy}-A_{yx}=\partial_{x}u_{y}-\partial_{y}u_{x}.
\end{eqnarray}

\subsection{Real Schur form and invariant vorticity decompositions}\label{IVD12}
For 2D compressible flow, the discriminant of $\mathbf{A}$ in~\eqref{VGT_matrix} can be written as
\begin{eqnarray}\label{discriminant1}
	\Delta\equiv4\det(\mathbf{A})-\vartheta^2=4\left(A_{xx}A_{yy}-A_{xy}A_{yx}\right)-(A_{xx}+A_{yy})^2.
\end{eqnarray}
At a point $P$ where $\Delta>0$, $\mathbf{A}$ has a pair of complex conjugate eigenvalues $\lambda_{1,2}=\lambda_{\rm cr}\pm{\rm i}\lambda_{\rm ci}$ where \( \chi \equiv \lambda_{\text{cr}} \) and \( \lambda_{\text{ci}}~(> 0)  \) denote their real and imaginary parts, respectively. The local rotation axis $\bm{e}_{3}$ at $P$ is equal to ${\rm sgn}(\omega_{z})\bm{e}_{z}$ where ${\rm sgn}$ is the sign function; the rotation-axis-normal plane is denoted by $\mathcal{P}\equiv\mathbb{R}^2$. According to the classical critical-point theory \citep{chong1990general}, a closed or spiral streamline pattern must appear in a small neighborhood of $P$. The theory of linear algebra shows that under an appropriate right-handed orthonormal basis $(\bm{e}_{1},\bm{e}_{2},\bm{e}_{3})$ spanning $\mathcal{P}$ (where $\bm{e}_{3}=\bm{e}_{1}\times\bm{e}_{2}$), $\mathbf{A}$ can be expressed as an real Schur form~\citep{schur1909uber,LiZhen2014,Liu2018,Chen2026operator} restricted on $\mathcal{P}$:
\begin{equation}\label{eqq13}
	\mathbf{A}=
	\begin{bmatrix}
		\chi & \psi+\gamma \\
		-\psi & \chi
	\end{bmatrix}.
\end{equation}
which is fully characterized by three rotational invariants \( (\chi,\psi,\gamma) \).
Here, the diagonal element $\chi$ represents the local stretching or contraction rate of a material line element along $\bm{e}_{1}$ or $\bm{e}_{2}$. The dilatation is given as $\vartheta=2\chi$. The off-diagonal element \( \psi \) quantifies the characteristic angular velocity associated with rigid rotation about the axis \( \bm{e}_3 \) within \( \mathscr{P} \), while $\gamma$ describes the shear strain rate (or the rate of angular deformation).
Note that the orthonormal basis \( (\bm{e}_1, \bm{e}_2, \bm{e}_3) \), which serves as the fundamental basis for the real Schur form and the NND (normal-nilpotent decomposition), is referred to as NND basis or NND triad. 
Using~\eqref{discriminant1} and~\eqref{eqq13}, $\Delta$ can be alternatively expressed as
\begin{equation}\label{discriminant2}
	\Delta= 4\lambda_{\text{ci}}^2= 4\psi(\psi+\gamma).
\end{equation}

From~\eqref{eqq13}, $\mathbf{D}$ and $\bm{\Omega}$ are evaluated as
\begin{equation}
	\mathbf{D} = \begin{bmatrix}
		\chi & \frac{1}{2}\gamma \\
		\frac{1}{2}\gamma & \chi
	\end{bmatrix}, \quad
	\bm{\Omega} = \begin{bmatrix}
		0 & \psi+\frac{1}{2}\gamma \\
		-\left(\psi+\frac{1}{2}\gamma\right) & 0
	\end{bmatrix}.
\end{equation}
Then, the dual vector of $\bm{\Omega}$ results in the following representations of the axial vorticity component $\omega_{3}\equiv\bm{\omega}\bm{\cdot}\bm{e}_{3}(>0)$ and the discriminant $\Delta$~\citep{chen2025kinematic}:
\begin{eqnarray}
	\omega_3 = 2\psi+\gamma,~~\Delta = \omega_3^2 - \gamma^2.
\end{eqnarray}
Mathematically, either \( \gamma > 0 \) (denoted by \( \gamma^+ \)) or \( \gamma < 0 \) (denoted by \( \gamma^- \)) can serve as an alternative uniqueness condition for determining the real Schur form~\citep{Chen2025arxiv,Chen2026operator}, leading to two distinct invariant vorticity decompositions (IVDs) as follows.

\textbf{Case I.} When $\gamma > 0$, $\Delta>0$ implies $\omega_3 > \gamma > 0$, and thus $\psi > 0$. Let $(\psi^+, \gamma^+)$ denote the corresponding invariants and $(\bm{e}_{1}^{+},\bm{e}_{2}^{+})$ the NND basis. The first IVD is given by
\begin{subequations}\label{CD1}
	\begin{eqnarray}\label{plus_expression0}
		\omega_{3}=2\psi^{+}+\gamma^{+},
	\end{eqnarray}
	\begin{eqnarray}\label{plus_expression1}
		R_{N}^{+}\equiv2\psi^{+}=\omega_{3}-\sqrt{\omega_{3}^{2}-4\lambda_{\rm ci}^{2}},~~s_{N}^{+}\equiv\gamma^{+}=\sqrt{\omega_{3}^{2}-4\lambda_{\rm ci}^{2}}.
	\end{eqnarray}
\end{subequations}
Note that both $R_{N}^{+}$ and $s_{N}^{+}$ are bivariate functions of the fundamental invariants $(\omega_{3},\lambda_{\rm ci})$. The characteristic rigid-rotation mode $\bm{R}_{N}^{+}\equiv{R}_{N}^{+}\bm{e}_{3}$ is precisely the Liutex (or Rortex) vector~\citep{Liu2018}, which has been widely adopted to extract vortex structures in complex flows.
Equation~\eqref{plus_expression0} represents the Liutex-shear decomposition proposed by~\citet{Liu2018}.
Meanwhile,~\eqref{plus_expression1} has previously been reported in studies such as~\citet{Liu2020} and~\citet{Xu2019}. This configuration aligns with the KKB mechanism: axial vortices form via the wrapping of shear layers, accompanied by a transfer from spin mode to rigid-rotation mode, ultimately resulting in vortex intensification~\citep{Klein1910,Kaden1931,Betz1950}.

\textbf{Case II.} When $\gamma<0$, there is $2\psi>\omega_{3}>-\gamma>0$, which implies $\psi>-\gamma>0$. Let $(\psi^-, \gamma^-)$ denote the corresponding invariants and $(\bm{e}_{1}^{-},\bm{e}_{2}^{-})$ the NND basis. The second IVD is given by
\begin{subequations}\label{CD2}
	\begin{eqnarray}\label{plus_expression2}
		\omega_{3}=2\psi^{-}+\gamma^{-},
	\end{eqnarray}
	\begin{eqnarray}\label{plus_expression3}
		R_{N}^{-}\equiv2\psi^{-}=\omega_{3}+\sqrt{\omega_{3}^{2}-4\lambda_{\rm ci}^{2}},~~s_{N}^{-}\equiv\gamma^{-}=-\sqrt{\omega_{3}^{2}-4\lambda_{\rm ci}^{2}}.
	\end{eqnarray}
\end{subequations}
The characteristic rigid-rotation mode $\bm{R}_{N}^{-} \equiv R_{N}^{-}\bm{e}_{3}$, formally introduced by \citet{Chen2025arxiv}, corresponds to the anti-KKB mechanism. This mechanism occurs when the axial vortex has sufficient swirling strength to maintain its coherence, such that opposing spin effects cannot significantly disrupt the primary vortex structure. A typical example is the Burgers vortex, an analytical solution of the NS equations~\citep{burgers1948mathematical,Chen2025arxiv}. Moreover, both the KKB and anti-KKB mechanisms coexist in the secondary hurricane-eye vortex at Saturn’s north pole~\citep{Chen2025arxiv}. During the formation of an axial vortex, the anti-KKB mechanism could be activated to suppress the unbounded growth of swirling strength in the inner region, while the KKB mechanism may continue to govern the roll-up of the outer shear layers.
Using~\eqref{CD1} and~\eqref{CD2}, four elegant identities are obtained for characteristic variables~\citep{Chen2025arxiv,Chen2026operator}:
\begin{eqnarray}\label{flag2}
	\psi^{+}=\psi^{-}+\gamma^{-},~\psi^{-}=\psi^{+}+\gamma^{+},~\gamma^{-}=-\gamma^{+},
	~\psi^{+}+\psi^{-}=\omega_{3}.
\end{eqnarray}

\section{Decomposition and estimation of total vorticity flux}\label{Decomposition and Estimate of total vorticity}
With the preparation in~\S\S~\ref{Streamline-based DVD} and~\ref{IVD12}, we consider the global integral properties of the vorticity field and its constituents.
 The vorticity flux $\Gamma_{\omega}$ across the domain $\mathcal{D}$, bounded by the closed curvilinear contour $\mathcal{L}$ oriented counterclockwise about the $z$-axis, can be decomposed exactly as
\begin{subequations}
	\begin{equation}\label{circ1}
		\Gamma_{\omega} \equiv \int_{\mathcal{D}} \omega_z \, d\sigma = \Gamma_{\rm{R}} + \Gamma_{\rm{s}}, 
	\end{equation}
	\begin{equation}\label{circ2}
		\Gamma_{\rm{R}} \equiv \int_{\mathcal{D}} \tilde{R}(\bm{t})d\sigma~~\text{and}~~\Gamma_{\rm{s}} \equiv \int_{\mathcal{D}} \tilde{s}(\bm{t}) d\sigma.
	\end{equation}
\end{subequations}
where $d\sigma=dxdy$ represents the area element. The integration domain  $\mathcal{D}$ in~\eqref{circ1} can be replaced by $\mathcal{D}^{\prime}\equiv\mathcal{D}\setminus\left\{\bm{x}\in\mathcal{D}\vert\bm{\omega}(\bm{x})=\bm{0}\right\}$ without changing the result. In practical evaluations of geometrically antisymmetric flow fields, $\mathcal{D}$ is often confined to the rectangular half-domain above the symmetry axis (i.e., $[0,L_{x}]\times[L_{y}/2,L_{y}]$) to prevent cancellation of opposing contributions.
In terms of the sign of $\omega_{z}$ in the region where $\Delta>0$., two fundamental inequalities are derived from the inequality in~\citet{Chen2025arxiv}:
\begin{subequations}
\begin{itemize}
	\item If $\omega_z < 0$ (in which $\bm{e}_3 = -\bm{e}_z$), then $\tilde{R}_N^{\pm} = -R_N^{\pm}$ and $\tilde{s}_N^{\pm} = -s_N^{\pm}=s_{N}^{\mp}$. Consequently,
	\begin{equation}\label{bound_neg}
		\tilde{R}_N^- \le \tilde{R}(\bm{t}) \le \tilde{R}_N^+, \quad s_{N}^{-}=\tilde{s}_N^+ \le \tilde{s}(\bm{t}) \le \tilde{s}_N^-=s_{N}^{+}.
	\end{equation}
	\item If $\omega_z > 0$ (in which $\bm{e}_3 = \bm{e}_z$), then $\tilde{R}_N^{\pm} = R_N^{\pm}$ and $\tilde{s}_N^{\pm} = s_N^{\pm}$. Consequently,
	\begin{equation}\label{bound_pos}
		\tilde{R}_N^+ \le \tilde{R}(\bm{t}) \le \tilde{R}_N^-, \quad s_{N}^{-}=\tilde{s}_N^- \le \tilde{s}(\bm{t}) \le \tilde{s}_N^+=s_{N}^{+}.
	\end{equation}
\end{itemize}
\end{subequations}

Now apply integration to both sides of the first equalities in~\eqref{bound_neg} and~\eqref{bound_pos}, we obtain
\begin{equation}\label{ttt1}
	I_{\rm L}[\Gamma_{\rm{R}}]\le \Gamma_{\rm{R}} \le I_{\rm U}[\Gamma_{\rm{R}}],
\end{equation}
\begin{subequations}
where the lower and upper bounds of $\Gamma_{\rm{R}}$ are respectively given by
\begin{eqnarray}\label{lll}
	I_{\rm{L}}[\Gamma_{\rm{R}}] =I_{\rm{L}}^{-}[\Gamma_{\rm{R}}]+I_{\rm{L}}^{+}[\Gamma_{\rm{R}}], %\int_{\mathcal{D} \cap \{\omega_z < 0\}} \tilde{R}_N^- \, d\sigma + \int_{\mathcal{D} \cap \{\omega_z > 0\}} \tilde{R}_N^+ \, d\sigma\nonumber\\
%	&=&-\int_{\mathcal{D} \cap \{\omega_z < 0\}} {R}_N^- \, d\sigma + \int_{\mathcal{D} \cap \{\omega_z > 0\}} {R}_N^+ \, d\sigma,
\end{eqnarray}
\begin{eqnarray}
	I_{\rm{L}}^{+}[\Gamma_{\rm{R}}]\equiv\int_{\mathcal{D} \cap \{\omega_z > 0\}} \tilde{R}_N^+ \, d\sigma,~~I_{\rm{L}}^{-}[\Gamma_{\rm{R}}]\equiv\int_{\mathcal{D} \cap \{\omega_z < 0\}} \tilde{R}_N^- \, d\sigma;
\end{eqnarray}
\begin{eqnarray}\label{rrr}
	I_{\rm{U}}[\Gamma_{\rm{R}}]=I_{\rm{U}}^{-}[\Gamma_{\rm{R}}]+I_{\rm{U}}^{+}[\Gamma_{\rm{R}}], %\int_{\mathcal{D} \cap \{\omega_z < 0\}} \tilde{R}_N^+ \, d\sigma + \int_{\mathcal{D} \cap \{\omega_z > 0\}} \tilde{R}_N^- \, d\sigma\nonumber\\
%	&=&-\int_{\mathcal{D} \cap \{\omega_z < 0\}} {R}_N^+ \, d\sigma + \int_{\mathcal{D} \cap \{\omega_z > 0\}} {R}_N^- \, d\sigma.
\end{eqnarray}
\begin{eqnarray}
	I_{\rm U}^{+}[\Gamma_{\rm{R}}]\equiv\int_{\mathcal{D} \cap \{\omega_z > 0\}} \tilde{R}_N^- \, d\sigma,~~I_{\rm U}^{-}[\Gamma_{\rm{R}}]\equiv\int_{\mathcal{D} \cap \{\omega_z < 0\}} \tilde{R}_N^+ \, d\sigma.
\end{eqnarray}
\end{subequations}
It is seen that $I_{\rm{L}}[\Gamma_{\rm{R}}]$ and $	I_{\rm{L}}[\Gamma_{\rm{s}}]$ are inherently asymmetric, which are determined by the piecewise integrations that switch between $R_N^-$ and $R_N^+$ depending on the local sign of $\omega_z$, and thus strictly reflect the spatial distribution and relative dominance of positive and negative vorticity regions within the flow field.
Similarly, from the second equalities in~\eqref{bound_neg} and~\eqref{bound_pos}, it follows that
\begin{equation}\label{ttt2}
	J_{\rm L}[\Gamma_{\rm{s}}]\le \Gamma_{\rm{s}} \le J_{\rm U}[\Gamma_{\rm{s}}],
\end{equation}
\begin{subequations}
where the lower and upper bounds of $\Gamma_{\rm{s}}$ are respectively given by
\begin{equation}\label{lllp}
	J_{\rm{L}}[\Gamma_{\rm{s}}] \equiv \int_{\mathcal{D} \cap \{\omega_z < 0\}} \tilde{s}_N^+ \, d\sigma + \int_{\mathcal{D} \cap \{\omega_z > 0\}} \tilde{s}_N^- \, d\sigma=\int_{\mathcal{D}^\prime}s_{N}^{-}d\sigma,
\end{equation}
\begin{equation}\label{rrrp}
	J_{\rm{U}}[\Gamma_{\rm{s}}] \equiv \int_{\mathcal{D} \cap \{\omega_z < 0\}} \tilde{s}_N^- \, d\sigma + \int_{\mathcal{D} \cap \{\omega_z > 0\}} \tilde{s}_N^+ \, d\sigma=\int_{\mathcal{D}^\prime}s_{N}^{+}d\sigma.
\end{equation}
Since $s_{N}^{-}=-s_{N}^{+}$, the bounds of $\Gamma_{\rm s}$ are perfectly symmetric, i.e., $J_{\rm{L}}[\Gamma_{\rm{s}}]=-J_{\rm{U}}[\Gamma_{\rm{s}}]$.
\end{subequations}

\section{Numerical aspects}\label{Numerical aspects}
\subsection{Governing equations for binary gas mixture}
Two-dimensional (2D) RMI simulations for a binary, ideal, non-reactive gas mixture system are performed by solving the compressible multicomponent Euler equations, which are capable of capturing the interface structures in reasonable agreement with experiments~\citep{Li2025,singh2025shock}. In the Cartesian coordinate system $(x,y)$, the conservative laws of mass, momentum, energy, and species transport are written in conservative form as
\begin{equation}\label{19eq}
	\frac{\partial \mathbf{U}}{\partial t} + \frac{\partial \mathbf{F(U)}}{\partial x} + \frac{\partial \mathbf{G(U)}}{\partial y} = \bm{0},
\end{equation}
where the conservative state vector $\mathbf{U}$ and the inviscid flux vectors ($\mathbf{F(U)}$ and $\mathbf{G(U)}$) in the $x$- and $y$-directions, respectively, are defined as
\begin{equation}\label{UFG}
	\mathbf{U} = \begin{bmatrix}
		\rho \\
		\rho u_{x} \\
		\rho u_{y} \\
		\rho E \\
		\rho Y_{k}
	\end{bmatrix}, \quad \mathbf{F}(\mathbf{U}) = \begin{bmatrix}
		\rho u_{x} \\
		\rho u_{x}^2 + p \\
		\rho u_{x}u_{y}\\
		(\rho E + p) u_{x} \\
		\rho Y_{k} u_{x}
	\end{bmatrix}, \quad \mathbf{G(U)} = \begin{bmatrix}
		\rho u_{y} \\
		\rho u_{x}u_{y} \\
		\rho u_{y}^2 + p \\
		(\rho E + p) u_{y} \\
		\rho Y_{k}u_{y}
	\end{bmatrix}.
\end{equation}
In this inviscid formulation, diffusive transport phenomena are neglected, and the evolution of material interfaces is tracked solely through the species mass fraction equation. 

In~\eqref{UFG}, using Dalton's law of partial pressures, the ideal equation of state is applied to the thermodynamic pressure $p$:
\begin{eqnarray}
	p=\rho RT,~\quad R=\sum_{k=1}^{2} Y_{k}R_{k},
\end{eqnarray}
where $\rho$ is the mixed density, $T$ is the temperature, $R$ is the specific gas constant of the mixture. $Y_{k}$ denotes the mass fraction of species $k~(k=1,2)$, satisfying $Y_{1}+Y_{2}=1$. $R_{k}\equiv{R}_{\rm u}/M_{k}$ is the specific gas constant of species $k$, with $M_{k}$ being the molar mass, and $R_{\rm u}=8.314{\rm J}/({\rm mol\cdot K})$ is the universal gas constant. The total energy per unit volume, $\rho{E}$ (where $E$ is the total energy per unit mass), is related to the pressure and kinetic energy as follows~\citep{Singh2025ccc}:
\begin{eqnarray}
		\rho{E}=\frac{p}{\gamma-1}+\frac{1}{2}\rho\left(u_{x}^{2}+u_{y}^{2}\right),
\end{eqnarray}
where the effective specific heat ratio of the gas mixture is given by
\begin{eqnarray}
	\gamma=\frac{\sum_{k=1}^{2} Y_{k}C_{p,k}}{\sum_{k=1}^{2} Y_{k}C_{v,k}},~~C_{p,k}=\frac{\gamma_{k}R_{k}}{\gamma_{k}-1}=C_{v,k}+R_{k},~~C_{v,k}=\frac{R_{k}}{\gamma_{k}-1}.
\end{eqnarray}
Here, both gas components are assumed to behave as calorically perfect gases, characterized by the specific heats at constant pressure $(C_{p,1},C_{p,2})$ and constant volume $(C_{v,1},C_{v,2})$, with the corresponding specific heat ratios $\gamma_{k}=C_{p,k}/C_{v,k}~(k=1,2)$. The above approach provides a general framework for modeling shock-interface interactions in multicomponent compressible flows, enabling the present investigation of RMI from a novel vorticty-dynamical perspective.

\subsection{Hybrid HOWD approach}
In this study, the hybrid HOWD approach~\citep{Ding2017JFM828} is employed to solve~\eqref{19eq}. This method integrates a high-order weighted essentially non-oscillatory (WENO) scheme~\citep{Jiang1996} with a double-flux algorithm for spatial discretisation. The reliability and high computational efficiency of the code has been thoroughly validated in the previous studies~\citep{Ding2017JFM828,LiJX2024compress,Li2025}; therefore, only a brief introduction is given here. The HOWD scheme achieves third-order temporal accuracy and fifth-order spatial accuracy. On the one hand, the fifth-order WENO scheme (WENO5) is implemented to accurately capture the shock wave propagation, interface deformation, and the ensuing instability dynamics. It effectively suppresses potential spurious oscillations near the interface, a capability that is also crucial for the robust calculation of baroclinic vorticity deposition and small-scale vortical structures at the interface.
On the other hand, a third-order total variation diminishing (TVD) Runge-Kutta scheme is utilized for time discretization, preserving both numerical stability and time accuracy:
\begin{subequations}
	\begin{equation}
		\mathbf{U}^{(1)} = \mathbf{U}^n + \Delta t \mathbf{L}(\mathbf{U}^n),
	\end{equation}
	\begin{equation}
		\mathbf{U}^{(2)} = \frac{3}{4} \mathbf{U}^n + \frac{1}{4} \mathbf{U}^{(1)} + \frac{1}{4} \Delta t  \mathbf{L}(\mathbf{U}^{(1)}),
	\end{equation}
	\begin{equation}
		\mathbf{U}^{n+1} = \frac{1}{3} \mathbf{U}^n + \frac{2}{3} \mathbf{U}^{(2)} + \frac{2}{3} \Delta t\mathbf{L}(\mathbf{U}^{(2)}),
	\end{equation}
\end{subequations}
where $\mathbf{L(U)}$ is the spatial discretization operator in the equation $\partial_{t}\mathbf{U}=\mathbf{L}(\mathbf{U})$, and the time step $\Delta{t}$ is determined by the Courant-Friedrichs-Lewy (CFL) condition
\begin{equation}
\Delta{t}= {\rm CFL} \cdot \min \left( \frac{h}{\lvert{u}_{x}\rvert + c}, \frac{h}{\lvert{u}_{y}\rvert + c} \right),
\end{equation}
where the CFL number is set to 0.5 for all simulations, $h=\Delta{x}=\Delta{y}$ denotes the uniform grid spacing, and $c\equiv\sqrt{\gamma RT}$ is the speed of sound.
To mitigate the discontinuity of the initially sharp interface, a numerical diffusion layer is introduced, providing a smooth transition of gas concentration from 0 to 1 over three cells. This softening strategy has been validated in previous research to ensure physical accuracy~\citep{Ding2017JFM828}. For all simulations presented below, symmetric boundary conditions are enforced on the top and bottom surfaces, which are oriented parallel to the flow direction. Non-reflecting outflow conditions are applied to the boundaries perpendicular to the streamwise direction, with zeroth-order extrapolation employed at both the left and right edges, which prevents spurious wave contamination. The numerical results will be presented in dimensionless forms, with the reference quantities summarized in Table~\ref{reference_quantities}. It is worth noting that the reference velocity is comparable in magnitude to the speed of sound, while the reference time is on the order of microseconds $(\mu\rm{s})$.

\begin{table}[t]
	\caption{Reference quantities for all simulations in the present study. $R_{\rm u}=8.314{\rm J}/({\rm mol\cdot K})$ is the universal gas constant.}
	\centering
	\vspace{1em}
	% 第一个 minipage（First Branch）
	\begin{minipage}[t]{\textwidth}
		\centering
		\setlength{\tabcolsep}{15pt}
	%	\textbf{First Primary Branch} \\[0.5em] % 手动添加标题
		\begin{tabular}{cccc}
		\toprule \toprule
		%	\multicolumn{4}{c}{\textbf{Dimensionless Reference Bases}} & \multicolumn{4}{c}{\textbf{Dimensionless Reference Bases (Cont.)}} \\
		%	\cmidrule(r){1-4} \cmidrule(l){5-8}
		\textbf{Name} & \textbf{Symbol} & \textbf{Value} & \textbf{SI Unit} \\
		\midrule		
		Air Molar Mass & $M_{0}$ & $28.97\times10^{-3}$ & $\text{kg}/\text{mol}$  \\
		Reference Length & $L_{0}$ & $10^{-3}$ & $\text{m}$  \\
		Reference Pressure & $p_{0}\equiv\rho_{0}R_{0}T_{0}$ & $1.01325\times10^5$ & ${\rm Pa}={\rm kg}/({\rm m}\cdot\rm{s^2})$  \\
		Reference Density & $\rho_{0}$ &$1.1966$ & ${\rm kg/m^3}$ \\
		\bottomrule\bottomrule
	\end{tabular}
	\end{minipage}
	
	\vspace{1.5em} % 垂直间距
	
	% 第二个 minipage（Second Branch）
	\begin{minipage}[t]{\textwidth}
	\centering
	\setlength{\tabcolsep}{15pt}
	%	\textbf{First Primary Branch} \\[0.5em] % 手动添加标题
	\begin{tabular}{cccc}
		\toprule \toprule
		%	\multicolumn{4}{c}{\textbf{Dimensionless Reference Bases}} & \multicolumn{4}{c}{\textbf{Dimensionless Reference Bases (Cont.)}} \\
		%	\cmidrule(r){1-4} \cmidrule(l){5-8}
		\textbf{Name} & \textbf{Symbol} & \textbf{Value} & \textbf{SI Unit} \\
		\midrule		

		Reference Gas Constant & $R_{0}\equiv R_{\rm u}/M_{0}$ & $286.9865$ & $\text{J}/(\text{kg}{\cdot}\text{K})={\rm m^2}/\rm{(s^2\cdot K)}$ \\
		Reference Temperature & $T_{0}$ & 295.05 & $\text{K}$ \\
		Reference Velocity & $U_{0}\equiv\sqrt{R_{0} T_{0}}$ & $290.9903$ & m/s \\ 
		Reference Time & $t_{0}\equiv L_{0}/U_{0}$ & $3.4365$ & ${\rm\mu s}$ \\
		\bottomrule\bottomrule
	\end{tabular}
\end{minipage}
	\label{reference_quantities}
\end{table}

\section{RMI in a single-mode perturbed interface}\label{RMI of a single-mode perturbed interface}
\subsection{Characteristics of interface morphology and vorticity modes}
\begin{figure}[h!]
	\centering
	\includegraphics[width=1.0\columnwidth,trim={0cm 0cm 0cm 0.1cm},clip]{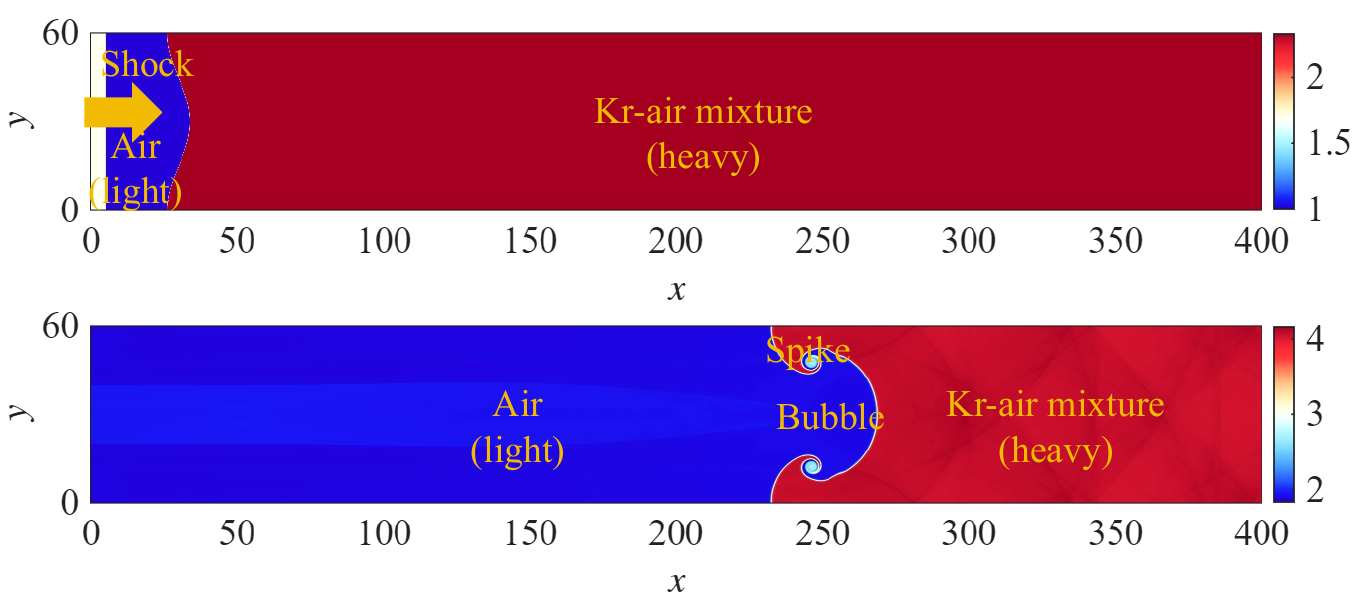}
	%\label{example}%    left down  right up
	\caption{Contour maps of the normalized density field for a shock-accelerated single-mode perturbed interface at $t^{*}=0$ and $442.31$. The Atwood number is $0.40$, and the shock Mach number is $1.40$. The incident shock wave propagates from left to right.} 
	\label{danmo}
\end{figure}
We simulate the RM instability on a single-mode perturbed interface separating two fluid components initially at rest.
The computational domain is rectangular with physical dimensions 
$L_{x}\times{L}_{y}=400~{\rm mm}\times60~{\rm mm}$ and discretized using uniform spatial grids of high resolution $4000\times600$. As illustrated in the upper panel of figure~\ref{danmo}, the upstream region to the left of the interface is initially occupied by air $(\rho_{1}=1.20~{\rm kg}/{\rm m}^3,~\gamma_{1}=1.40,~R_{1}=R_{0}=286.99~{\rm J}/({\rm kg}\cdot{\rm K}),~T_{1}=T_{0})$, whereas the downstream region to the right is filled with a mixture of 30\% air and 70\% Krypton (Kr) by volume fraction $(\rho_{2}=2.78~{\rm kg}/{\rm m}^3,~\gamma_{2}=1.56,~R_{2}=123.44~{\rm J}/({\rm kg}\cdot{\rm K}),~T_{2}=T_{0})$. The corresponding Atwood number is $At\equiv(\rho_{2}-\rho_{1})/(\rho_{1}+\rho_{2})=0.40$.
The initially sinuisoidal disturbed interface has an amplitude $a_{0}=4.0~{\rm mm}$ and wavelength $\lambda=60.0~{\rm mm}$, yielding a dimensionless amplitude $ka_{0}=2\pi{a}_{0}/\lambda=0.42$.
At the temperature $T_{0}$, the speeds of sound are $c_{1}=\sqrt{\gamma_{1}R_{1}T_0}=344.3~{\rm m/s}$ upstream and $c_{2}=\sqrt{\gamma_{2}R_{2}T_0}=238.35~{\rm m/s}$ downstream. The interface is impulsively accelerated by an incident shock wave with the shock Mach number $Ma_{s}\equiv{u}_{s}/c_{1}=1.40$, where $u_{s}=482~{\rm m/s}$ is the incident shock speed. The lower panel of figure~\ref{danmo}
displays the developed spike (heavy fluid penetrating into the light fluid) and bubble (light fluid penetrating into the heavy fluid) at $t^{*}=442.31$ (physically corresponding to $t=1520~{\rm \mu s}$), indicating the nonlinear growth of the interface perturbation magnitude and the formation of the mushroom-like structures.
\begin{figure}[t]
	\centering
	\includegraphics[width=1.0\columnwidth,trim={1.0cm 1.6cm 1cm 2.4cm},clip]{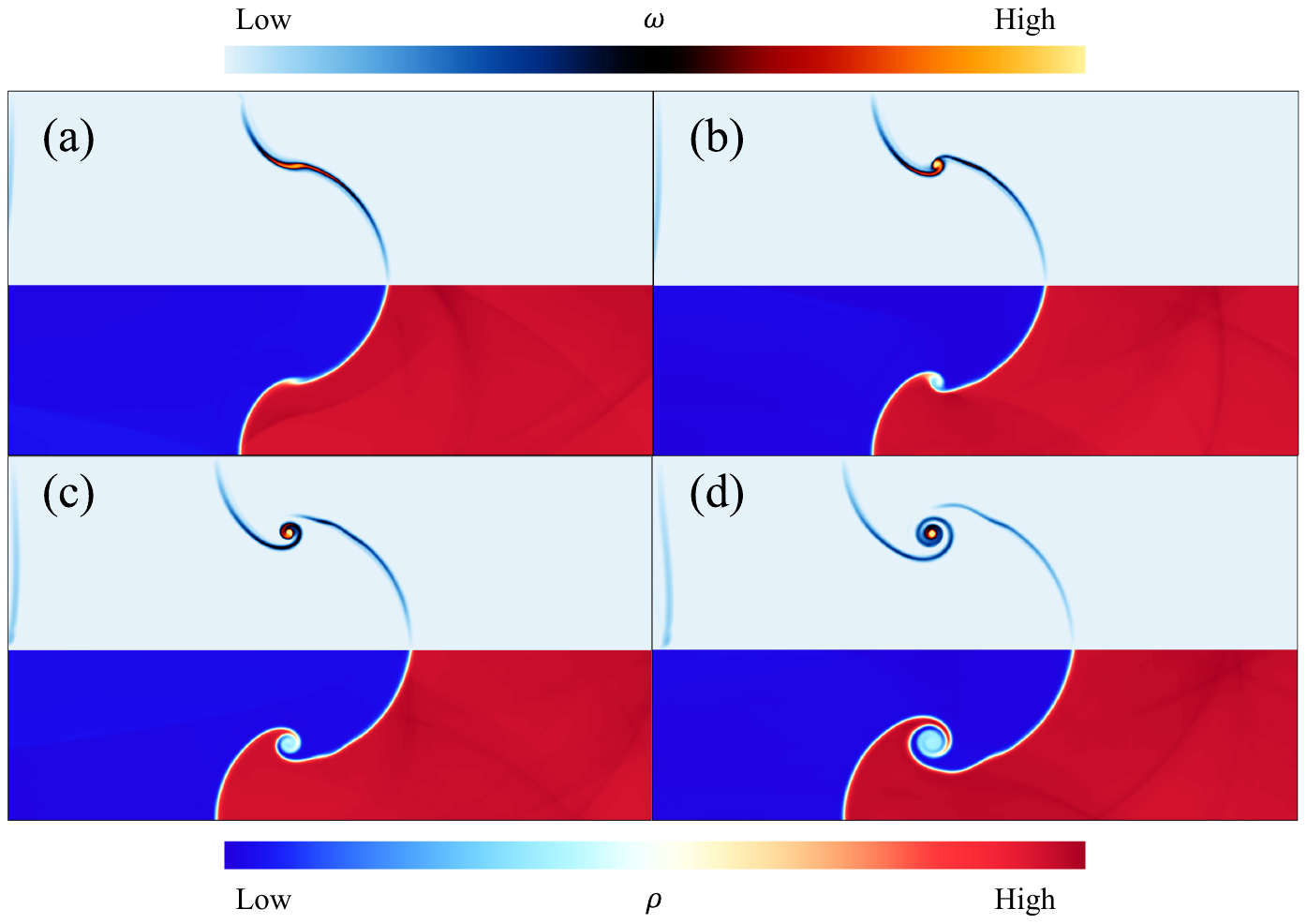}
	\caption{Contour maps of the vorticity magnitude $\lVert\bm{\omega}\rVert$ (top panels) and the density $\rho$ (bottom panels) at various instants: (a) $t^{*} = 87.3$, (b) $t^{*} = 203.7$, (c) $t^{*} =320$, and (d) $t^{*} = 436.5$. } 
	\label{evolution_density_vorticity}
\end{figure}

Figure~\ref{evolution_density_vorticity} demonstrates the spatiotemporal evolution of density (lower panel) and vorticity (upper panel) patterns as interface perturbations grow following the shock impact. Upon interaction with the incident normal shock wave, baroclinic vorticity is deposited at the interface due to the misalignment between the local pressure and density gradients (i.e., the baroclinic torque). This critical mechanism drives an initial linear growth phase over a short period and the subsequent nonlinear regime characterized by the rolling-up of the stretching spiral arms into strong vortex cores. The heavy fluid intrudes into the light fluid, forming a `spike' structure; meanwhile, the light fluid is entrained into the heavy fluid, generating a `bubble' structure. At this stage, the interfacial dynamics are dominated by nonlinear growth and the mutual competition between `spike' and `bubble'. However, vorticity is the directional average of angular velocities of material line elements originating from a spatial point, which cannot distinguish between rigid rotation (swirl-dominant) and shearing motion (shear-dominant), nor capture their dynamic transformation and interaction. Therefore, the streamline-based DVD~\citep{chen2025kinematic,Chen2025arxiv} is employed to investigate the general features of vortical structures associated with RMI.
\begin{figure}[t]
	\centering
	\subfigure[$\bm{R}(\bm{t})\cdot\bm{e}_z$]{
		\begin{minipage}[t]{0.49\linewidth}
			\centering
			\includegraphics[width=1.0\columnwidth,trim={0cm 0cm 0cm 0cm},clip]{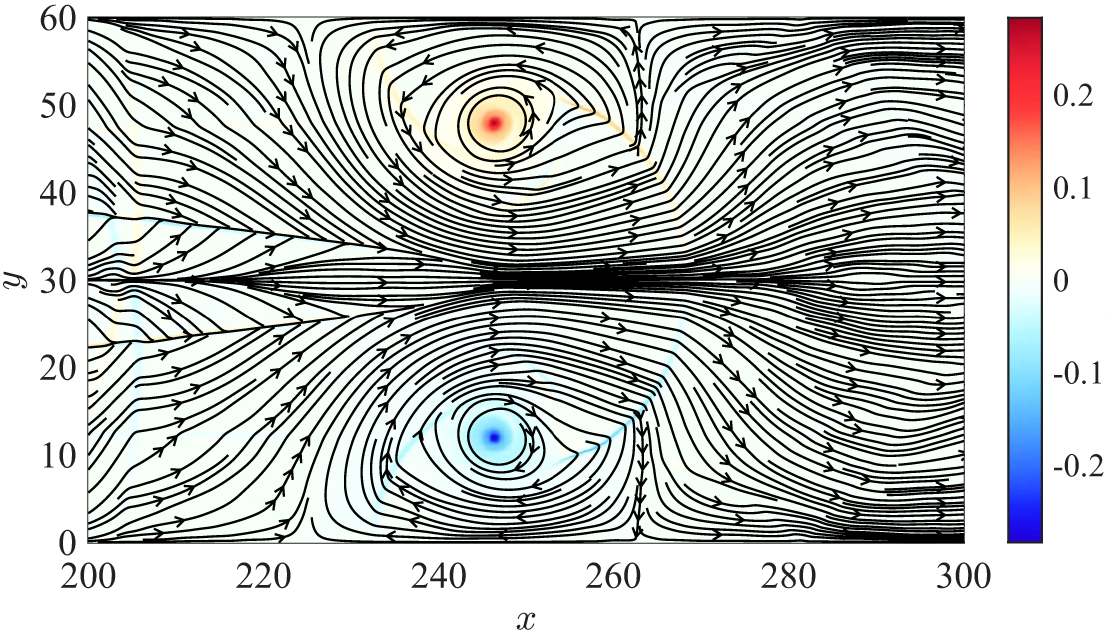}
			%	\vspace{-0.6cm} 
			\label{kk1}
		\end{minipage}%
	}
	\subfigure[$\bm{s}(\bm{t})\cdot\bm{e}_z$]{
		\begin{minipage}[t]{0.49\linewidth}
			\centering
			\includegraphics[width=1.0\columnwidth,trim={0cm 0cm 0cm 0cm},clip]{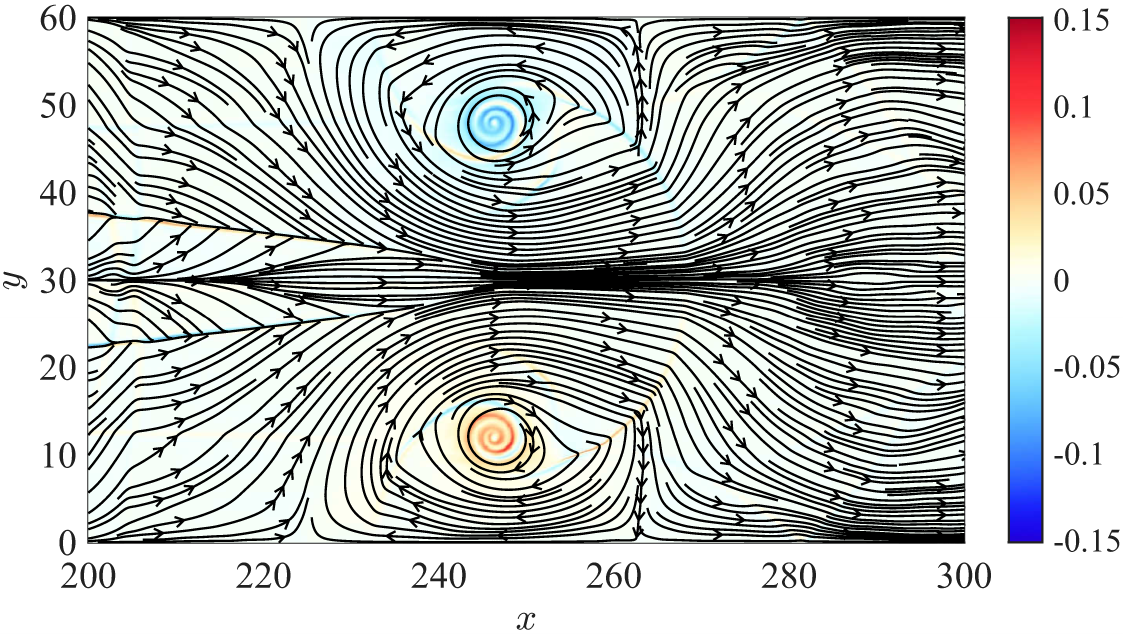}
			%\vspace{-19.6cm} 
			\label{kk2}
		\end{minipage}%
	}	
	\subfigure[${\omega}_z$]{
		\begin{minipage}[t]{0.49\linewidth}
			\centering
			\includegraphics[width=1.0\columnwidth,trim={0cm 0cm 0cm 0cm},clip]{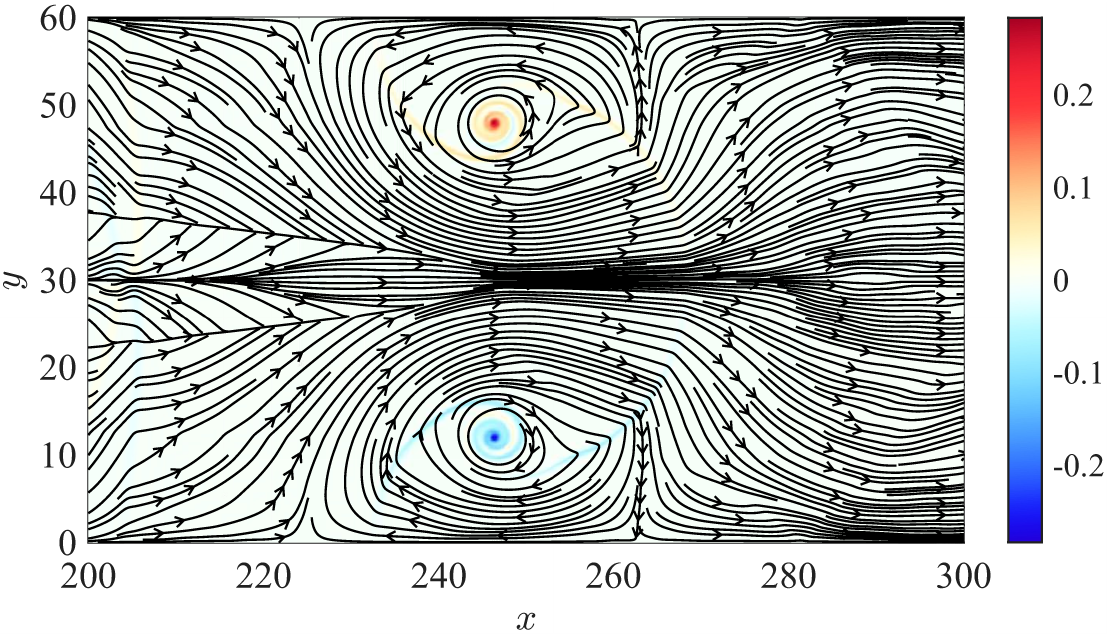}
			%   \vspace{-6.6cm} 
			\label{kk3}
		\end{minipage}%
	}	
	\subfigure[$\bm{R}_N^+\cdot\bm{e}_z$]{
		\begin{minipage}[t]{0.49\linewidth}
			\centering
			\includegraphics[width=1.0\columnwidth,trim={0cm 0cm 0cm 0cm},clip]{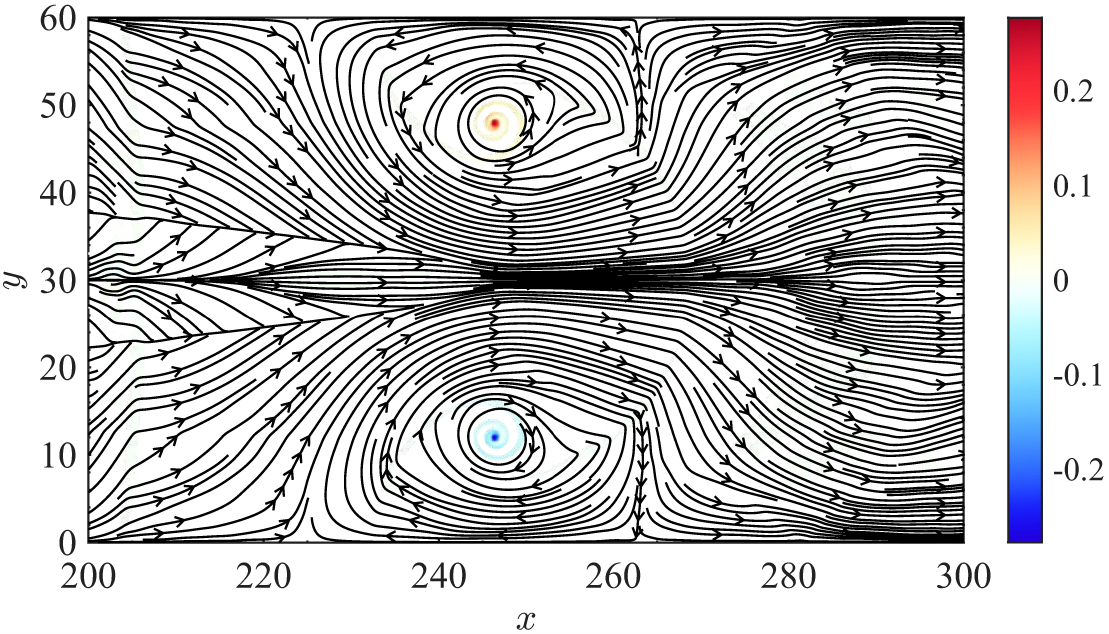}
			%   \vspace{-6.6cm} 
			\label{kk4}
		\end{minipage}%
	}	
	
	\subfigure[$\bm{R}_N^{-}\cdot\bm{e}_z$]{
		\begin{minipage}[t]{0.49\linewidth}
			\centering
			\includegraphics[width=1.0\columnwidth,trim={0cm 0cm 0cm 0cm},clip]{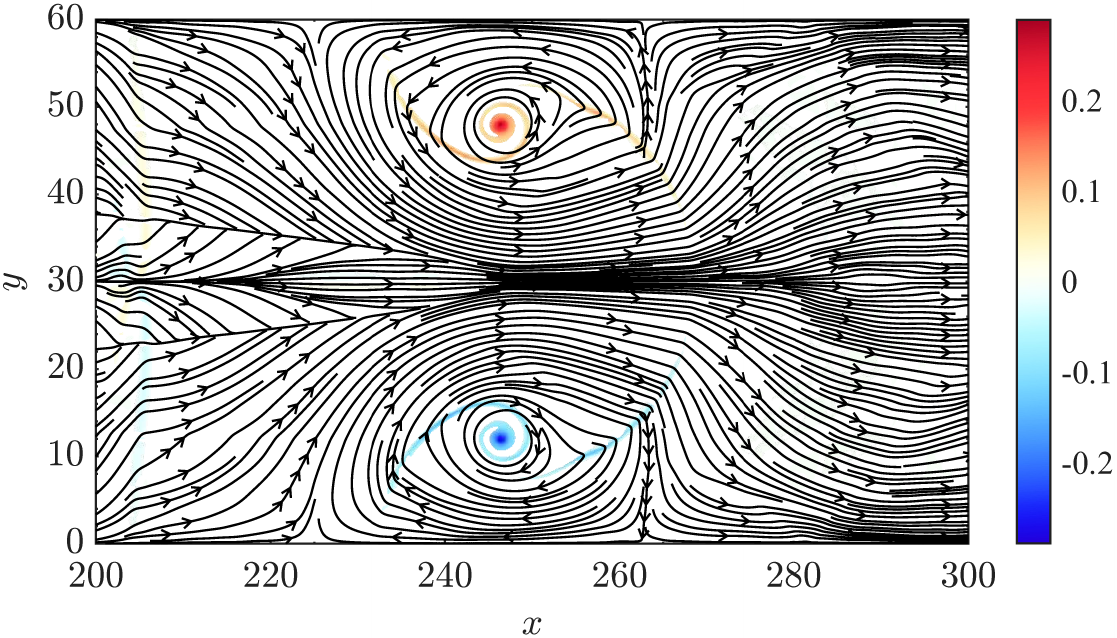}
			%   \vspace{-6.6cm} 
			\label{kk5}
		\end{minipage}%
	}	
	\caption{Normalized snapshots of the $\bm{e}_z$-components of the DVD and IVD vorticity modes at $t^{*}=442.31$: (a) the orbital-rotation mode
		$\bm{R}(\bm{t})$, (b) the spin mode $\bm{s}(\bm{t})$, (c) the total vorticity $ \bm{\omega}$, (d) $\bm{R}_{N}^{+}$ (Liutex)~\citep{Liu2018}, and (e) $\bm{R}_{N}^{-}$~\citep{Chen2025arxiv,Chen2026operator}.
		The reference frame translates at the vortex-pair propagation velocity $\mathbf{U}_{1}= 0.512{\rm U}_{0}\bm{e}_{x}$.} 
	\label{333333}
\end{figure}
\begin{figure}[t]
	\centering
	\subfigure[ $\bm{R}(\bm{t})\cdot\bm{e}_z$]{
		\begin{minipage}[t]{0.49\linewidth}
			\centering
			\includegraphics[width=1.0\columnwidth,trim={0cm 0cm 0cm 0cm},clip]{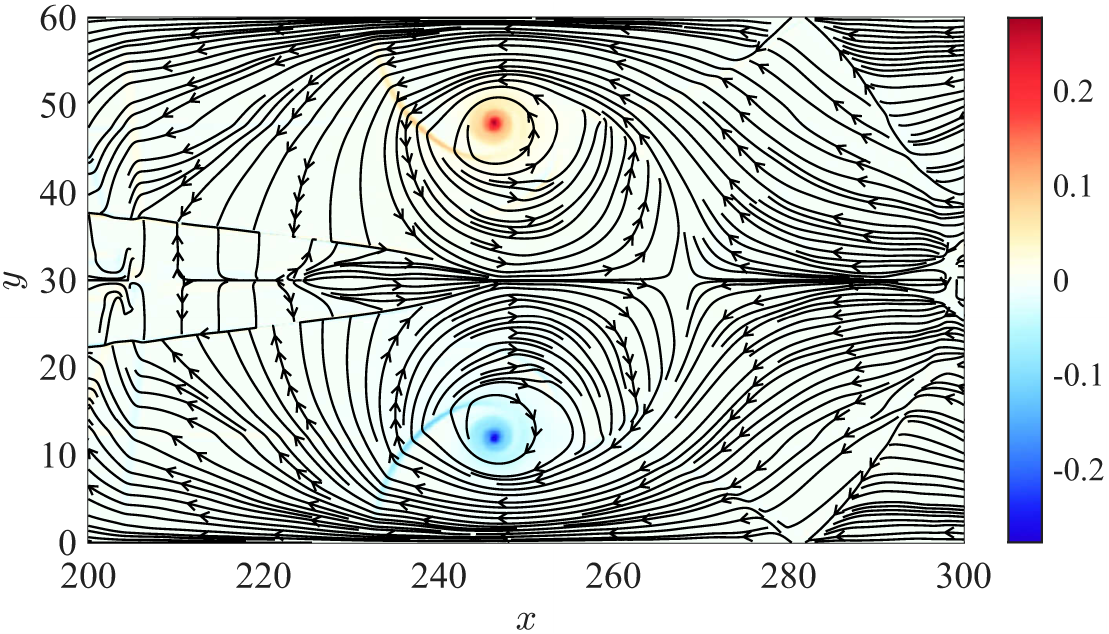}
			% \vspace{-19.6cm} %    left down  right up
			\label{212}
		\end{minipage}%
	}%    left down  right up
	\subfigure[$\bm{s}(\bm{t})\cdot\bm{e}_z$]{
		\begin{minipage}[t]{0.49\linewidth}
			\centering
			\includegraphics[width=1.0\columnwidth,trim={0cm 0cm 0cm 0cm},clip]{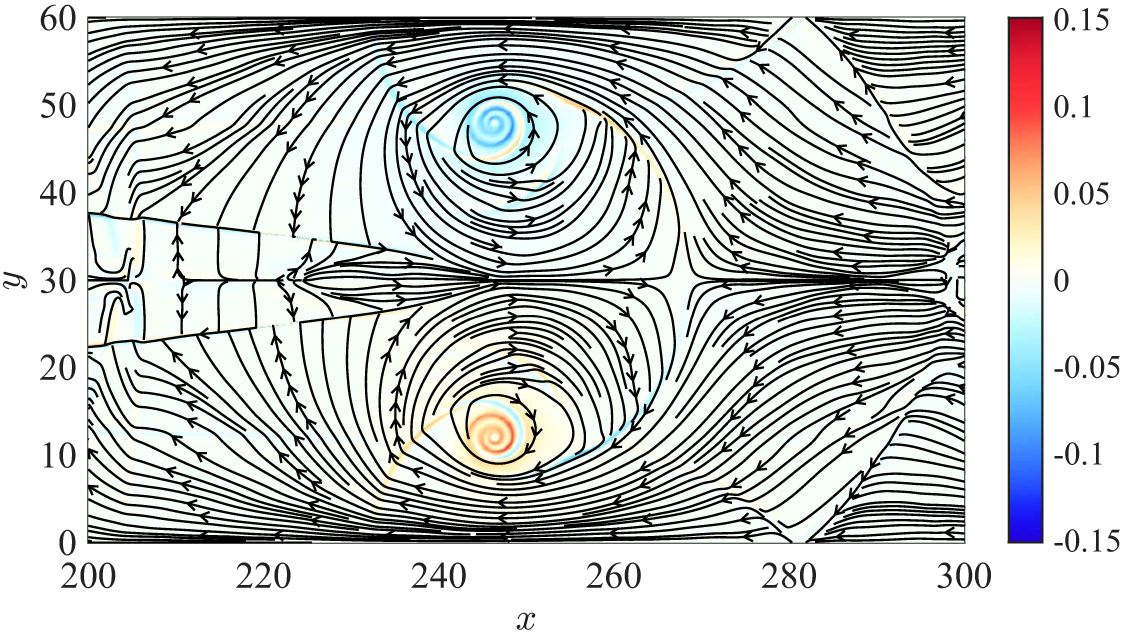}
			%\vspace{-19.6cm} 
			\label{22}
		\end{minipage}%
	}	
	\subfigure[${\omega}_z$]{
		\begin{minipage}[t]{0.49\linewidth}
			\centering
			\includegraphics[width=1.0\columnwidth,trim={0cm 0cm 0cm 0cm},clip]{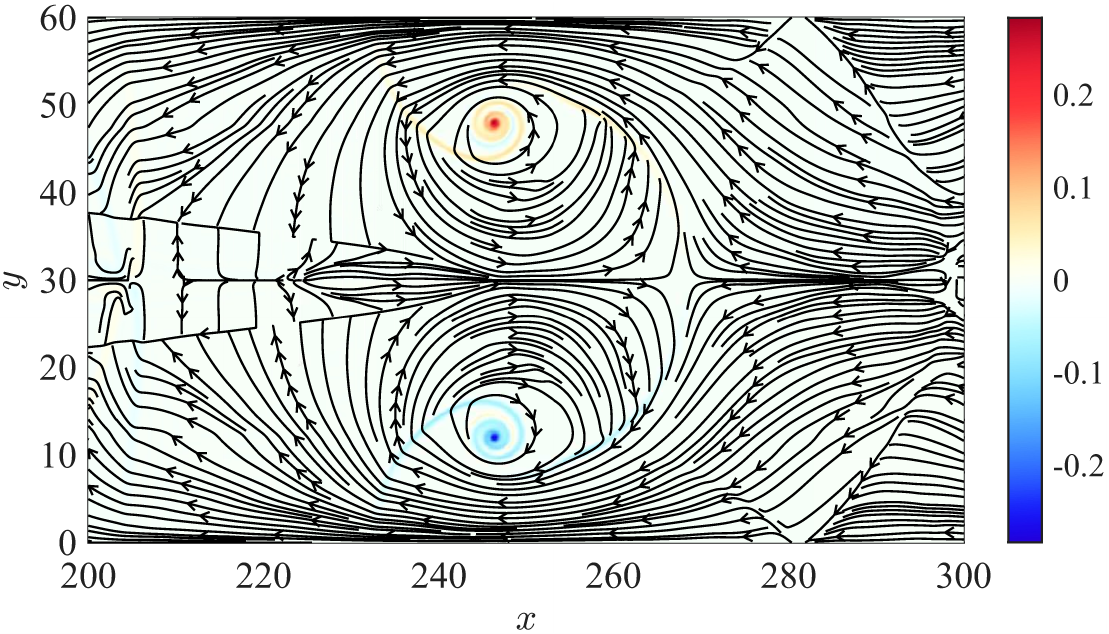}
			%   \vspace{-6.6cm} 
			\label{231}
		\end{minipage}%
	}	
	\subfigure[$\bm{R}_N^+\cdot\bm{e}_z$]{
		\begin{minipage}[t]{0.49\linewidth}
			\centering
			\includegraphics[width=1.0\columnwidth,trim={0cm 0cm 0cm 0cm},clip]{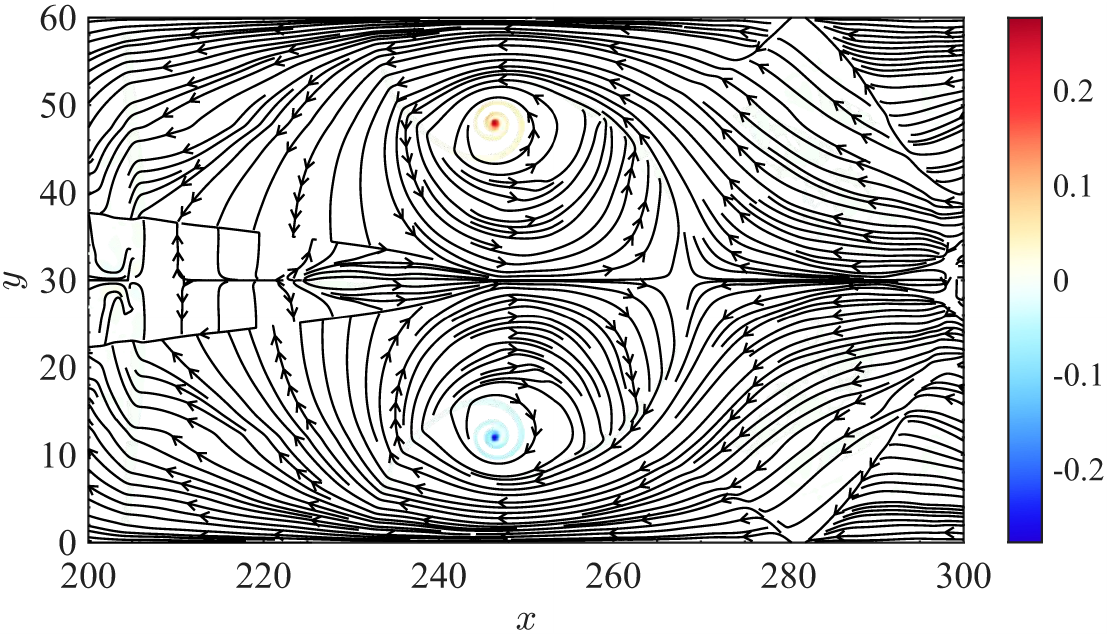}
			%   \vspace{-6.6cm} 
			\label{2332}
		\end{minipage}%
	}	
	
	\subfigure[$\bm{R}_{N}^{-}\cdot\bm{e}_z$]{
		\begin{minipage}[t]{0.49\linewidth}
			\centering
			\includegraphics[width=1.0\columnwidth,trim={0cm 0cm 0cm 0cm},clip]{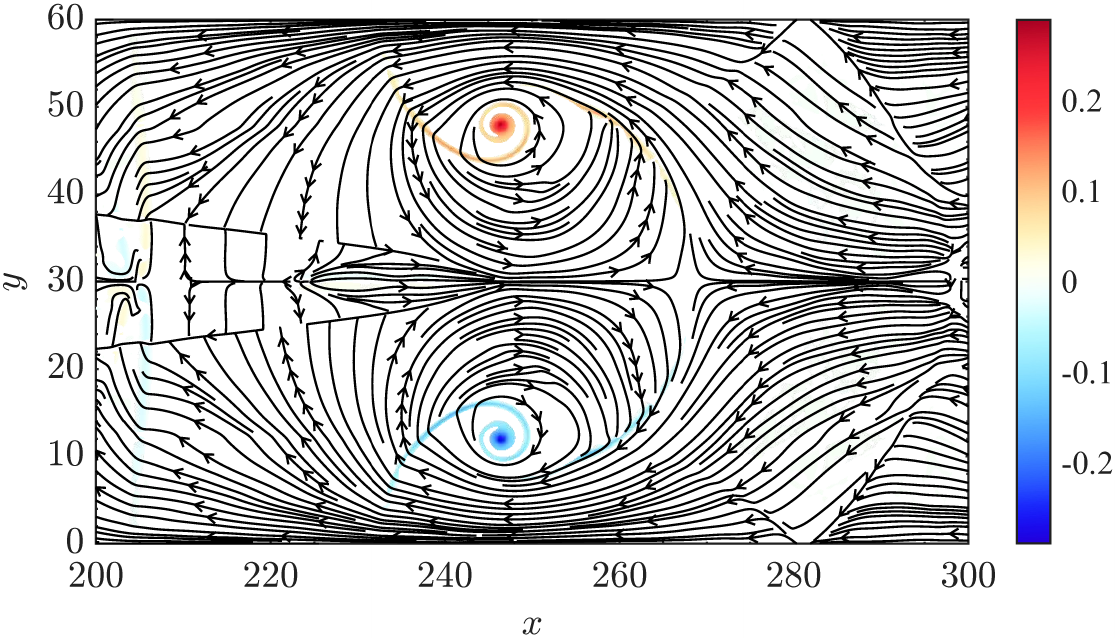}
			%   \vspace{-6.6cm} 
			\label{2332minus}
		\end{minipage}%
	}	
	\caption{Normalized snapshots of the $\bm{e}_z$-components of the DVD and IVD vorticity modes at $t^{*}=442.31$: (a) the orbital-rotation mode
		$\bm{R}(\bm{t})$, (b) the spin mode $\bm{s}(\bm{t})$, (c) the total vorticity $ \bm{\omega}$, (d) $\bm{R}_{N}^{+}$ (Liutex)~\citep{Liu2018}, and (e) $\bm{R}_{N}^{-}$~\citep{Chen2025arxiv,Chen2026operator}.
		The reference frame translates at the incoming flow velocity $\mathbf{U}_{2}= 0.5334{\rm U}_{0}\bm{e}_{x}$.} 
	\label{3045334}
\end{figure}
\begin{figure}[t]
	\centering
	\subfigure[]{
		\begin{minipage}[t]{0.5\linewidth}
			\centering
			\includegraphics[width=1.0\columnwidth,trim={0cm 0.0cm 0.0cm 0.0cm},clip]{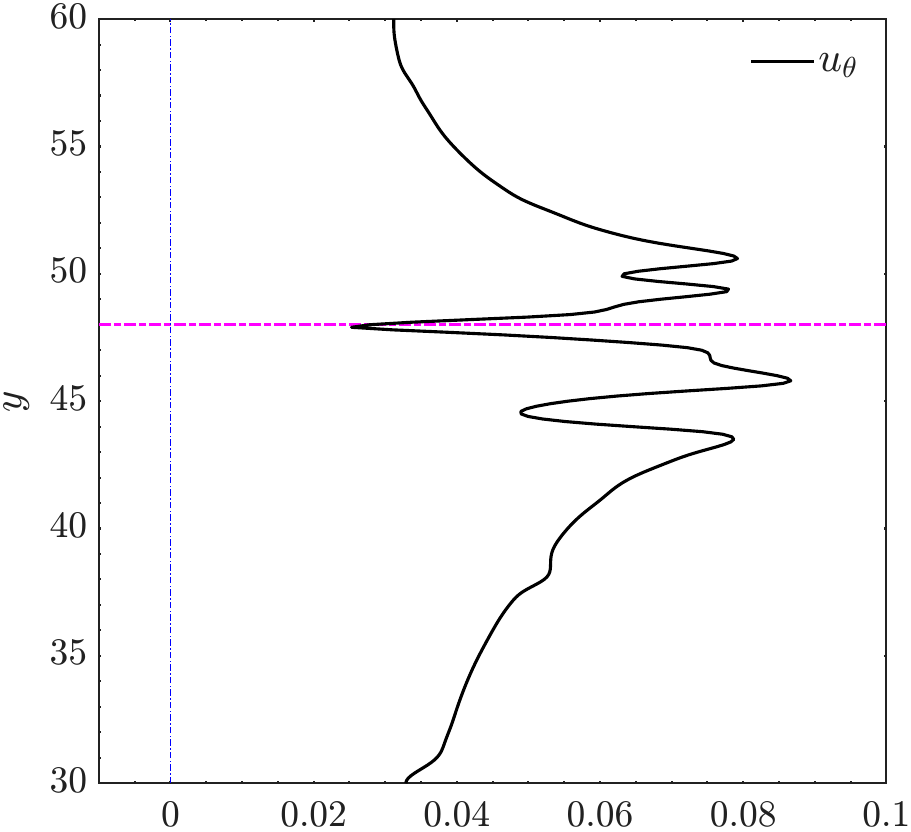}
			\label{compare2}
		\end{minipage}%
	}%	
	\subfigure[]{
		\begin{minipage}[t]{0.5\linewidth}
			\centering
			\includegraphics[width=1.0\columnwidth,trim={0cm 0.0cm 0.0cm 0.0cm},clip]{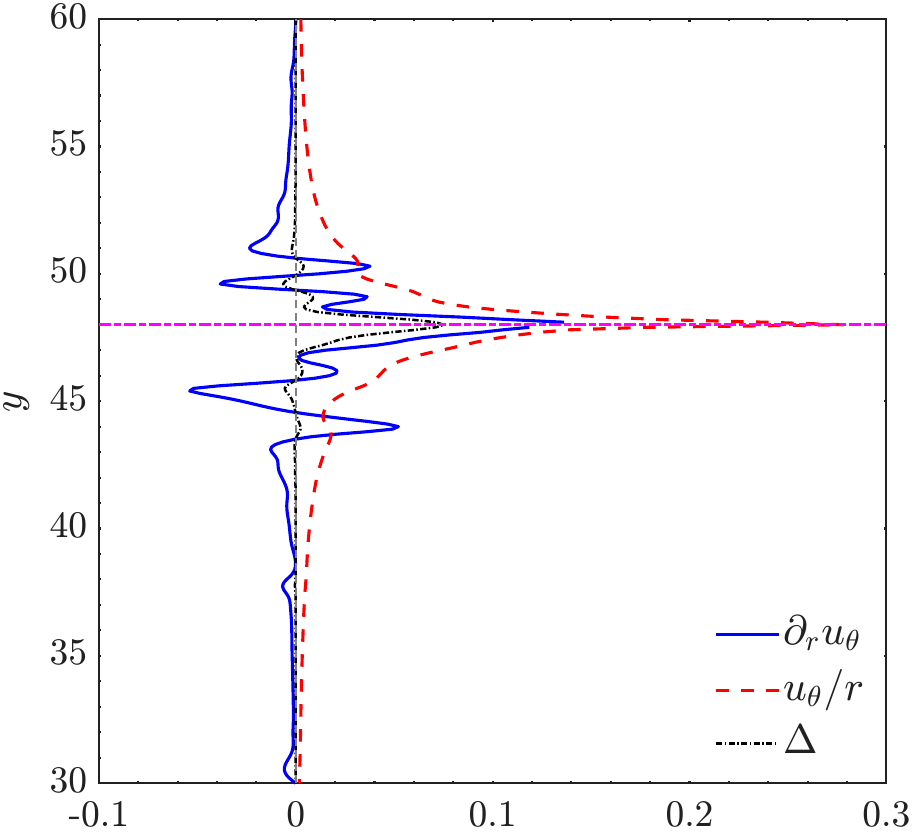}
			\label{compare1}
		\end{minipage}%
	}%	
	\caption{Normalized radial profiles along the line $x^{*}=246.3$: (a) the azimuthal velocity component $u_{\theta}$, (b) $(\partial_{r}u_{\theta},u_{\theta}/r,\Delta)$.}
	\label{fig:4-3}
\end{figure}
\begin{figure}[t]
	\centering
	\subfigure[$(\bm{R}(\bm{t}),\bm{R}^+_N,\bm{R}^{-}_N)$]{
		\begin{minipage}[t]{0.49\linewidth}
			\centering
			\includegraphics[width=1.0\columnwidth,trim={0cm 0cm 0.0cm 0cm},clip]{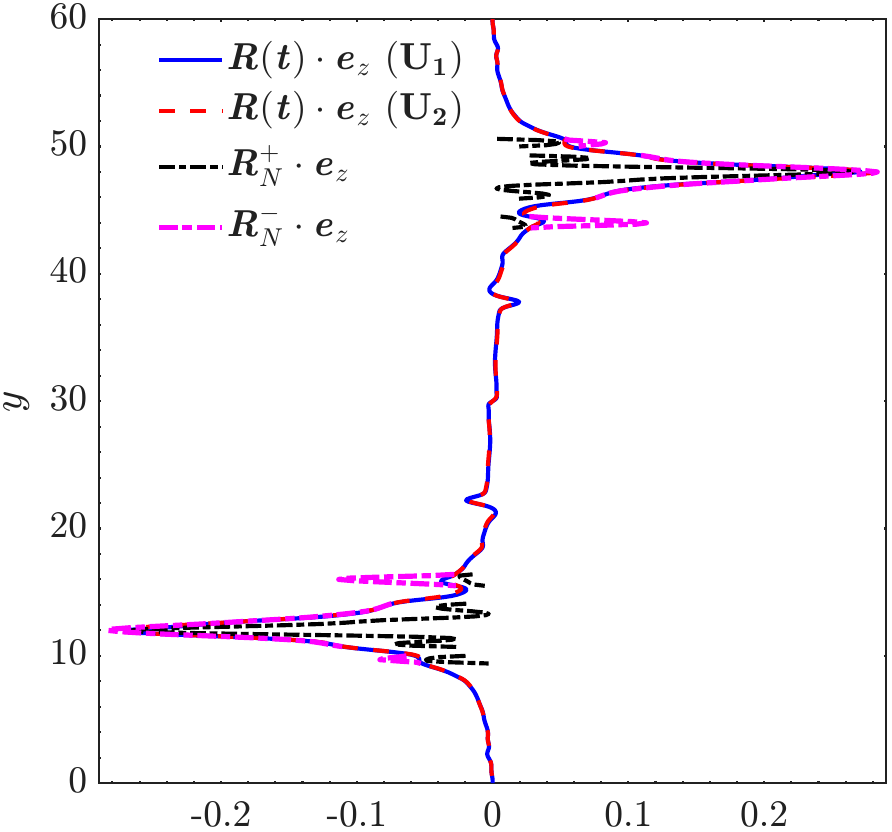}
			% \vspace{-19.6cm} 
			\label{hhhh1}
		\end{minipage}%
	}
	\subfigure[$(\bm{s}(\bm{t}),\bm{s}_{N}^{+},\bm{s}_{N}^{-})$]{
		\begin{minipage}[t]{0.49\linewidth}
			\centering
			\includegraphics[width=1.0\columnwidth,trim={0cm 0.0cm 0.0cm 0cm},clip]{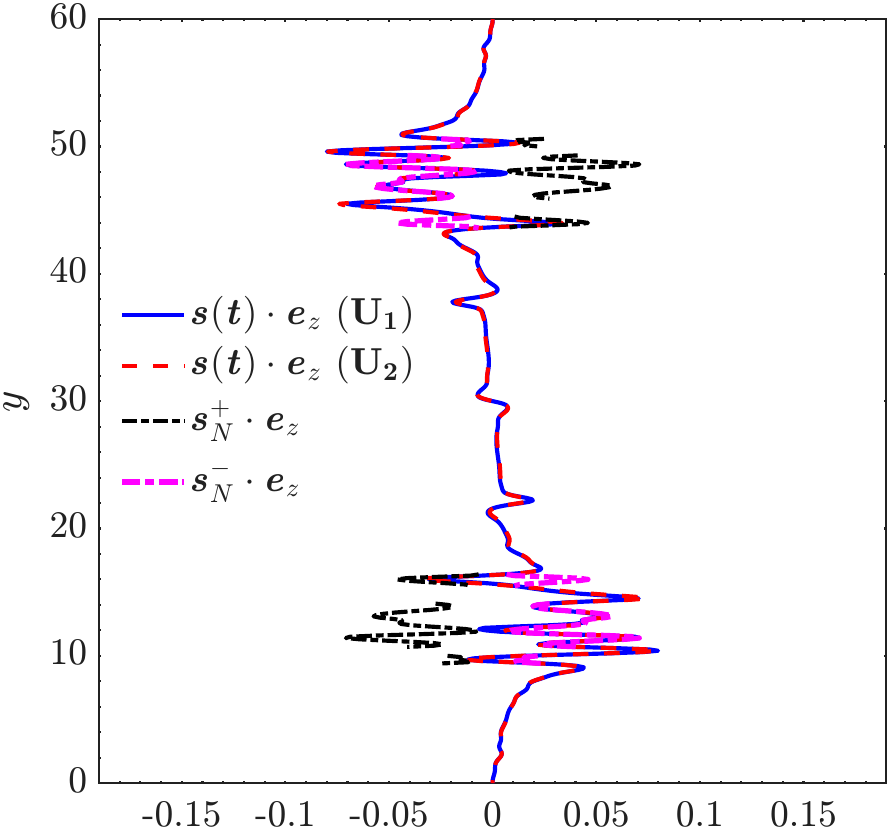}
			%\vspace{-19.6cm} 
			\label{hhhh2}
		\end{minipage}%
	}	
	
	\subfigure[$(\bm{R}(\bm{t}),\bm{R}^+_N,\bm{R}^{-}_N)$ (zoom-in view)]{
		\begin{minipage}[t]{0.49\linewidth}
			\centering
			\includegraphics[width=1.0\columnwidth,trim={0cm 0.0cm 0.0cm 0cm},clip]{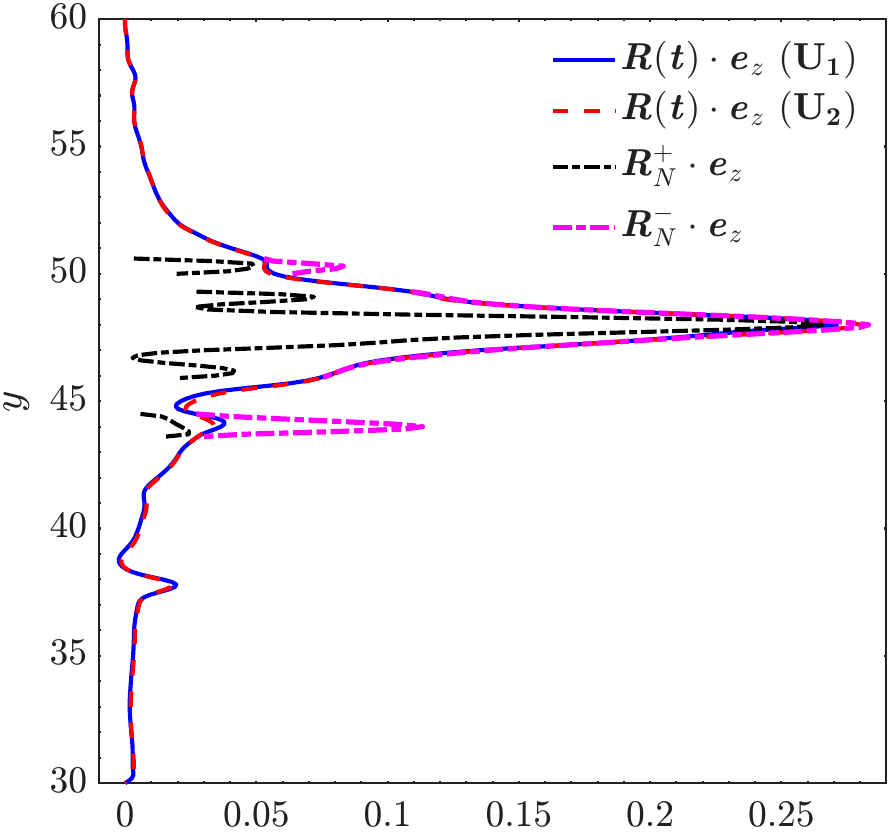}
			%   \vspace{-6.6cm} 
			\label{hhhh3}
		\end{minipage}%
	}	
	\subfigure[$(\bm{s}(\bm{t}),\bm{s}_{N}^{+},\bm{s}_{N}^{-})$ (zoom-in view)]{
		\begin{minipage}[t]{0.49\linewidth}
			\centering
			\includegraphics[width=1.0\columnwidth,trim={0cm 0.0cm 0.0cm 0cm},clip]{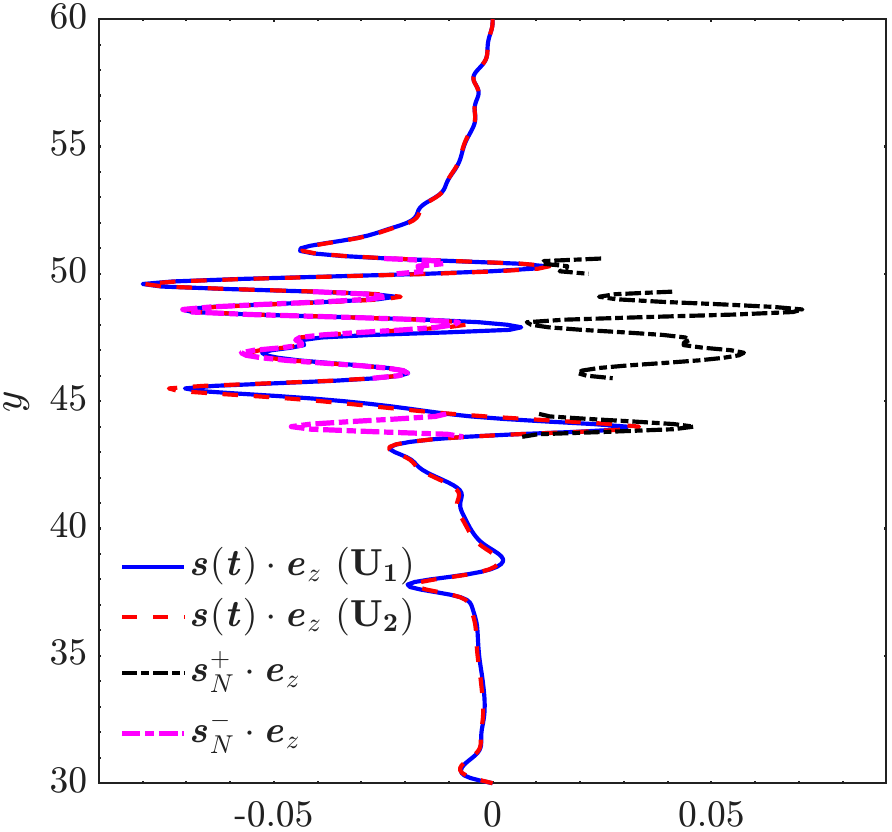}
			%   \vspace{-6.6cm} 
			\label{hhhh4}
		\end{minipage}%
	}	
	\caption{Comparison of the DVD and IVD vorticity modes along the line $x^{*} = 246.3$ connecting the vortex-pair centers for different translation velocities of the reference frame $(\mathbf{U}_{1},\mathbf{U}_{2})=(0.512{\rm U}_{0}\bm{e}_{x},0.5334{\rm U}_{0}\bm{e}_{x})$. Note the IVD vorticity modes is defined in the region where $\Delta>0$.} 
	\label{hhhh1234}
\end{figure}

Applying the streamline-based DVD requires selecting an appropriate reference frame to adequately capture the interface rolling-up process induced by vorticity deposition. We consider two reference frames moving at constant translational velocities: (i) the vortex-pair propagation velocity $\mathbf{U}_{1}=0.512{\rm U}_{0}\bm{e}_{x}$ (figure~\ref{333333}), and (ii) the incoming freestream velocity $\mathbf{U}_{2}=0.5334{\rm U}_{0}\bm{e}_{x}$ (figure~\ref{3045334}). Figure~\ref{333333} presents normalized contour maps of the orbital-rotation mode $\bm{R}(\bm{t})\bm{\cdot}\bm{e}_{z}$ (figure~\ref{kk1}), the spin mode $\bm{s}(\bm{t})\bm{\cdot}\bm{e}_{z}$ (figure~\ref{kk2}), and the total vorticity $\omega_{z}$ (figure~\ref{kk3}) for $\mathbf{U}_{1}=0.512{\rm U}_{0}\bm{e}_{x}$. For comparison, the contour maps of the characteristic rigid-rotation modes, $\bm{R}_N^+$ and $\bm{R}_N^-$, are also shown in figures~\ref{kk4} and~\ref{kk5}. In figure~\ref{kk1}, the rigid-rotation contribution driving interface deformation is isolated from shear contamination via $\bm{R}(\bm{t})\bm{\cdot}\bm{e}_{z}$, which exhibits antisymmetric pattern with positive and negative peaks in the upper and lower vortex cores, respectively. Notably, the spin mode $\bm{s}(\bm{t})\bm{\cdot}\bm{e}_{z}$ (figure~\ref{kk2}), representing the rate of angular deformation, has opposite sign to $\bm{R}(\bm{t})\bm{\cdot}\bm{e}_{z}$ throughout most of the vortex core, whereas both positive and negative signs appear in the outer regions and spiral arms. Unlike the conventional Klein-Kaden-Betz (KKB) configuration, where $\bm{R}(\bm{t})$ and $\bm{s}(\bm{t})$ share the same sign~\citep{Klein1910,Kaden1931,Betz1950}, the observed spatial distribution resembles the Burgers vortex solution~\citep{burgers1948mathematical} and is dominated by the anti-KKB configuration (opposite signs of $\bm{R}(\bm{t})$ and $\bm{s}(\bm{t})$), as previously proposed by~\citet{Chen2025arxiv}. This is further supported by the radial profile of the azimuthal velocity component $u_{\theta}$ in figure~\ref{compare2}, which resembles the Burgers vortex solution but exhibits more fluctuations due to increased complexity. The detailed analysis of the Burgers vortex can be found in~\citet{Chen2025arxiv}. It is noted that this is the first observation of the anti-KKB configuration in a shock–interface interaction. The potential coexistence of both KKB and anti-KKB patterns in a swirling system under certain conditions has been shown to arise directly from the commutativity of a pair of vorticity operators~\citep{Chen2026operator}.

The inner region of the vortex core is clearly captured by the Liutex 
$\bm{R}_N^{+}$, which exhibits a more concentrated peak region compared to $\bm{R}(\bm{t})$ (figure~\ref{kk4}). In contrast, $\bm{R}(\bm{t})$ resembles the characteristic rigid-rotation mode $\bm{R}_N^{-}$ serving as the upper bound (figure~\ref{kk5}), indicating a larger vortex core size. This observation can be interpreted as follows. As shown in figure~\ref{compare1}, it holds that $\gamma_{0}\equiv\partial_{r}u_{\theta}-u_{\theta}/r<0$. Under this condition, the relationships between $(R(\bm{t}),s(\bm{t}))$ and $(R_{N}^{-},s_{N}^{-})$ are derived as $R(\bm{t})=R_{N}^{-}+2s_{N}^{-}\cos\Theta$ and $s(\bm{t})=s_{N}^{+}\cos2\Theta$, where $R_{N}^{-}=2\psi^{-}=2u_{\theta}/r+\mathcal{O}(\varepsilon^2)$, $s_{N}^{-}=\gamma^{-}=\partial_{r}u_{\theta}-u_{\theta}/r+\mathcal{O}(\varepsilon^2)$, and $\varepsilon$ is a small dimensionless parameter~\citep{Chen2026operator}. Here, in the cylindrical coordinate system $(r,\Theta)$ with the basis $(\bm{e}_{r},\bm{e}_{\theta})$, it holds that $\bm{u}=\bm{u}_{r}\bm{e}_{r}+u_{\theta}\bm{e}_{\theta}$ and $(\cos\Theta,\sin\Theta)=(u_{r}/q,u_{\theta}/q)+\mathcal{O}(\varepsilon)$, where $r$ is the radial distance from the vortex center, $\Theta$ the azimuthal angle, and $q=\sqrt{u_{\theta}^{2}+u_{r}^{2}}$ the velocity magnitude. Since the streamline direction is approximately aligned with the basis vector $\bm{e}_{\theta}$ in the neighborhood of the vortex core, we obtain $\Theta\approx\pi/2$, and hence $(R(\bm{t}),s(\bm{t}))\approx({R}_{N}^{-},s_{N}^{-})$. Consequently, the differences between $(R(\bm{t}),s(\bm{t}))$ and $(R_{N}^{+},s_{N}^{+})$ are determined as $R(\bm{t})-R_{N}^{+}=2\gamma^{+}=2s_{N}^{+}$ and $s(\bm{t})-s_{N}^{+}=2\gamma^{-}=2s_{N}^{-}$. A comparison of figures~\ref{333333} and~\ref{3045334} reveals that the streamline patterns around the vortex pair appear largely similar in both cases, albeit with some visible differences which do not alter the essential features nearby the vortex cores.
\begin{figure}[t]
	\centering
	\includegraphics[width=0.5\columnwidth,trim={0cm 0cm 0cm 0cm},clip]{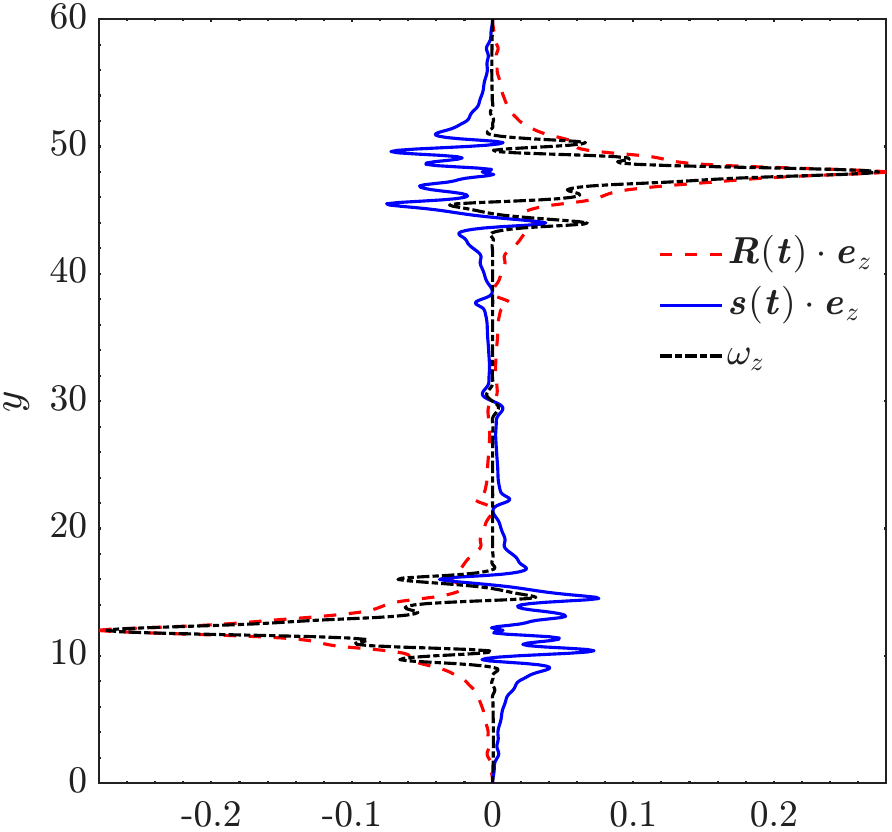}
	\caption{Comparison of $(\bm{R}(\bm{t}),\bm{s}(\bm{t}),\bm{\omega})$ along the line $x^{*} = 246.3$ connecting the vortex-pair centers moving at $\mathbf{U}_{1}= 0.512{\rm U}_{0}\bm{e}_{x}$.} 
	\label{compare_vertical_Rsomega}
\end{figure}

In figures~\ref{hhhh1} and~\ref{hhhh2}, the streamline-based DVD vorticity modes $(\bm{R}(\bm{t}),\bm{s}(\bm{t}))$ are quantitatively compared along the line $x^*=246.3$ connecting the centers of the vortex pair for different translation velocities of the reference frame $(\mathbf{U}_{1},\mathbf{U}_{2})$. Zoom-in views around the upper vortex are shown in figures~\ref{hhhh3} and~\ref{hhhh4}. It is observed that the DVD vorticity modes $(\bm{R}(\bm{t}),\bm{s}(\bm{t}))$ are rigorously bounded by the IVD vorticity modes $(\bm{R}_{N}^{\pm},\bm{s}_{N}^{\pm})$ in the region where $\Delta>0$, being consistent with the theoretical analysis of~\citet{Chen2025arxiv}. High peaks of $\bm{R}(\bm{t})$ and high-frequency oscillations of $\bm{s}(\bm{t})$ are primarily concentrated nearby the vortex pair, and the distributions of these DVD modes are insensitive to the choice of the reference frame. As shown in figure~\ref{compare_vertical_Rsomega}, vorticity is dominated by the orbital-rotation and essentially attenuated by the spin mode inside the inner vortex cores. Interestingly, during the vorticity-driven interface evolution, the dominated pattern of the DVD vorticity modes exhibits a dynamic transition from a KKB to an anti-KKB configuration. In the early stage, the KKB mechanism is activated, promoting the formation of axial vortices through strong entrainment of the sheet-like spiral arms generated by the baroclinic torque, where ${R}(\bm{t})$ and ${s}(\bm{t})$ share the same sign (i.e., the KKB configuration). In contrast, as the vortex pair matures, the vorticity modes display an anti-KKB pattern with opposite signs of $\bm{R}(\bm{t})$ and $\bm{s}(\bm{t})$ in most regions, which tends to suppress further increases in the swirling intensity of the vortex.

\subsection{Integral diagnostics and interface-vortex-wave interaction}
\begin{figure}[t]
	\centering
	\subfigure[$\mathcal{D}\cap\left\{\bm{x}\in\mathbb{R}^{2}|\Delta(\bm{x})>0\right\}$]{
		\begin{minipage}[t]{0.49\linewidth}
			\centering
			\includegraphics[width=1.0\columnwidth,trim={0cm 0cm 0cm 0cm},clip]{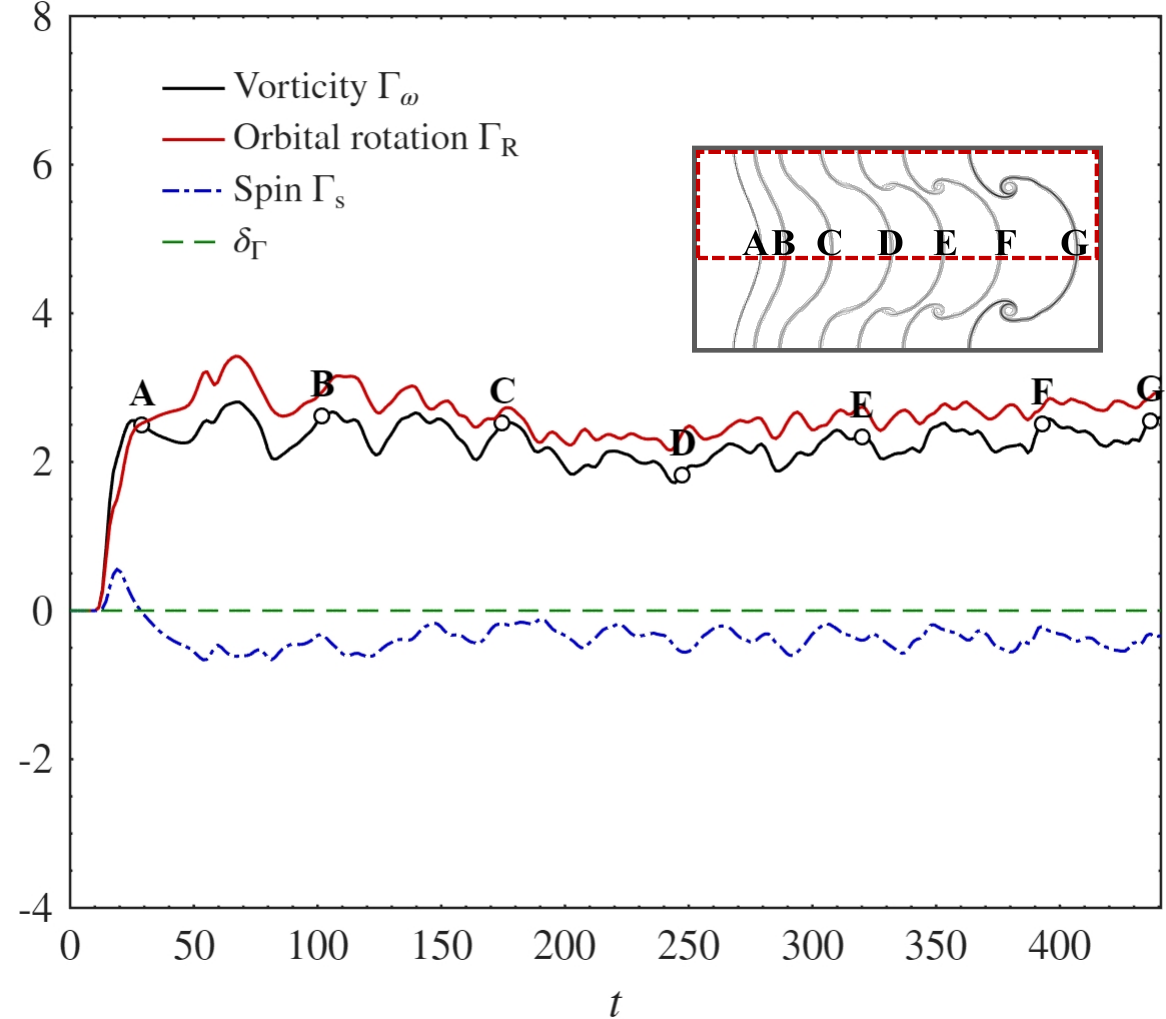}
			%\vspace{-0.6cm} 
			\label{Delta0}
		\end{minipage}%
	}
			\subfigure[$\mathcal{D}$]{
		\begin{minipage}[t]{0.49\linewidth}
			\centering
			\includegraphics[width=1.0\columnwidth,trim={0cm 0cm 0cm 0cm},clip]{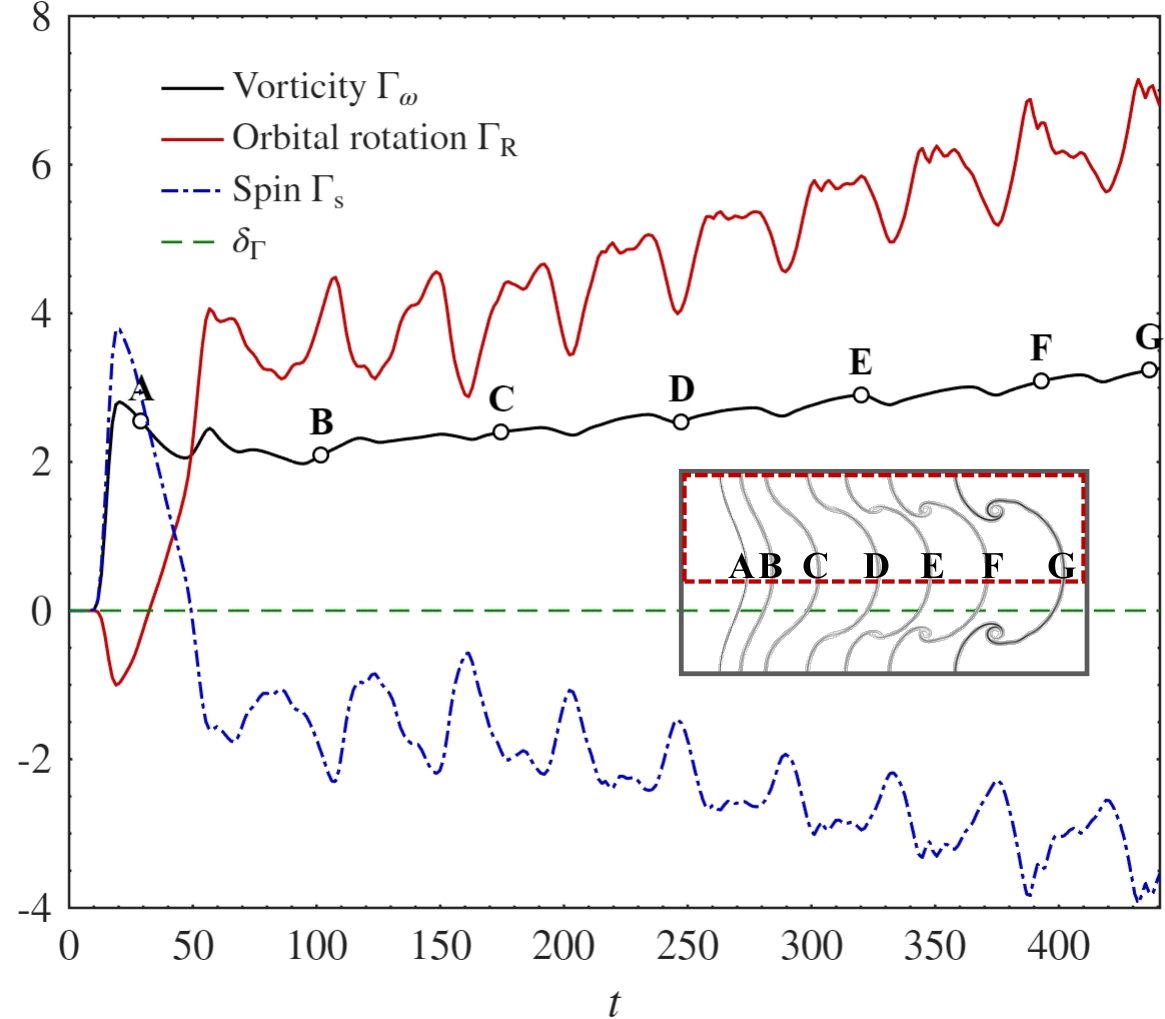}
			%\vspace{-0.6cm} 
			\label{Delta1}
		\end{minipage}%
	}	
	
%	\subfigure[$\mathcal{D}\cap\left\{\bm{x}\in\mathbb{R}^{2}|\Delta(\bm{x})>0\right\}$ (zoom-in view)]{
%	\begin{minipage}[t]{0.49\linewidth}
%		\centering
%		\includegraphics[width=1.0\columnwidth,trim={0cm 0cm 0cm 0cm},clip]{whole1.pdf}
%		%\vspace{-0.6cm} 
%		\label{Delta00}
%	\end{minipage}%
%}
		\subfigure[$\mathcal{D}\cap\left\{\bm{x}\in\mathbb{R}^{2}|\Delta(\bm{x})>0\right\}$ (the early stage)]{
		\begin{minipage}[t]{0.49\linewidth}
			\centering
			\includegraphics[width=1.0\columnwidth,trim={0cm 0cm 0cm 0cm},clip]{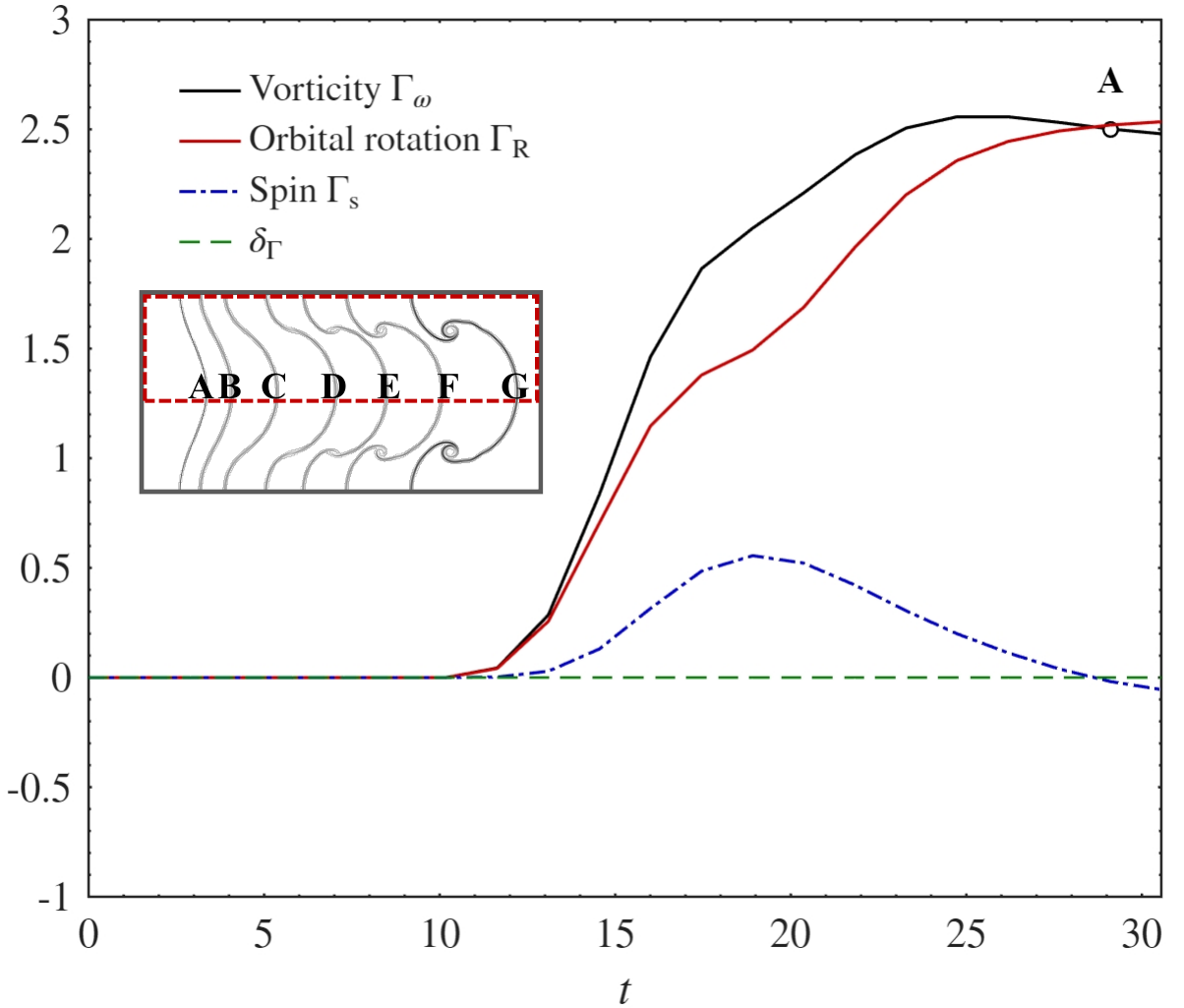}
			%\vspace{-0.6cm} 
			\label{Delta2}
		\end{minipage}%
	}	
			\subfigure[$\mathcal{D}$ (the early stage)]{
		\begin{minipage}[t]{0.49\linewidth}
			\centering
			\includegraphics[width=1.0\columnwidth,trim={0cm 0cm 0cm 0cm},clip]{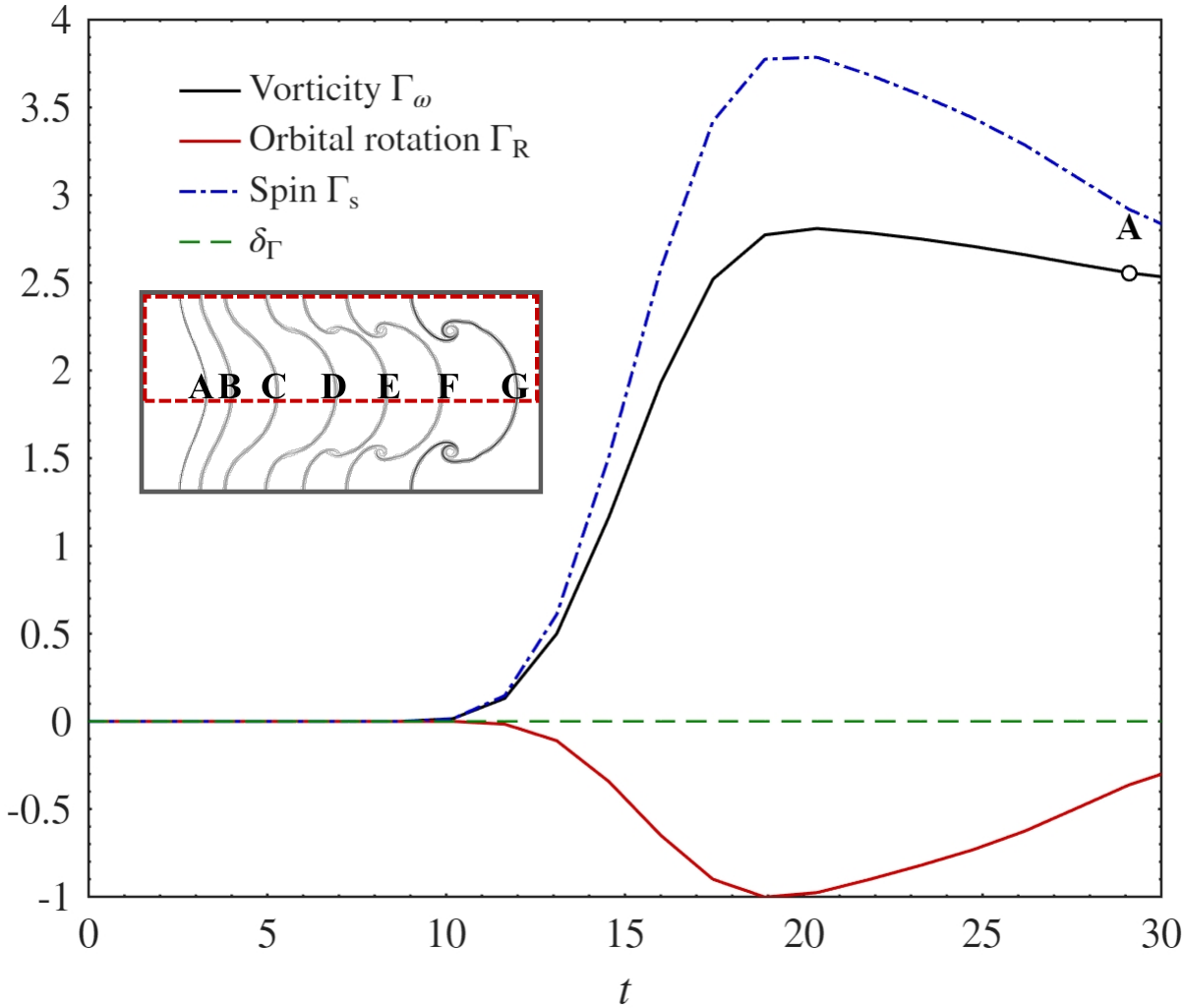}
			%\vspace{-0.6cm} 
			\label{Delta3}
		\end{minipage}%
	}	
	\caption{Spatiotemporal evolution of the vorticity flux $\Gamma_{\omega}$, the orbital-rotation flux $\Gamma_{\rm R}$, and the spin flux  $\Gamma_{\rm s}$ during the vorticity-driven interface evolution. The integration domain is (a,c) $\mathcal{D}\cap\left\{\bm{x}\in\mathbb{R}^{2}|\Delta(\bm{x})>0\right\}$, and (b,d) $\mathcal{D}$ (the rectangular upper half domain). Note that the residual error is $\delta_{\Gamma}\equiv\Gamma_{\omega}-\Gamma_{\rm{R}}-\Gamma_{\rm{s}}$.} 
	\label{DeltaGamma}
\end{figure}
\begin{figure}[t]
	\centering
	\subfigure[$A$, $t^{*}=29.099$]{
		\begin{minipage}[t]{0.49\linewidth}
			\centering
			\includegraphics[width=1.0\columnwidth,trim={0cm 0.0cm 0cm 0.0cm},clip]{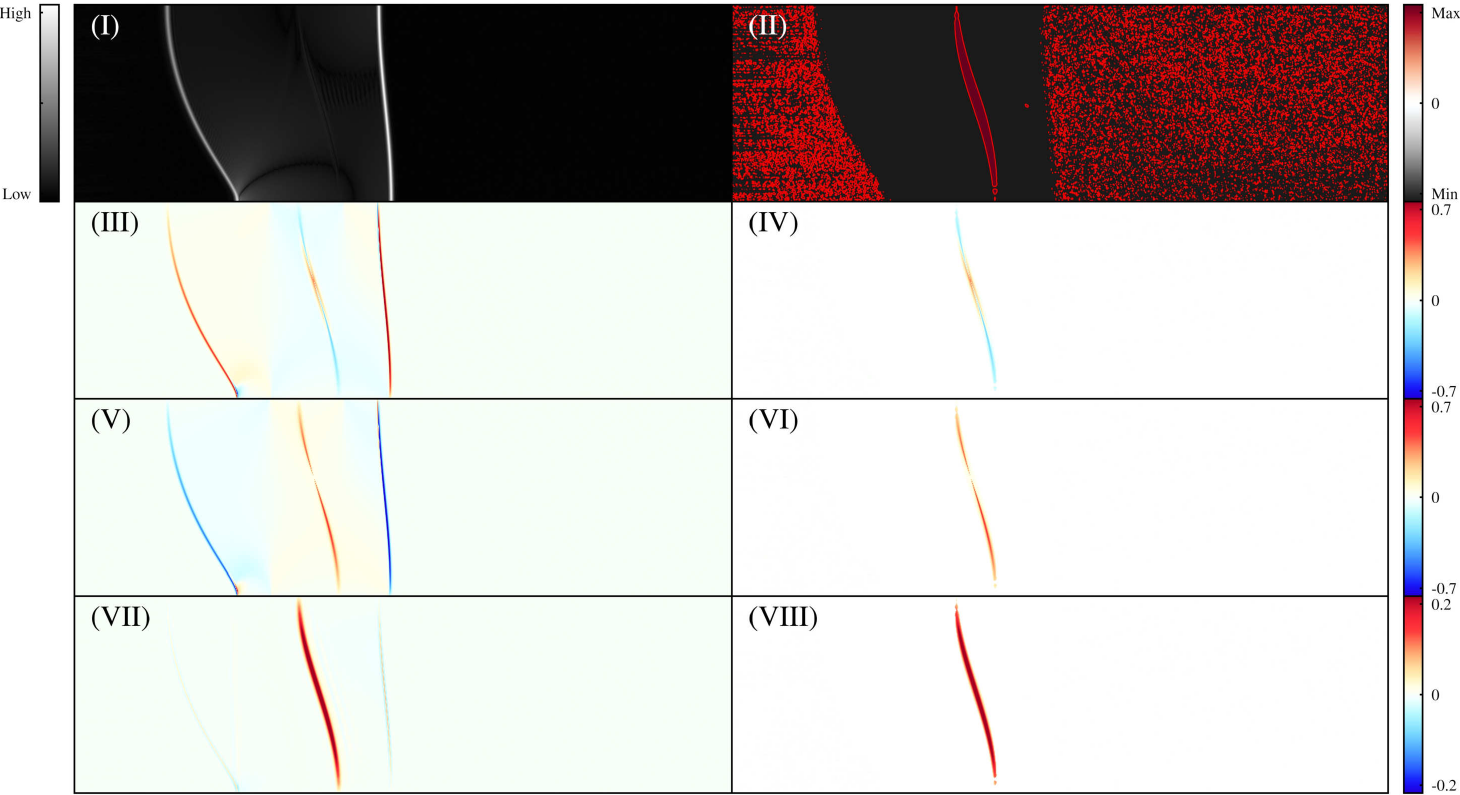}
			%\vspace{-0.6cm} 
			\label{and1}
		\end{minipage}%
	}
	\subfigure[$B$, $t^{*}=101.847$]{
		\begin{minipage}[t]{0.49\linewidth}
			\centering
			\includegraphics[width=1.0\columnwidth,trim={0cm 0.0cm 0cm 0.0cm},clip]{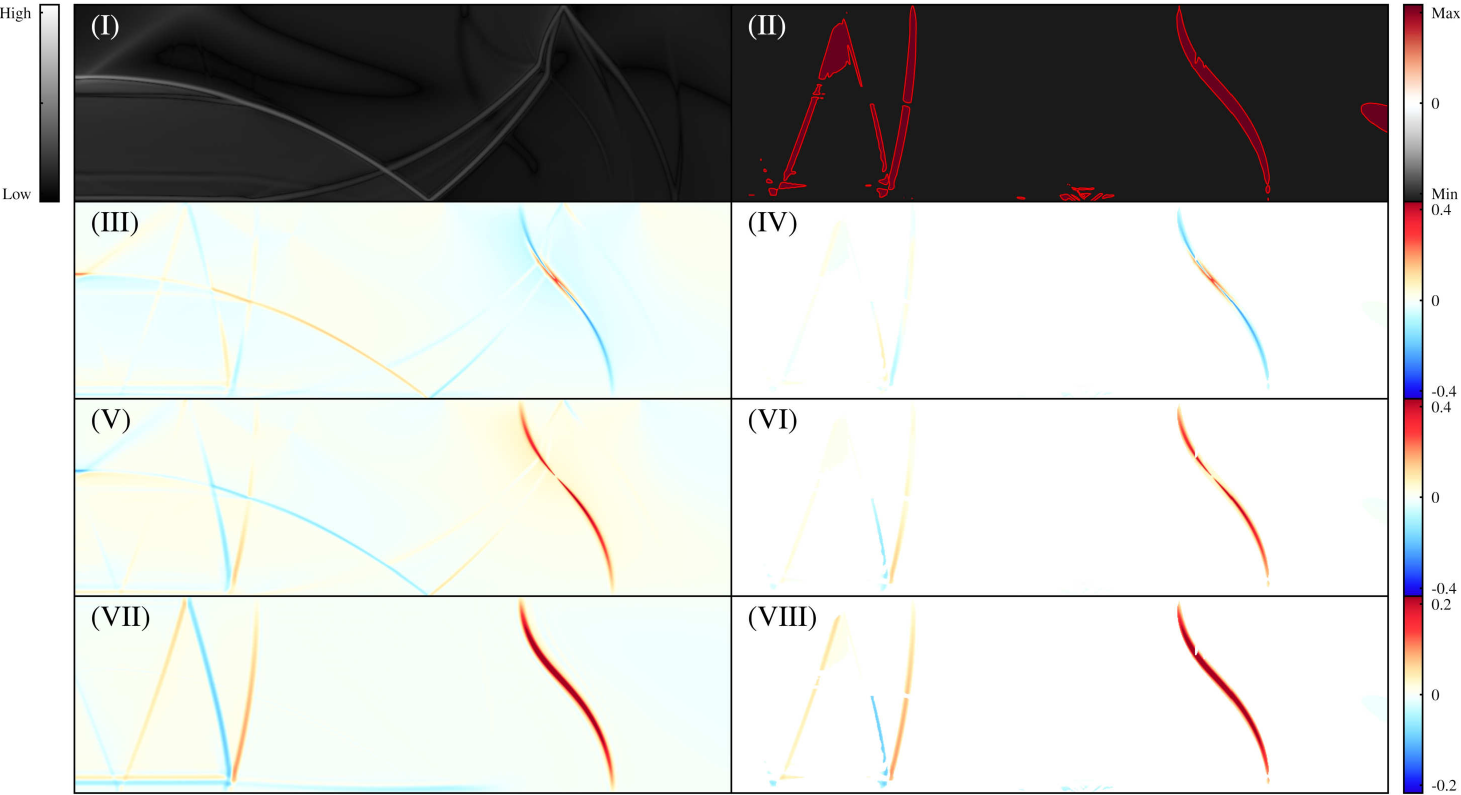}
			%\vspace{-0.6cm} %    left down  right up
			\label{and2}
		\end{minipage}%
	}	
	
	\subfigure[$C$, $t^{*}=174.594$]{
		\begin{minipage}[t]{0.49\linewidth}
			\centering
			\includegraphics[width=1.0\columnwidth,trim={0cm 0.0cm 0cm 0.0cm},clip]{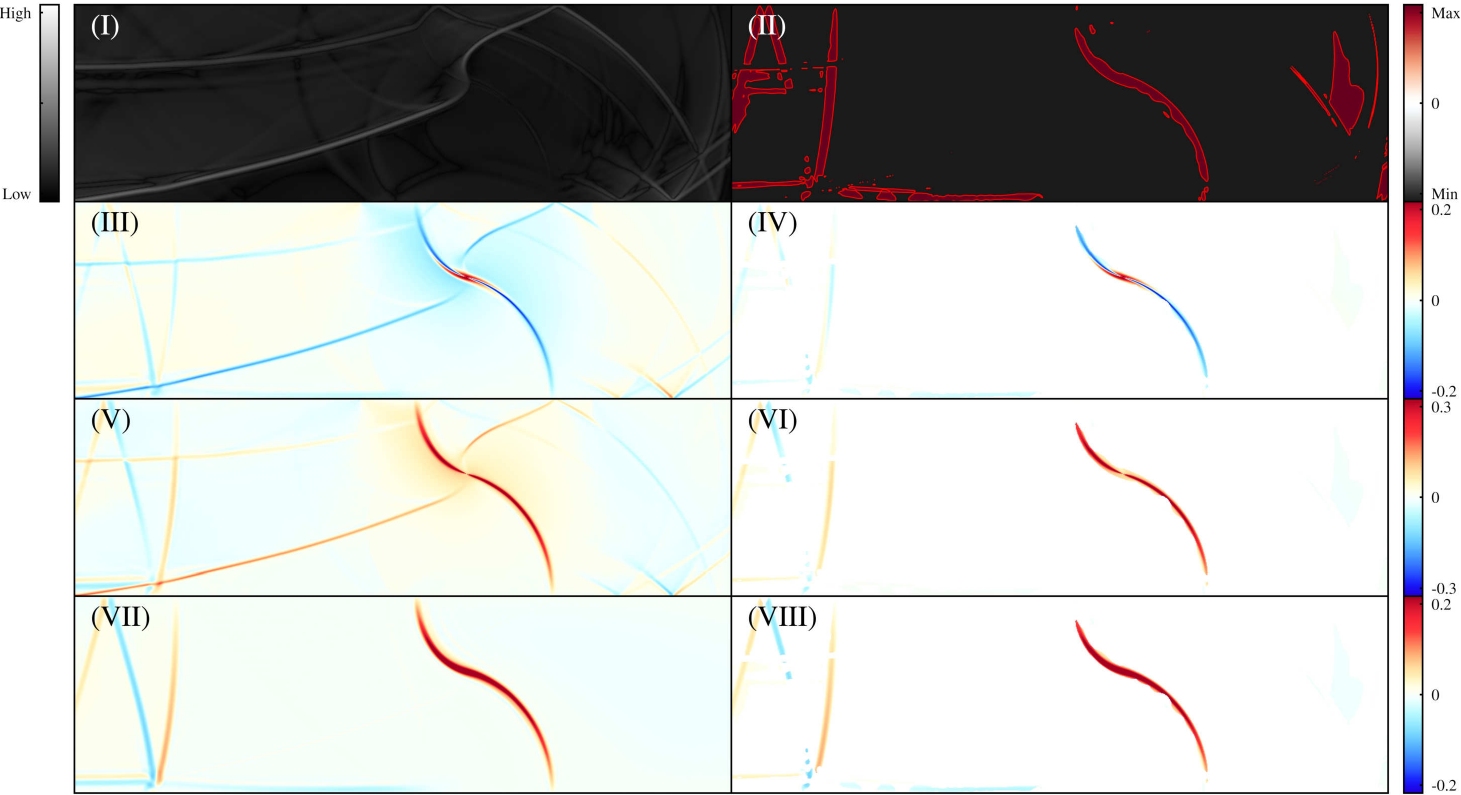}
			%\vspace{-0.6cm} 
			\label{and3}
		\end{minipage}%
	}
	\subfigure[$D$, $t^{*}=247.342$]{
		\begin{minipage}[t]{0.49\linewidth}
			\centering
			\includegraphics[width=1.0\columnwidth,trim={0cm 0.0cm 0cm 0.0cm},clip]{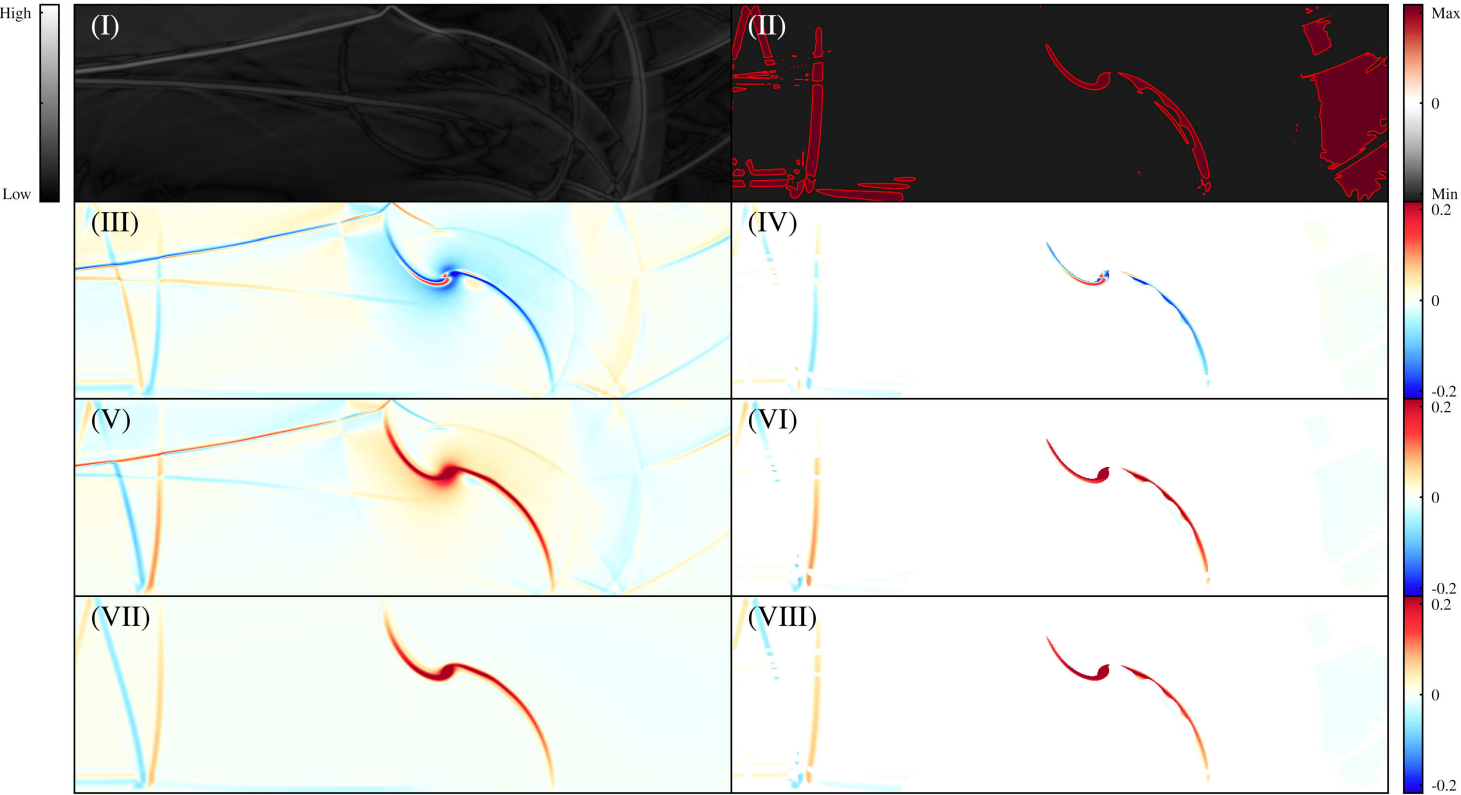}
			%\vspace{-0.6cm} 
			\label{and4}
		\end{minipage}%
	}	
	
	\subfigure[$E$, $t^{*}=320.089$]{
		\begin{minipage}[t]{0.49\linewidth}
			\centering
			\includegraphics[width=1.0\columnwidth,trim={0cm 0.0cm 0cm 0.0cm},clip]{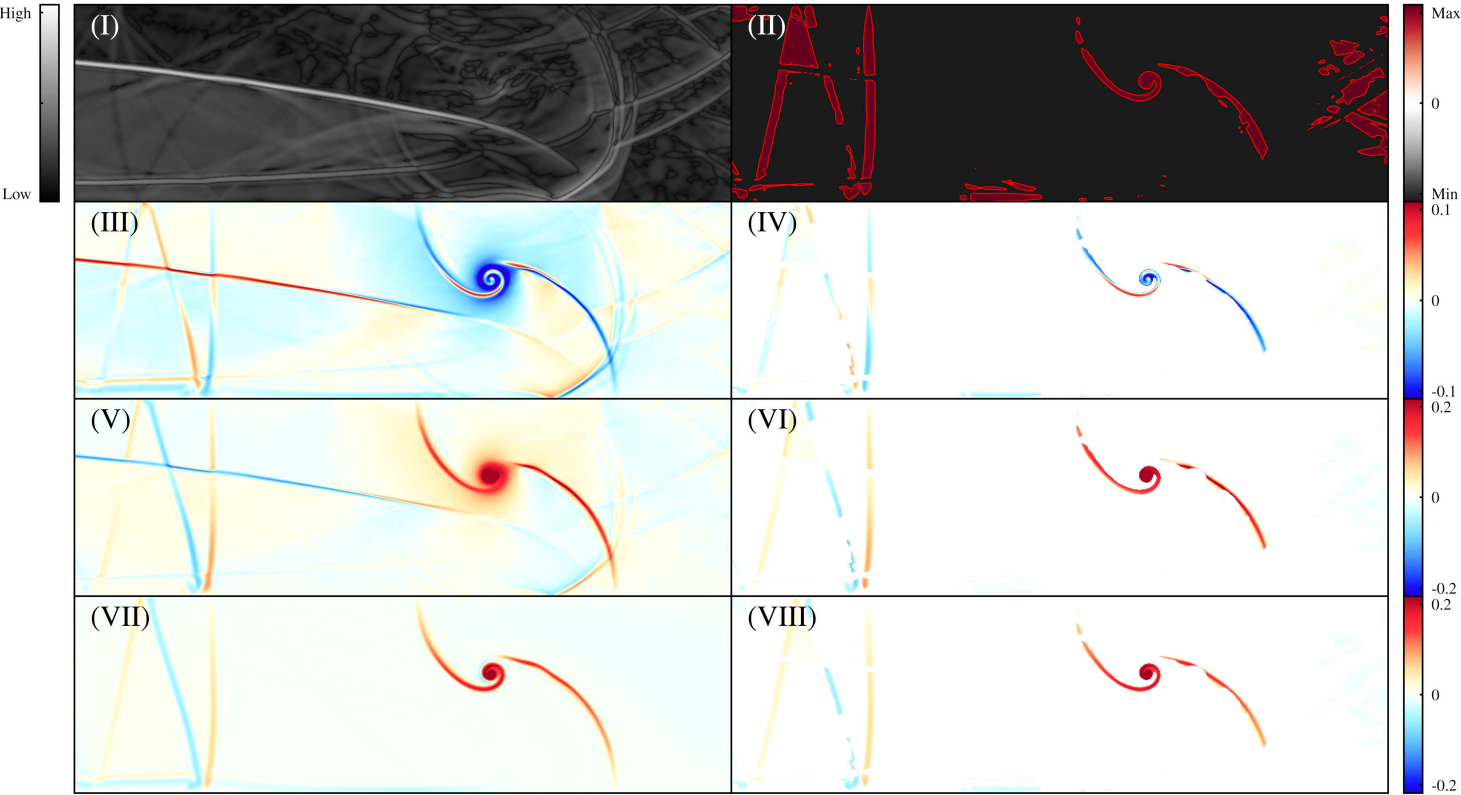}
			%\vspace{-0.6cm} 
			\label{and5}
		\end{minipage}%
	}
	\subfigure[$F$, $t^{*}=392.837$]{
		\begin{minipage}[t]{0.49\linewidth}
			\centering
			\includegraphics[width=1.0\columnwidth,trim={0cm 0.0cm 0cm 0.0cm},clip]{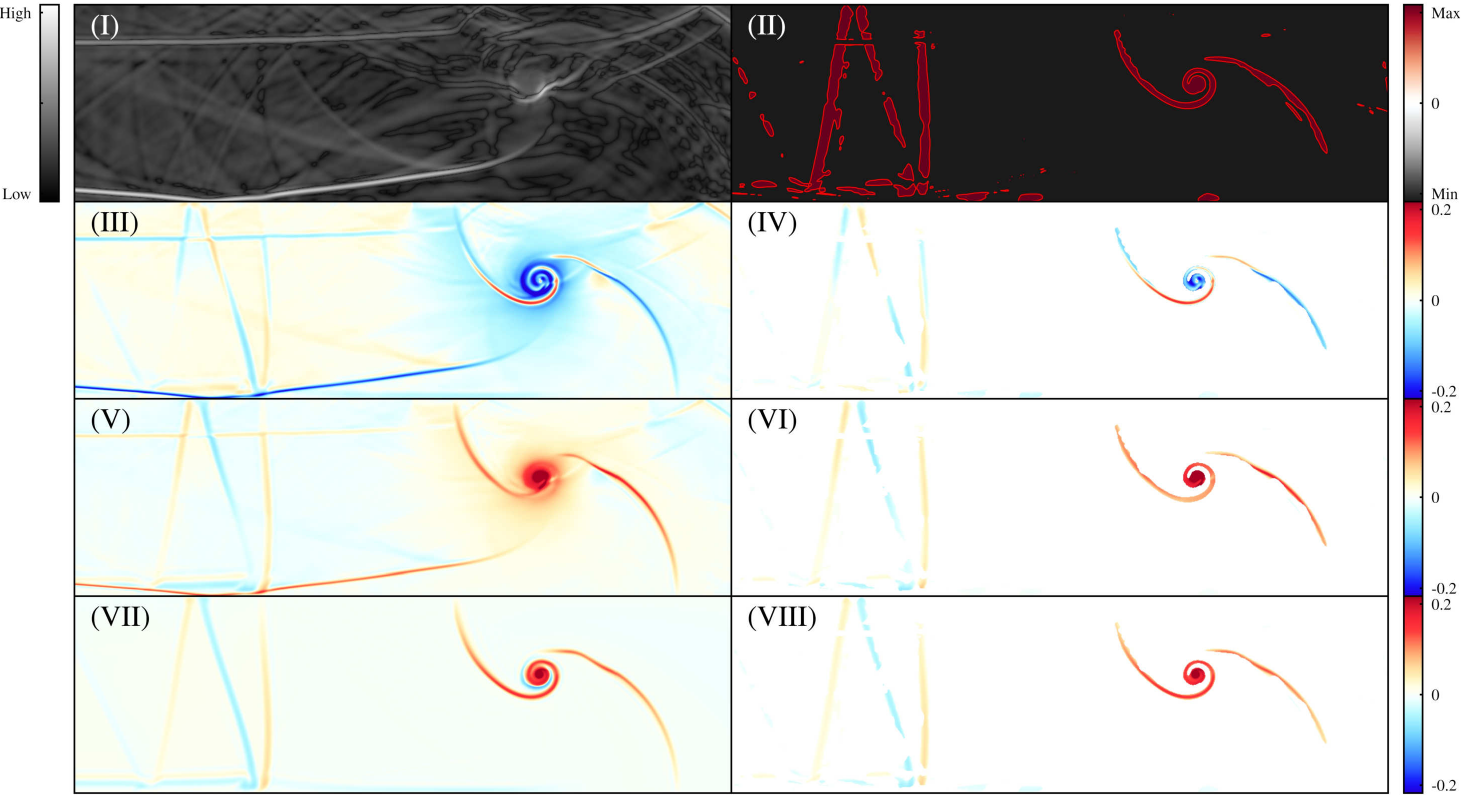}
			%\vspace{-0.6cm} 
			\label{and6}
		\end{minipage}%
	}	
	\caption{Evolutionary characteristics of interface morphology, vorticity/vortex structures, and wave fields at six representative instants, which correspond to points $A$~--~$F$ marked in figure~\ref{DeltaGamma}. For each instant, eight subpanels (I)~--~(VIII) are defined as follows: (I) the dilatation $\vartheta\equiv\bm{\nabla}\bm{\cdot}\bm{u}$; (II) the discriminant $\Delta$ (red for $\Delta > 0$ and black for $\Delta < 0$, with the bright red solid line representing the isoline $\Delta = 0$); (III) the unconditional spin mode $\bm{s}(\bm{t}) \cdot \bm{e}_z$; (IV) the spin mode $\bm{s}(\bm{t})\cdot \bm{e}_z$ conditioned on $\Delta > 0$; (V) the unconditional orbital-rotation mode $\bm{R}(\bm{t}) \cdot \bm{e}_z$; (VI) the orbital-rotation mode $\bm{R}(\bm{t}) \cdot \bm{e}_z$ conditioned on $\Delta > 0$; (VII) the unconditional vorticity $\omega_{z}$; (VIII) the vorticity $\omega_z$ conditioned on $\Delta>0$.} 
	\label{and}
\end{figure}

The global integral properties are further examined based on the general theory proposed in \S~\ref{Decomposition and Estimate of total vorticity}. Figure~\ref{DeltaGamma} illustrates the temporal evolution histories of the total vorticity flux ($\Gamma_{\rm{\omega}}$), and its orbital-rotation and spin components ($\Gamma_{\rm{R}}$ and $\Gamma_{\rm{s}}$), along with the corresponding evolution of interfacial morphology shown in insets $A$~--~$G$. The integration domains are chosen as the subregion $\mathcal{D}\cap\left\{\bm{x}\in\mathbb{R}^{2}|\Delta(\bm{x})>0\right\}$ in figure~\ref{Delta0} (being consistent with the conventional vortical-flow region), and the rectangular upper half-domain $\mathcal{D}$ (figure~\ref{Delta1}). Zoom-in views of the early-stage evolution are specifically displayed in figures~\ref{Delta2} and~\ref{Delta3}. Figure~\ref{and} presents snapshots of the wave and vortex structure distributions at six representative instants, corresponding to the characteristic points $A$~--~$F$ marked in figure~\ref{DeltaGamma}.

In the early stage of the shock-interface interaction ($t^{*}<30$, prior to point $A$), the misalignment between density and pressure gradients generates
baroclinic torque, leading to rapid accumulation of vorticity on the perturbed interface. This physical process is intuitively reflected in the rapid initial growth of all three vorticity fluxes within the region where $\Delta>0$ (figures~\ref{Delta0} and~\ref{Delta2}). Before point $A$, the simulation results show that a strong spin layer with positive spin mode $(\bm{s}(\bm{t})\bm{\cdot}\bm{e}_{z})$ develops along the interface during the time interval $t^{*}\in(10,30)$, keeping $\Gamma_{s}$ positive. During this period, the spin flux $\Gamma_{\rm s}$ (or the spin mode $\bm{s}(\bm{t})\bm{\cdot}\bm{e}_{z}$) is continuously transformed into the orbital-rotation flux $\Gamma_{\rm R}$ (or the orbital-rotation mode $\bm{R}(\bm{t})\bm{\cdot}\bm{e}_{z}$), resulting in a downward trend of $\Gamma_{\rm s}$ over $t^{*}\in(20,30)$ after a distinct positive peak. Locally, both $\bm{R}(\bm{t})\bm{\cdot}\bm{e}_{z}$ and $\bm{s}(\bm{t})\bm{\cdot}\bm{e}_{z}$ are positive, which is identified as KKB configuration (i.e., KKB synergy effect)~\citep{Chen2025arxiv}. The total vorticity flux $\Gamma_{\omega}$ is dominated by $\Gamma_{\rm R}$ and partially modulated by $\Gamma_{\rm s}$. However, when integrated over the upper half domain $\mathcal{D}$, the early-stage behavior in figures~\ref{Delta1} and~\ref{Delta3} exhibits a completely opposite trend, where the two vorticity modes exhibit opposite signs (negative $\bm{R}(\bm{t})\bm{\cdot}\bm{e}_{z}$) and positive $\bm{s}(\bm{t})\bm{\cdot}\bm{e}_{z}$), following the anti-KKB configuration (antagonistic effect)~\citep{Chen2025arxiv,Chen2026operator}. As visualized in the subpanel (I) of figure~\ref{and1} using the dilatation $\vartheta\equiv\bm{\nabla}\bm{\cdot}\bm{u}$, this observation is attributed to the high concentration of the two vorticity modes associated with strong reflected and transmitted shock waves in the region where $\Delta<0$. By comparing the left and right columns in figure~\ref{and1}, it is found that restricting to $\Delta>0$ attenuates shock-wave interference while retaining relatively pure vortex dynamics related to the interface evolution. The influence of shocks is also evident in the pronounced
resonance amplification observed during the evolution of the two vorticity modes, as seen from figures~\ref{Delta0} and~\ref{Delta1}. Notably, vorticity is primarily generated during the interaction between the incident shock and the interface and carried by the transmitted and reflected shock waves, so that the total vorticity flux $\Gamma_{\omega}$ remains approximately conserved during the system evolution, exhibiting a relatively smooth curve. In contrast, $\Gamma_{\omega}$ in the region $\Delta>0$ shows more obvious oscillations with low amplitude, partly due to the variation of the boundary $\Delta=0$ of the conventional vortical-flow domain and the relatively weak dilatation-vorticity coupling and baroclinic torque in the intermediate and late times (figure~\ref{Delta0}).

\begin{figure}[t]
	\centering
	% 第一行
	\subfigure[$\bm{R}(\bm{t})\cdot\bm{e}_z$ (point B) \label{5a}]{
		\includegraphics[width=0.315\linewidth, trim={0cm 0cm 0cm 0cm}, clip]{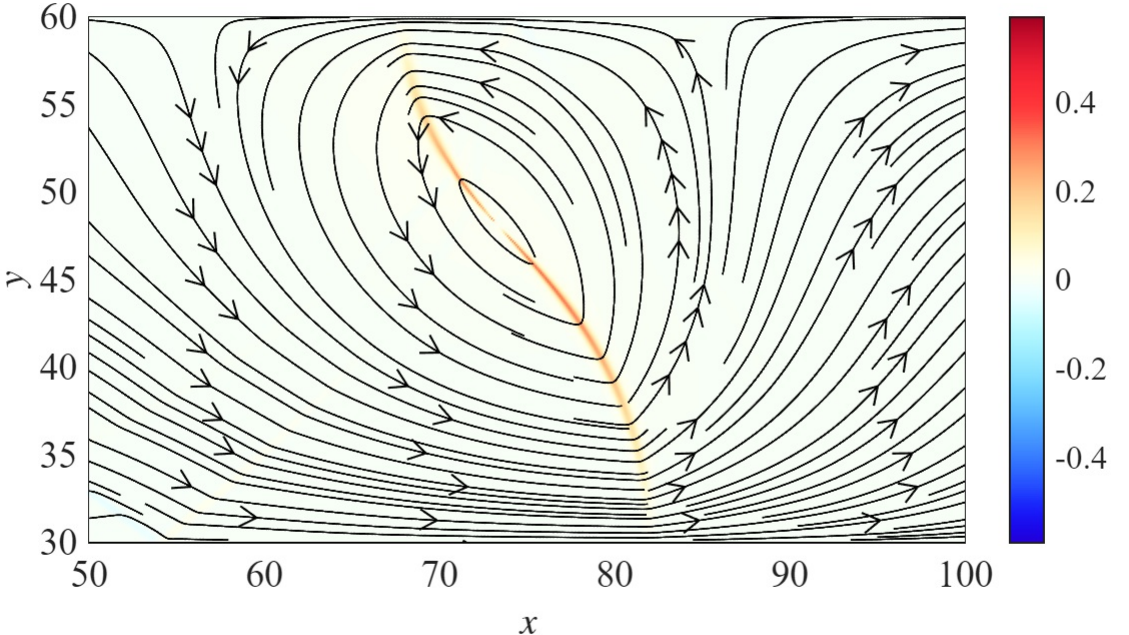}
	}
	\hfill
	\subfigure[$\bm{s}(\bm{t})\cdot\bm{e}_z$ (point B)\label{5b}]{
		\includegraphics[width=0.315\linewidth,trim={0cm 0cm 0cm 0cm}, clip]{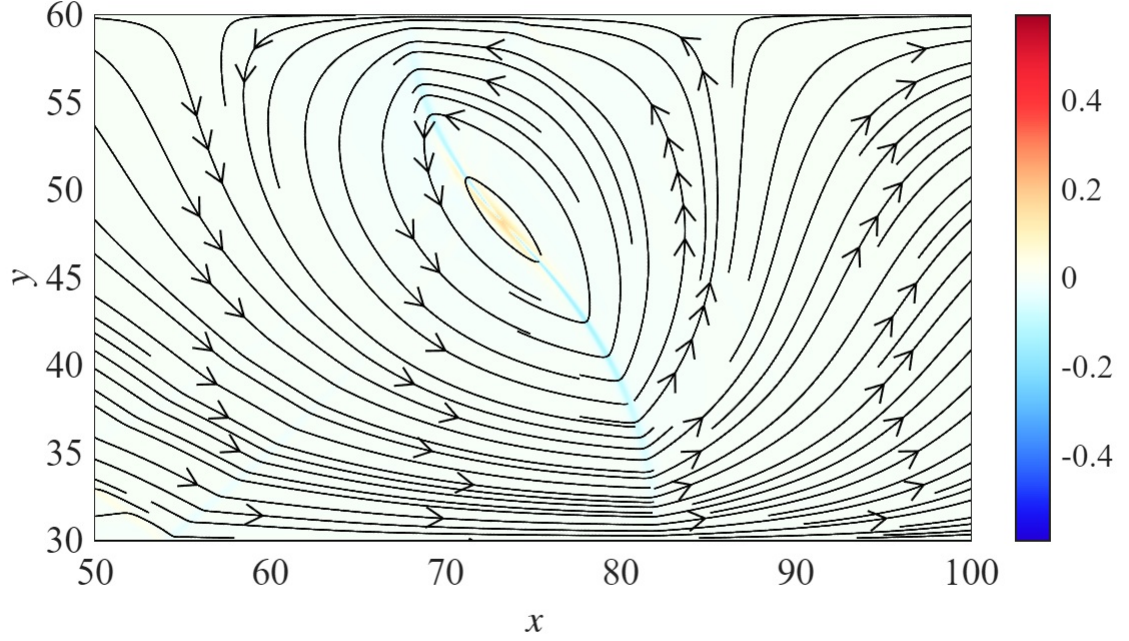}
	}
	\hfill
	\subfigure[${\omega}_z$ (point B) \label{5c}]{
		\includegraphics[width=0.315\linewidth, trim={0cm 0cm 0cm 0cm}, clip]{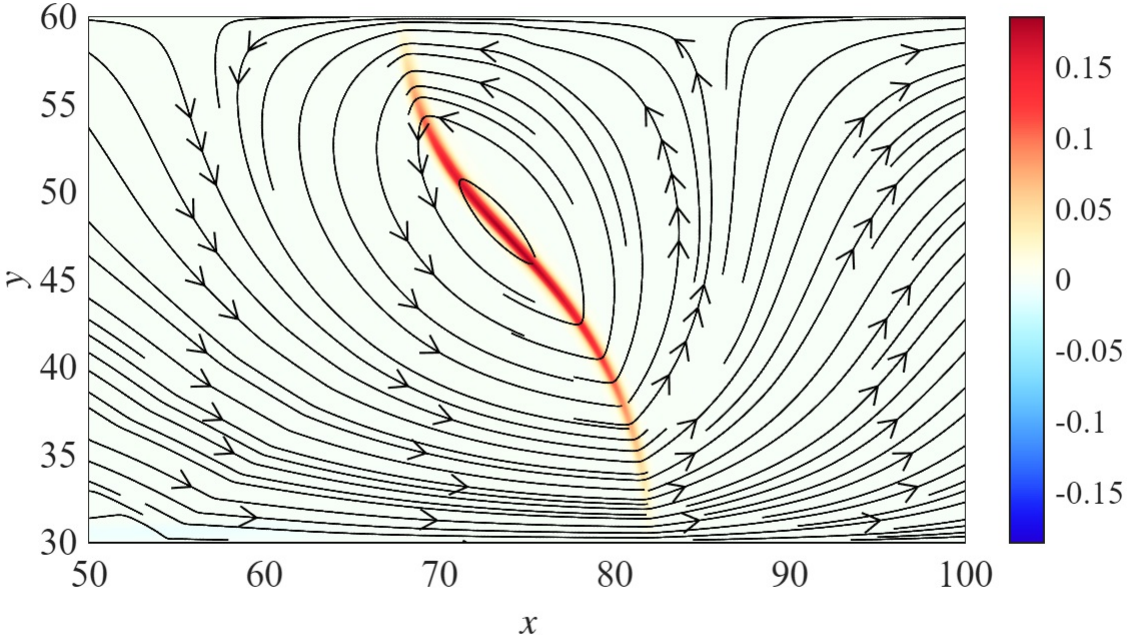}
	}
	
	%\vspace{1cm}
	% 第二行
	\subfigure[$\bm{R}(\bm{t})\cdot\bm{e}_z$ (point E) \label{5d}]{
		\includegraphics[width=0.315\linewidth, trim={0cm 0cm 0cm 0cm}, clip]{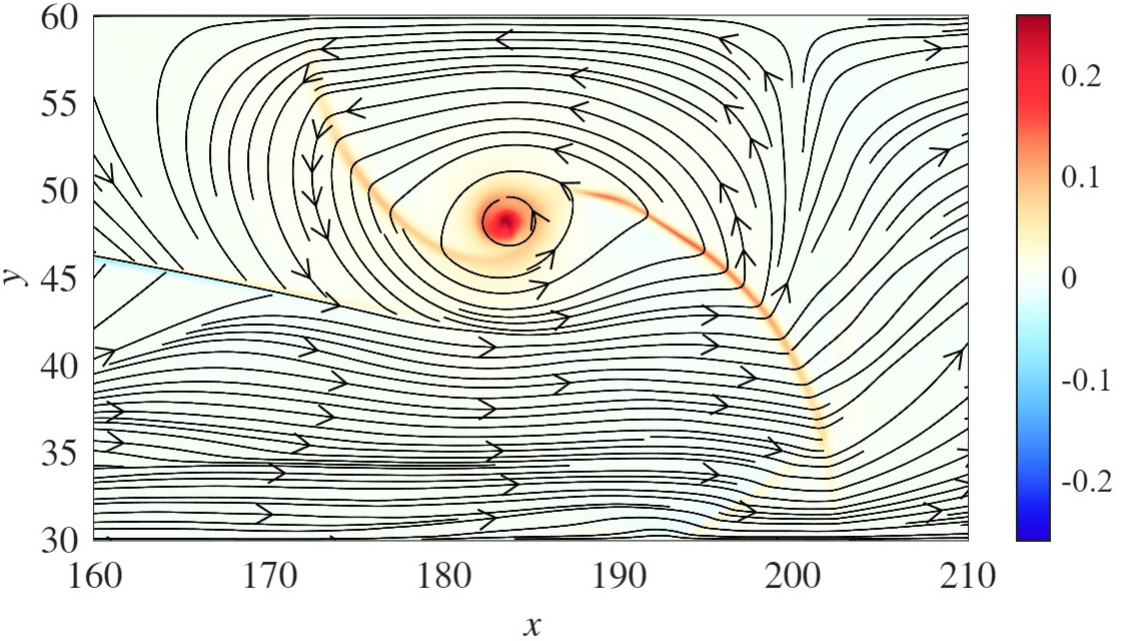}
	}
	\hfill
	\subfigure[$\bm{s}(\bm{t})\cdot\bm{e}_z$ (point E) \label{5e}]{
		\includegraphics[width=0.315\linewidth,trim={0cm 0cm 0cm 0cm}, clip]{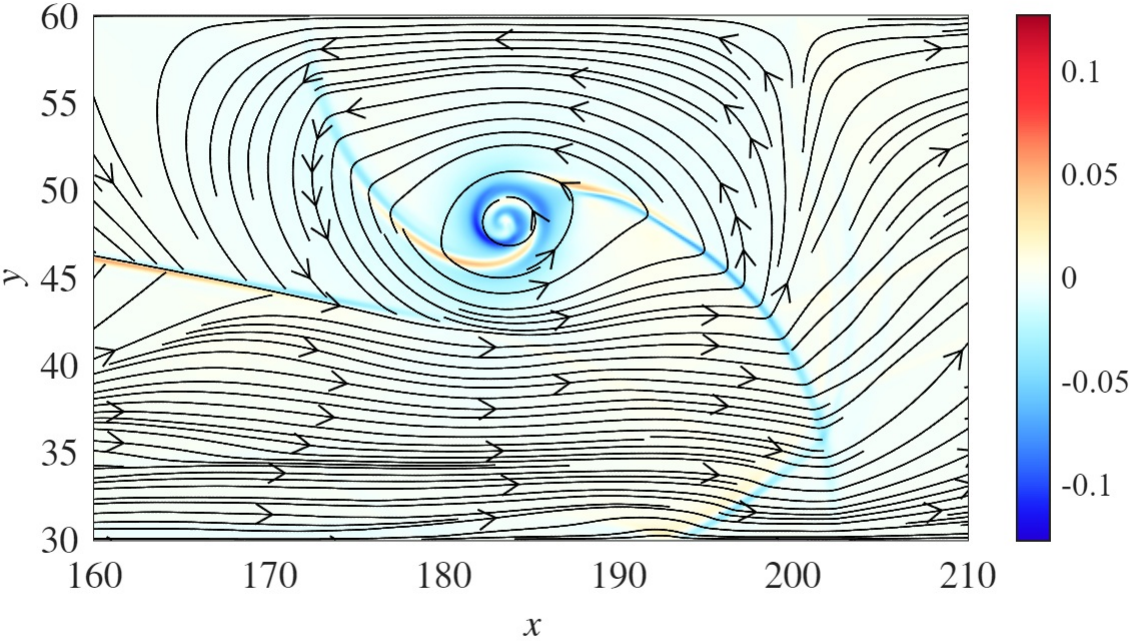}
	}
	\hfill
	\subfigure[${\omega}_z$ (point E) \label{5f}]{
		\includegraphics[width=0.315\linewidth, trim={0cm 0cm 0cm 0cm}, clip]{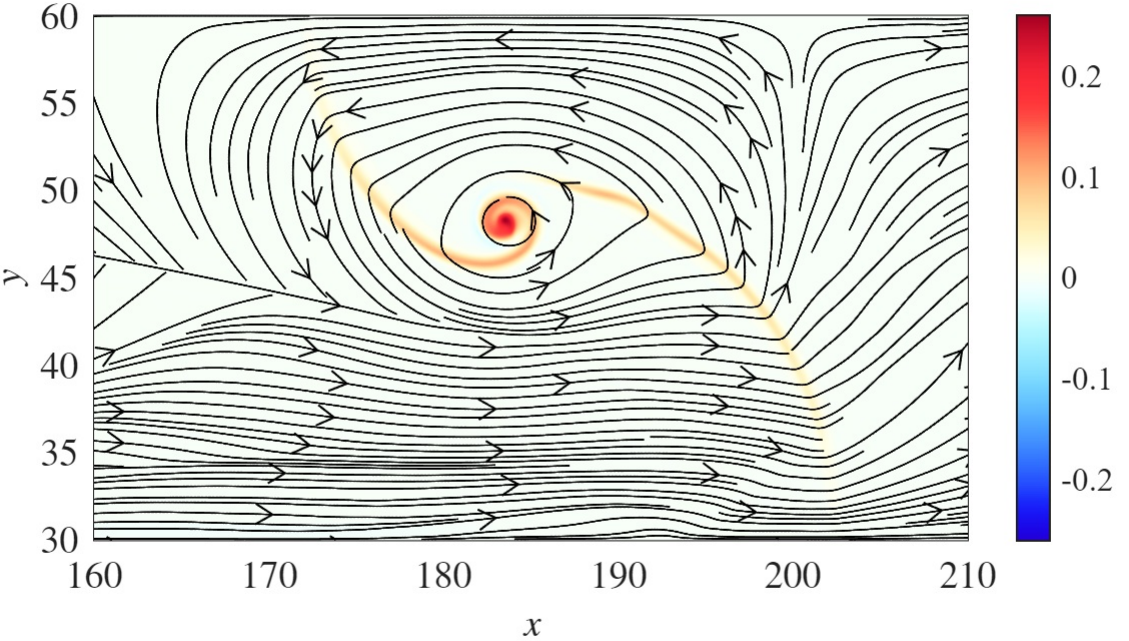}
	}
	
	% \vspace{1cm}
	% 第三行
	\subfigure[$\bm{R}(\bm{t})\cdot\bm{e}_z$ (point G) \label{5g}]{
		\includegraphics[width=0.315\linewidth, trim={0cm 0cm 0cm 0cm}, clip]{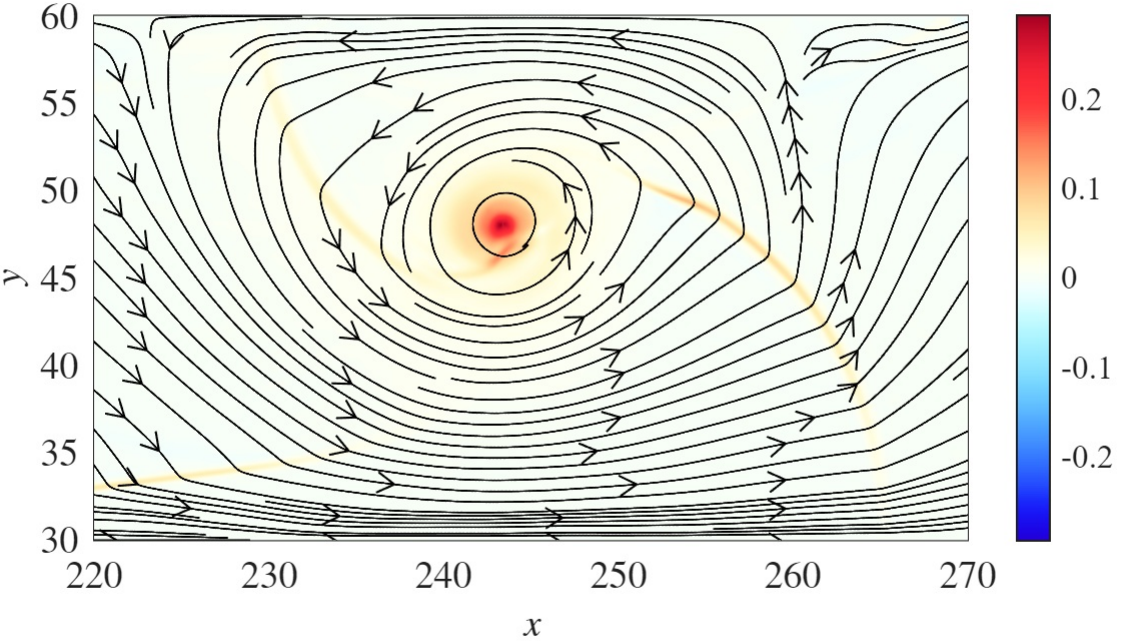}
	}
	\hfill
	\subfigure[$\bm{s}(\bm{t})\cdot\bm{e}_z$ (point G) \label{5h}]{
		\includegraphics[width=0.315\linewidth, trim={0cm 0cm 0cm 0cm}, clip]{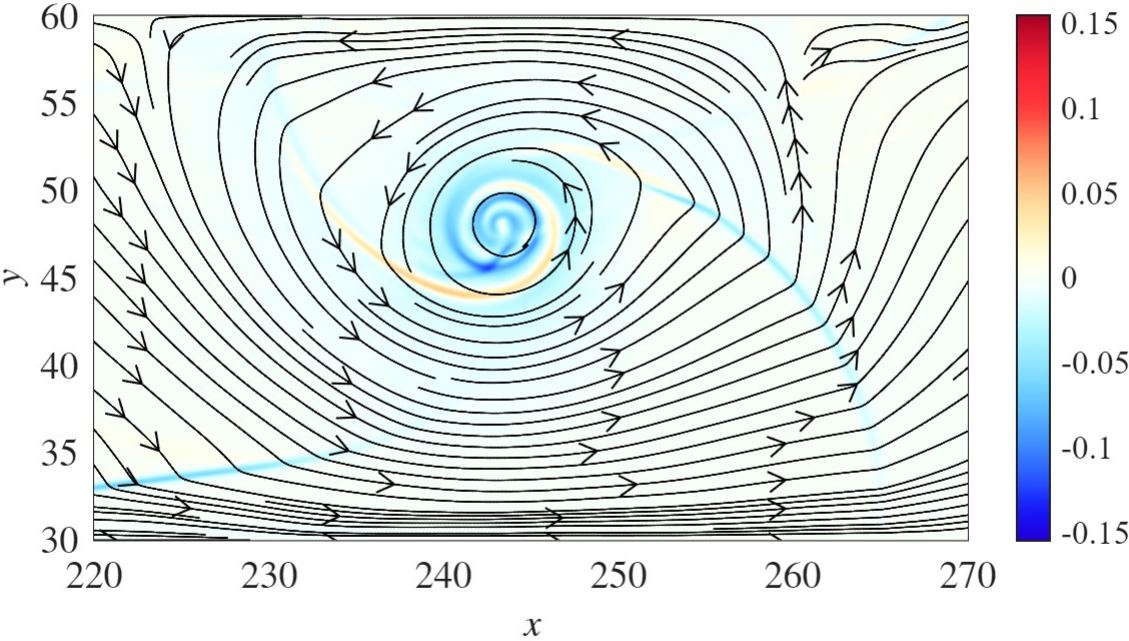}
	}
	\hfill
	\subfigure[${\omega}_z$ (point G) \label{5i}]{
		\includegraphics[width=0.315\linewidth,trim={0cm 0cm 0cm 0cm}, clip]{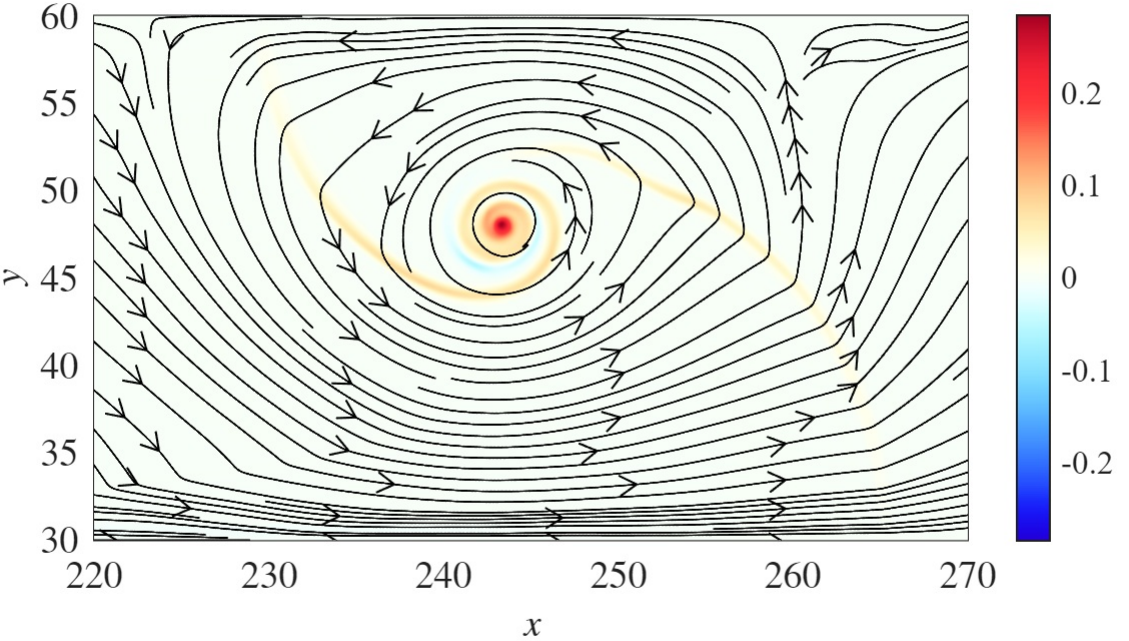}
	}
	
	\caption{Normalized snapshots of the $\bm{e}_z$-components during the interface evolution: (a,d,g) the orbital-rotation mode $\bm{R}(\bm{t})$, (b,e,h) the spin mode $\bm{s}(\bm{t})$, and (c,f,i) the vorticity $\bm{\omega}$. }
	\label{vvvv}
\end{figure}

\begin{figure}[t]
	\centering
	\subfigure[]{
		\begin{minipage}[t]{0.49\linewidth}
			\centering
			\includegraphics[width=1.0\columnwidth,trim={0cm 0.0cm 0.0cm 0.0cm},clip]{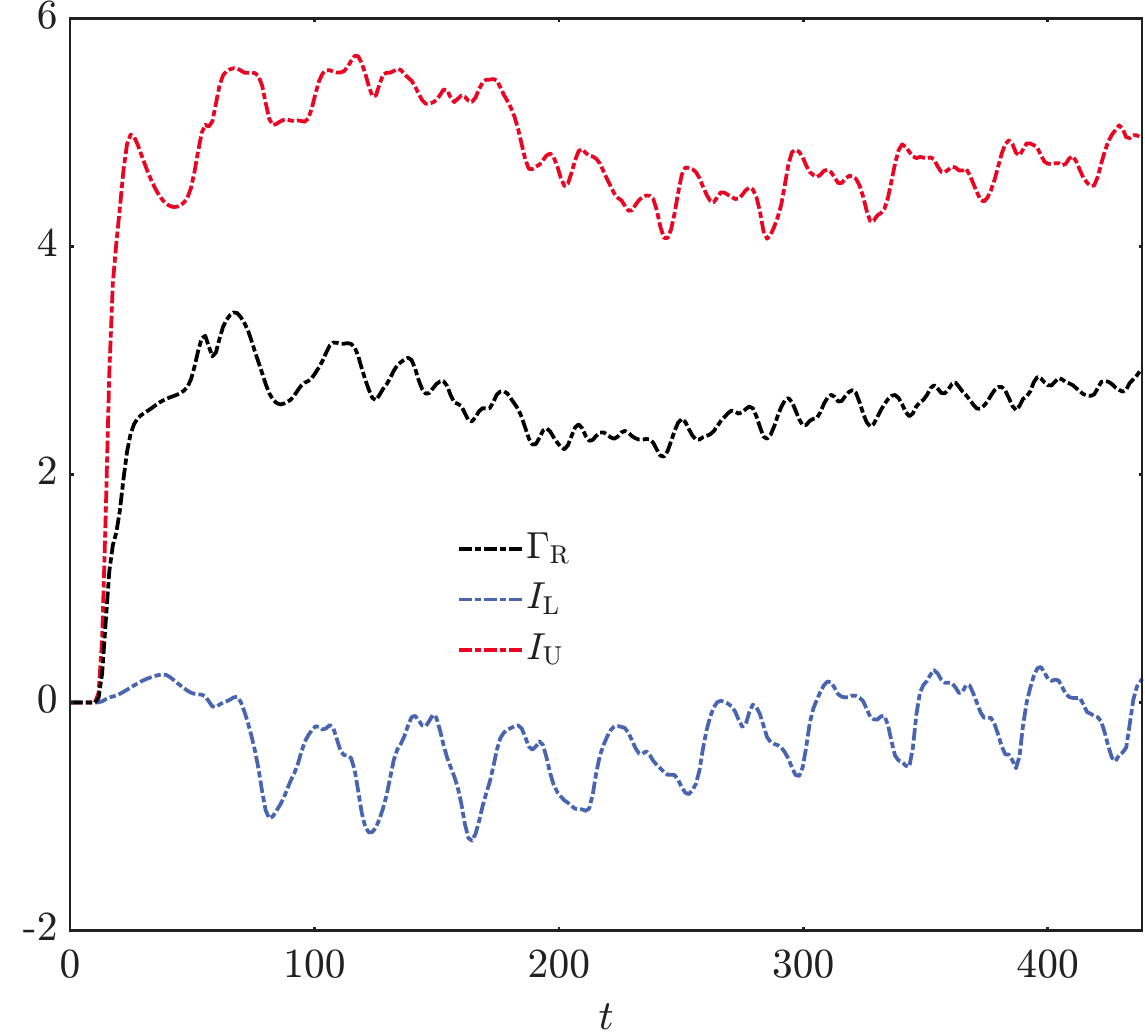}
			%	\vspace{-0.6cm} 
			\label{uu1}
		\end{minipage}%
	}
	\subfigure[]{
		\begin{minipage}[t]{0.49\linewidth}
			\centering
			\includegraphics[width=1.0\columnwidth,trim={0cm 0.0cm 0.0cm 0.0cm},clip]{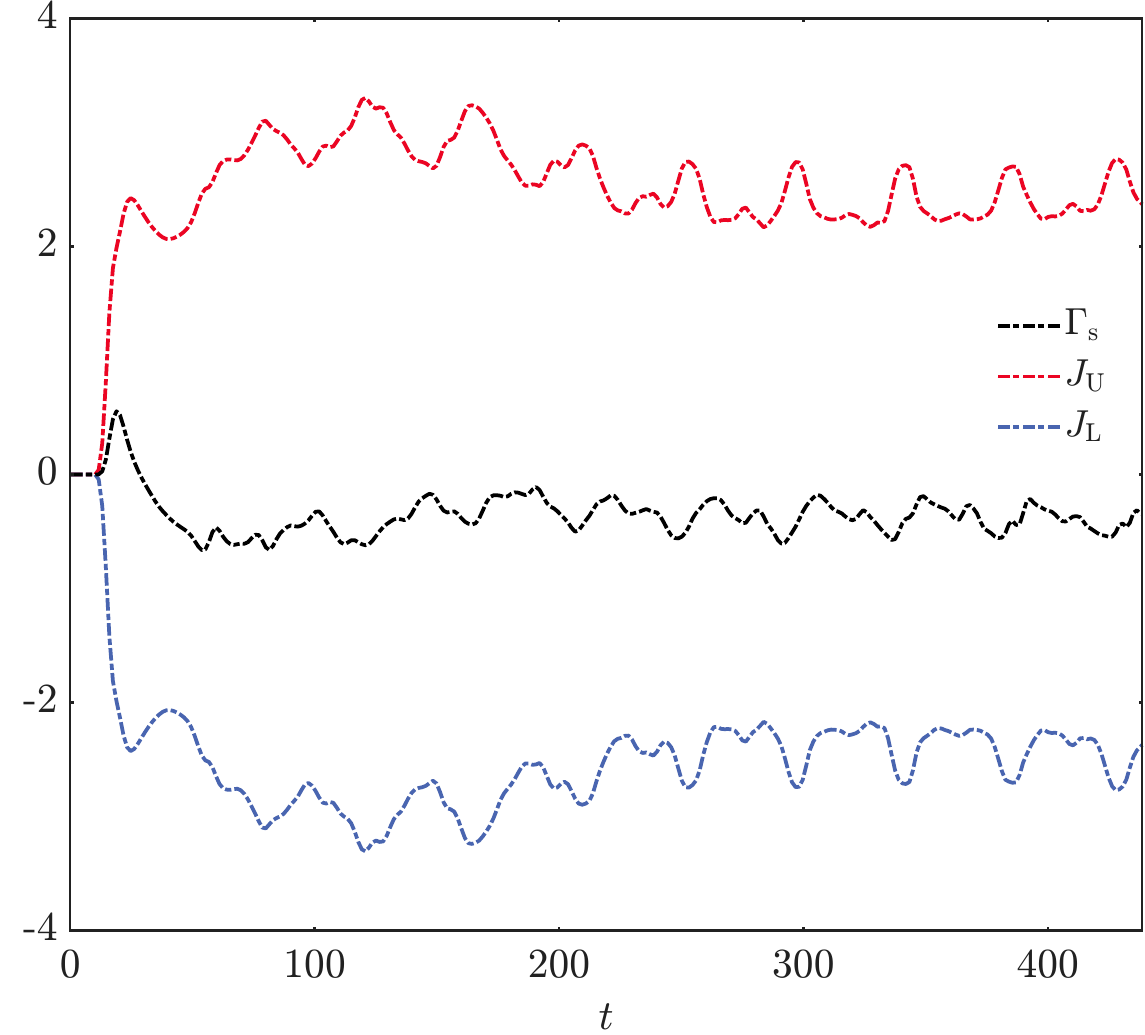}
			%	\vspace{-0.6cm} 
			\label{uu2}
		\end{minipage}%
	}	
	\subfigure[]{
		\begin{minipage}[t]{0.49\linewidth}
			\centering
			\includegraphics[width=1.0\columnwidth,trim={0cm 0.0cm 0.0cm 0.0cm},clip]{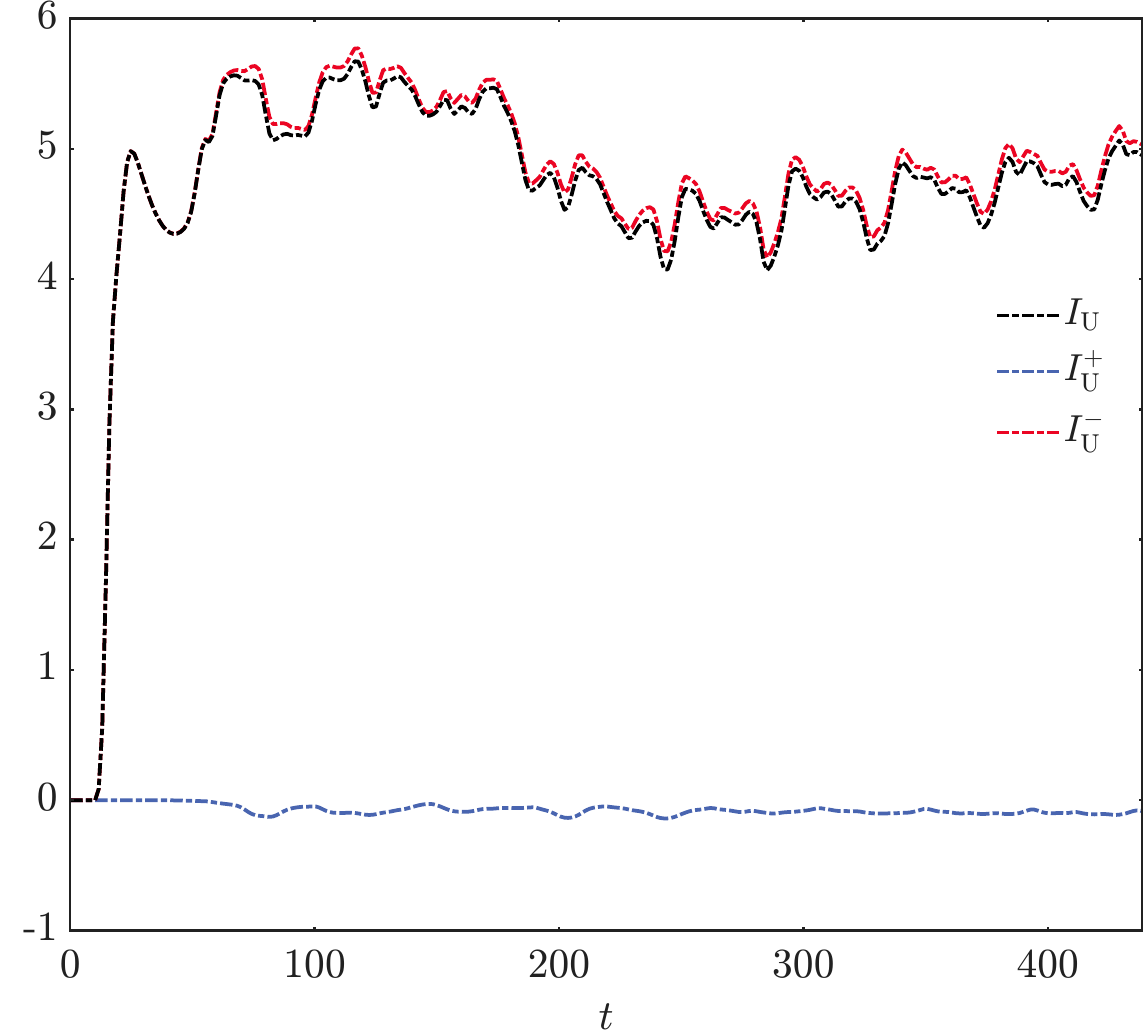}
			%	\vspace{-0.6cm} 
			\label{uu3}
		\end{minipage}%
	}
	\subfigure[]{
		\begin{minipage}[t]{0.49\linewidth}
			\centering
			\includegraphics[width=1.0\columnwidth,trim={0cm 0.0cm 0.0cm 0.0cm},clip]{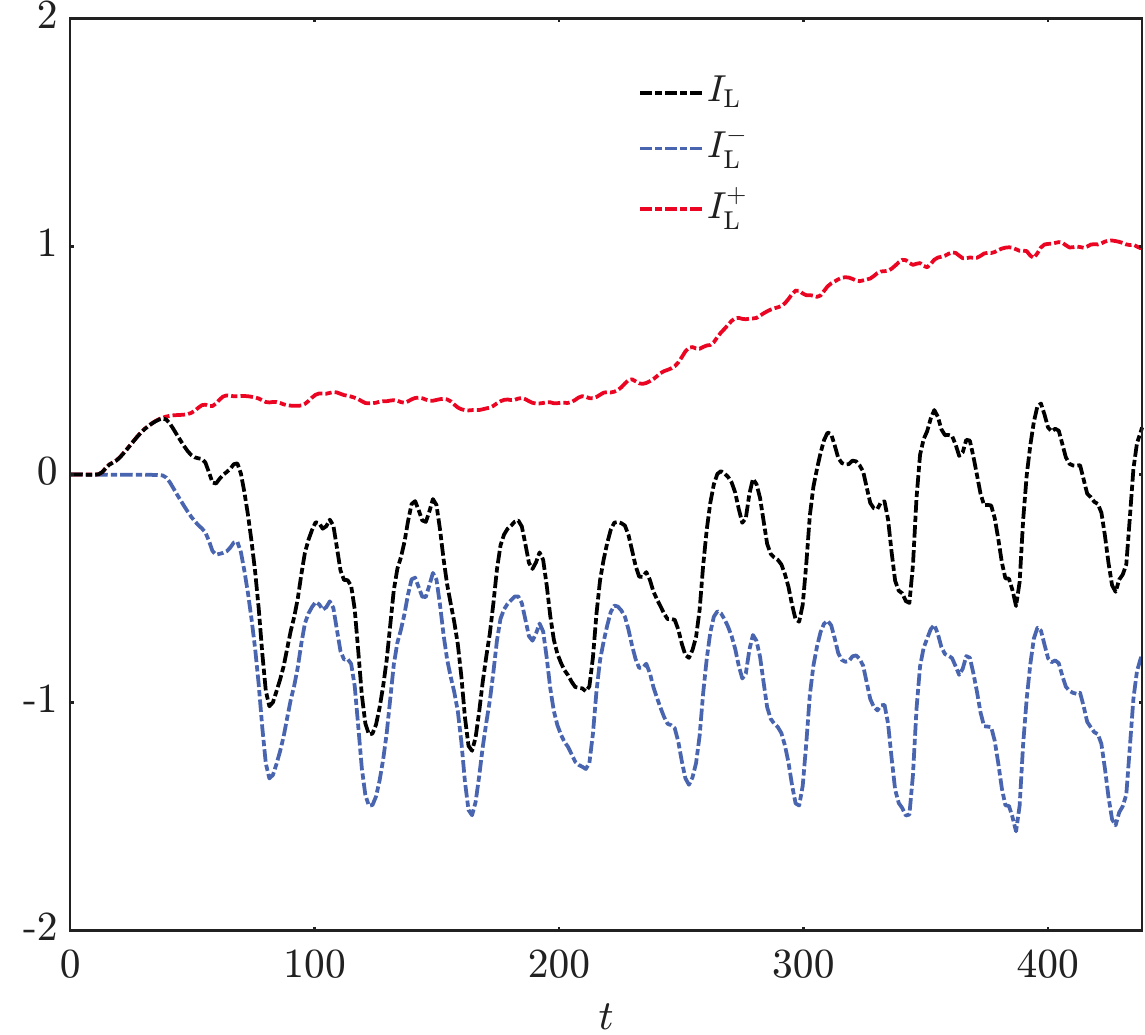}
			%	\vspace{-0.6cm} 
			\label{uu4}
		\end{minipage}%
	}	
	\caption{Evolution of the upper and lower bounds of the orbital-rotation flux $\Gamma_{\rm{R}}$ and the spin flux $\Gamma_{\rm{s}}$. (a) $(\Gamma_{\rm R},I_{\rm L},I_{\rm U})$, (b) $(\Gamma_{s},J_{\rm L},J_{\rm U})$, (c) $(I_{\rm U},I_{\rm U}^{+},I_{\rm U}^{-})$, and (d) $(I_{\rm L},I_{\rm L}^{+},I_{\rm L}^{-})$.} 
	\label{fug121}
\end{figure}

The spatial characteristics of the vorticity modes at points B and C (figures~\ref{and2} and~\ref{and3}) resemble those at point A (figure~\ref{and1}), with sheet-like vortices showing distinct rigid-rotation motion and higher interface curvature.
As the secondary instability (KHI) gradually develops (points D, E, F, and G in figures~\ref{Delta0}), the mushroom structures are observed owing to the rolling-up of the perturbed interface, where the vorticity modes evolve into anti-KKB configuration with opposite signs of $\bm{R}(\bm{t})$ and $\bm{s}(\bm{t})$ inside the vortex cores (figures~\ref{and4},~\ref{and5}, and~\ref{and6}). Correspondingly, $\Gamma_{\rm R}$ and $\Gamma_{\rm{s}}$ remain on a plateau in the intervals $(2,3)$ and $(-0.5,0)$, respectively, which together yield a stable level of $\Gamma_{\omega}$ that is slightly lower than $\Gamma_{\rm R}$ (figure~\ref{Delta0}). Interestingly, as the interface evolves, the vorticity modes are influenced by the reflected shock waves and exhibit approximately an anti-KKB pattern in the region $\Delta<0$ dominated by the potential flow (figures~\ref{and4},~\ref{and5}, and~\ref{and6}). Being distinguished from the single-vortex winding system, the extending spiral arms are trapped by adjacent primary vortices, which display anti-KKB patterns in the tip regions of the spike and bubble structures (figures~\ref{and} and~\ref{vvvv}). Figure~\ref{fug121} shows the evolution histories of the orbital-rotation flux $\Gamma_{\rm{R}}$ and the spin flux $\Gamma_{\rm{s}}$, along with their upper and lower bounds.  Figures~\ref{uu1} and~\ref{uu2} validate the inequalities proposed in~\eqref{ttt1} and~\eqref{ttt2}. It is observed that the evolution trend of $\Gamma_{\rm{R}}$ is very similar to the upper bound $I_{\rm U}$, although with different magnitudes. $I_{\rm U}^{-}$ dominates $I_{\rm U}$ while the contribution from $I_{\rm U}^{+}$ is negligible (figure~\ref{uu3}). In contrast, $I_{\rm L}$ is dominated by $I_{\rm L}^{-}$ before $t=300$, and modulated by $I_{\rm L}^{\pm}$ in the later evolution (figure~\ref{uu4}).

\subsection{Decomposition of $Q$-criterion based on orbital-spin decomposition}
For compressible flow, the second principal invariant of the VGT $\mathbf{A}\equiv\bm{\nabla  u}$ is defined as~\citep{hunt1988eddies,Jeong1995}
\begin{eqnarray}
	Q\equiv\frac{1}{2}\left[{\rm tr}(\mathbf{A})^{2}-{\rm tr}(\mathbf{A}^{2})\right]=\frac{1}{2}\left[\vartheta^2-{\rm tr}(\bm{\Omega}^2)-{\rm tr}(\mathbf{D}^2)\right]=\frac{1}{2}\left[\vartheta^2+\frac{1}{2}\omega^2-\mathbf{D}\bm{:}\mathbf{D}\right],
\end{eqnarray}
where $-{\rm tr}(\bm{\Omega}^2)=\bm{\Omega}\bm{:}\bm{\Omega}=\omega^2/2$ is the enstrophy, and ${\rm tr}(\mathbf{D}^2)=\mathbf{D}\bm{:}\mathbf{D}$ is the squared strain rate. Therefore, $Q$ represents the local balance among the squared dilatation, enstrophy, and the squared strain rate. For incompressible flow $(\vartheta=0)$, the pressure can be written as a weighted integral of $Q$ over the entire space via the pressure Poisson equation $\nabla^{2}p=2\rho Q$. Although there is no explicit connection between $Q>0$ and local pressure minimum, the isosurface of $Q$ typically identifies a low-pressure region associated with a strong vortex core.

\begin{figure}[t]
	\centering
	\subfigure[$Q$]{
		\begin{minipage}[t]{0.49\linewidth}
			\centering
			\includegraphics[width=1.0\columnwidth,trim={0cm 0cm 0cm 0cm},clip]{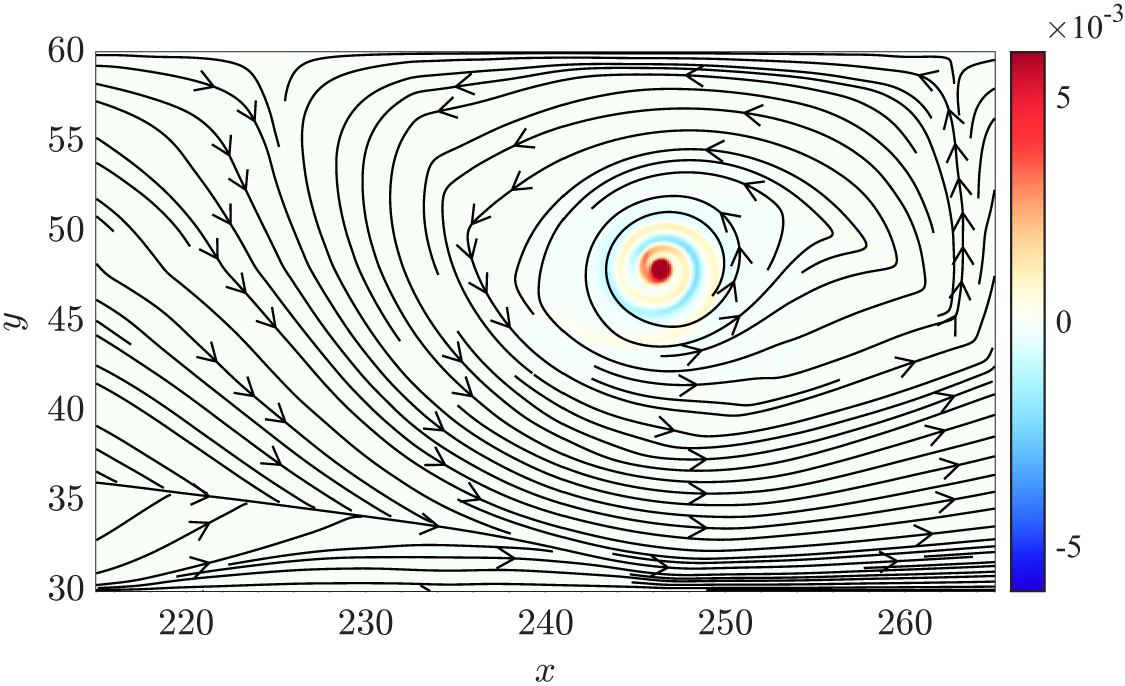}
			% \vspace{-19.6cm} 
			\label{curvaturt}
		\end{minipage}%
	}
	\subfigure[$\frac{1}{4}{R}^2(\bm{t})$]{
		\begin{minipage}[t]{0.49\linewidth}
			\centering
			\includegraphics[width=1.0\columnwidth,trim={0cm 0cm 0cm 0cm},clip]{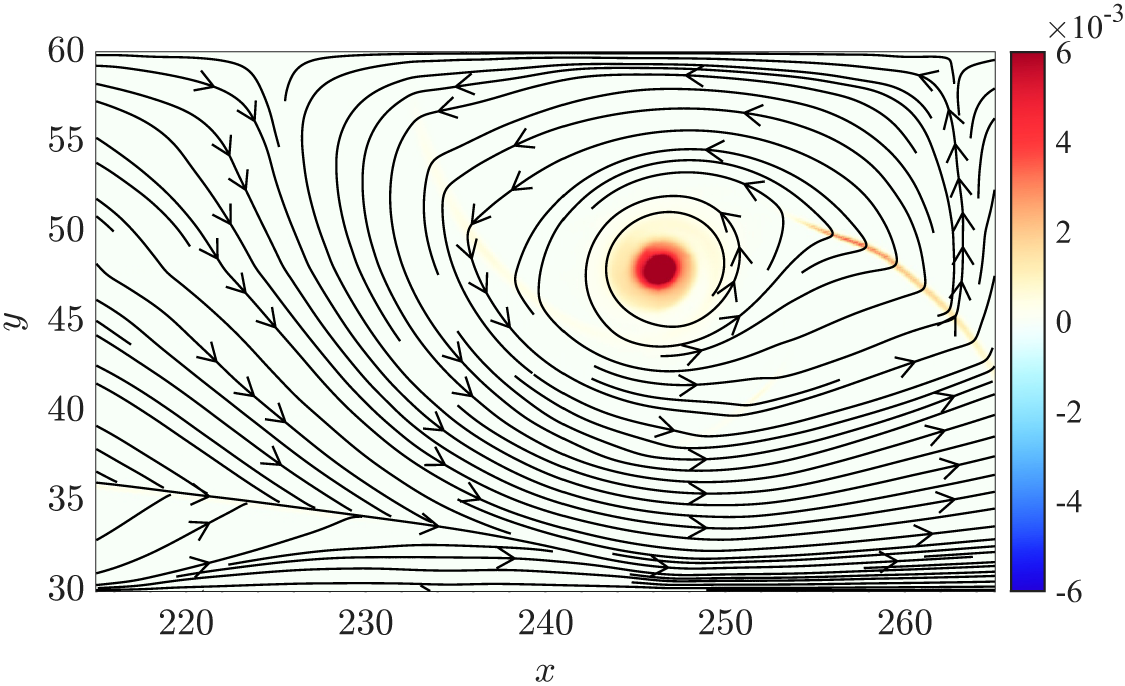}
			%\vspace{-19.6cm} 
			\label{curvbda_zz}
		\end{minipage}%
	}	
	\subfigure[$\frac{1}{2}\bm{R}(\bm{t})\bm{\cdot}\bm{s}(\bm{t})$]{
		\begin{minipage}[t]{0.49\linewidth}
			\centering
			\includegraphics[width=1.0\columnwidth,trim={0cm 0cm 0cm 0cm},clip]{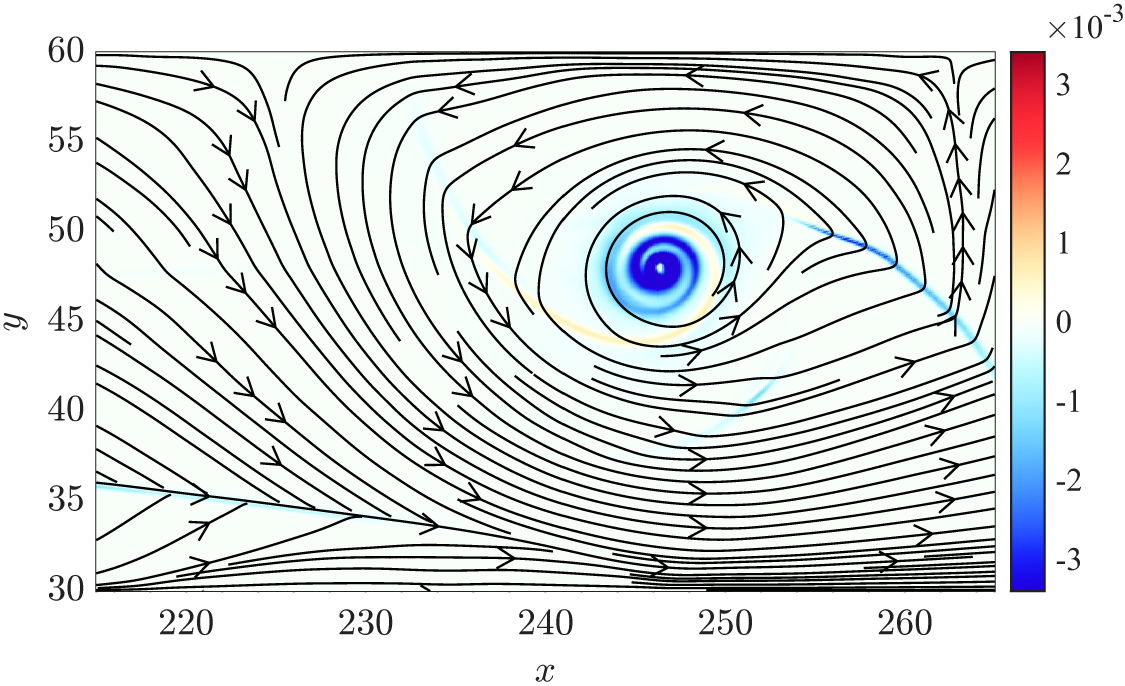}
			%   \vspace{-6.6cm} 
			\label{curvzz}
		\end{minipage}%
	}	
	\subfigure[${\chi}(\bm{t}){\chi}(\bm{n})$]{
		\begin{minipage}[t]{0.49\linewidth}
			\centering
			\includegraphics[width=1.0\columnwidth,trim={0cm 0cm 0cm 0cm},clip]{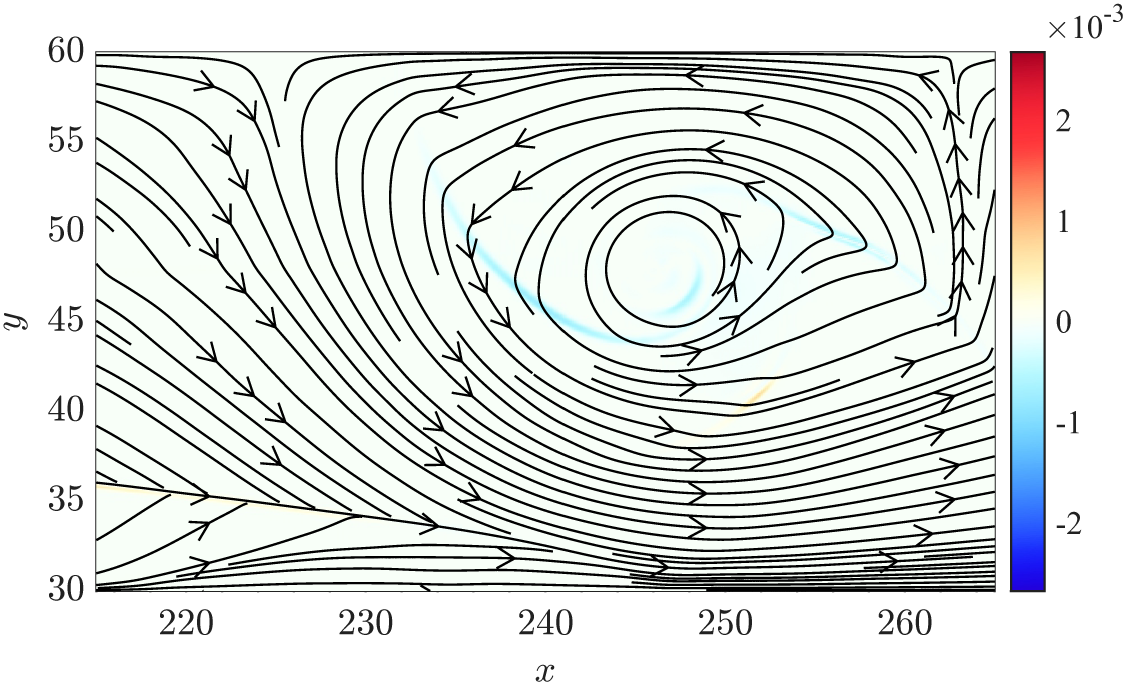}
			%   \vspace{-6.6cm} 
			\label{cz}
		\end{minipage}%
	}	
	\caption{Normalized snapshots of (a) $Q$, (b) $\frac{1}{4}{R}^2(\bm{t})$, (c) $\frac{1}{2}\bm{R}(\bm{t})\bm{\cdot}\bm{s}(\bm{t})$, and (d) ${\chi}(\bm{t}){\chi}(\bm{n})$ for a shock-accelerated single-mode perturbed interface at $t^{*}=442.31$.} 
	\label{Q_decomposition}
\end{figure}

Under the Frenet-Serret orthonormal frame $(\bm{t},\bm{n},\bm{b})$, the restricted VGT on $\mathbb{R}^{2}$ can be expressed in its irreducible real Schur form~\citep{chen2025kinematic}:
\begin{equation}\label{AFS}
	\mathbf{A}=
	\begin{bmatrix}
		\chi(\bm{t}) & W_{L}(\bm{t}) \\
		-W_{L}(\bm{n}) & \chi(\bm{n})
	\end{bmatrix},
\end{equation}
where $\chi(\bm{t})$ and $\chi(\bm{n})$ denote the relative stretching rates of the material line elements along $\bm{t}$ and $\bm{n}$, respectively. The dilatation is given by $\vartheta=\chi(\bm{t})+\chi(\bm{n})$. Here, the $\bm{b}$-components of the total angular velocities of the material line elements along $\bm{t}$ and $\bm{n}$ are $W_{L}(\bm{t})=R(\bm{t})/2$ and $W_{L}(\bm{n})=R(\bm{t})/2+s(\bm{t})$, respectively. Using~\eqref{AFS}, the strain-rate and rotation-rate tensors are derived as
\begin{equation}\label{DFS}
	\mathbf{D}=
	\begin{bmatrix}
		\chi(\bm{t}) & -\frac{1}{2}s(\bm{t}) \\
		-\frac{1}{2}s(\bm{t}) & \chi(\bm{n})
	\end{bmatrix},~~
	\bm{\Omega}=
	\begin{bmatrix}
		0 & \frac{1}{2}(R(\bm{t})+s(\bm{t})) \\
		-\frac{1}{2}(R(\bm{t})+s(\bm{t})) & 0
	\end{bmatrix}.
\end{equation}
Consequently, we obtain
\begin{subequations}\label{FS37}
\begin{eqnarray}\label{FS37a}
	\vartheta^2=\chi^2(\bm{t})+\chi^2(\bm{n})+2\chi(\bm{t})\chi(\bm{n}),
\end{eqnarray}
\begin{eqnarray}\label{FS37b}
	\frac{1}{2}\omega^2=\frac{1}{2}R^2(\bm{t})+\frac{1}{2}s^{2}(\bm{t})+\bm{R}(\bm{t})\bm{\cdot}\bm{s}(\bm{t}),
\end{eqnarray}
\begin{eqnarray}\label{FS37c}
	\mathbf{D}\bm{:}\mathbf{D}=\chi^2(\bm{t})+\chi^2(\bm{n})+\frac{1}{2}s^2(\bm{t}).
\end{eqnarray}
\end{subequations}
Substituting~\eqref{FS37} into~\eqref{AFS} yields the decomposition of $Q$:
\begin{eqnarray}\label{Q38}
	Q=\frac{1}{4}R^2(\bm{t})+\frac{1}{2}\bm{R}(\bm{t})\bm{\cdot}\bm{s}(\bm{t})+\chi(\bm{t})\chi(\bm{n}),
\end{eqnarray}
where on the right-hand side of~\eqref{Q38}, the first term represents the enstrophy contribution solely from $\bm{R}(\bm{t})$; the second term represents the cross-coupling between $\bm{R}(\bm{t})$ and $\bm{s}(\bm{t})$; and the third term arises from the product of the relative stretching rates, i.e., the diagonal elements of $\mathbf{D}$. Note that the terms, $\chi^2(\bm{t})+\chi^2(\bm{n})$ and $s^2(\bm{t})/2$ in~\eqref{FS37c}, are  canceled by their counterparts in $\vartheta^2$~\eqref{FS37a} and $\omega^2/2$~\eqref{FS37b}, respectively.
Figure~\ref{Q_decomposition} visualizes the invariant $Q$ and different contributions in~\eqref{Q38}. It is observed that the positive term $R^2(\bm{t})/4$ can well capture the vortex core region. However, the size of the identified vortex core is significantly reduced due to the influence of the negative coupling term $\bm{R}(\bm{t})\bm{\cdot}\bm{s}(\bm{t})/2$, which originates from $\omega^2/2$ in~\eqref{FS37b}. This reduction is essentially caused by the dominance of the anti-KKB configuration (i.e., opposite signs of $\bm{R}(\bm{t})$ and $\bm{s}(\bm{t})$) within the vortex core region. By contrast, the stretching effect $\chi(\bm{t})\chi(\bm{n})$, which comes from $\vartheta^2$ in~\eqref{FS37a}, is largely confined to the peripheral region near the vortex core, whereas its influence on the inner region of the vortex is negligible.
\begin{figure}[t]
	\centering
	\includegraphics[width=0.9\columnwidth,trim={0cm 0.0cm 0cm 0cm},clip]{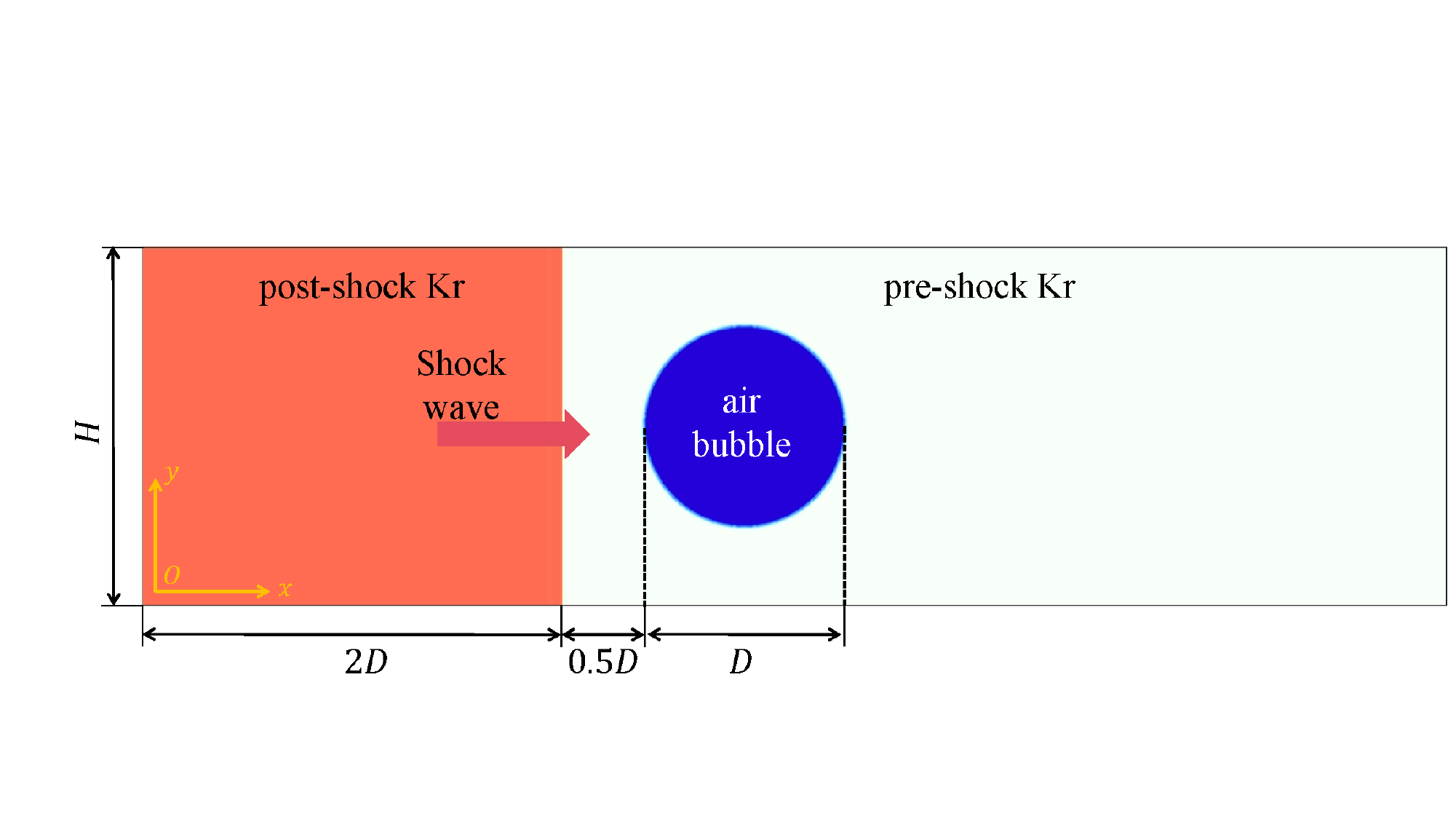}
	\caption{Schematic of the initial 2D configuration for a single cylindrical air bubble immersed in the pre-shock Krypton (Kr). The incident shock wave propagates from left to right, compressing the post-shock Kr. $D$ is the diameter of the bubble, and $H$ is the height of the computational domain.} 
	\label{bubble}
\end{figure}
\begin{figure}[h!]
	\centering
	\includegraphics[width=0.8\columnwidth,trim={4.5cm 0cm 4.0cm 1.2cm},clip]{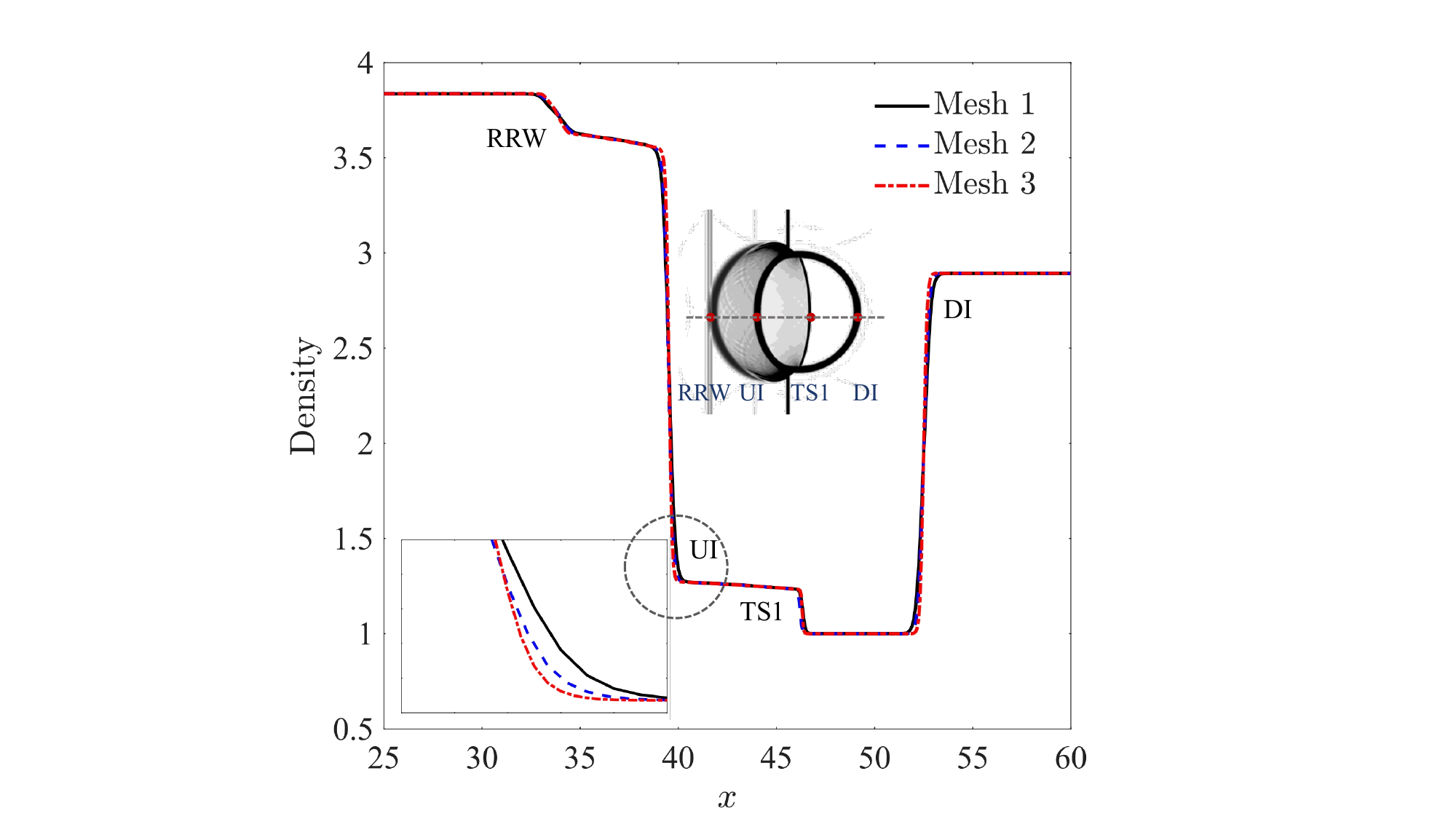}
	\caption{Grid refinement study: normalized density profiles along the centerline ($y^{*} = 13.35$) of the cylindrical air bubble at $t^{*} = 14.5$ for three different meshes. RRW, UI, DI, and TS1 represent the reflected rarefaction wave, upstream interface, downstream interface, and transmitted shock wave, respectively, with qualitative agreement with~\citet{singh2023shock}.} 
	\label{wangge}
\end{figure}
\begin{figure}[h!]
	\centering
	% 第一行
	\subfigure[Mesh-1 ($975 \times 267$) \label{12a}]{
		\includegraphics[width=0.315\linewidth, trim={6.5cm 0cm 6.5cm 0cm}, clip]{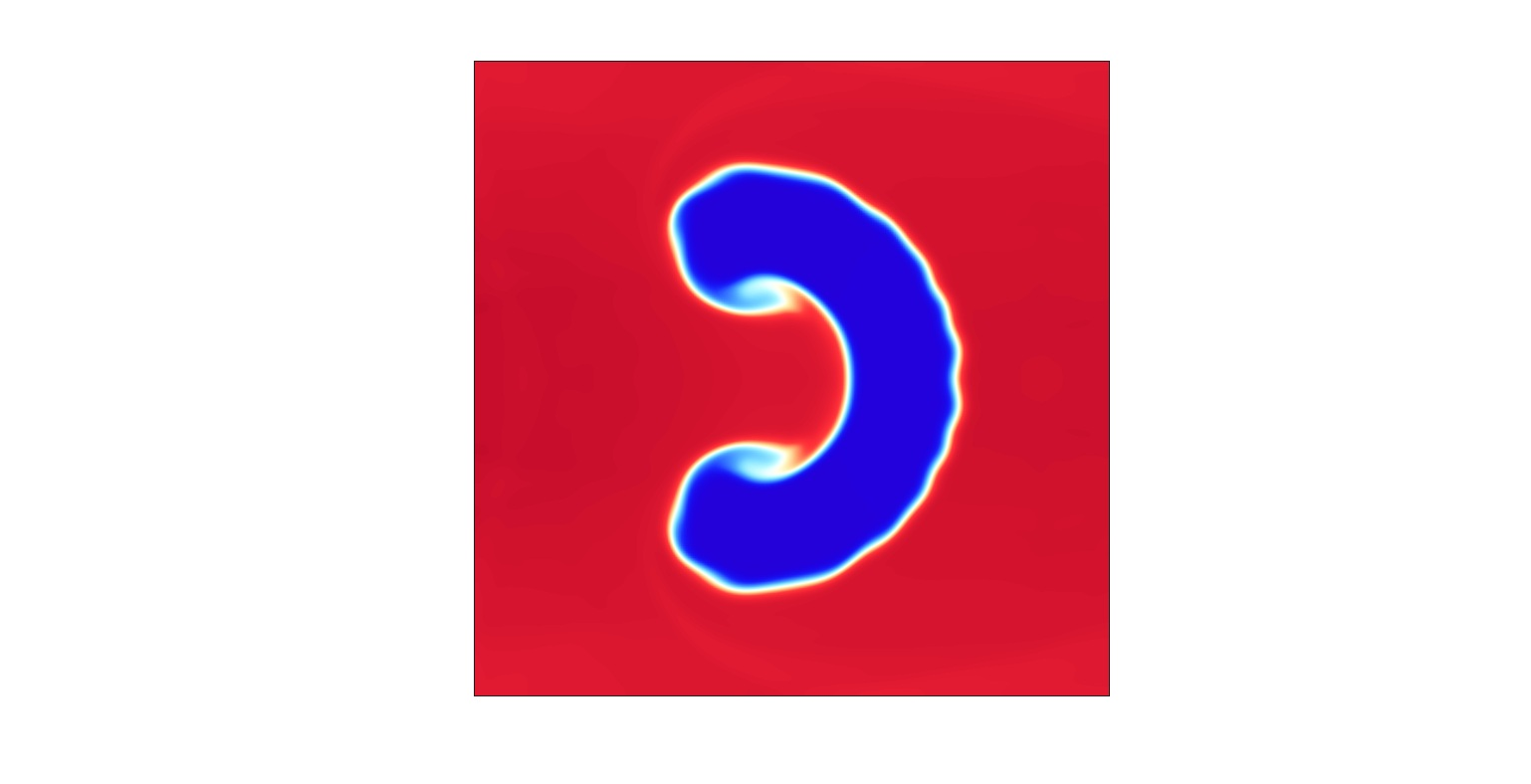}
	}%left down  right up
	\hfill
	\subfigure[Mesh-2 ($1300 \times 356$) \label{12b}]{
		\includegraphics[width=0.315\linewidth, trim={6.5cm 0cm 6.5cm 0cm}, clip]{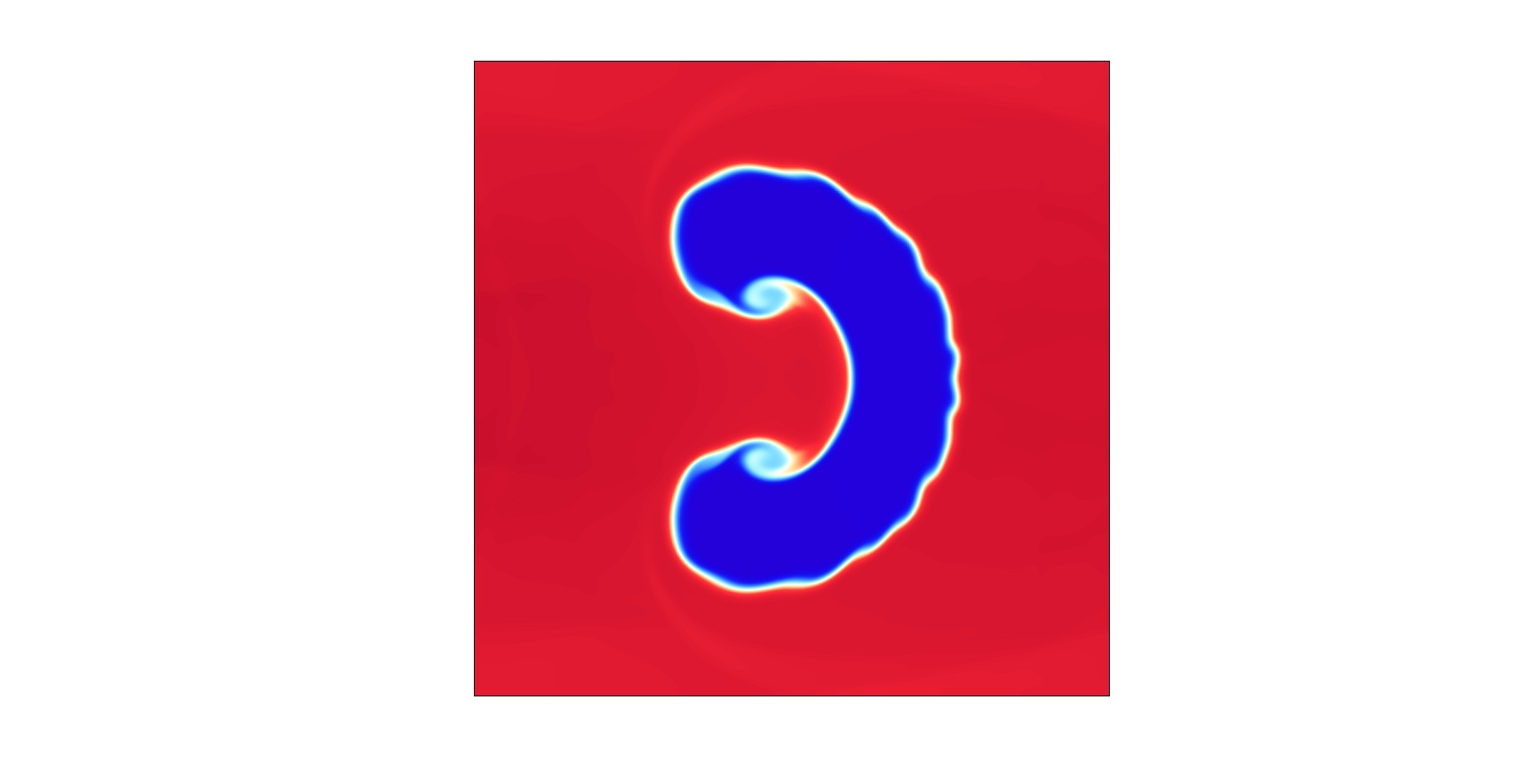}
	}
	\hfill
	\subfigure[Mesh-3 ($1950 \times 534$) \label{12c}]{
		\includegraphics[width=0.315\linewidth, trim={6.5cm 0cm 6.5cm 0cm}, clip]{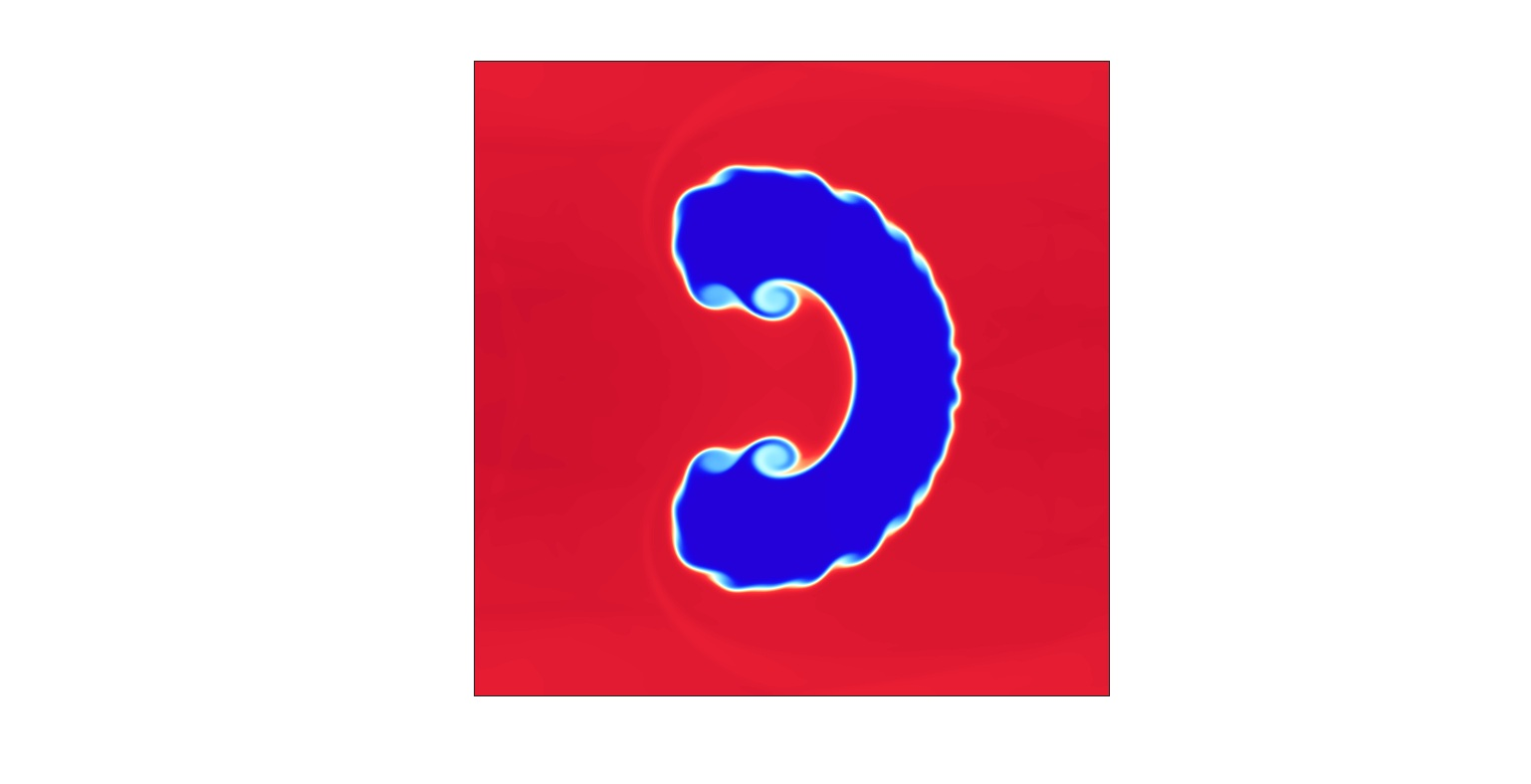}
	}
	\caption{Grid refinement study for a single shock-accelerated light cylindrical air bubble surrounded by Krypton at $t^{*}=87.2$. Density contour maps are shown for three grid resolutions: (a) Mesh-1 ($975 \times 267$), (b) Mesh-2 ($1300 \times 356$), and (c) Mesh-3 ($1950 \times 534$).}
	\label{300o}
\end{figure}

\section{RMI in shock-accelerated cylindrical air bubble in Krypton}\label{RMI in shock-accelerated cylindrical air bubble in Krypton}
\subsection{Initialization and grid refinement study}
The interaction between a planar shock wave and a 2D stationary bubble has been extensively studied in the existing literature~\citep{shankar2010numerical,singh2023shock}. As shown in figure~\ref{bubble}, the computation domain spans $[x/D,y/D]=[0,6.5]\times[0,1.78]$, where $D=300\Delta{x}=15{\rm~mm}$ is the diameter of the initial air bubble immersed in pre-shock Krypton (Kr). The right edge of the air cylinder corresponds to $x/D=3.5$, with its center located at $(x/D,y/D)=(3,0.89)$. The shape factor of the bubble is given by $D/H=0.56$, with the height $H=1.78D$. At the beginning, an incident shock with a prescribed pre-shock Mach number $Ma_{s}=1.22$ is imposed at $x/D=2$ and propagates from left to right. The specific heat ratios of air and Kr are $\gamma_{\rm air}=1.40$ and $\gamma_{\rm Kr}=1.667$, respectively. Using subscripts $(-)$ and $(+)$ to denote the pre-shock and post-shock quantities, respectively, the post-shock variables in Kr are determined via the one-dimensional (1D) Rankine-Hugoniot (R-H) jump relations across a gas discontinuity:
\begin{eqnarray*}
	Ma_{+}^{2}=\frac{1+\frac{1}{2}(\gamma_{\rm Kr}-1)Ma_{s}^{2}}{\gamma_{\rm Kr} Ma_{s}^{2}-\frac{1}{2}(\gamma_{\rm Kr}-1)},
\end{eqnarray*}
\begin{eqnarray}
	\frac{p_{+}}{p_{-}}=\frac{1+\gamma_{\rm Kr} Ma_{s}^{2}}{1+\gamma_{\rm Kr} Ma_{+}^{2}},~~\frac{\rho_{+}}{\rho_{-}}=\frac{(\gamma_{\rm Kr}+1)p_{+}/p_{-}+(\gamma_{\rm Kr}-1)}{(\gamma_{\rm Kr}-1)p_{+}/p_{-}+(\gamma_{\rm Kr}+1)},
\end{eqnarray}
where $Ma_{+}$ is the post-shock Mach number, from which the dimensionless initial quantities in the laboratory reference frame are evaluated as
\begin{eqnarray}
	(\rho, u_{x}, u_{y}, p) = 
	\begin{cases} 
		(2.8926,0,0,1), & \text{for pre-shock Kr,} \\
		(3.8368,0.2279,0,1.6106), & \text{for post-shock Kr,} \\
		(1,0,0,1), & \text{for air cylinder.}
	\end{cases}
\end{eqnarray}
The Atwood number, based on the pre-shock density values, is $At=0.4862$.
\begin{figure}[t]
	\centering
	\subfigure[$t^{*}=13.09$]{
		\begin{minipage}[t]{0.45\linewidth}
			\centering
			\includegraphics[width=1.0\columnwidth,trim={5cm 0.1cm 6cm 1.5cm},clip]{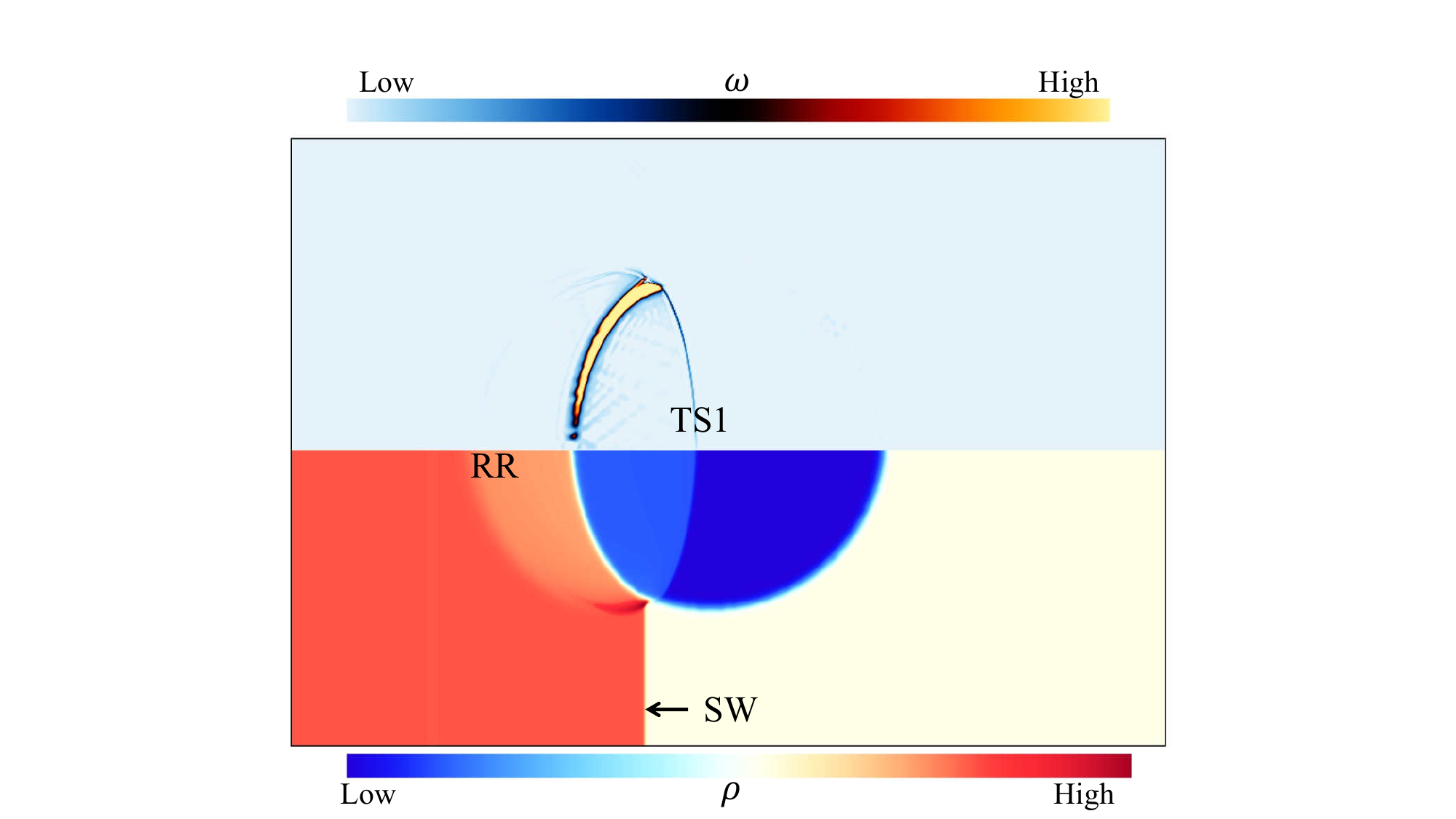}
			% \vspace{-19.6cm} 
		\end{minipage}%
	}
	\subfigure[$t^{*}=14.50$]{
		\begin{minipage}[t]{0.45\linewidth}
			\centering
			\includegraphics[width=1.0\columnwidth,trim={5cm 0.1cm 6cm 1.5cm},clip]{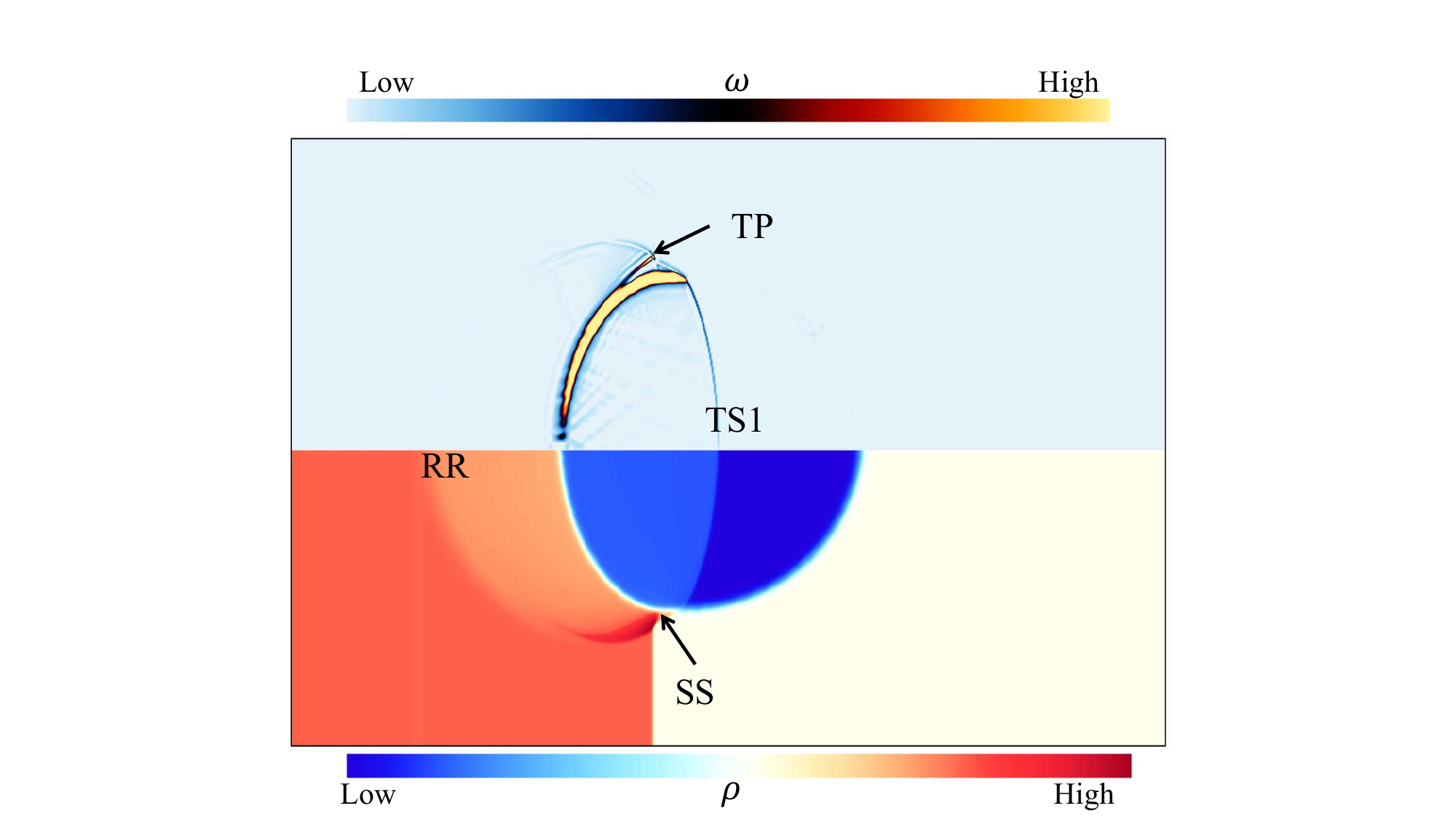}
			%\vspace{-19.6cm} 
			%\label{22}
		\end{minipage}%
	}	
	\subfigure[$t^{*}=17.46$]{
		\begin{minipage}[t]{0.45\linewidth}
			\centering
			\includegraphics[width=1.0\columnwidth,trim={5cm 0.1cm 6cm 1.5cm},clip]{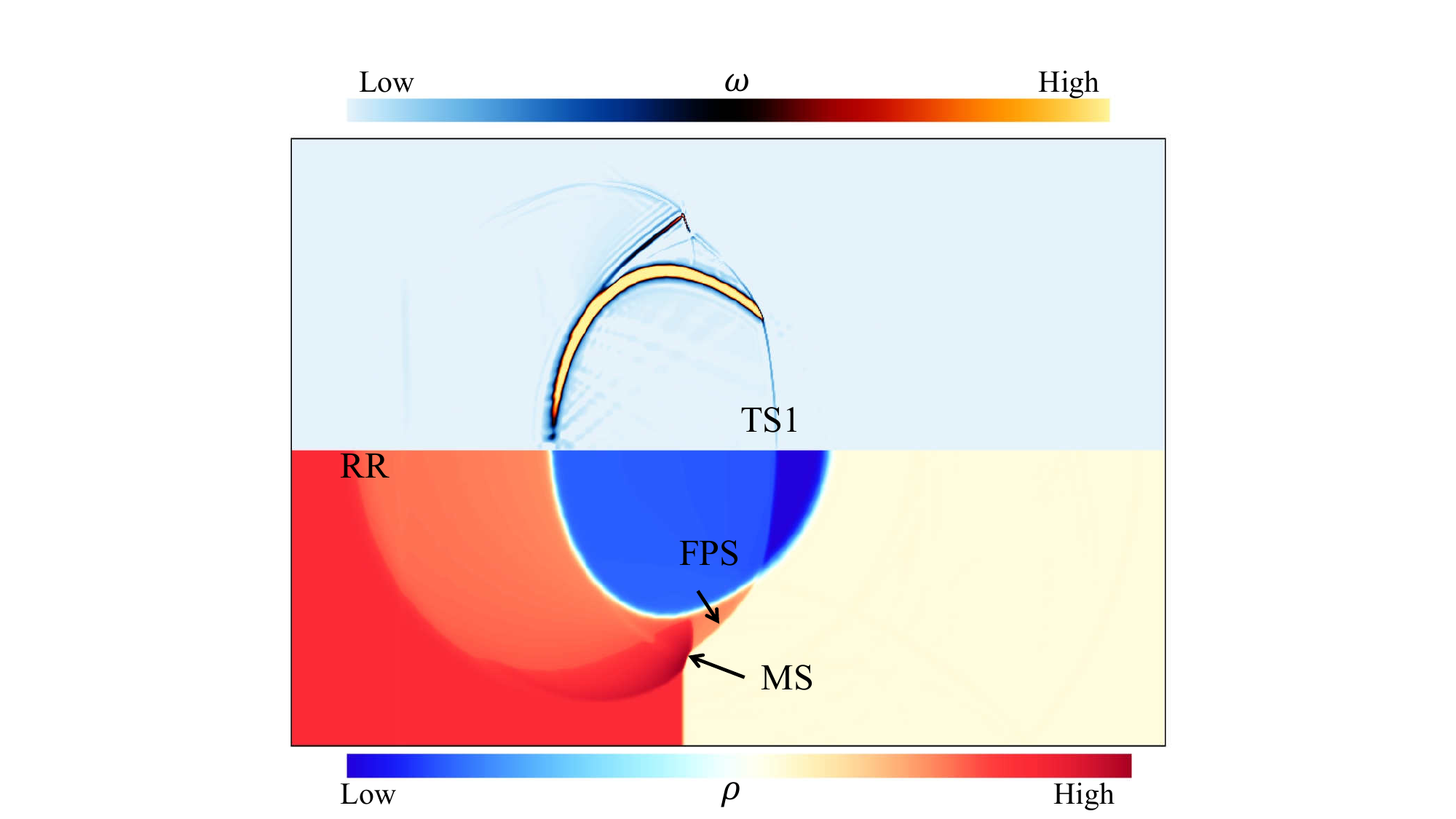}
			%   \vspace{-6.6cm} 
			%	\label{231}
		\end{minipage}%
	}	
	\subfigure[$t^{*}=20.37$]{
		\begin{minipage}[t]{0.45\linewidth}
			\centering
			\includegraphics[width=1.0\columnwidth,trim={5cm 0.1cm 6cm 1.5cm},clip]{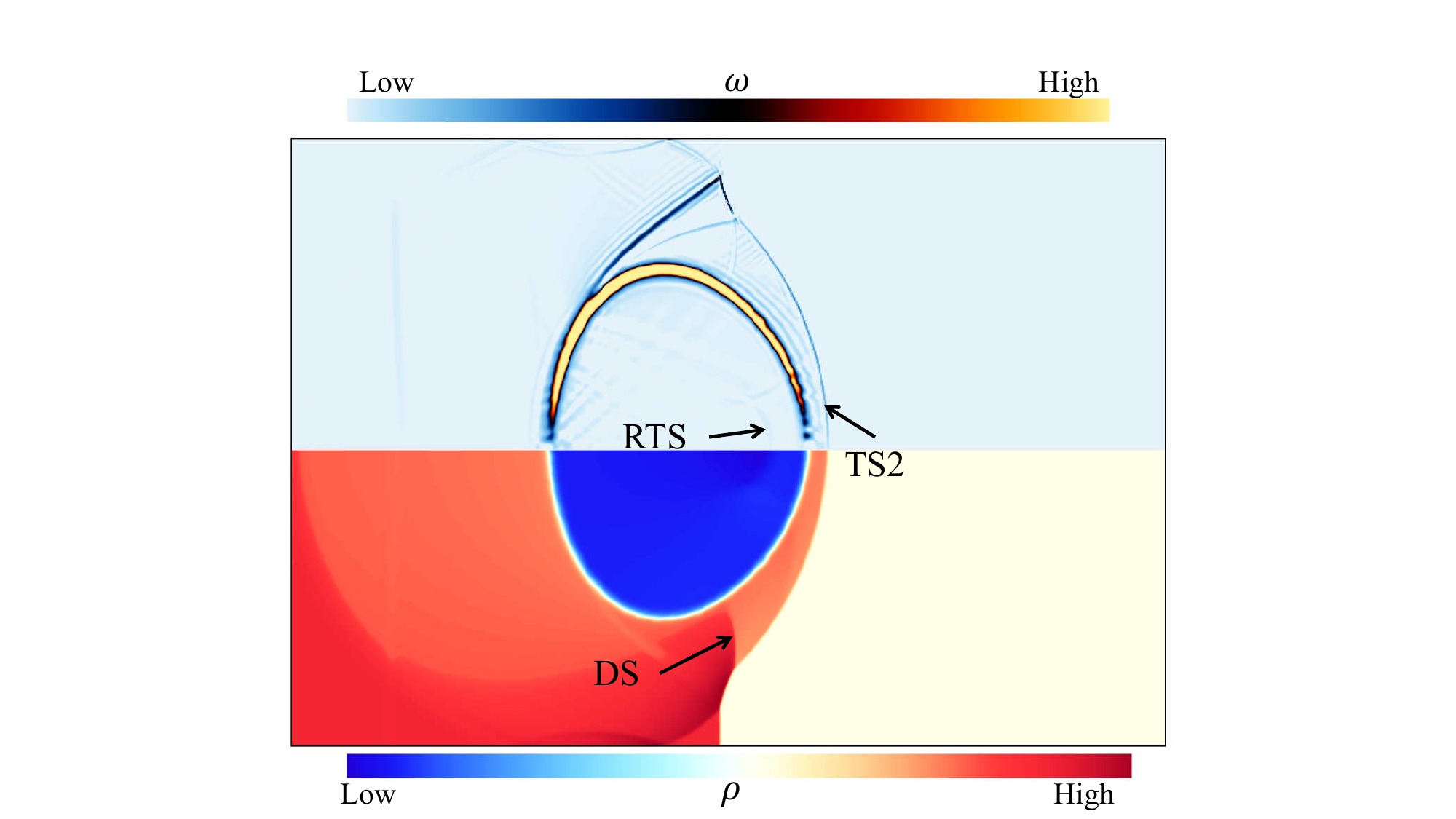}
			%   \vspace{-6.6cm} 
			%\label{curvature_lambda_zz}
		\end{minipage}%
	}	
	% left down  right up
	\caption{Contour maps of the vorticity magnitude $\lVert\bm{\omega}\rVert$ (upper panels) and the density $\rho$ (lower panels) at different dimensionless instants during early stages of shock-bubble interaction: (a) $t^* = 13.09$, (b) $t^* = 14.50$, (c) $t^* =17.46$, and (d) $t^* = 20.37$. See the descriptions in~\S\ref{MV} for full definitions of abbreviations.} 
	\label{152}
\end{figure}
\begin{figure}[t]
	\centering
	\includegraphics[width=1.0\columnwidth,trim={0.0cm 0cm 0.0cm 0cm},clip]{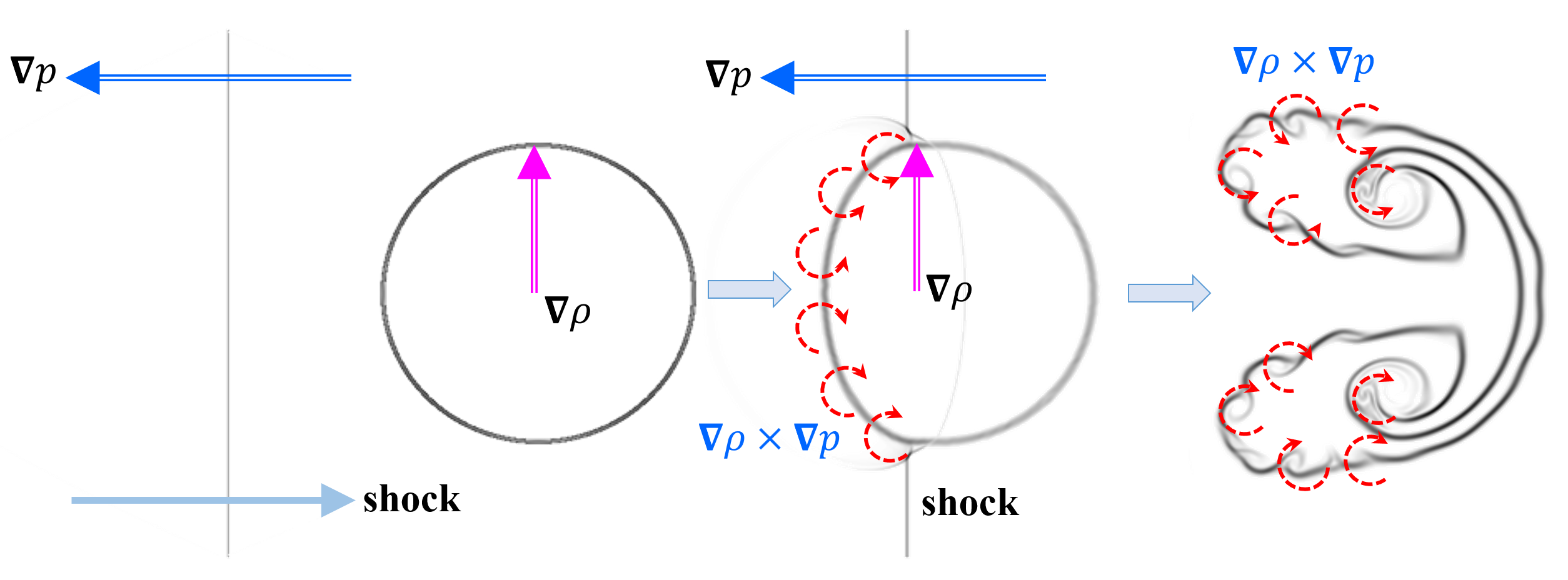}
	\caption{Schematic of vorticity deposition mechanism due to the baroclinic torque $\bm{\nabla}\rho\times\bm{\nabla}p$ along the shock-driven air bubble interface. The shock wave propagates from left to right.} 
	\label{Baroclinic_torque}
\end{figure}
\begin{figure}[h!]
	\centering
	\includegraphics[width=1.0\columnwidth,trim={0.0cm 0cm 0.0cm 0cm},clip]{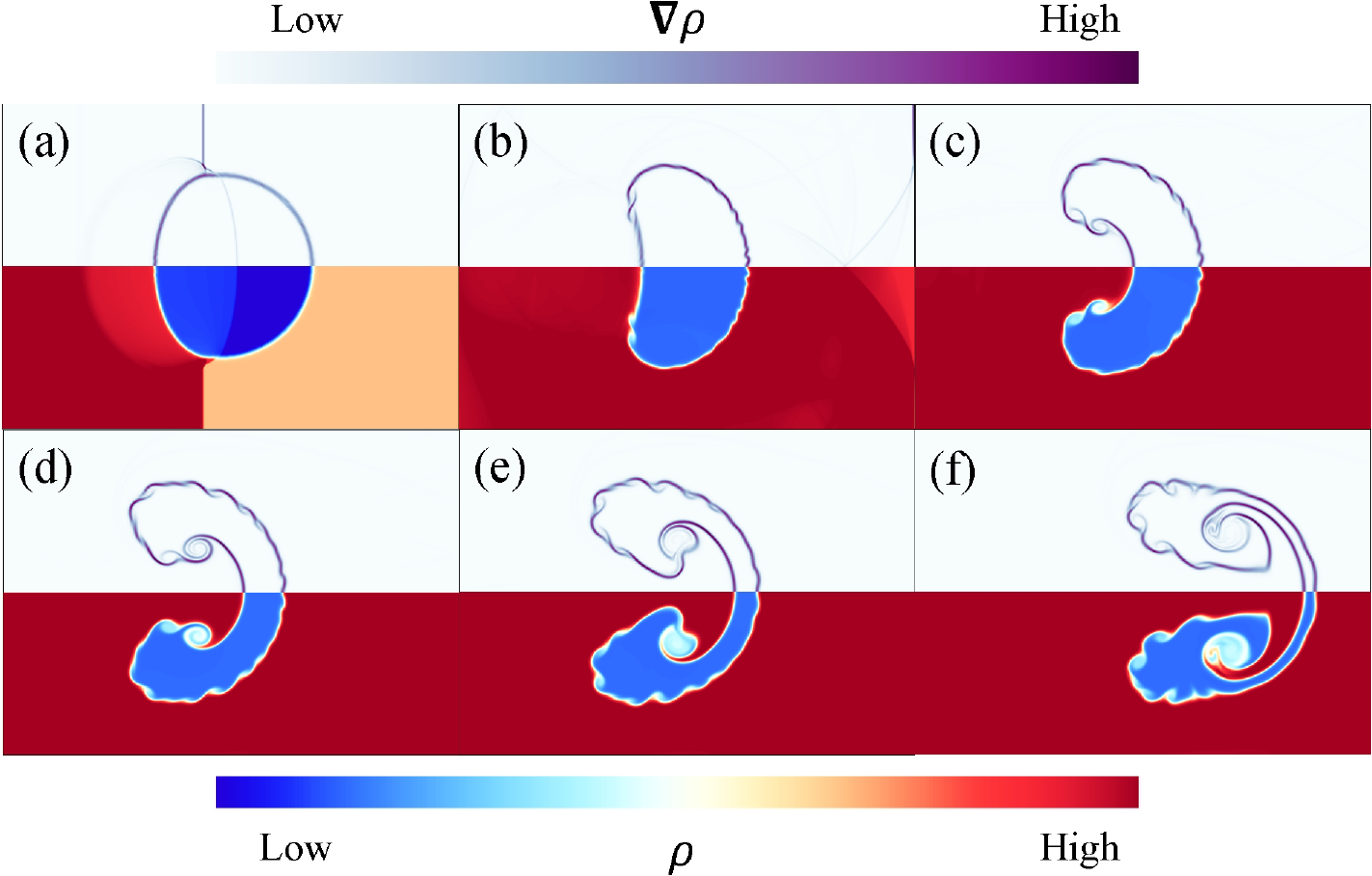}
	\caption{Evolution of the density $\rho$ (lower panels) and the magnitude of the density gradient $\lVert\bm{\nabla}\rho\rVert$ (upper panels) : (a) $t^*  = 14.5$, (b) $t^* = 29.1$,
		(c) $t^* =58.2$, (d) $t^* = 87.3$, (e) $t^* = 116.4$, and (f) $t^* = 174.6$.} 
	\label{history}
\end{figure}

To ensure grid-independent results, a grid refinement study was conducted for the simulated problem of shock-bubble interaction.
Three structured meshes were considered: Mesh-1 ($975 \times 267$), Mesh-2 ($1300 \times 356$), and Mesh-3 ($1950 \times 534$). Figure~\ref{wangge} compares the density profiles along the centerline $y^{*}=13.35$ at $t^{*}=14.5$, a relatively early stage of transient evolution.
While all meshes capture the global positions of the shock and interfaces, Mesh-1 exhibits noticeable smearing. Mesh-2 partially improves the sharpness to an acceptable level.
In contrast, Mesh-3 best resolves the steep density gradients at both the leading and trailing interfaces, demonstrating superior fidelity.
Figure~\ref{300o} presents the contour maps of density at $t^{*}=87.2$ (a relatively later transient stage), where secondary KH instabilities are resolved with varying levels of detail in terms of small-scale vortical roll-ups along the interface.
The coarsest mesh (Mesh-1) shows significant numerical diffusion, resulting in a smeared bubble interface and  partially obscured small vortical structures.
Mesh-2 improves interface sharpness and primary deformation, but still underestimates the development of small-scale interfacial vortices.
The finest mesh (Mesh-3) produces the steepest gradients, well-defined roll-up patterns, and the richest post-shock surface curvature, indicating that high resolution is essential for accurately capturing RMI and interface evolution. Based on these observations, Mesh-3 was selected for the subsequent physical analysis in this study.
\begin{figure}[h!]
	\centering
	\subfigure[$\bm{R}(\bm{t})\cdot\bm{e}_z$]{
		\begin{minipage}[t]{0.4\linewidth}
			\centering
			\includegraphics[width=1.0\columnwidth,trim={0.0cm 0cm 0.0cm 0cm},clip]{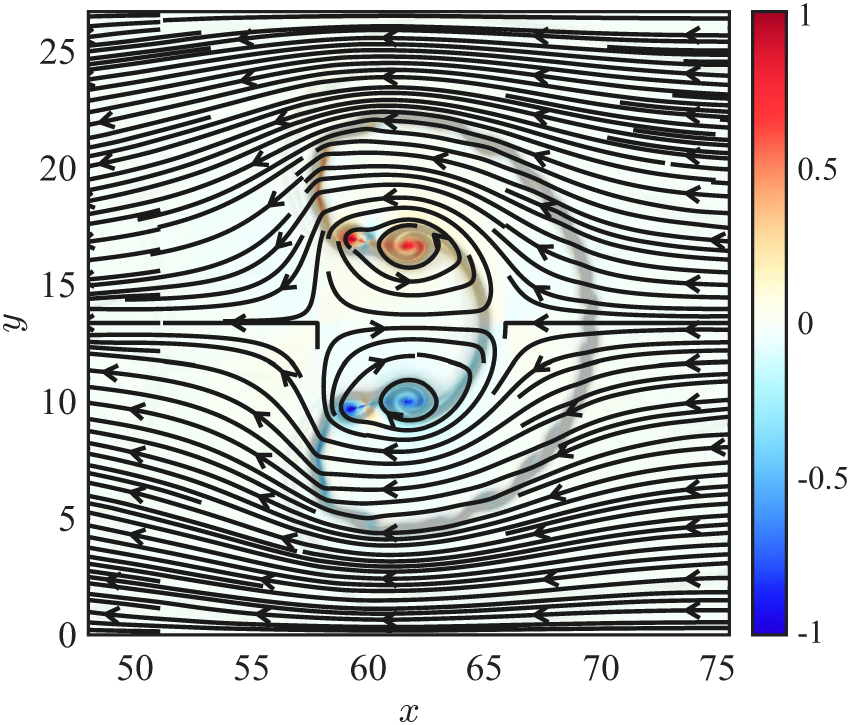}
			% \vspace{-19.6cm} 
			\label{Rotational_Vorticity_R1}
		\end{minipage}%
	}
	\subfigure[$\bm{s}(\bm{t})\cdot\bm{e}_z$]{
		\begin{minipage}[t]{0.4\linewidth}
			\centering
			\includegraphics[width=1.0\columnwidth,trim={0.0cm 0cm 0.0cm 0cm},clip]{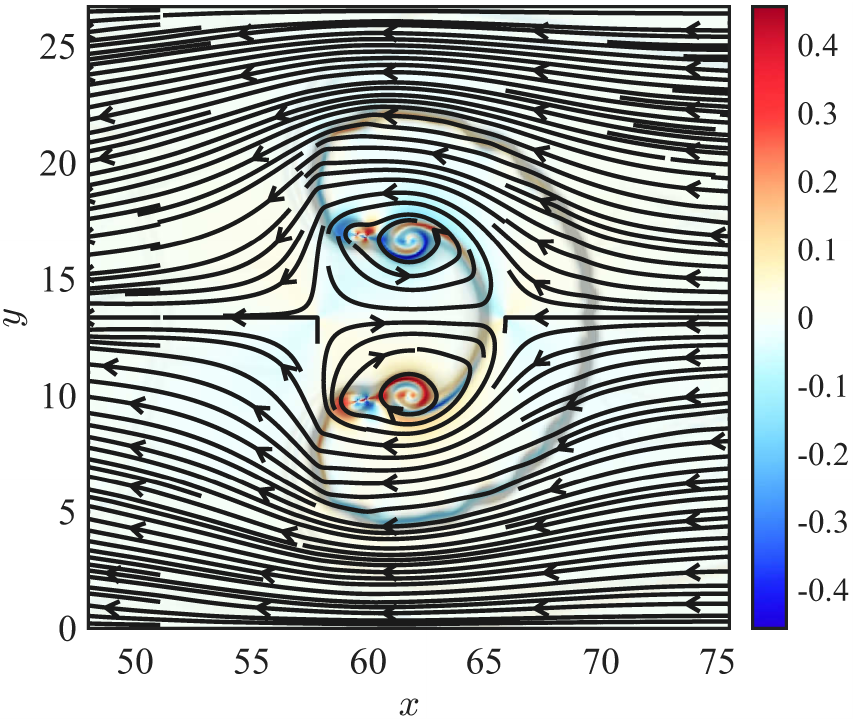}
			%\vspace{-19.6cm} 
			\label{Shear_Vorticity_s1}
		\end{minipage}%
	}	
	
	\subfigure[${\omega}_z$]{
		\begin{minipage}[t]{0.4\linewidth}
			\centering
			\includegraphics[width=1.0\columnwidth,trim={0.0cm 0cm 0.0cm 0cm},clip]{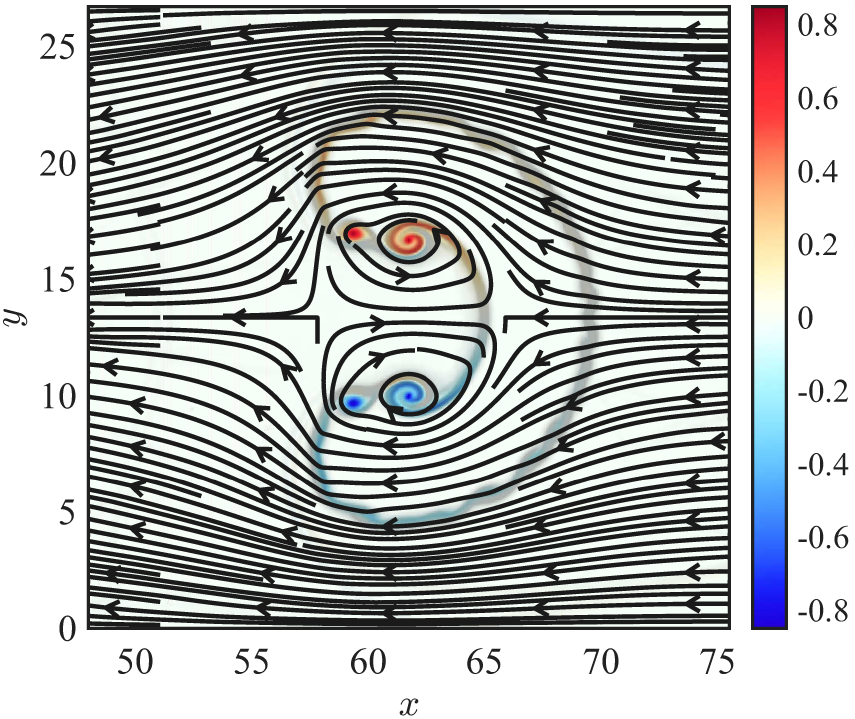}
			%   \vspace{-6.6cm} 
			\label{Total_Vorticity_omega_z1}
		\end{minipage}%
	}	% left down  right up
	\subfigure[$\bm{R}_N^+\cdot\bm{e}_z$ (Liutex)]{
		\begin{minipage}[t]{0.4\linewidth}
			\centering
			\includegraphics[width=1.0\columnwidth,trim={0.0cm 0cm 0.0cm 0cm},clip]{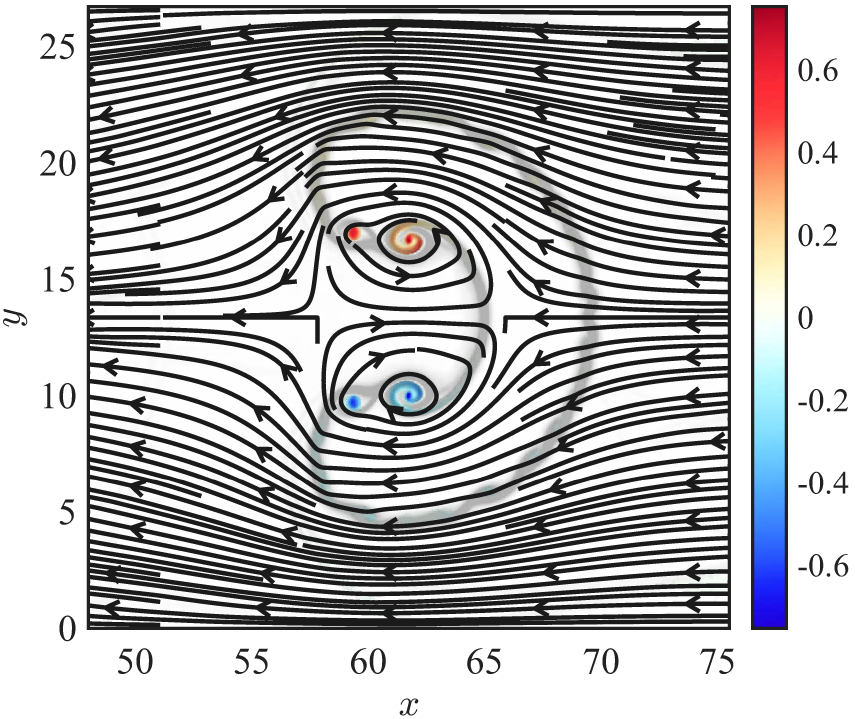}
			%   \vspace{-6.6cm} 
			\label{Liutex_Magnitude_R_Liutex1}
		\end{minipage}%
	}	
	
	\subfigure[$\bm{R}_N^-\cdot\bm{e}_z$]{
		\begin{minipage}[t]{0.4\linewidth}
			\centering
			\includegraphics[width=1.0\columnwidth,trim={0.0cm 0cm 0.0cm 0cm},clip]{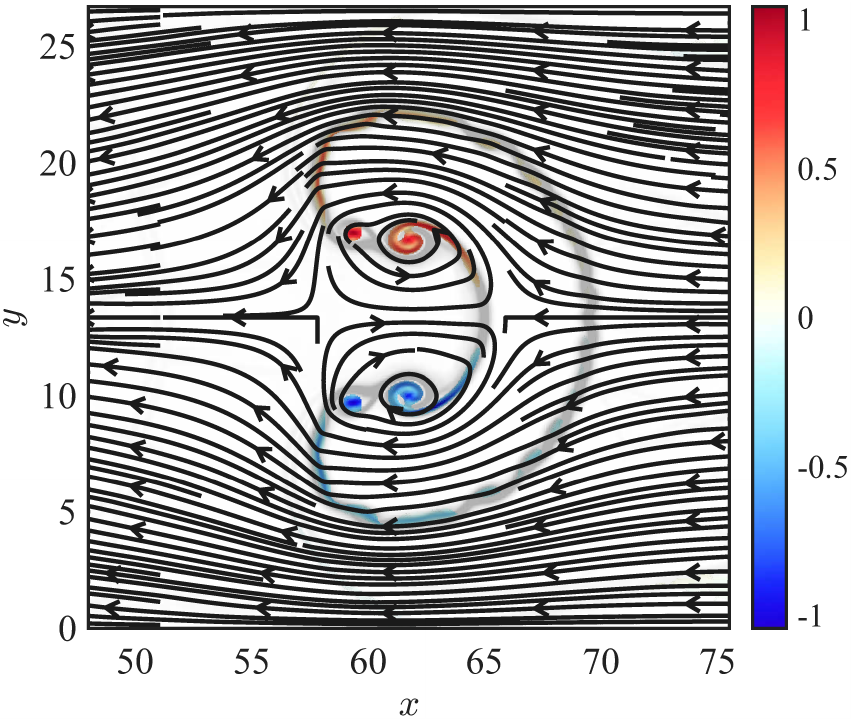}
			%   \vspace{-6.6cm} 
			\label{R_N_minus1}
		\end{minipage}%
	}	
	\caption{Normalized snapshots of the $\bm{e}_z$-components of the DVD and IVD vorticity modes at the instant $t^{*}=84.39$ (point C): (a) the orbital-rotation mode
		$\bm{R}(\bm{t})$, (b) the spin mode $\bm{s}(\bm{t})$, (c) the total vorticity $\bm{\omega}$, (d)
		$\bm{R}_{N}^{+}$ (Liutex)~\citep{Liu2018}, and (e) $\bm{R}_{N}^{-}$~\citep{Chen2025arxiv,Chen2026operator}.
		The reference frame translates at the transient vortex-pair propagation velocity $\mathbf{U}_{\rm f}={\rm 0.3093}{\rm U}_{0}\bm{e}_{x}$.} 
	\label{fug9a}
\end{figure}
\begin{figure}[h!]
	\centering
	\subfigure[$\bm{R}(\bm{t})\cdot\bm{e}_z$]{
		\begin{minipage}[t]{0.4\linewidth}
			\centering
			\includegraphics[width=1.0\columnwidth,trim={0.0cm 0cm 0.0cm 0cm},clip]{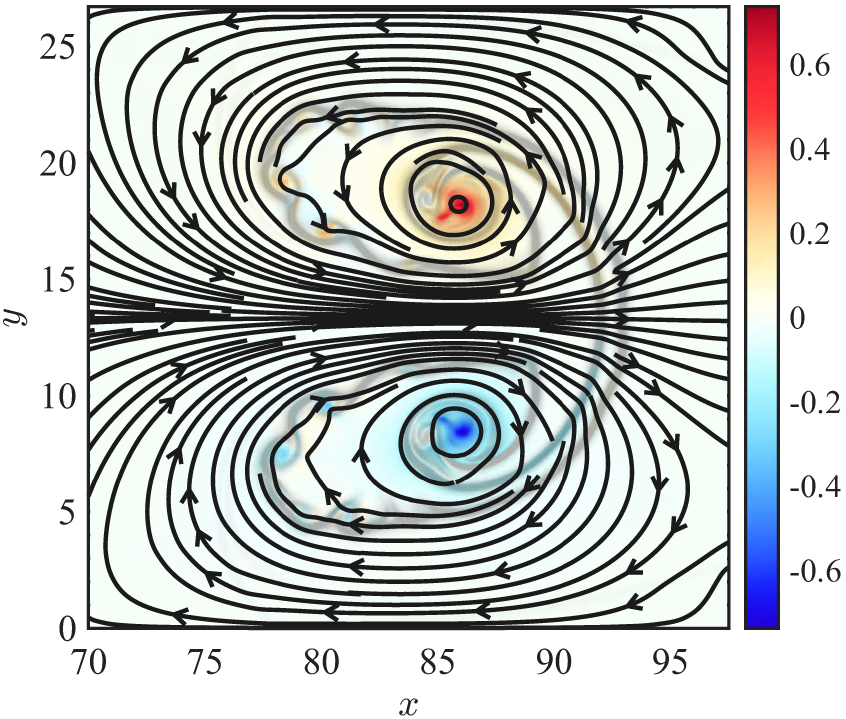}
			% \vspace{-19.6cm} 
			\label{Rotational_Vorticity_R}
		\end{minipage}%
	}
	\subfigure[$\bm{s}(\bm{t})\cdot\bm{e}_z$]{
		\begin{minipage}[t]{0.4\linewidth}
			\centering
			\includegraphics[width=1.0\columnwidth,trim={0.0cm 0cm 0.0cm 0cm},clip]{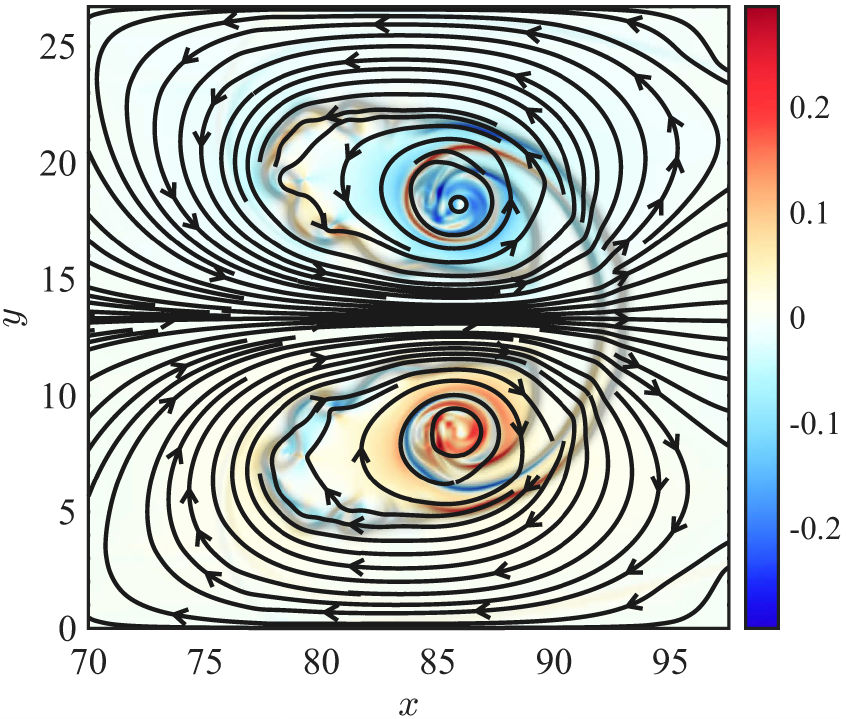}
			%\vspace{-19.6cm} 
			\label{Shear_Vorticity_s}
		\end{minipage}%
	}	
	
	\subfigure[${\omega}_z$]{
		\begin{minipage}[t]{0.4\linewidth}
			\centering
			\includegraphics[width=1.0\columnwidth,trim={0.0cm 0cm 0.0cm 0cm},clip]{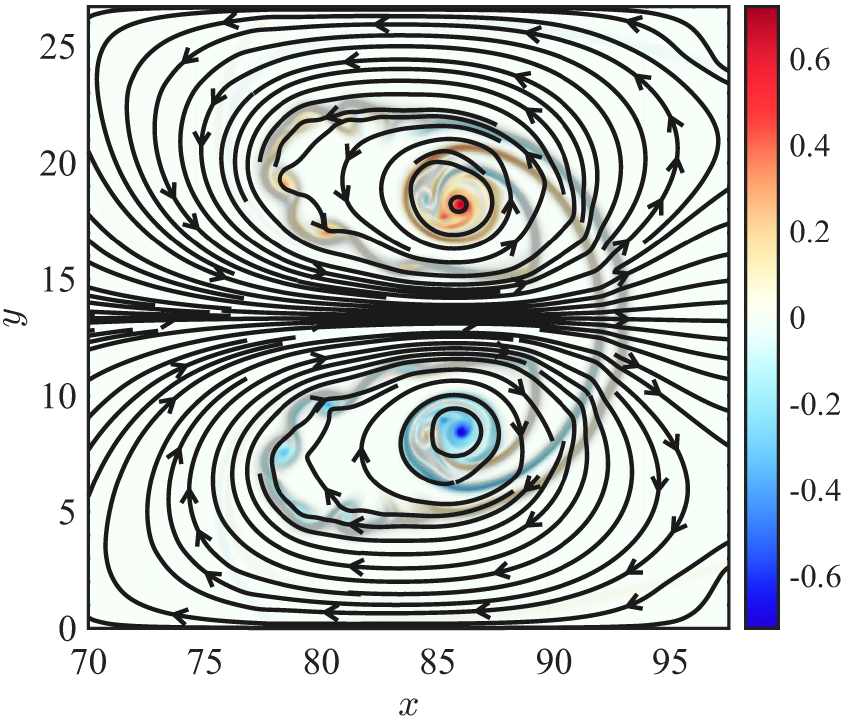}
			%   \vspace{-6.6cm} 
			\label{Total_Vorticity_omega_z}
		\end{minipage}%
	}	% left down  right up
	\subfigure[$\bm{R}_N^+\cdot\bm{e}_z$ (Liutex)]{
		\begin{minipage}[t]{0.4\linewidth}
			\centering
			\includegraphics[width=1.0\columnwidth,trim={0.0cm 0cm 0.0cm 0cm},clip]{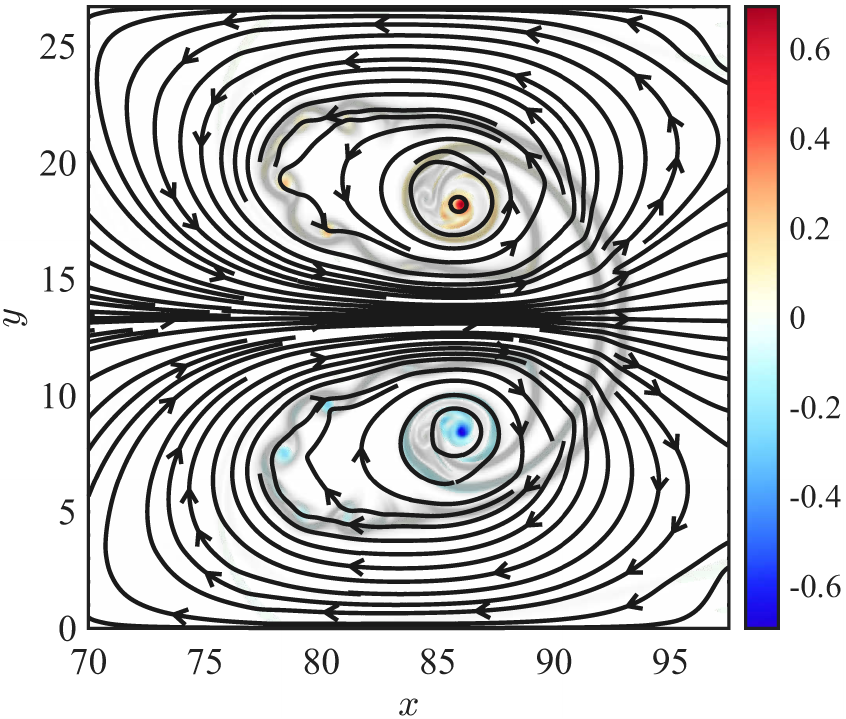}
			%   \vspace{-6.6cm} 
			\label{Liutex_Magnitude_R_Liutex}
		\end{minipage}%
	}	
	
	\subfigure[$\bm{R}_N^-\cdot\bm{e}_z$]{
		\begin{minipage}[t]{0.4\linewidth}
			\centering
			\includegraphics[width=1.0\columnwidth,trim={0.0cm 0cm 0.0cm 0cm},clip]{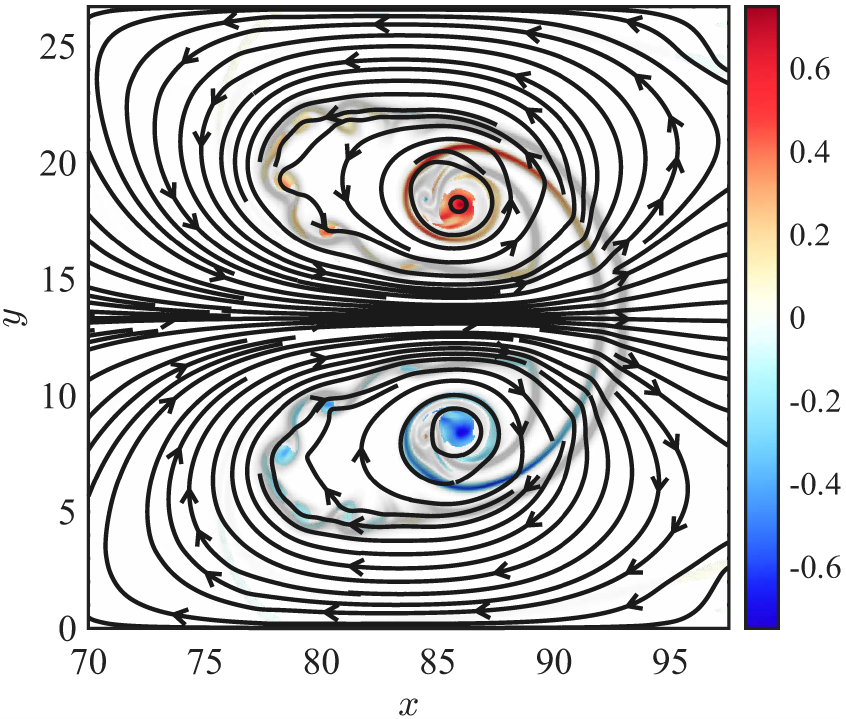}
			%   \vspace{-6.6cm} 
			\label{R_N_minus}
		\end{minipage}%
	}	
	\caption{Normalized snapshots of the $\bm{e}_z$-components of the DVD and IVD vorticity modes at the instant $t^{*}=174.59$ (point E): (a) the orbital-rotation mode
		$\bm{R}(\bm{t})$, (b) the spin mode $\bm{s}(\bm{t})$, (c) the total vorticity $\bm{\omega}$, (d)
		$\bm{R}_{N}^{+}$ (Liutex)~\citep{Liu2018}, and (e) $\bm{R}_{N}^{-}$~\citep{Chen2025arxiv,Chen2026operator}.
		The reference frame translates at the transient vortex-pair propagation velocity $\mathbf{U}_{\rm f}=0.223{\rm U}_{0}\bm{e}_{x}$.} 
	\label{fug9}
\end{figure}

\subsection{Baroclinic vorticity generation and interface morphology evolution}\label{MV}

%\begin{figure}[t]
%	\centering
%	\subfigure[$\bm{R}_N^+\cdot\bm{e}_z$ (Liutex)]{
%		\begin{minipage}[t]{0.49\linewidth}
%			\centering
%			\includegraphics[width=1.0\columnwidth,trim={0.0cm 0cm 0.0cm 0cm},clip]{Liutex_Magnitude_R_Liutex.pdf}
%			%   \vspace{-6.6cm} 
%			\label{Liutex_Magnitude_R_Liutex}
%		\end{minipage}%
%	}	
%	\subfigure[$\bm{R}_N^-\cdot\bm{e}_z$]{
%		\begin{minipage}[t]{0.49\linewidth}
%			\centering
%			\includegraphics[width=1.0\columnwidth,trim={0.0cm 0cm 0.0cm 0cm},clip]{R_N_minus.pdf}
%			%   \vspace{-6.6cm} 
%			\label{R_N_minus}
%		\end{minipage}%
%	}	
%	\caption{Normalized snapshots of the $\bm{e}_z$-components of the IVD vorticity modes at the instant $t=174.5$: .
%		The reference frame translates at the vortex-pair propagation velocity $\mathbf{U}_{\rm f}=0.223\sqrt{RT_0}\bm{e}_{x}$.} 
%	\label{fug10}
%\end{figure}
\begin{figure}[t]
	\centering
	\subfigure[$(\bm{R}(\bm{t}),\bm{R}^+_N,\bm{R}^{-}_N)$]{
		\begin{minipage}[t]{0.476\linewidth}
			\centering
			\includegraphics[width=1.0\columnwidth,trim={0cm 0cm 0.0cm 0cm},clip]{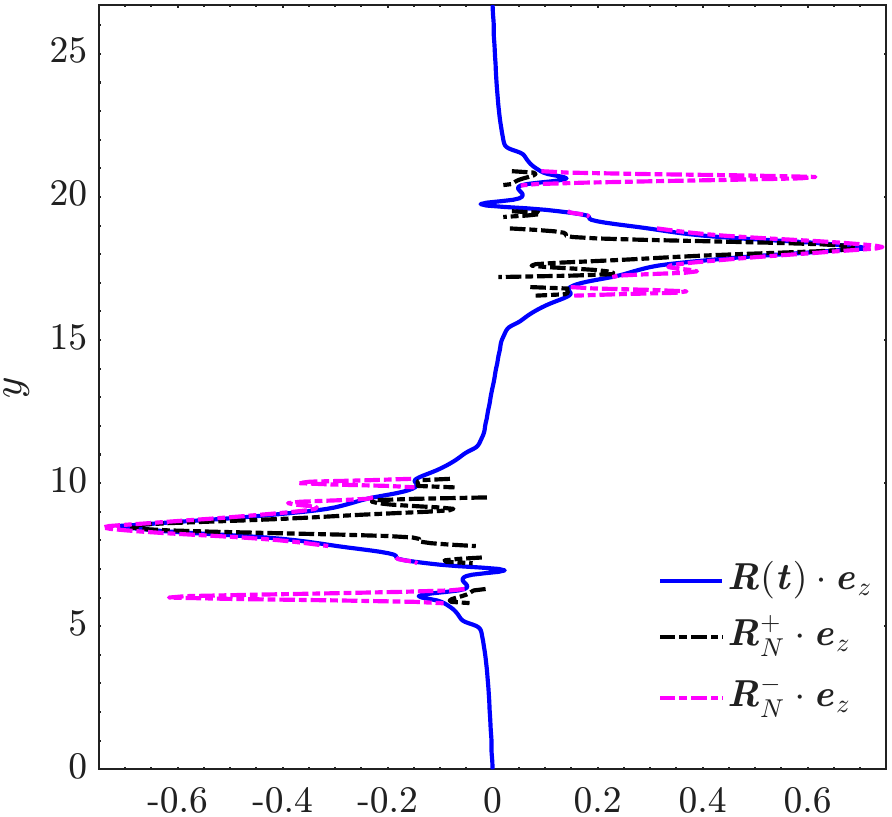}
			% \vspace{-19.6cm} 
			\label{R compare}
		\end{minipage}%
	}
	\subfigure[$(\bm{s}(\bm{t}),\bm{s}^+_N,\bm{s}^{-}_N)$]{
		\begin{minipage}[t]{0.49\linewidth}
			\centering
			\includegraphics[width=1.0\columnwidth,trim={0cm 0.0cm 0.0cm 0cm},clip]{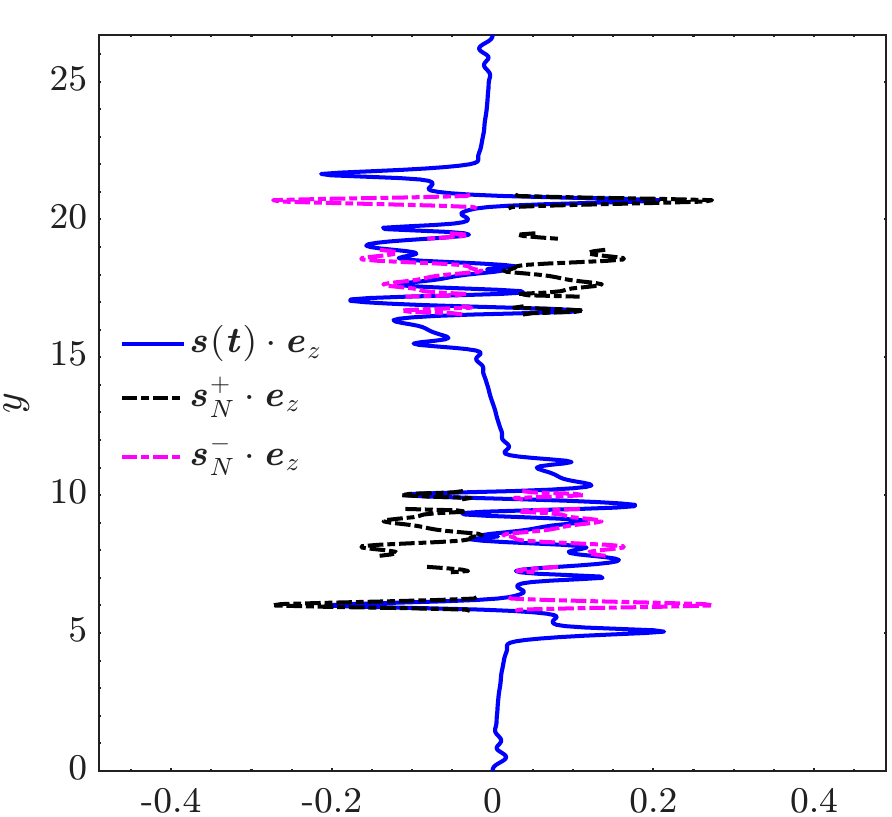}
			%\vspace{-19.6cm} 
			\label{s compare}
		\end{minipage}%
	}	
	
		\subfigure[$(\bm{R}(\bm{t}),\bm{s}(\bm{t}),\bm{\omega})$]{
		\begin{minipage}[t]{0.49\linewidth}
			\centering
			\includegraphics[width=1.0\columnwidth,trim={0cm 0.0cm 0.0cm 0cm},clip]{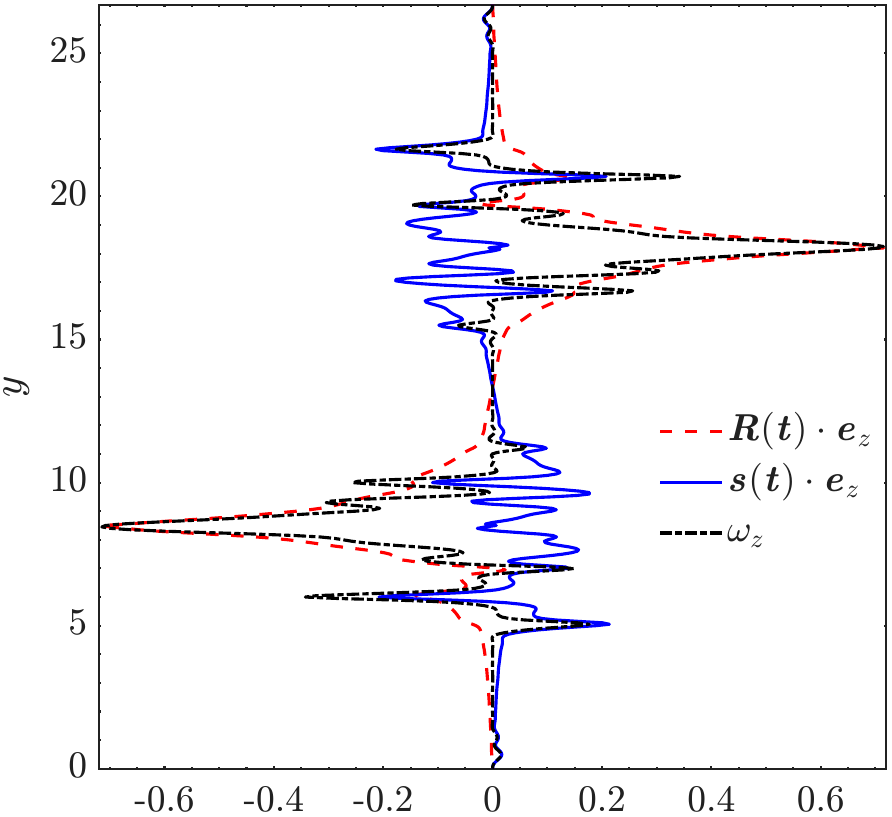}
			%\vspace{-19.6cm} 
			\label{R_S_Omega_z_bubble}
		\end{minipage}%
	}	
	\caption{Comparison of the DVD and IVD (defined in the region $\Delta>0$) vorticity modes, along the line $x^{*} = 86$ connecting the vortex-pair centers moving at $\mathbf{U}_{\rm f}=0.223{\rm U}_{0}\bm{e}_{x}$. (a) $(\bm{R}(\bm{t}),\bm{R}^+_N,\bm{R}^{-}_N)$, (b) $(\bm{s}(\bm{t}),\bm{s}^+_N,\bm{s}^{-}_N)$, and (c) $(\bm{R}(\bm{t}),\bm{s}(\bm{t}),\bm{\omega})$. The dimensionless time is $t^{*}=174.59$.} 
	\label{fug11}
\end{figure}
\begin{figure}[h!]
	\centering
	\subfigure[$\mathcal{D}\cap\left\{\bm{x}\in\mathbb{R}^{2}|\Delta(\bm{x})>0\right\}$]{
		\begin{minipage}[t]{0.49\linewidth}
			\centering
			\includegraphics[width=1.0\columnwidth,trim={0cm 0cm 0cm 0cm},clip]{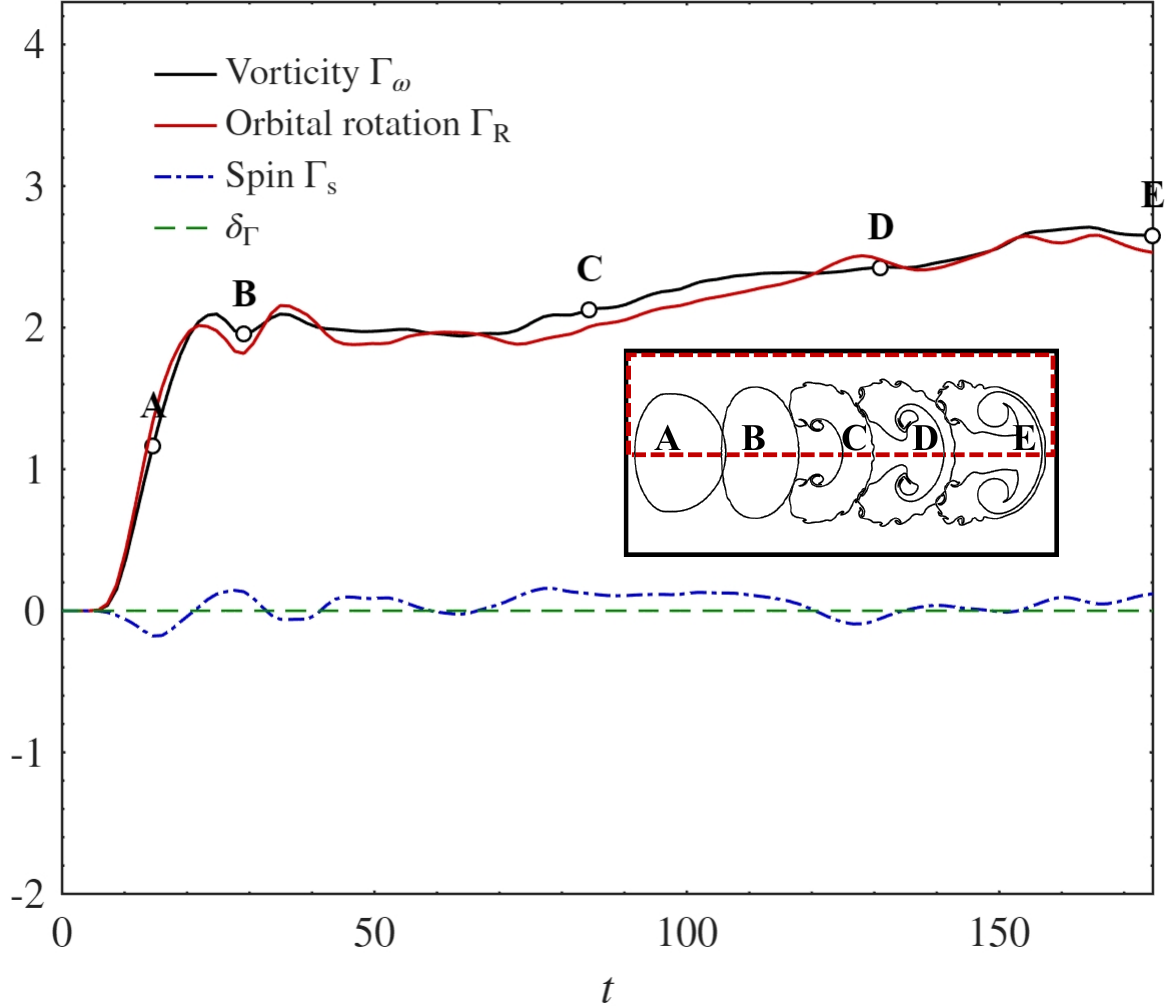}
			\label{Delta0-1}
		\end{minipage}%
	}
	\subfigure[$\mathcal{D}$]{
		\begin{minipage}[t]{0.49\linewidth}
			\centering
			\includegraphics[width=1.0\columnwidth,trim={0cm 0cm 0cm 0cm},clip]{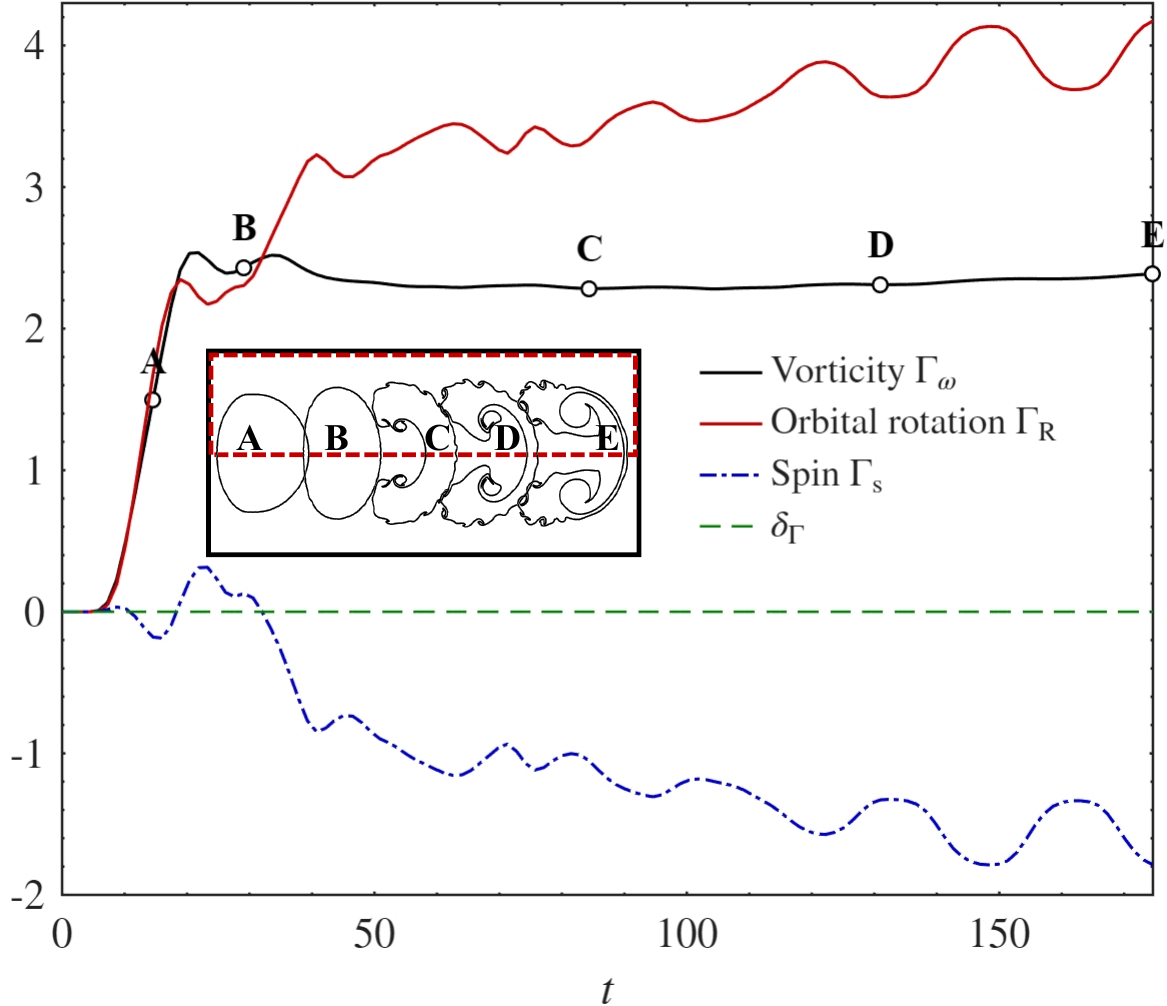}			\label{Delta1-1}
		\end{minipage}%
	}	
	\caption{Spatiotemporal evolution of the vorticity flux $\Gamma_{\omega}$, the orbital-rotation flux $\Gamma_{\rm R}$, and the spin flux  $\Gamma_{\rm s}$ during the shock-driven bubble evolution. The integration domain is (a) $\mathcal{D}\cap\left\{\bm{x}\in\mathbb{R}^{2}|\Delta(\bm{x})>0\right\}$ (restricted to the conventional vortical-flow region), and (b) $\mathcal{D}$ (the rectangular upper half domain). Note that the residual error is defined as $\delta_{\Gamma}\equiv\Gamma_{\omega}-\Gamma_{\rm{R}}-\Gamma_{\rm{s}}$.} 
	\label{DeltaGamma-1}
\end{figure}
\begin{figure}[h!]
	\centering
	\subfigure[point ${\rm A}$, $t^{*}=14.55$]{
		\begin{minipage}[t]{1.0\linewidth}
			\centering
			\includegraphics[width=1.0\columnwidth,trim={0cm 1cm 0cm 1cm},clip]{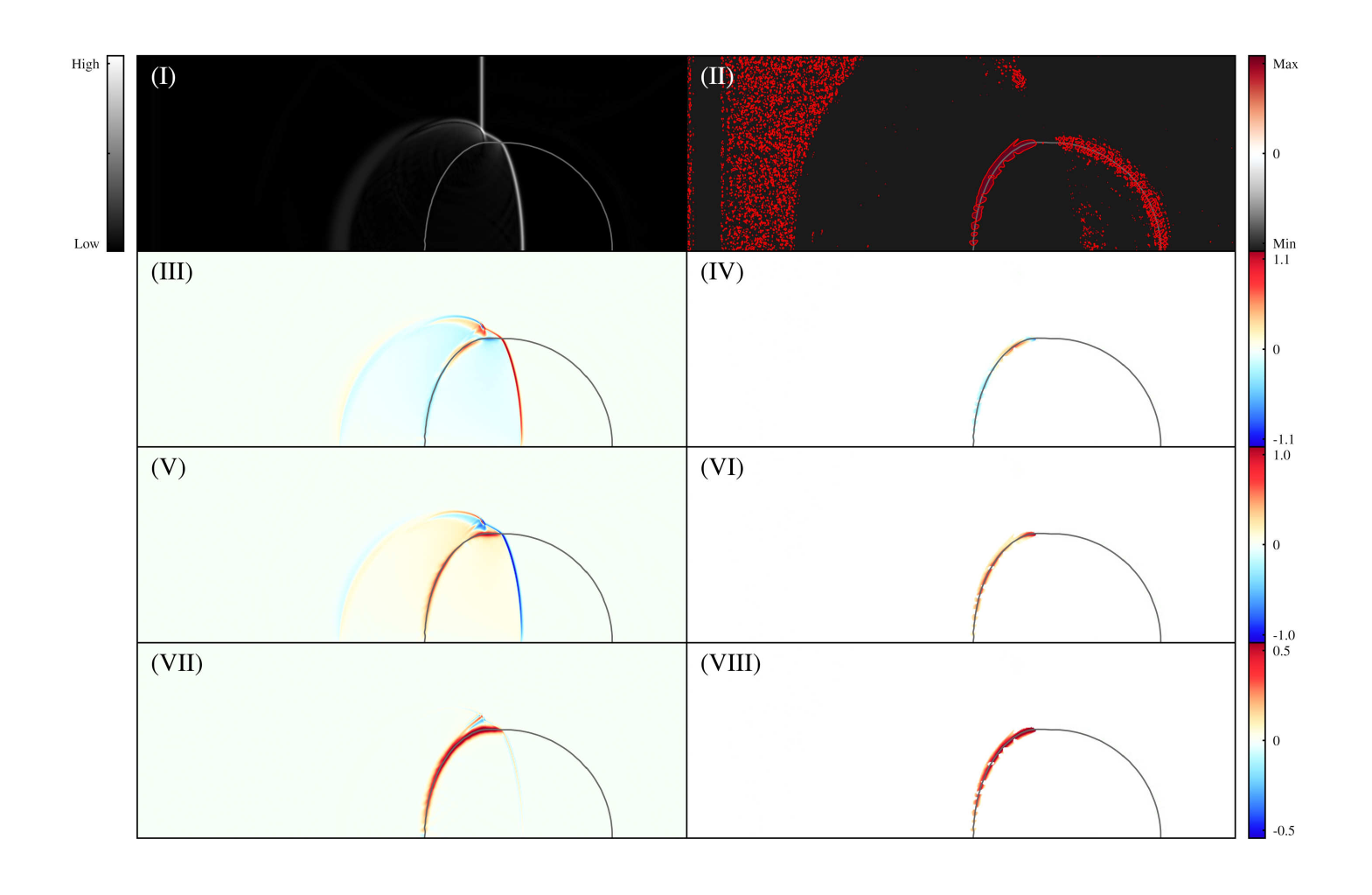}
			\label{and t14_55}
		\end{minipage}%
	}
		\subfigure[point ${\rm B}$, $t^{*}=29.10$]{
		\begin{minipage}[t]{1.0\linewidth}
			\centering
			\includegraphics[width=1.0\columnwidth,trim={0cm 1cm 0cm 1cm},clip]{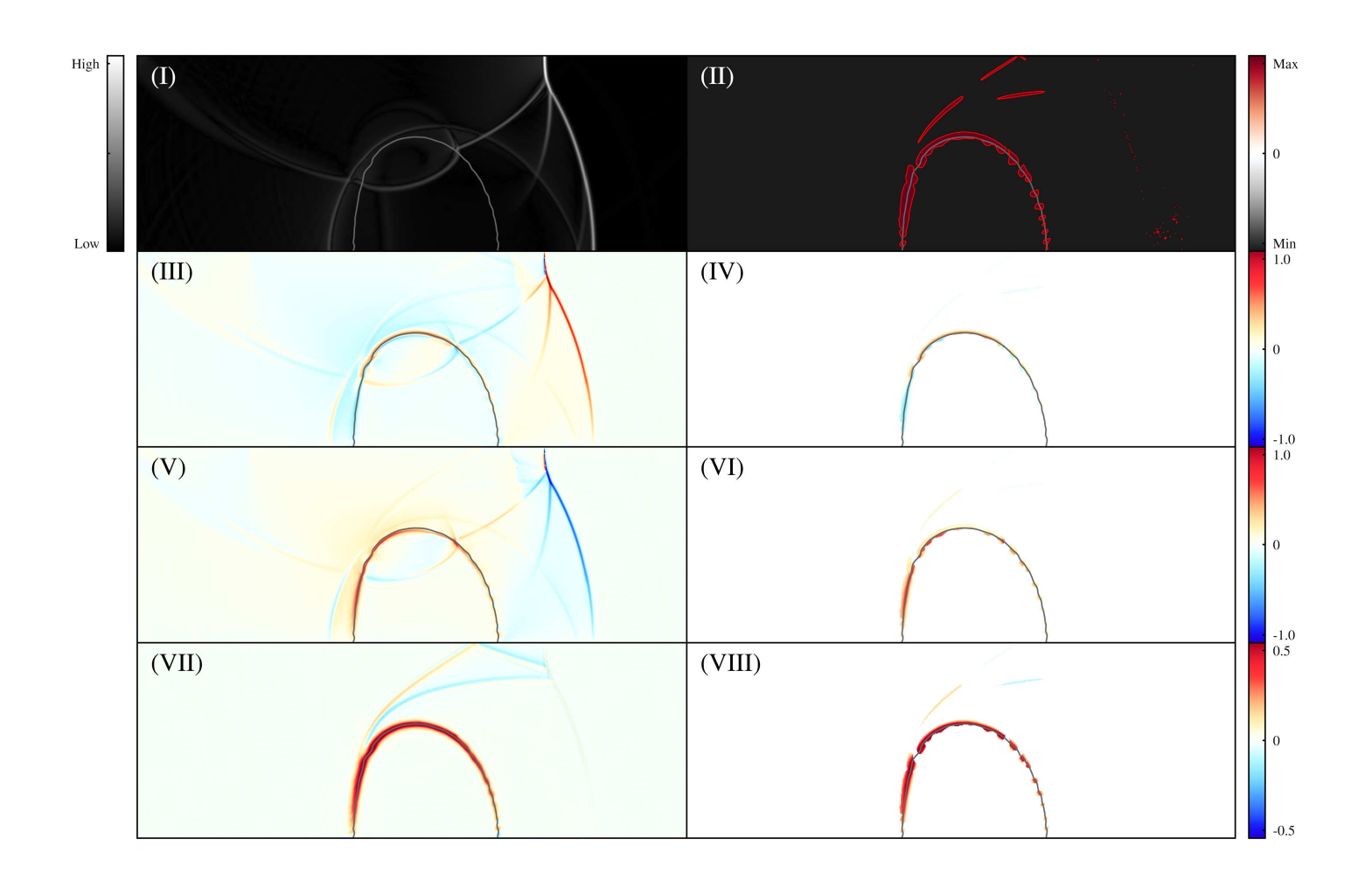}
			\label{and t29_10}
		\end{minipage}%
	}
	\caption{Evolutionary characteristics of interface morphology, vorticity/vortex structures, and dilatational wave fields at representative instants (points A--E in figure~\ref{DeltaGamma-1}). The definitions of subpanels (I) -- (VIII) follow those in figure~\ref{and}.} 
	\label{air_bubble_AB}
\end{figure}
\begin{figure}[h!]
	\centering
	\subfigure[point ${\rm C}$, $t^{*}=84.39$]{
		\begin{minipage}[t]{1.0\linewidth}
			\centering
			\includegraphics[width=1.0\columnwidth,trim={0cm 1cm 0cm 1cm},clip]{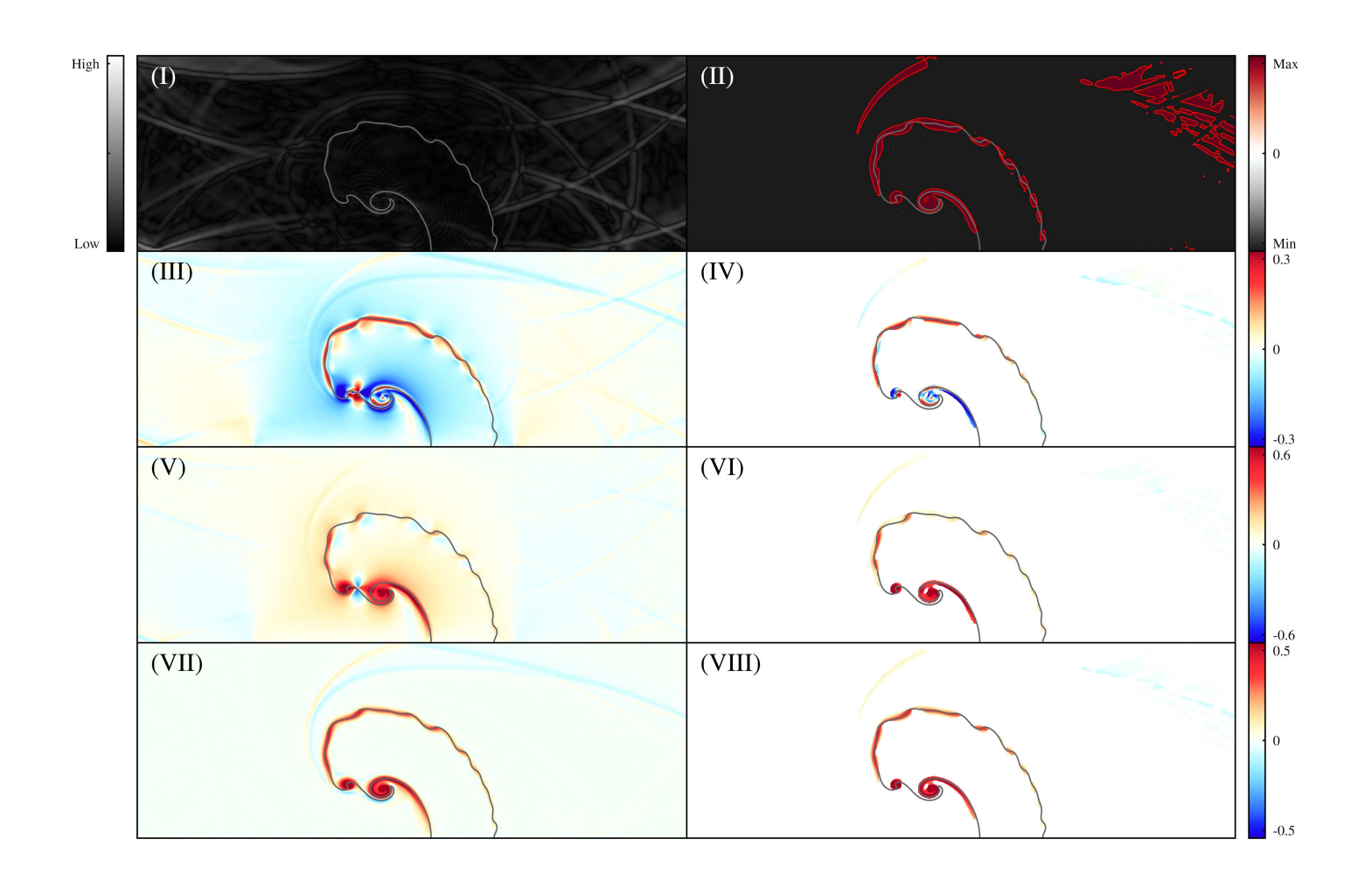}
			\label{and t84_39}
		\end{minipage}%
	}
	\subfigure[point ${\rm D}$, $t^{*}=130.95$]{
		\begin{minipage}[t]{1.0\linewidth}
			\centering
			\includegraphics[width=1.0\columnwidth,trim={0cm 1cm 0cm 1cm},clip]{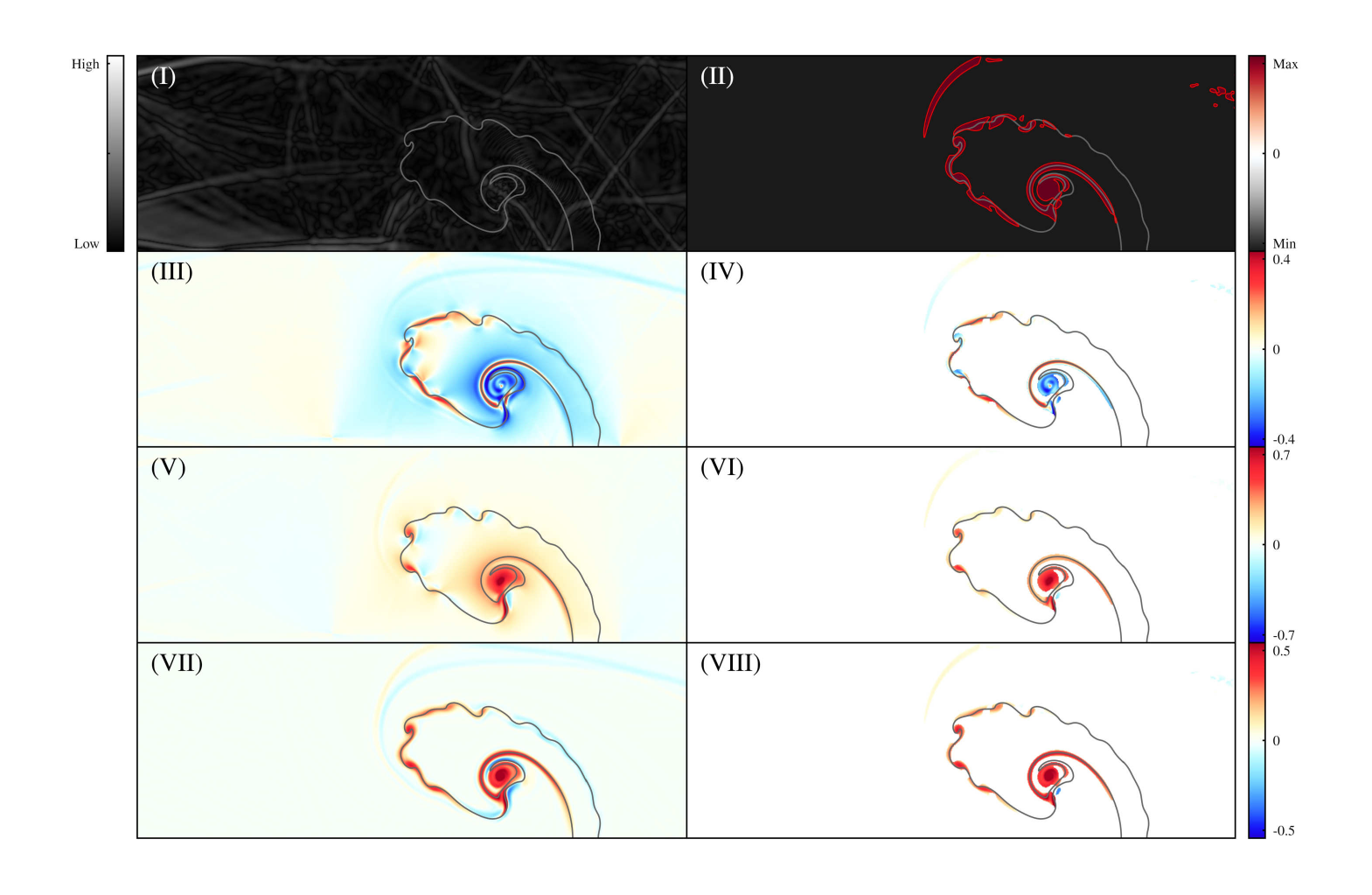}
			\label{and t130_95}
		\end{minipage}%
	}
	\caption{Continuation of figure~\ref{air_bubble_AB}.} 
	\label{air_bubble_CD}
\end{figure}
\begin{figure}[t]
	\centering
			\includegraphics[width=1.0\columnwidth,trim={0cm 1cm 0cm 1cm},clip]{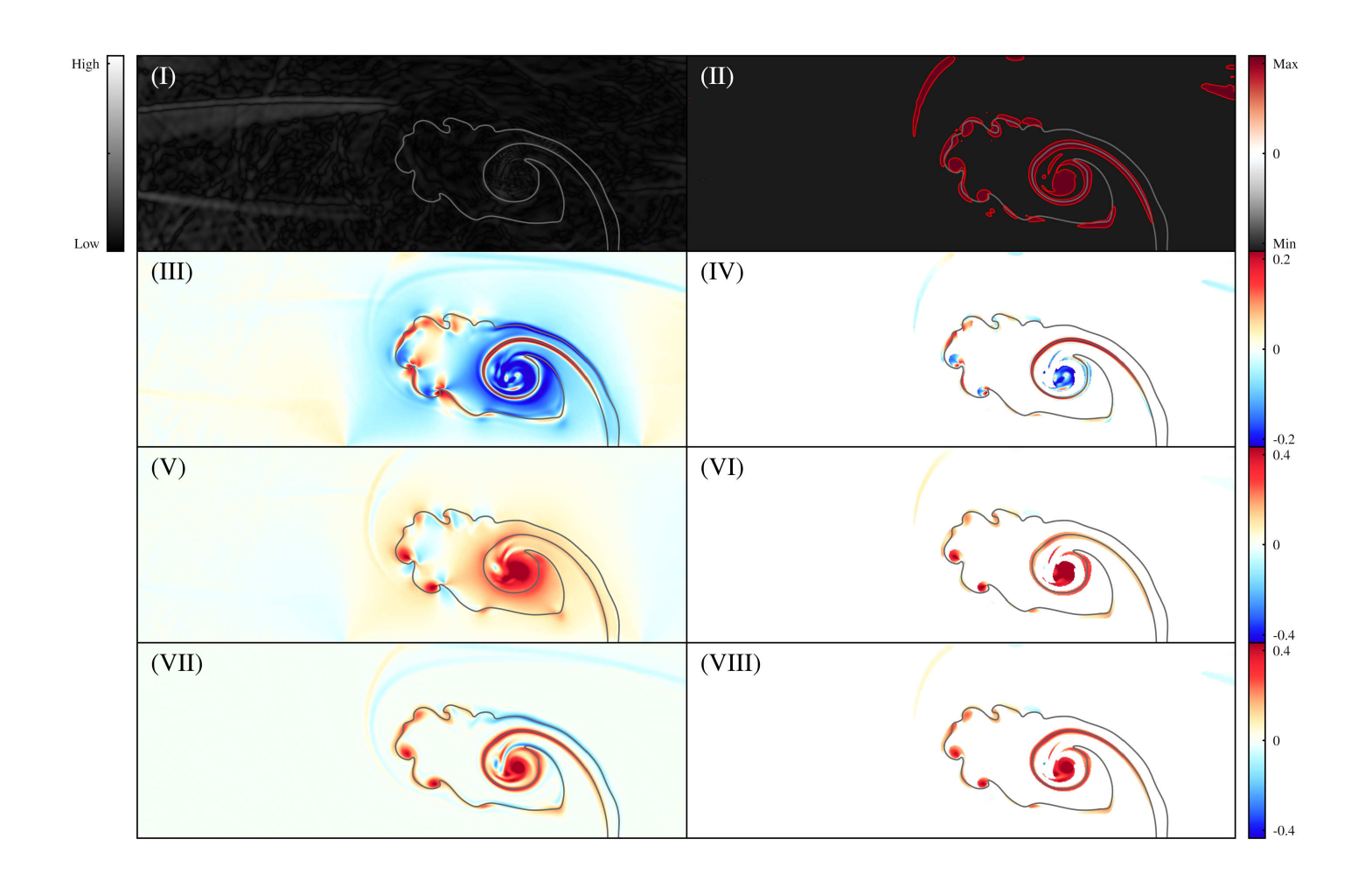}
	\caption{Continuation of figure~\ref{air_bubble_CD}. Point ${\rm E}$, $t^{*}=174.59$.} 
			\label{and t174_5}
\end{figure}

Figure~\ref{152} demonstrates the contour maps of the vorticity magnitude (upper half) and  the density $\rho$ (lower half)
during the early stages of the shock-bubble interaction, where the annotated flow structures, such as the shock wave (SW), transmitted shocks (TS1, TS2), reflected rarefaction wave (RR), free precursor shock (FPS), triple points (TP), Mach stem (MS), reflected transmitted shock (RTS), and slip surface (SS), qualitatively correspond with the visualization reported in~\citet{Alsaeed2025}.
The shock wave first acts on the front of the cylindrical bubble; on the one hand, it propagates into the light gas, forming a transmitted shock wave, and on the other hand, it reflects in the surrounding ambient medium, generating a reflected rarefaction wave. Due to the curved geometric characteristics of the bubble interface, this interaction process further induces the generation of secondary transmitted shock waves and secondary reflected shock waves, accompanied by the formation of local triple points and Mach stems, thereby significantly enhancing the generation effect of baroclinic vorticity, which can clearly be seen in the upper half of figure~\ref{152}.

Figure~\ref{Baroclinic_torque} provides a schematic illustration of the baroclinic vorticity deposition mechanism along the shock-driven air bubble interface, which arises whenever the density and pressure gradients are not aligned. The incident shock induces a strong pressure gradient $\bm{\nabla} p$ oriented opposite to the propagation direction, while the density gradient $\bm{\nabla}\rho$ points across the bubble interface from the light interior (air) to the heavy surrounding medium (Kr). This misalignment generates a localized baroclinic torque (proportional to $\bm{\nabla}\rho\times\bm{\nabla}p$), depositing vorticity of opposite signs with respect to the axis of symmetry. This deposited vorticity drives the rolling-up of the interface and promotes the nonlinear growth of the secondary Kelvin-Helmholtz instabilities (KHIs).
As further shown in figure~\ref{history}, the baroclinic term in~\eqref{jianh} serves as the primary vorticity source for the shock-driven bubble interaction problem. The small-scale vortices distributed along the interface are clearly identifiable in the Schlieren images of density $\rho$ and density-gradient magnitude $\lVert\bm{\nabla}\rho\rVert$. During the evolution of the shock–bubble interaction, both the shock wave and the material interfaces are well resolved without significant spurious oscillations. Owing to the difference in sound speed between the interior gas (air) and the ambient gas (Kr), the shock wave propagates faster within the cylinder. This can be understood from gas dynamics that the shock propagation speed scales with the speed of sound in the pre-shock region, modified by an appropriate factor. Upon impact, the cylindrical interface flattens, and the baroclinic torque generates a jet that subsequently pierces through the cylinder, as revealed by the streamlines in figures~\ref{fug9a} and~\ref{fug9}.

\subsection{Analyzing vortex structures based on kinematic vorticity decompositions}
\begin{figure}[t]
	\centering
	\subfigure[$Q$]{
		\begin{minipage}[t]{0.49\linewidth}
			\centering
			\includegraphics[width=1.0\columnwidth,trim={0cm 0cm 0cm 0cm},clip]{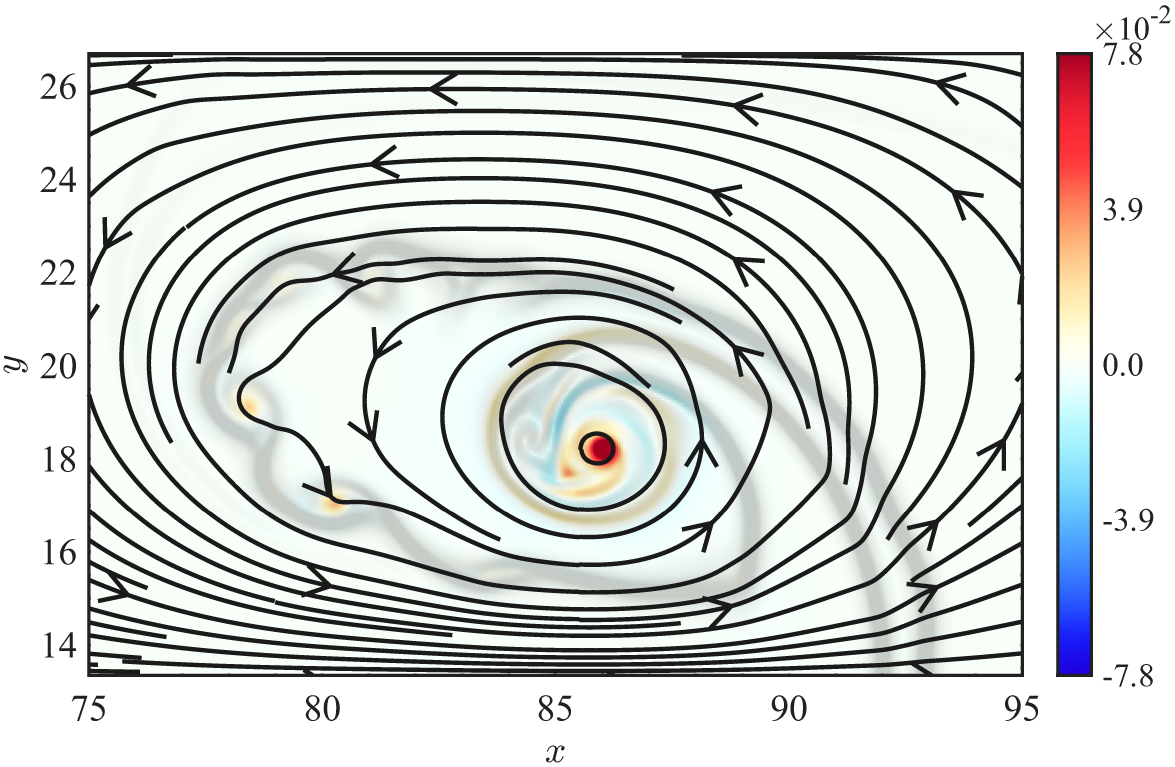}%
			% \vspace{-19.6cm}  left down  right up
			\label{221}
		\end{minipage}%
	}
	\subfigure[$\frac{1}{4}{R}^2(\bm{t})$]{
		\begin{minipage}[t]{0.49\linewidth}
			\centering
			\includegraphics[width=1.0\columnwidth,trim={0cm 0cm 0cm 0cm},clip]{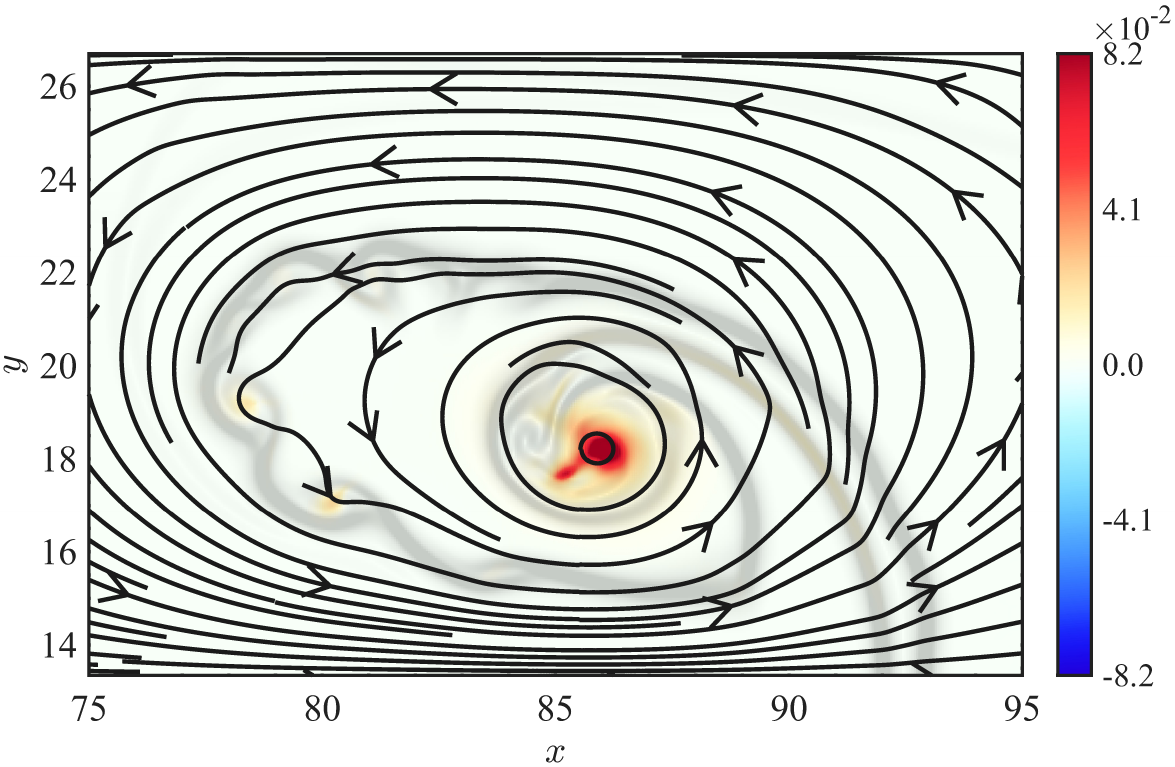}
			%\vspace{-19.6cm} 
			\label{222}
		\end{minipage}%
	}	
	
	\subfigure[$\frac{1}{2}\bm{R}(\bm{t})\cdot \bm{s}(\bm{t})$]{
		\begin{minipage}[t]{0.49\linewidth}
			\centering
			\includegraphics[width=1.0\columnwidth,trim={0cm 0cm 0cm 0cm},clip]{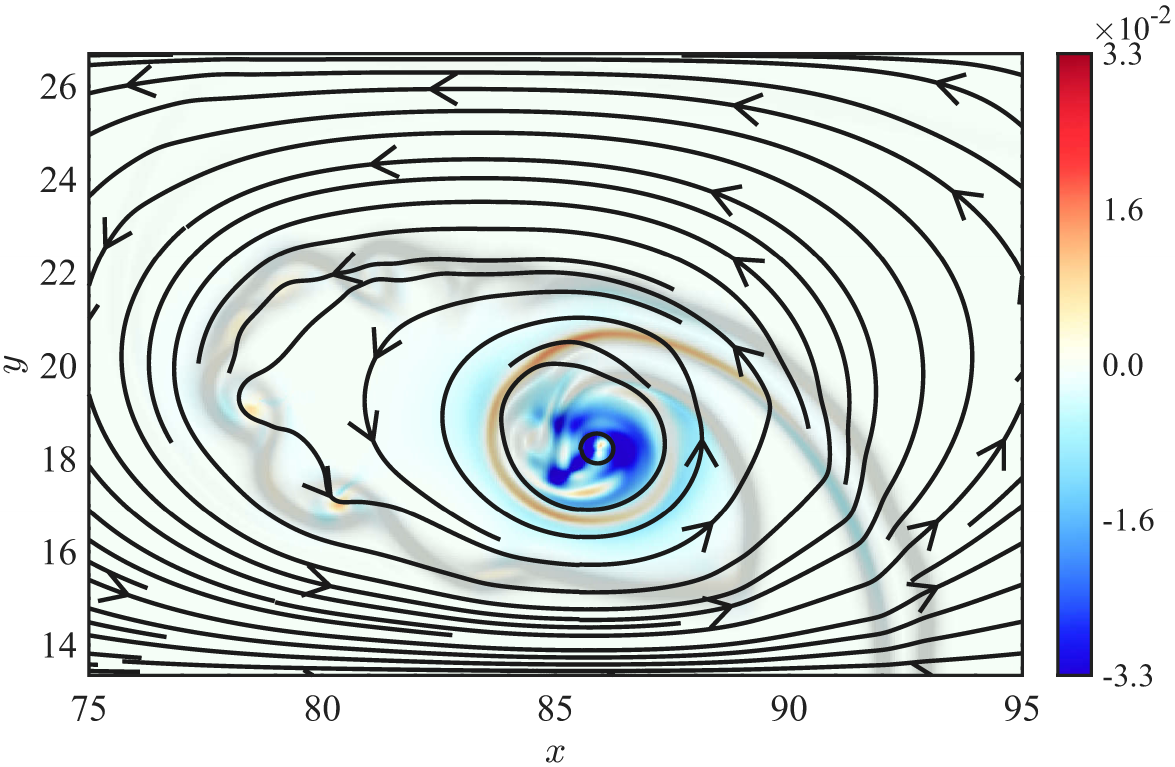}
			%   \vspace{-6.6cm} 
			\label{223}
		\end{minipage}%
	}	% left down  right up
	\subfigure[${\chi}(\bm{t}){\chi}(\bm{n})$]{
		\begin{minipage}[t]{0.49\linewidth}
			\centering
			\includegraphics[width=1.0\columnwidth,trim={0cm 0cm 0cm 0cm},clip]{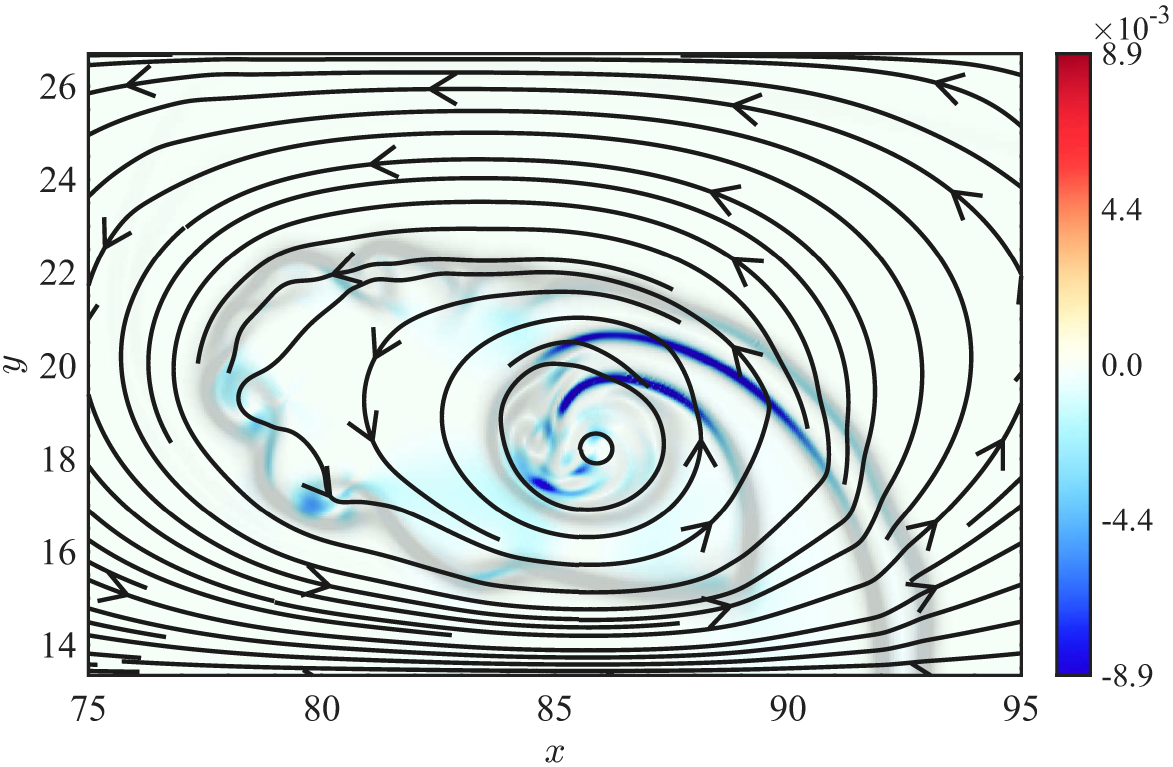}
			%   \vspace{-6.6cm} 
			\label{224}
		\end{minipage}%
	}	
	\caption{Normalized snapshots of (a) $Q$, (b) ${R}^2(\bm{t})/4$, (c) $\bm{R}(\bm{t})\bm{\cdot}\bm{s}(\bm{t})/2$, and (d) ${\chi}(\bm{t}){\chi}(\bm{n})$ for shock-accelerated air bubble evolution at $t^{*}=174.59$ (point E).} 
	\label{Q_decomposition_2}
\end{figure}
\begin{figure}[t]
	\centering
		\subfigure[]{
		\begin{minipage}[t]{0.49\linewidth}
			\centering
			\includegraphics[width=1.0\columnwidth,trim={0cm 0cm 0cm 0cm},clip]{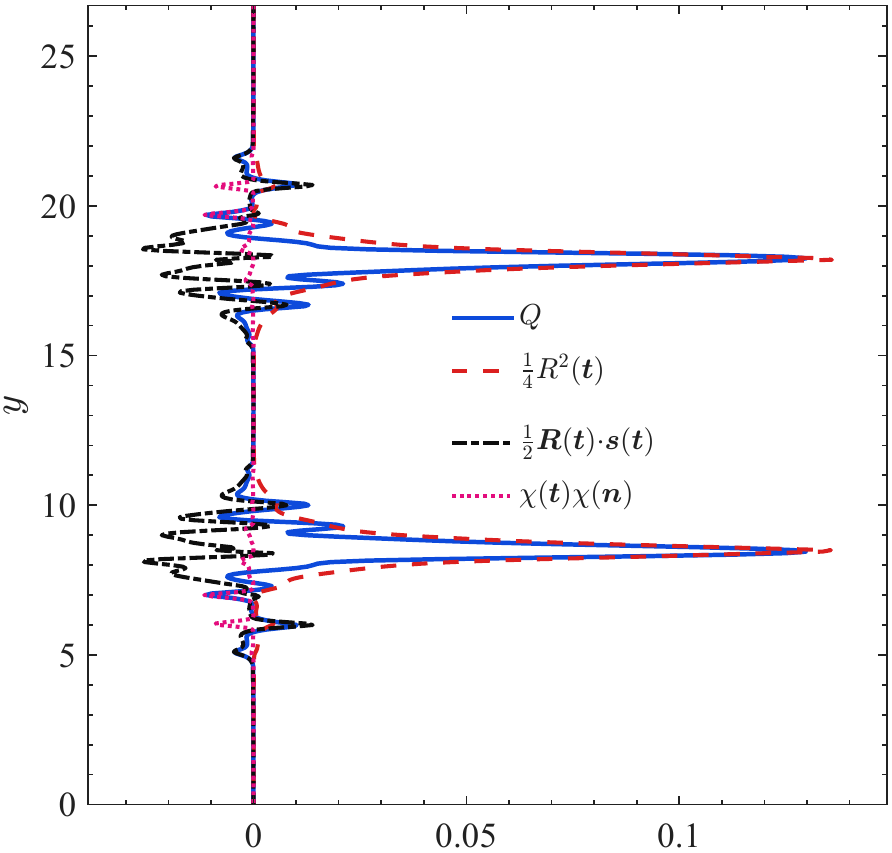}			\label{Q_profile_comparison}
		\end{minipage}%
	}	
	
	\subfigure[
	]{
		\begin{minipage}[t]{0.49\linewidth}
			\centering
			\includegraphics[width=1.0\columnwidth,trim={0cm 0cm 0cm 0cm},clip]{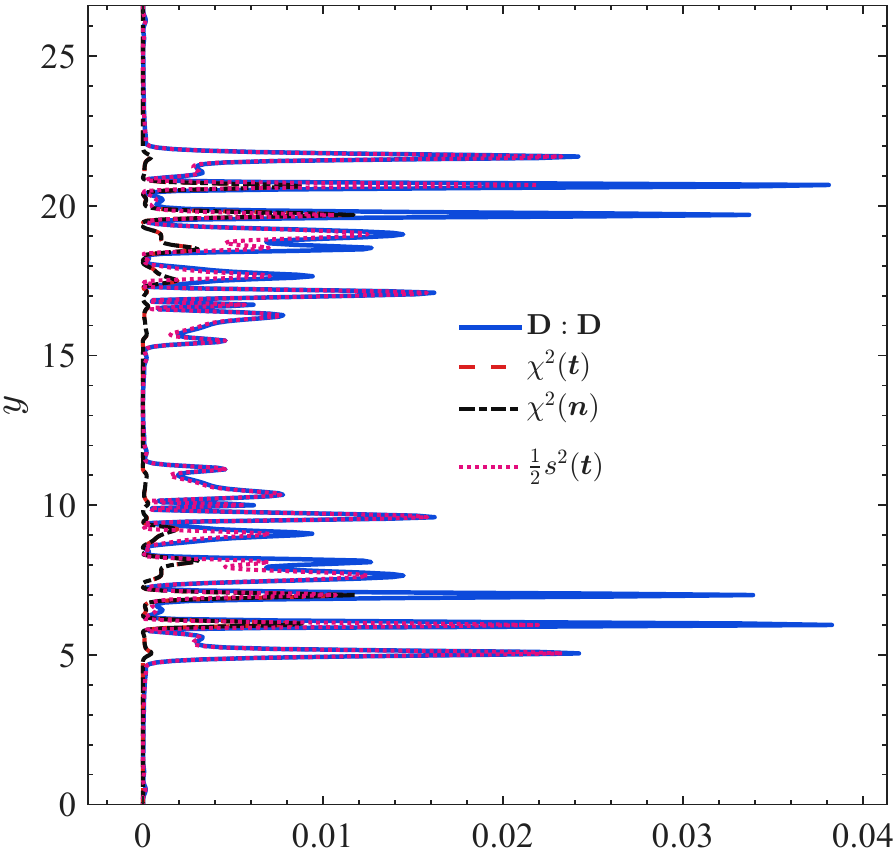}
			\label{DD_profile_comparison}
		\end{minipage}%
	}
	\subfigure[]{
		\begin{minipage}[t]{0.49\linewidth}
			\centering
			\includegraphics[width=1.0\columnwidth,trim={0cm 0cm 0cm 0cm},clip]{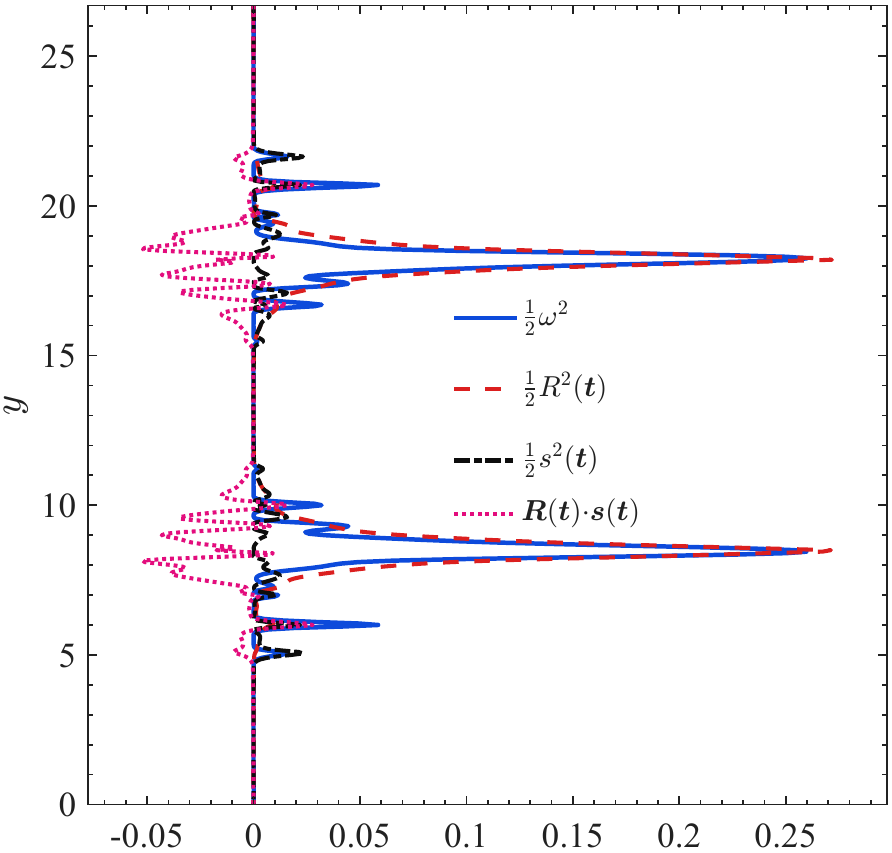}			\label{enstro_profile_comparison}
		\end{minipage}%
	}	
	\caption{Decompositions of (a) $Q$-criterion, (b) squared strain rate $\mathbf{D}\bm{:}\mathbf{D}$, and (c) enstrophy $\omega^2/2$ along the line $x^{*}=86$ for the shock-accelerated air bubble evolution at $t^{*}=174.59$ (point E).} 
	\label{Decomposition_Q_enstrophy_squared_strain_rate}
\end{figure}

Figures~\ref{fug9a} and~\ref{fug9} demonstrate the normalized snapshots of the $\bm{e}_{z}$-components of the DVD modes $(\bm{R}(\bm{t}),\bm{s}(\bm{t}))$, the vorticity $\omega_{z}$, and the IVD modes $(\bm{R}_{N}^{+},\bm{R}_{N}^{-})$ in a reference frame translating with instantaneous velocity of the vortex pair, at the relatively late times $t^{*}=84.39$ and $174.59$, respectively. At each instant, the orbital-rotation mode $\bm{R}(\bm{t})\bm{\cdot}\bm{e}_{z}$ exhibits strong swirling concentration inside the distinct cores of the primary vortex pair and within the secondary smaller vortices distributed along the deformed bubble interface (figures~\ref{Rotational_Vorticity_R1}
and \ref{Rotational_Vorticity_R}). Inside the inner region of the primary vortex, the vorticity-mode pattern is dominated by opposite signs of $\bm{R}(\bm{t})\bm{\cdot}\bm{e}_{z}$ and $\bm{s}(\bm{t})\bm{\cdot}\bm{e}_{z}$, which is identified as an anti-KKB configuration (figures~\ref{Shear_Vorticity_s1} and~\ref{Shear_Vorticity_s}). This behavior resembles that observed in a planar interface subjected to a single-mode initial perturbation.  The sum of these two vorticity modes yields $\omega_{z}$, which is predominantly governed by $\bm{R}(\bm{t})\bm{\cdot}\bm{e}_{z}$ (figures~\ref{Total_Vorticity_omega_z1} and~\ref{Total_Vorticity_omega_z}). Notably, the KKB configuration, where $\bm{R}(\bm{t})\bm{\cdot}\bm{e}_{z}$ and $\bm{s}(\bm{t})\bm{\cdot}\bm{e}_{z}$ share the same sign, is observed at the periphery of the primary vortex core and inside the secondary rolling-up vortices. The formation of small axial vortices distributed along the interface is driven by the ongoing KHI. As shown in figures~\ref{Liutex_Magnitude_R_Liutex1},~\ref{R_N_minus1},~\ref{Liutex_Magnitude_R_Liutex}, and~\ref{R_N_minus}, the cores of axial vortices are well captured by both Liutex $\bm{R}_{N}^{+}$~\citep{Liu2018}, and $\bm{R}_{N}^{-}$~\citep{Chen2025arxiv} in regions where $\Delta>0$. However, the characteristic rigid-rotation mode could also exist in sheet-like vortices attached to the interface, which is captured only by $\bm{R}_{N}^{-}\bm{\cdot}\bm{e}_{z}$ (figures~\ref{R_N_minus1} and~\ref{R_N_minus}).
Similar to the single-mode perturbed interface, $\bm{R}_{N}^{-}\bm{\cdot}\bm{e}_{z}$ yields larger vortex-core sizes than $\bm{R}_{N}^{+}\bm{\cdot}\bm{e}_{z}$, whose pattern is more comparable to $\bm{R}(\bm{t})\bm{\cdot}\bm{e}_{z}$.
Moreover, as shown in figures~\ref{R compare} and~\ref{s compare}, a quantitative comparison along the line $x^{*}=86$ at $t^*=174.59$ further verifies that the DVD vorticity modes $(\bm{R}(\bm{t}),\bm{s}(\bm{t}))$ are bounded by the IVD vorticity modes $(\bm{R}_{N}^{\pm},\bm{s}_{N}^{\pm})$ in the conventional vortical-flow region $\Delta>0$. Overall, the distribution of $\bm{R}(\bm{t})\bm{\cdot}\bm{e}_{z}$ is smoother than that of the IVD modes $\bm{R}_{N}^{\pm}\bm{\cdot}\bm{e}_{z}$ , with no sharp peaks appearing outside the main vortex core. As shown in figure~\ref{R_S_Omega_z_bubble}, the vorticity $\omega_{z}$ is dominated by $\bm{R}(\bm{t})\bm{\cdot}\bm{e}_{z}$ and essentially attenuated by $\bm{s}(\bm{t})\bm{\cdot}\bm{e}_{z}$. The IVD and DVD vorticity modes provide complementary descriptions of vorticity kinematics at different levels for a generic flow field~\citep{Chen2025arxiv}. The IVD modes, representing vorticity extrema in the plane normal to the vortex axis, can be evaluated to identify the primary features of flow structures at the characteristic algebraic level, albeit with some loss of flow details. In contrast, the DVD modes offer a more detailed description based on material and field perspectives, offering deeper physical insight into the flow details.

Figures~\ref{Delta0-1} and~\ref{Delta1-1} present the evolution histories of the vorticity fluxes $(\Gamma_{\rm{\omega}},\Gamma_{\rm{R}},\Gamma_{\rm{s}})$ during the shock-driven evolution of an air bubble. In contrast to the case of a single-mode perturbed interface (\S\ref{RMI of a single-mode perturbed interface}), where vorticity deposition is primarily induced by RMI during the shock-bubble interaction, the present scenario shows distinct characteristics dominated by shear-driven KHI in the late stages.
As illustrated in figure~\ref{Delta0-1}, when restricted to regions where $\Delta>0$, $\Gamma_{\rm{\omega}}$ is dominated by $\Gamma_{\rm{R}}$ and only weakly modulated by $\Gamma_{\rm{s}}$. Early in the evolution, sheet-like vortices are created and accumulate along the interface due to the baroclinic torque $\bm{\nabla}\rho\times\bm{\nabla}p$, which results in a rapid increase of 
$\Gamma_{\rm{\omega}}$ and $\Gamma_{\rm{R}}$ (figures~\ref {and t14_55} and~\ref{and t29_10}). The magnitude of $\Gamma_{\rm{s}}$ is obviously lower than that of $\Gamma_{\rm{R}}$, owing to mutual cancellation between positive and negative spin layers attached to the interface. Both orbital-rotation and spin modes are also observed in the wave structures near the bubble where $\Delta<0$. Unlike the single-mode interface case, the overall evolution at this stage is essentially unaffected by the modulation of the shock wave (figure~\ref{Delta1-1}). Following the onset of KHI, the initially attached sheet-like vortices gradually roll up into many secondary vortices along the interface, as well as into mushroom-shaped structures. This process redistributes the pre-existing vorticity between the two vorticity modes, and simultaneously promotes the generation of new vorticity along the interface, as illustrated in figures~\ref{and t84_39},~\ref{and t130_95}, and~\ref{and t174_5}. Compared with figure~\ref{Delta0-1}, $\Gamma_{\omega}$ in figure~\ref{Delta1-1} exhibits the same evolutionary trend but with a higher amplitude. This is because some strong interface vorticity exists in the region $\Delta<0$, which is excluded under the restriction $\Delta>0$. The total vorticity is generated primarily before point B and remains conserved thereafter. Notably, at points $C$, $D$, and $E$, both $\Gamma_{\rm{R}}$ and $\Gamma_{\rm{s}}$ exhibit obvious anti-KKB configurations over most regions external to the vortex core. In fact, owing to the very weak vorticity there, these regions can be approximately treated as a potential-flow region, where the associated velocity field is induced by the interface-deposited vorticity via the Biot-Savart law. Equivalently, the induced velocity field outside the interface can be approximated as the superposition of point-vortex solutions distributed along the interface~\citep{jacobs1996experimental}. It then follows that $\bm{R}(\bm{t})\approx-\bm{s}(\bm{t})$, which corresponds to an anti-KKB configuration but is fundamentally different from that observed inside the primary vortex cores. This induction mechanism becomes increasingly evident after point $B$ where the variations of $\Gamma_{\rm{R}}$ and $\Gamma_{\rm{s}}$ are approximately anti-symmetric.

Figure~\ref{Q_decomposition_2} illustrates the normalized snapshots of the second principal invariant $Q$ of the VGT, along with its physical constituents as defined in~\eqref{Q38}.
Compared to the vortex core identified using $R^2(\bm{t})/4$ (figure~\ref{222}), the region with positive $Q$ is notably smaller, indicating that the $Q$-criterion yields a more compact representation of the coherent vortex core. As displayed in figure~\ref{223}, the orbital-spin coupling term $\bm{R}(\bm{t})\bm{\cdot}\bm{s}(\bm{t})/2$ exhibits a spatially alternating positive-negative distribution, suggesting the coexistence of both KKB and anti-KKB configurations for the DVD vorticity modes. Specifically, within the inner region of the vortex core, this coupling term is predominantly negative, implying a locally dominant anti-KKB configuration characterized by opposite signs of $\bm{R}(\bm{t})$ and $\bm{s}(\bm{t})$. In contrast, an interface-deposited annular vortex band with a distinctly positive coupling term emerges clearly at the periphery of the vortex core, which can be attributed to the presence of KKB configuration where $\bm{R}(\bm{t})$ and $\bm{s}(\bm{t})$ share the same sign.
Meanwhile, the product of the relative stretching rates, $\chi(\bm{t})\chi(\bm{n})$, contributes negatively in the peripheral sheet-like spin layer, as depicted in figure~\ref{224}. Figure~\ref{Decomposition_Q_enstrophy_squared_strain_rate} provides quantitative comparisons of various contributions to enstrophy $\omega^{2}/2$ (equation~\eqref{FS37b}), squared strain rate $\mathbf{D}\bm{:}\mathbf{D}$ (equation~\eqref{FS37c}), and $Q$ (equation~\eqref{Q38}) along the line $x^{*}=86$ connecting the vortex pair. In figure~\ref{Q_profile_comparison}, $Q$ is dominated by $R^{2}(\bm{t})/4$ and attenuated by $\bm{R}(\bm{t})\bm{\cdot}\bm{s}(\bm{t})/2$ in the vortex-pair region, while the contribution of $\chi(\bm{t})\chi(\bm{n})$ to $Q$ is negligible. The squared strain rate $\mathbf{D}\bm{:}\mathbf{D}$ is primarily governed by $s^{2}(\bm{t})/2$ (figure~\ref{DD_profile_comparison}), whereas the contribution of $s^{2}(\bm{t})/2$ to $\omega^2/2$ is insignificant when compared to  $R^{2}(\bm{t})/2$ and $\bm{R}(\bm{t})\bm{\cdot}\bm{s}(\bm{t})$ (figure~\ref{enstro_profile_comparison}).

\section{Conclusions and discussions}\label{Conclusions and discussions}
In this study, direct numerical simulations of two canonical Richtmyer-Meshkov instability (RMI) flows are carried out using a hybrid HOWD approach: shock interaction with a single-mode perturbed interface (Example Flow I), and a cylindrical air bubble immersed in Krypton (Example Flow II). The key contributions that extend beyond the existing vorticity-based paradigm in the community are summarized as follows.
\subsection{Vorticity and vortex dynamics: interface, vortex, and shock wave}
\begin{itemize}
	\item For both example flows, vorticity deposition due to the baroclinic torque $(\bm{\nabla}\rho\times\bm{\nabla}p)$ occurs primarily during the period of shock-interface interaction, which arises from the misalignment of density and pressure gradients. This serves as the primary mechanism driving the evolution and formation of mushroom-shaped roll-up structures. The total vorticity flux $\Gamma_{\omega}$ in the upper half-domain remains approximately conserved after the initial generation. In the region $\Delta>0$, the fluctuation of $\Gamma_{\omega}$ originates from the relatively weak dilatation-vorticity coupling and baroclinic torque in the intermediate and late times. In contrast to Example Flow I, Example Flow II exhibits secondary vortices distributed along the interface, driven by the shear-induced Kelvin–Helmholtz instability (KHI).
	\item For the first time, the recently proposed streamline-based direction-dependent vorticity decomposition (DVD)~\citep{chen2025kinematic}
	is applied to analyze the shock-driven flow physics in canonical RMI flows, where the vorticity $\bm{\omega}$ is decomposed into the sum of the orbital-rotation mode $\bm{R}(\bm{t})$ and the spin mode $\bm{s}(\bm{t})$. 
    Both KKB and anti-KKB configurations of the vorticity modes are observed within a single flow field, where $\bm{R}(\bm{t})$ and $\bm{s}(\bm{t})$ exhibit the same sign and opposite signs, respectively. The DVD vorticity modes are quantitatively compared with the IVD vorticity modes, namely, $\bm{R}_{N}^{+}$ (Liutex)~\citep{Liu2018} and $\bm{R}_{N}^{-}$~\citep{Chen2025arxiv}, in the conventional vortical-flow region satisfying $\Delta>0$. The cores of axial vortices deposited at interface are well captured by both $\bm{R}_{N}^{+}$ and $\bm{R}_{N}^{-}$. Nevertheless, the spatial distribution of $\bm{R}_{N}^{-}$ is closer to that of $\bm{R}(\bm{t})$, which shows larger size of vortex cores than Liutex $\bm{R}_{N}^{+}$. The sheet-like vortices at the periphery of the primary vortices are captured only by $\bm{R}_{N}^{-}$ and $\bm{R}(\bm{t})$. The DVD vorticity modes are bounded by the IVD vorticity modes, which agrees with the theoretical analysis in~\citet{Chen2025arxiv}.
	\item To quantify the global flow behaviors, we propose a double decomposition of the vorticity flux  $\Gamma_{\omega}=\Gamma_{\rm{R}}+\Gamma_{\rm{s}}$, where $\Gamma_{\rm{R}}$ is the orbital-rotation flux, and $\Gamma_{\rm{s}}$ is the spin flux. The upper and lower bounds of $\Gamma_{\rm{R}}$ and $\Gamma_{\rm{s}}$ are explicitly determined using the IVD vorticity modes. In the early evolution of Example Flow I, when restricting the integration domain within $\Delta>0$, it is found that $\Gamma_{\omega}$ is dominated by the positive orbital-rotation flux $\Gamma_{\rm{R}}$ and enhanced by the positive $\Gamma_{\rm{s}}$ (a behavior termed KKB synergy effect) which arises from the rolling-up of the sheet-like spin layer attached to the interface.
	However, without the restriction of $\Delta>0$, we observe a completely opposite anti-KKB antagonistic effect, characterized by positive $\Gamma_{\rm{s}}$ and negative $\Gamma_{\rm{R}}$. This configuration is attributed to the interference from high concentrations of the two vorticity modes associated with the reflected and transmitted shock waves. In contrast, Example Flow II remains largely unaffected by shock-wave modulation and is governed primarily by the coupling-driven mechanism of RMI and subsequent KHI.
	\item The analysis reveals a three-layer structure of the pair of primary vortices formed at the interface. For each primary vortex, the inner core region (where $\Delta>0$) is identified as anti-KKB configuration, which resembles the classical Burgers vortex solution in terms of both the cross-sectional velocity field and the distributions of vorticity modes. In this region, vorticity $\bm{\omega}$ is dominated by $\bm{R}(\bm{t})$ and essentially attenuated by $\bm{s}(\bm{t})$. Notably, the KKB configuration is observed at the annular vortex band attached to the air-bubble interface, severing as the periphery of the primary vortex core, which drives further rolling up of the air-bubble interface. External to this region, the velocity field approaches the potential flow regime, induced by the interface-deposited vorticity via the Biot-Savart law. In this outer region, the vorticity modes exhibit an anti-KKB configuration satisfying  $\bm{R}(\bm{t})\approx-\bm{s}(\bm{t})$.
	\item Using the DVD framework, the well known $Q$-criterion is decomposed into three contributions: $R^{2}(\bm{t})/4$ (half the  Using the DVD, the well known $Q$-criterion is decomposed as the sum of $R^{2}(\bm{t})/4$ (half of the enstrophy contributed by the orbital-rotation mode), $\bm{R}(\bm{t})\bm{\cdot}\bm{s}(\bm{t})/2$ (the orbital-spin coupling), and $\chi(\bm{t})\chi(\bm{n})$ (the product of the relative stretching rates). Consistent with the three-layer structure of the primary vortex pair, $\bm{R}(\bm{t})\bm{\cdot}\bm{s}(\bm{t})/2$ is negative within the inner vortex cores following the anti-KKB configuration, thereby attenuating the positive contribution from $R^{2}(\bm{t})/4$. This coupling term becomes distinctly positive in the interface-deposited annular vortex band at the edge of the inner core due to the KKB configuration, and then negative outside this region. The stretching term $\chi(\bm{t})\chi(\bm{n})$ is negligible in the inner vortex cores, which however, becomes significant in the spiral arms of the annular vortex band.
\end{itemize}

\subsection{Outlook and future work}
In future work, the proposed approach will be extended to investigate vorticity and vortex dynamics associated with RMI and coupling-driven shock-interface interaction in more complex systems. These include single gas cylinders with varying geometries~\citep{singh2023shock}, multiple-interface configurations involving light/heavy bubble sequences~\citep{Alsaeed2025,Singh2024ScienceChina}, the effects of 3D interface curvature and rippled shock wave~\citep{zhai2017review}, and the mixing mechanism of compressible turbulence~\citep{ranjan2011shock,zhou2017rayleigh,Zhou2024}.
The influence of key dimensionless parameters on the spatiotemporal evolution of the two vorticity modes will be systematically examined from both structural and statistical perspectives. In particular, we aim to elucidate how variations in interface geometry, shock strength, and Atwood number modulate the competition and synergy between the orbital-rotation and spin modes. This line of research will advance the fundamental understanding and modeling of the interplay among interface geometry, vortex organization, and background wave fields, while also providing robust diagnostic tools for practical engineering applications, such as inertial confinement fusion and RMI-induced supersonic mixing~\citep{zhou2025instabilities}.

\section*{Acknowledgements}
This work was funded by the National Natural Science
Foundation of China (Grant No. 12402262). The authors are grateful to Zhigang Zhai from University of Science and Technology of China for valuable discussions.

\section*{AUTHOR DECLARATIONS Conflict of Interest} 
The authors have no conflicts to disclose.

\section*{Author Contributions} 
\textbf{Xi Chen}: Formal analysis; Investigation; 
Visualization; Writing – original draft; Writing – review
\& editing.
\textbf{Tao Chen}: Conceptualization; Formal analysis; Funding
acquisition; Investigation; Methodology; Resources; Supervision; Writing – original draft; Writing – review \& editing. \textbf{Tianshu Liu}: Writing – review \& editing. 

\section*{Data availability} 
Data will be made available on request. 

\appendix
%\section{Chapman-Enskog analysis of the lattice Boltzmann model}\label{Chapman-Enskog analysis}

%\newpage
\bibliography{ChenPOFref}% Produces the bibliography via BibTeX.

%aipnauth4-2.bst 2018-12-27 (MD) hand-edited version of apsauth4-1.bst
%Control: author (9) reversed initials
\begin{thebibliography}{72}%
\makeatletter
\providecommand \@ifxundefined [1]{%
 \@ifx{#1\undefined}
}%
\providecommand \@ifnum [1]{%
 \ifnum #1\expandafter \@firstoftwo
 \else \expandafter \@secondoftwo
 \fi
}%
\providecommand \@ifx [1]{%
 \ifx #1\expandafter \@firstoftwo
 \else \expandafter \@secondoftwo
 \fi
}%
\providecommand \natexlab [1]{#1}%
\providecommand \enquote  [1]{``#1''}%
\providecommand \bibnamefont  [1]{#1}%
\providecommand \bibfnamefont [1]{#1}%
\providecommand \citenamefont [1]{#1}%
\providecommand \href@noop [0]{\@secondoftwo}%
\providecommand \href [0]{\begingroup \@sanitize@url \@href}%
\providecommand \@href[1]{\@@startlink{#1}\@@href}%
\providecommand \@@href[1]{\endgroup#1\@@endlink}%
\providecommand \@sanitize@url [0]{\catcode `\\12\catcode `\$12\catcode
  `\&12\catcode `\#12\catcode `\^12\catcode `\_12\catcode `\%12\relax}%
\providecommand \@@startlink[1]{}%
\providecommand \@@endlink[0]{}%
\providecommand \url  [0]{\begingroup\@sanitize@url \@url }%
\providecommand \@url [1]{\endgroup\@href {#1}{\urlprefix }}%
\providecommand \urlprefix  [0]{URL }%
\providecommand \Eprint [0]{\href }%
\providecommand \doibase [0]{https://doi.org/}%
\providecommand \selectlanguage [0]{\@gobble}%
\providecommand \bibinfo  [0]{\@secondoftwo}%
\providecommand \bibfield  [0]{\@secondoftwo}%
\providecommand \translation [1]{[#1]}%
\providecommand \BibitemOpen [0]{}%
\providecommand \bibitemStop [0]{}%
\providecommand \bibitemNoStop [0]{.\EOS\space}%
\providecommand \EOS [0]{\spacefactor3000\relax}%
\providecommand \BibitemShut  [1]{\csname bibitem#1\endcsname}%
\let\auto@bib@innerbib\@empty
%</preamble>
\bibitem [{\citenamefont {Alsaeed}\ and\ \citenamefont
  {Singh}(2025{\natexlab{a}})}]{Alsaeed_Singh_PhysicaD2025}%
  \BibitemOpen
  \bibfield  {author} {\bibinfo {author} {\bibnamefont {Alsaeed}, \bibfnamefont
  {S.~S.}}and\ \bibinfo {author} {\bibnamefont {Singh}, \bibfnamefont {S.}},\
  }\bibfield  {title} {\enquote {\bibinfo {title} {Insights into coupling
  effects of double light square bubbles on shocked hydrodynamic
  instability},}\ }\href@noop {} {\bibfield  {journal} {\bibinfo  {journal}
  {Physica D}\ }\textbf {\bibinfo {volume} {476}},\ \bibinfo {pages} {134646}
  (\bibinfo {year} {2025}{\natexlab{a}})}\BibitemShut {NoStop}%
\bibitem [{\citenamefont {Alsaeed}\ and\ \citenamefont
  {Singh}(2025{\natexlab{b}})}]{Alsaeed2025}%
  \BibitemOpen
  \bibfield  {author} {\bibinfo {author} {\bibnamefont {Alsaeed}, \bibfnamefont
  {S.~S.}}and\ \bibinfo {author} {\bibnamefont {Singh}, \bibfnamefont {S.}},\
  }\bibfield  {title} {\enquote {\bibinfo {title} {{Shock-driven flow physics
  of Richtmyer–Meshkov instability in tandem light cylindrical
  interfaces}},}\ }\href@noop {} {\bibfield  {journal} {\bibinfo  {journal}
  {Physics of Fluids}\ }\textbf {\bibinfo {volume} {37}},\ \bibinfo {pages}
  {122116} (\bibinfo {year} {2025}{\natexlab{b}})}\BibitemShut {NoStop}%
\bibitem [{\citenamefont {Balick}\ and\ \citenamefont
  {Frank}(2002)}]{balick2002shapes}%
  \BibitemOpen
  \bibfield  {author} {\bibinfo {author} {\bibnamefont {Balick}, \bibfnamefont
  {B.}}and\ \bibinfo {author} {\bibnamefont {Frank}, \bibfnamefont {A.}},\
  }\bibfield  {title} {\enquote {\bibinfo {title} {Shapes and shaping of
  {Planetary} {Nebulae}},}\ }\href@noop {} {\bibfield  {journal} {\bibinfo
  {journal} {Annual Review of Astronomy and Astrophysics}\ }\textbf {\bibinfo
  {volume} {40}},\ \bibinfo {pages} {439--486} (\bibinfo {year}
  {2002})}\BibitemShut {NoStop}%
\bibitem [{\citenamefont {Betz}(1950)}]{Betz1950}%
  \BibitemOpen
  \bibfield  {author} {\bibinfo {author} {\bibnamefont {Betz}, \bibfnamefont
  {A.}},\ }\bibfield  {title} {\enquote {\bibinfo {title} {{Wie entsteht ein
  Wirbel in einer wenig zähen Flüssigkeit}},}\ }\href@noop {} {\bibfield
  {journal} {\bibinfo  {journal} {Naturwissenschaften}\ }\textbf {\bibinfo
  {volume} {37}},\ \bibinfo {pages} {193--196} (\bibinfo {year}
  {1950})}\BibitemShut {NoStop}%
\bibitem [{\citenamefont {Burgers}(1948)}]{burgers1948mathematical}%
  \BibitemOpen
  \bibfield  {author} {\bibinfo {author} {\bibnamefont {Burgers}, \bibfnamefont
  {J.~M.}},\ }\bibfield  {title} {\enquote {\bibinfo {title} {A mathematical
  model illustrating the theory of turbulence},}\ }\href@noop {} {\bibfield
  {journal} {\bibinfo  {journal} {Advances in Applied Mechanics}\ }\textbf
  {\bibinfo {volume} {1}},\ \bibinfo {pages} {171--199} (\bibinfo {year}
  {1948})}\BibitemShut {NoStop}%
\bibitem [{\citenamefont {Chapman}\ and\ \citenamefont
  {Jacobs}(2006)}]{Chapman2006}%
  \BibitemOpen
  \bibfield  {author} {\bibinfo {author} {\bibnamefont {Chapman}, \bibfnamefont
  {P.~R.}}and\ \bibinfo {author} {\bibnamefont {Jacobs}, \bibfnamefont
  {J.~W.}},\ }\bibfield  {title} {\enquote {\bibinfo {title} {{Experiments on
  the three-dimensional incompressible Richtmyer-Meshkov instability}},}\
  }\href@noop {} {\bibfield  {journal} {\bibinfo  {journal} {Physics of
  Fluids}\ }\textbf {\bibinfo {volume} {18}},\ \bibinfo {pages} {074101}
  (\bibinfo {year} {2006})}\BibitemShut {NoStop}%
\bibitem [{\citenamefont {Chen}(2026)}]{Chen2026operator}%
  \BibitemOpen
  \bibfield  {author} {\bibinfo {author} {\bibnamefont {Chen}, \bibfnamefont
  {T.}},\ }\bibfield  {title} {\enquote {\bibinfo {title} {Kinematic vorticity
  decompositions and operator commutativity},}\ }\href@noop {} {\bibfield
  {journal} {\bibinfo  {journal} {Physics of Fluids}\ }\textbf {\bibinfo
  {volume} {38}},\ \bibinfo {pages} {027140} (\bibinfo {year}
  {2026})}\BibitemShut {NoStop}%
\bibitem [{\citenamefont {Chen}\ and\ \citenamefont
  {Liu}(2025)}]{chen2025kinematic}%
  \BibitemOpen
  \bibfield  {author} {\bibinfo {author} {\bibnamefont {Chen}, \bibfnamefont
  {T.}}and\ \bibinfo {author} {\bibnamefont {Liu}, \bibfnamefont {T.}},\
  }\bibfield  {title} {\enquote {\bibinfo {title} {Kinematic vorticity
  decomposition for generic two-dimensional flow},}\ }\href@noop {} {\bibfield
  {journal} {\bibinfo  {journal} {Physics of Fluids}\ }\textbf {\bibinfo
  {volume} {37}} (\bibinfo {year} {2025})}\BibitemShut {NoStop}%
\bibitem [{\citenamefont {Chen}\ \emph {et~al.}(2026)\citenamefont {Chen},
  \citenamefont {Wu}, \citenamefont {Mao},\ and\ \citenamefont
  {Liu}}]{Chen2025arxiv}%
  \BibitemOpen
  \bibfield  {author} {\bibinfo {author} {\bibnamefont {Chen}, \bibfnamefont
  {T.}}, \bibinfo {author} {\bibnamefont {Wu}, \bibfnamefont {J.-Z.}}, \bibinfo
  {author} {\bibnamefont {Mao}, \bibfnamefont {F.}}, and\ \bibinfo {author}
  {\bibnamefont {Liu}, \bibfnamefont {T.}},\ }\bibfield  {title} {\enquote
  {\bibinfo {title} {A general kinematic theory of fluid-element rotation and
  intrinsic vorticity decompositions},}\ }\href@noop {} {\bibfield  {journal}
  {\bibinfo  {journal} {Journal of Fluid Mechanics}\ }\textbf {\bibinfo
  {volume} {1032}},\ \bibinfo {pages} {A55} (\bibinfo {year}
  {2026})}\BibitemShut {NoStop}%
\bibitem [{\citenamefont {Chong}, \citenamefont {Perry},\ and\ \citenamefont
  {Cantwell}(1990)}]{chong1990general}%
  \BibitemOpen
  \bibfield  {author} {\bibinfo {author} {\bibnamefont {Chong}, \bibfnamefont
  {M.~S.}}, \bibinfo {author} {\bibnamefont {Perry}, \bibfnamefont {A.~E.}},
  and\ \bibinfo {author} {\bibnamefont {Cantwell}, \bibfnamefont {B.~J.}},\
  }\bibfield  {title} {\enquote {\bibinfo {title} {A general classification of
  three-dimensional flow field},}\ }\href@noop {} {\bibfield  {journal}
  {\bibinfo  {journal} {Physics of Fluids}\ }\textbf {\bibinfo {volume} {2}},\
  \bibinfo {pages} {765} (\bibinfo {year} {1990})}\BibitemShut {NoStop}%
\bibitem [{\citenamefont {Ding}\ \emph {et~al.}(2017)\citenamefont {Ding},
  \citenamefont {Si}, \citenamefont {Chen}, \citenamefont {Zhai}, \citenamefont
  {Lu},\ and\ \citenamefont {Luo}}]{Ding2017JFM828}%
  \BibitemOpen
  \bibfield  {author} {\bibinfo {author} {\bibnamefont {Ding}, \bibfnamefont
  {J.}}, \bibinfo {author} {\bibnamefont {Si}, \bibfnamefont {T.}}, \bibinfo
  {author} {\bibnamefont {Chen}, \bibfnamefont {M.}}, \bibinfo {author}
  {\bibnamefont {Zhai}, \bibfnamefont {Z.}}, \bibinfo {author} {\bibnamefont
  {Lu}, \bibfnamefont {X.}}, and\ \bibinfo {author} {\bibnamefont {Luo},
  \bibfnamefont {X.}},\ }\bibfield  {title} {\enquote {\bibinfo {title} {On the
  interaction of a planar shock with a three-dimensional light gas cylinder},}\
  }\href@noop {} {\bibfield  {journal} {\bibinfo  {journal} {Journal of Fluid
  Mechanics}\ }\textbf {\bibinfo {volume} {828}},\ \bibinfo {pages} {289--317}
  (\bibinfo {year} {2017})}\BibitemShut {NoStop}%
\bibitem [{\citenamefont {Fan}\ \emph {et~al.}(2012)\citenamefont {Fan},
  \citenamefont {Zhai}, \citenamefont {Si}, \citenamefont {Luo}, \citenamefont
  {Zou},\ and\ \citenamefont {Tan}}]{fan2012numerical}%
  \BibitemOpen
  \bibfield  {author} {\bibinfo {author} {\bibnamefont {Fan}, \bibfnamefont
  {M.}}, \bibinfo {author} {\bibnamefont {Zhai}, \bibfnamefont {Z.}}, \bibinfo
  {author} {\bibnamefont {Si}, \bibfnamefont {T.}}, \bibinfo {author}
  {\bibnamefont {Luo}, \bibfnamefont {X.}}, \bibinfo {author} {\bibnamefont
  {Zou}, \bibfnamefont {L.}}, and\ \bibinfo {author} {\bibnamefont {Tan},
  \bibfnamefont {D.}},\ }\bibfield  {title} {\enquote {\bibinfo {title}
  {{Numerical study on the evolution of the shock-accelerated SF6 interface:
  Influence of the interface shape}},}\ }\href@noop {} {\bibfield  {journal}
  {\bibinfo  {journal} {Science China Physics, Mechanics and Astronomy}\
  }\textbf {\bibinfo {volume} {55}},\ \bibinfo {pages} {284--296} (\bibinfo
  {year} {2012})}\BibitemShut {NoStop}%
\bibitem [{\citenamefont {Fraley}(1986)}]{Fraley1986}%
  \BibitemOpen
  \bibfield  {author} {\bibinfo {author} {\bibnamefont {Fraley}, \bibfnamefont
  {G.}},\ }\bibfield  {title} {\enquote {\bibinfo {title} {{Rayleigh-Taylor
  stability for a normal shock wave-density discontinuity interaction}},}\
  }\href@noop {} {\bibfield  {journal} {\bibinfo  {journal} {Phys. Fluids}\
  }\textbf {\bibinfo {volume} {29}},\ \bibinfo {pages} {376--386} (\bibinfo
  {year} {1986})}\BibitemShut {NoStop}%
\bibitem [{\citenamefont {Haines}(2016)}]{haines2016detailed}%
  \BibitemOpen
  \bibfield  {author} {\bibinfo {author} {\bibnamefont {Haines}, \bibfnamefont
  {B.~M.}},\ }\bibfield  {title} {\enquote {\bibinfo {title} {{Detailed
  high-resolution three-dimensional simulations of OMEGA separated reactants
  inertial confinement fusion experiments}},}\ }\href@noop {} {\bibfield
  {journal} {\bibinfo  {journal} {Physics of Plasmas}\ }\textbf {\bibinfo
  {volume} {23}},\ \bibinfo {pages} {072709} (\bibinfo {year}
  {2016})}\BibitemShut {NoStop}%
\bibitem [{\citenamefont {Hejazialhosseini}, \citenamefont {Rossinelli},\ and\
  \citenamefont {Koumoutsakos}(2013)}]{Hejazialhosseini2013}%
  \BibitemOpen
  \bibfield  {author} {\bibinfo {author} {\bibnamefont {Hejazialhosseini},
  \bibfnamefont {B.}}, \bibinfo {author} {\bibnamefont {Rossinelli},
  \bibfnamefont {D.}}, and\ \bibinfo {author} {\bibnamefont {Koumoutsakos},
  \bibfnamefont {P.}},\ }\bibfield  {title} {\enquote {\bibinfo {title}
  {{Vortex dynamics in 3D shock-bubble interaction}},}\ }\href@noop {}
  {\bibfield  {journal} {\bibinfo  {journal} {Physics of Fluids}\ }\textbf
  {\bibinfo {volume} {25}},\ \bibinfo {pages} {110816} (\bibinfo {year}
  {2013})}\BibitemShut {NoStop}%
\bibitem [{\citenamefont {von Helmholtz}(1868)}]{Helmholtz1868}%
  \BibitemOpen
  \bibfield  {author} {\bibinfo {author} {\bibnamefont {von Helmholtz},
  \bibfnamefont {H.}},\ }\bibfield  {title} {\enquote {\bibinfo {title} {On
  discontinuous movements of fluid},}\ }\href@noop {} {\bibfield  {journal}
  {\bibinfo  {journal} {Philosophical Magazine}\ }\textbf {\bibinfo {volume}
  {36}},\ \bibinfo {pages} {337} (\bibinfo {year} {1868})}\BibitemShut
  {NoStop}%
\bibitem [{\citenamefont {Hunt}, \citenamefont {Wray},\ and\ \citenamefont
  {Moin}(1988)}]{hunt1988eddies}%
  \BibitemOpen
  \bibfield  {author} {\bibinfo {author} {\bibnamefont {Hunt}, \bibfnamefont
  {J.~C.~R.}}, \bibinfo {author} {\bibnamefont {Wray}, \bibfnamefont {A.~A.}},
  and\ \bibinfo {author} {\bibnamefont {Moin}, \bibfnamefont {P.}},\
  }\href@noop {} {\enquote {\bibinfo {title} {Eddies, stream, and convergence
  zones in turbulent flows},}\ }\bibinfo {type} {Tech. Rep.}\ \bibinfo {number}
  {CTR-S88}\ (\bibinfo  {institution} {Center for Turbulence Research, Stanford
  University},\ \bibinfo {year} {1988})\BibitemShut {NoStop}%
\bibitem [{\citenamefont {Jacobs}\ and\ \citenamefont
  {Sheeley}(1996)}]{jacobs1996experimental}%
  \BibitemOpen
  \bibfield  {author} {\bibinfo {author} {\bibnamefont {Jacobs}, \bibfnamefont
  {J.~W.}}and\ \bibinfo {author} {\bibnamefont {Sheeley}, \bibfnamefont
  {J.~M.}},\ }\bibfield  {title} {\enquote {\bibinfo {title} {{Experimental
  study of incompressible Richtmyer-Meshkov instability}},}\ }\href@noop {}
  {\bibfield  {journal} {\bibinfo  {journal} {Physics of Fluids}\ }\textbf
  {\bibinfo {volume} {8}},\ \bibinfo {pages} {405--415} (\bibinfo {year}
  {1996})}\BibitemShut {NoStop}%
\bibitem [{\citenamefont {Jeong}\ and\ \citenamefont
  {Hussain}(1995)}]{Jeong1995}%
  \BibitemOpen
  \bibfield  {author} {\bibinfo {author} {\bibnamefont {Jeong}, \bibfnamefont
  {J.}}and\ \bibinfo {author} {\bibnamefont {Hussain}, \bibfnamefont {F.}},\
  }\bibfield  {title} {\enquote {\bibinfo {title} {On the identification of a
  vortex},}\ }\href@noop {} {\bibfield  {journal} {\bibinfo  {journal} {Journal
  of Fluid Mechanics}\ }\textbf {\bibinfo {volume} {285}},\ \bibinfo {pages}
  {69--94} (\bibinfo {year} {1995})}\BibitemShut {NoStop}%
\bibitem [{\citenamefont {Jiang}\ and\ \citenamefont {Shu}(1996)}]{Jiang1996}%
  \BibitemOpen
  \bibfield  {author} {\bibinfo {author} {\bibnamefont {Jiang}, \bibfnamefont
  {G.-S.}}and\ \bibinfo {author} {\bibnamefont {Shu}, \bibfnamefont {C.-W.}},\
  }\bibfield  {title} {\enquote {\bibinfo {title} {Efficient implementation of
  weighted eno schemes},}\ }\href@noop {} {\bibfield  {journal} {\bibinfo
  {journal} {Journal of Computational Physics}\ }\textbf {\bibinfo {volume}
  {126}},\ \bibinfo {pages} {202--228} (\bibinfo {year} {1996})}\BibitemShut
  {NoStop}%
\bibitem [{\citenamefont {Kaden}(1931)}]{Kaden1931}%
  \BibitemOpen
  \bibfield  {author} {\bibinfo {author} {\bibnamefont {Kaden}, \bibfnamefont
  {H.}},\ }\bibfield  {title} {\enquote {\bibinfo {title} {{Aufwicklung einer
  unstabilen unstetigkeitsfl \"{a}che}},}\ }\href@noop {} {\bibfield  {journal}
  {\bibinfo  {journal} {Ingenieur Archive}\ }\textbf {\bibinfo {volume} {2}},\
  \bibinfo {pages} {140--168} (\bibinfo {year} {1931})}\BibitemShut {NoStop}%
\bibitem [{\citenamefont {Klein}(1910)}]{Klein1910}%
  \BibitemOpen
  \bibfield  {author} {\bibinfo {author} {\bibnamefont {Klein}, \bibfnamefont
  {F.}},\ }\bibfield  {title} {\enquote {\bibinfo {title} {{\"{U}ber die
  Bildung von Wirbeln in reibungslosen Fl\"{u}ssigkeiten}},}\ }\href@noop {}
  {\bibfield  {journal} {\bibinfo  {journal} {Zeitschrift f\"{u}r Angewandte
  Mathematik und Physik}\ }\textbf {\bibinfo {volume} {59}},\ \bibinfo {pages}
  {259--262} (\bibinfo {year} {1910})}\BibitemShut {NoStop}%
\bibitem [{\citenamefont {Kol{\'a}\v{r}}(2007)}]{Kolar2007IJHMF}%
  \BibitemOpen
  \bibfield  {author} {\bibinfo {author} {\bibnamefont {Kol{\'a}\v{r}},
  \bibfnamefont {V.}},\ }\bibfield  {title} {\enquote {\bibinfo {title} {Vortex
  identification: {N}ew requirements and limitations},}\ }\href@noop {}
  {\bibfield  {journal} {\bibinfo  {journal} {International Journal of Heat and
  Fluid Flow}\ }\textbf {\bibinfo {volume} {28}},\ \bibinfo {pages} {638--652}
  (\bibinfo {year} {2007})}\BibitemShut {NoStop}%
\bibitem [{\citenamefont {Li}\ \emph {et~al.}(2024)\citenamefont {Li},
  \citenamefont {Chen}, \citenamefont {Zhai},\ and\ \citenamefont
  {Luo}}]{LiJX2024compress}%
  \BibitemOpen
  \bibfield  {author} {\bibinfo {author} {\bibnamefont {Li}, \bibfnamefont
  {J.}}, \bibinfo {author} {\bibnamefont {Chen}, \bibfnamefont {C.}}, \bibinfo
  {author} {\bibnamefont {Zhai}, \bibfnamefont {Z.}}, and\ \bibinfo {author}
  {\bibnamefont {Luo}, \bibfnamefont {X.}},\ }\bibfield  {title} {\enquote
  {\bibinfo {title} {{Effects of compressibility on Richtmyer–Meshkov
  instability of heavy/light interface}},}\ }\href@noop {} {\bibfield
  {journal} {\bibinfo  {journal} {Phys. Fluids}\ }\textbf {\bibinfo {volume}
  {36}},\ \bibinfo {pages} {056104} (\bibinfo {year} {2024})}\BibitemShut
  {NoStop}%
\bibitem [{\citenamefont {Li}\ and\ \citenamefont {Zhai}(2025)}]{Li2025}%
  \BibitemOpen
  \bibfield  {author} {\bibinfo {author} {\bibnamefont {Li}, \bibfnamefont
  {J.}}and\ \bibinfo {author} {\bibnamefont {Zhai}, \bibfnamefont {Z.}},\
  }\bibfield  {title} {\enquote {\bibinfo {title} {{Modelling and mechanism of
  non-standard Richtmyer–Meshkov instability at heavy-light interfaces under
  moderate Mach numbers}},}\ }\href@noop {} {\bibfield  {journal} {\bibinfo
  {journal} {Journal of Fluid Mechanics}\ }\textbf {\bibinfo {volume} {1023}},\
  \bibinfo {pages} {A6} (\bibinfo {year} {2025})}\BibitemShut {NoStop}%
\bibitem [{\citenamefont {Li}\ \emph {et~al.}(2026)\citenamefont {Li},
  \citenamefont {Ding}, \citenamefont {Zhou}, \citenamefont {He}, \citenamefont
  {Leng},\ and\ \citenamefont {Gao}}]{LiXuan2026reactive}%
  \BibitemOpen
  \bibfield  {author} {\bibinfo {author} {\bibnamefont {Li}, \bibfnamefont
  {X.}}, \bibinfo {author} {\bibnamefont {Ding}, \bibfnamefont {J.}}, \bibinfo
  {author} {\bibnamefont {Zhou}, \bibfnamefont {Z.}}, \bibinfo {author}
  {\bibnamefont {He}, \bibfnamefont {D.}}, \bibinfo {author} {\bibnamefont
  {Leng}, \bibfnamefont {Y.}}, and\ \bibinfo {author} {\bibnamefont {Gao},
  \bibfnamefont {L.}},\ }\bibfield  {title} {\enquote {\bibinfo {title}
  {{Reactive Richtmyer--Meshkov instability under various Mach numbers}},}\
  }\href@noop {} {\bibfield  {journal} {\bibinfo  {journal} {Journal of Fluid
  Mechanics}\ }\textbf {\bibinfo {volume} {1032}},\ \bibinfo {pages} {A8}
  (\bibinfo {year} {2026})}\BibitemShut {NoStop}%
\bibitem [{\citenamefont {Li}, \citenamefont {Zhang},\ and\ \citenamefont
  {He}(2014)}]{LiZhen2014}%
  \BibitemOpen
  \bibfield  {author} {\bibinfo {author} {\bibnamefont {Li}, \bibfnamefont
  {Z.}}, \bibinfo {author} {\bibnamefont {Zhang}, \bibfnamefont {X.-W.}}, and\
  \bibinfo {author} {\bibnamefont {He}, \bibfnamefont {F.}},\ }\bibfield
  {title} {\enquote {\bibinfo {title} {Evaluation of vortex criteria by virtue
  of the quadruple decomposition of velocity gradient tensor},}\ }\href@noop {}
  {\bibfield  {journal} {\bibinfo  {journal} {Acta Physica Sinica}\ }\textbf
  {\bibinfo {volume} {63}},\ \bibinfo {pages} {054704} (\bibinfo {year}
  {2014})}\BibitemShut {NoStop}%
\bibitem [{\citenamefont {Liang}\ \emph {et~al.}(2021)\citenamefont {Liang},
  \citenamefont {Liu}, \citenamefont {Zhai}, \citenamefont {Ding},
  \citenamefont {Si},\ and\ \citenamefont {Luo}}]{Liang2021}%
  \BibitemOpen
  \bibfield  {author} {\bibinfo {author} {\bibnamefont {Liang}, \bibfnamefont
  {Y.}}, \bibinfo {author} {\bibnamefont {Liu}, \bibfnamefont {L.}}, \bibinfo
  {author} {\bibnamefont {Zhai}, \bibfnamefont {Z.}}, \bibinfo {author}
  {\bibnamefont {Ding}, \bibfnamefont {J.}}, \bibinfo {author} {\bibnamefont
  {Si}, \bibfnamefont {T.}}, and\ \bibinfo {author} {\bibnamefont {Luo},
  \bibfnamefont {X.}},\ }\bibfield  {title} {\enquote {\bibinfo {title}
  {{Richtmyer–Meshkov instability on two-dimensional multi-mode
  interfaces}},}\ }\href@noop {} {\bibfield  {journal} {\bibinfo  {journal}
  {Journal of Fluid Mechanics}\ }\textbf {\bibinfo {volume} {928}},\ \bibinfo
  {pages} {A37} (\bibinfo {year} {2021})}\BibitemShut {NoStop}%
\bibitem [{\citenamefont {Liang}\ \emph {et~al.}(2019)\citenamefont {Liang},
  \citenamefont {Zhai}, \citenamefont {Ding},\ and\ \citenamefont
  {Luo}}]{Liang2019}%
  \BibitemOpen
  \bibfield  {author} {\bibinfo {author} {\bibnamefont {Liang}, \bibfnamefont
  {Y.}}, \bibinfo {author} {\bibnamefont {Zhai}, \bibfnamefont {Z.}}, \bibinfo
  {author} {\bibnamefont {Ding}, \bibfnamefont {J.}}, and\ \bibinfo {author}
  {\bibnamefont {Luo}, \bibfnamefont {X.}},\ }\bibfield  {title} {\enquote
  {\bibinfo {title} {{Richtmyer–Meshkov instability on a quasi-single-mode
  interface}},}\ }\href@noop {} {\bibfield  {journal} {\bibinfo  {journal}
  {Journal of Fluid Mechanics}\ }\textbf {\bibinfo {volume} {872}},\ \bibinfo
  {pages} {729–751} (\bibinfo {year} {2019})}\BibitemShut {NoStop}%
\bibitem [{\citenamefont {Lighthill}(1956)}]{lighthill1956viscosity}%
  \BibitemOpen
  \bibfield  {author} {\bibinfo {author} {\bibnamefont {Lighthill},
  \bibfnamefont {M.~J.}},\ }\bibfield  {title} {\enquote {\bibinfo {title}
  {Viscosity effects in sound waves of finite amplitude},}\ }in\ \href@noop {}
  {\emph {\bibinfo {booktitle} {Surveys in Mechanics}}},\ \bibinfo {editor}
  {edited by\ \bibinfo {editor} {\bibfnamefont {G.~K.}\ \bibnamefont
  {Batchelor}}\ and\ \bibinfo {editor} {\bibfnamefont {R.~M.}\ \bibnamefont
  {Davies}}}\ (\bibinfo  {publisher} {Cambridge University Press},\ \bibinfo
  {year} {1956})\ pp.\ \bibinfo {pages} {250--351}\BibitemShut {NoStop}%
\bibitem [{\citenamefont {Liu}\ \emph {et~al.}(2018)\citenamefont {Liu},
  \citenamefont {Gao}, \citenamefont {Tian},\ and\ \citenamefont
  {Dong}}]{Liu2018}%
  \BibitemOpen
  \bibfield  {author} {\bibinfo {author} {\bibnamefont {Liu}, \bibfnamefont
  {C.}}, \bibinfo {author} {\bibnamefont {Gao}, \bibfnamefont {Y.}}, \bibinfo
  {author} {\bibnamefont {Tian}, \bibfnamefont {S.}}, and\ \bibinfo {author}
  {\bibnamefont {Dong}, \bibfnamefont {X.}},\ }\bibfield  {title} {\enquote
  {\bibinfo {title} {Rortex—a new vortex vector definition and vorticity
  tensor and vector decompositions},}\ }\href@noop {} {\bibfield  {journal}
  {\bibinfo  {journal} {Physics of Fluids}\ }\textbf {\bibinfo {volume} {30}},\
  \bibinfo {pages} {035103} (\bibinfo {year} {2018})}\BibitemShut {NoStop}%
\bibitem [{\citenamefont {Liu}\ and\ \citenamefont {Wang}(2020)}]{Liu2020}%
  \BibitemOpen
  \bibfield  {author} {\bibinfo {author} {\bibnamefont {Liu}, \bibfnamefont
  {C.}}and\ \bibinfo {author} {\bibnamefont {Wang}, \bibfnamefont {Y.}},\
  }\bibfield  {title} {\enquote {\bibinfo {title} {Liutex and third generation
  of vortex definition and identification},}\ }in\ \href@noop {} {\emph
  {\bibinfo {booktitle} {An Invited Workshop from Chaos 2020}}},\ \bibinfo
  {editor} {edited by\ \bibinfo {editor} {\bibfnamefont {C.}~\bibnamefont
  {Liu}}\ and\ \bibinfo {editor} {\bibfnamefont {Y.}~\bibnamefont {Wang}}}\
  (\bibinfo  {publisher} {Springer Nature Switzerland AG},\ \bibinfo {address}
  {Gewerbestrasse 11, 6330 Cham, Switzerland},\ \bibinfo {year} {2020})\ pp.\
  \bibinfo {pages} {1--479}\BibitemShut {NoStop}%
\bibitem [{\citenamefont {Markstein}(1957)}]{Markstein1957}%
  \BibitemOpen
  \bibfield  {author} {\bibinfo {author} {\bibnamefont {Markstein},
  \bibfnamefont {G.~H.}},\ }\bibfield  {title} {\enquote {\bibinfo {title}
  {Flow disturbances induced near a slightly wavy contact surface, or flame
  front, traversed by shock wave},}\ }\href@noop {} {\bibfield  {journal}
  {\bibinfo  {journal} {Journal of Aerosol Science}\ }\textbf {\bibinfo
  {volume} {24}},\ \bibinfo {pages} {238--239} (\bibinfo {year}
  {1957})}\BibitemShut {NoStop}%
\bibitem [{\citenamefont {Meshkov}(1969)}]{Meshkov1969}%
  \BibitemOpen
  \bibfield  {author} {\bibinfo {author} {\bibnamefont {Meshkov}, \bibfnamefont
  {E.~E.}},\ }\bibfield  {title} {\enquote {\bibinfo {title} {Instability of
  the interface of two gases accelerated by a shock wave},}\ }\href@noop {}
  {\bibfield  {journal} {\bibinfo  {journal} {Fluid Dynamics}\ }\textbf
  {\bibinfo {volume} {4}},\ \bibinfo {pages} {101} (\bibinfo {year}
  {1969})}\BibitemShut {NoStop}%
\bibitem [{\citenamefont {Meyer}\ and\ \citenamefont
  {Blewett}(1972)}]{meyer1972numerical}%
  \BibitemOpen
  \bibfield  {author} {\bibinfo {author} {\bibnamefont {Meyer}, \bibfnamefont
  {K.~A.}}and\ \bibinfo {author} {\bibnamefont {Blewett}, \bibfnamefont
  {P.~J.}},\ }\bibfield  {title} {\enquote {\bibinfo {title} {Numerical
  investigation of the stability of a shock-accelerated interface between two
  fluids},}\ }\href@noop {} {\bibfield  {journal} {\bibinfo  {journal} {Physics
  of Fluids}\ }\textbf {\bibinfo {volume} {15}},\ \bibinfo {pages} {753}
  (\bibinfo {year} {1972})}\BibitemShut {NoStop}%
\bibitem [{\citenamefont {Mikaelian}(2003)}]{Mikaelian2003}%
  \BibitemOpen
  \bibfield  {author} {\bibinfo {author} {\bibnamefont {Mikaelian},
  \bibfnamefont {K.~O.}},\ }\bibfield  {title} {\enquote {\bibinfo {title}
  {{Explicit expressions for the evolution of single-mode Rayleigh-Taylor and
  Richtmyer-Meshkov instabilities at arbitrary Atwood numbers}},}\ }\href@noop
  {} {\bibfield  {journal} {\bibinfo  {journal} {Physical Review E}\ }\textbf
  {\bibinfo {volume} {67}},\ \bibinfo {pages} {026319} (\bibinfo {year}
  {2003})}\BibitemShut {NoStop}%
\bibitem [{\citenamefont {Morgan}\ \emph {et~al.}(2012)\citenamefont {Morgan},
  \citenamefont {Aure}, \citenamefont {Stockero}, \citenamefont {Greenough},
  \citenamefont {Cabot}, \citenamefont {Likhachev},\ and\ \citenamefont
  {Jacobs}}]{Morgan2012}%
  \BibitemOpen
  \bibfield  {author} {\bibinfo {author} {\bibnamefont {Morgan}, \bibfnamefont
  {R.~V.}}, \bibinfo {author} {\bibnamefont {Aure}, \bibfnamefont {R.}},
  \bibinfo {author} {\bibnamefont {Stockero}, \bibfnamefont {J.~D.}}, \bibinfo
  {author} {\bibnamefont {Greenough}, \bibfnamefont {J.~A.}}, \bibinfo {author}
  {\bibnamefont {Cabot}, \bibfnamefont {W.}}, \bibinfo {author} {\bibnamefont
  {Likhachev}, \bibfnamefont {O.~A.}}, and\ \bibinfo {author} {\bibnamefont
  {Jacobs}, \bibfnamefont {J.~W.}},\ }\bibfield  {title} {\enquote {\bibinfo
  {title} {{On the late-time growth of the two-dimensional Richtmyer–Meshkov
  instability in shock tube experiments}},}\ }\href@noop {} {\bibfield
  {journal} {\bibinfo  {journal} {Journal of Fluid Mechanics}\ }\textbf
  {\bibinfo {volume} {712}},\ \bibinfo {pages} {354--383} (\bibinfo {year}
  {2012})}\BibitemShut {NoStop}%
\bibitem [{\citenamefont {Ranjan}, \citenamefont {Oakley},\ and\ \citenamefont
  {Bonazza}(2011)}]{ranjan2011shock}%
  \BibitemOpen
  \bibfield  {author} {\bibinfo {author} {\bibnamefont {Ranjan}, \bibfnamefont
  {D.}}, \bibinfo {author} {\bibnamefont {Oakley}, \bibfnamefont {J.}}, and\
  \bibinfo {author} {\bibnamefont {Bonazza}, \bibfnamefont {R.}},\ }\bibfield
  {title} {\enquote {\bibinfo {title} {{Shock-Bubble Interactions}},}\
  }\href@noop {} {\bibfield  {journal} {\bibinfo  {journal} {Annual Review of
  Fluid Mechanics}\ }\textbf {\bibinfo {volume} {43}},\ \bibinfo {pages}
  {117--140} (\bibinfo {year} {2011})}\BibitemShut {NoStop}%
\bibitem [{\citenamefont {Rayleigh}(1883)}]{Rayleigh1883}%
  \BibitemOpen
  \bibfield  {author} {\bibinfo {author} {\bibnamefont {Rayleigh},
  \bibfnamefont {L.}},\ }\bibfield  {title} {\enquote {\bibinfo {title}
  {Investigation of the character of the equilibrium of an incompressible heavy
  fluid of variable density},}\ }\href@noop {} {\bibfield  {journal} {\bibinfo
  {journal} {Proceedings of the London Mathematical Society}\ }\textbf
  {\bibinfo {volume} {s1-14}},\ \bibinfo {pages} {170--177} (\bibinfo {year}
  {1883})}\BibitemShut {NoStop}%
\bibitem [{\citenamefont {Richtmyer}(1960)}]{richtmyer1960taylor}%
  \BibitemOpen
  \bibfield  {author} {\bibinfo {author} {\bibnamefont {Richtmyer},
  \bibfnamefont {R.~D.}},\ }\bibfield  {title} {\enquote {\bibinfo {title}
  {{Taylor Instability in Shock Acceleration of Compressible Fluids}},}\
  }\href@noop {} {\bibfield  {journal} {\bibinfo  {journal} {Communications on
  Pure and Applied Mathematics}\ }\textbf {\bibinfo {volume} {13}},\ \bibinfo
  {pages} {297--319} (\bibinfo {year} {1960})}\BibitemShut {NoStop}%
\bibitem [{\citenamefont {Roberts}\ and\ \citenamefont
  {Jacobs}(2016)}]{RobertsJacobbs2016}%
  \BibitemOpen
  \bibfield  {author} {\bibinfo {author} {\bibnamefont {Roberts}, \bibfnamefont
  {M.~S.}}and\ \bibinfo {author} {\bibnamefont {Jacobs}, \bibfnamefont
  {J.~W.}},\ }\bibfield  {title} {\enquote {\bibinfo {title} {The effects of
  forced small-wavelength, finite-bandwidth initial perturbations and
  miscibility on the turbulent {Rayleigh}–{Taylor} instability},}\
  }\href@noop {} {\bibfield  {journal} {\bibinfo  {journal} {Journal of Fluid
  Mechanics}\ }\textbf {\bibinfo {volume} {787}},\ \bibinfo {pages} {50–83}
  (\bibinfo {year} {2016})}\BibitemShut {NoStop}%
\bibitem [{\citenamefont {Sadot}\ \emph {et~al.}(1998)\citenamefont {Sadot},
  \citenamefont {Erez}, \citenamefont {Alon}, \citenamefont {Oron},
  \citenamefont {Levin}, \citenamefont {Erez}, \citenamefont {Ben-Dor},\ and\
  \citenamefont {Shvarts}}]{Sadot1998}%
  \BibitemOpen
  \bibfield  {author} {\bibinfo {author} {\bibnamefont {Sadot}, \bibfnamefont
  {O.}}, \bibinfo {author} {\bibnamefont {Erez}, \bibfnamefont {L.}}, \bibinfo
  {author} {\bibnamefont {Alon}, \bibfnamefont {U.}}, \bibinfo {author}
  {\bibnamefont {Oron}, \bibfnamefont {D.}}, \bibinfo {author} {\bibnamefont
  {Levin}, \bibfnamefont {L.~A.}}, \bibinfo {author} {\bibnamefont {Erez},
  \bibfnamefont {G.}}, \bibinfo {author} {\bibnamefont {Ben-Dor}, \bibfnamefont
  {G.}}, and\ \bibinfo {author} {\bibnamefont {Shvarts}, \bibfnamefont {D.}},\
  }\bibfield  {title} {\enquote {\bibinfo {title} {{Study of Nonlinear
  Evolution of Single-Mode and Two-Bubble Interaction under Richtmyer-Meshkov
  Instability}},}\ }\href@noop {} {\bibfield  {journal} {\bibinfo  {journal}
  {Physical Review Letters}\ }\textbf {\bibinfo {volume} {80}},\ \bibinfo
  {pages} {1791--1794} (\bibinfo {year} {1998})}\BibitemShut {NoStop}%
\bibitem [{\citenamefont {Samtaney}\ and\ \citenamefont
  {Zabusky}(1994)}]{samtaney1994circulation}%
  \BibitemOpen
  \bibfield  {author} {\bibinfo {author} {\bibnamefont {Samtaney},
  \bibfnamefont {R.}}and\ \bibinfo {author} {\bibnamefont {Zabusky},
  \bibfnamefont {N.~J.}},\ }\bibfield  {title} {\enquote {\bibinfo {title}
  {Circulation deposition on shock-accelerated planar and curved
  density-stratified interfaces: models and scaling laws},}\ }\href@noop {}
  {\bibfield  {journal} {\bibinfo  {journal} {Journal of Fluid Mechanics}\
  }\textbf {\bibinfo {volume} {269}},\ \bibinfo {pages} {45--78} (\bibinfo
  {year} {1994})}\BibitemShut {NoStop}%
\bibitem [{\citenamefont {Schur}(1909)}]{schur1909uber}%
  \BibitemOpen
  \bibfield  {author} {\bibinfo {author} {\bibnamefont {Schur}, \bibfnamefont
  {I.}},\ }\bibfield  {title} {\enquote {\bibinfo {title} {{\"U}ber die
  charakteristischen wurzeln einer linearen substitution mit einer anwendung
  auf die theorie der integralgleichungen},}\ }\href@noop {} {\bibfield
  {journal} {\bibinfo  {journal} {Mathematische Annalen}\ }\textbf {\bibinfo
  {volume} {66}},\ \bibinfo {pages} {488--510} (\bibinfo {year}
  {1909})}\BibitemShut {NoStop}%
\bibitem [{\citenamefont {Shankar}, \citenamefont {Kawai},\ and\ \citenamefont
  {Lele}(2010)}]{shankar2010numerical}%
  \BibitemOpen
  \bibfield  {author} {\bibinfo {author} {\bibnamefont {Shankar}, \bibfnamefont
  {S.~K.}}, \bibinfo {author} {\bibnamefont {Kawai}, \bibfnamefont {S.}}, and\
  \bibinfo {author} {\bibnamefont {Lele}, \bibfnamefont {S.~K.}},\ }\bibfield
  {title} {\enquote {\bibinfo {title} {{Numerical Simulation of Multicomponent
  Shock Accelerated Flows and Mixing using Localized Artificial Diffusivity
  Method}},}\ }in\ \href@noop {} {\emph {\bibinfo {booktitle} {48th AIAA
  Aerospace Sciences Meeting Including the New Horizons Forum and Aerospace
  Exposition}}},\ \bibinfo {series and number} {\bibinfo {number} {AIAA
  2010-352}}\ (\bibinfo {address} {Orlando, Florida},\ \bibinfo {year}
  {2010})\BibitemShut {NoStop}%
\bibitem [{\citenamefont {Singh}(2023)}]{singh2023investigation}%
  \BibitemOpen
  \bibfield  {author} {\bibinfo {author} {\bibnamefont {Singh}, \bibfnamefont
  {S.}},\ }\bibfield  {title} {\enquote {\bibinfo {title} {Investigation of
  aspect ratio effects on flow characteristics and vorticity generation in
  shock-induced rectangular bubble},}\ }\href@noop {} {\bibfield  {journal}
  {\bibinfo  {journal} {European Journal of Mechanics/B Fluids}\ }\textbf
  {\bibinfo {volume} {101}},\ \bibinfo {pages} {131--148} (\bibinfo {year}
  {2023})}\BibitemShut {NoStop}%
\bibitem [{\citenamefont {Singh}, \citenamefont {Battiato},\ and\ \citenamefont
  {Myong}(2021)}]{Singh2021impact}%
  \BibitemOpen
  \bibfield  {author} {\bibinfo {author} {\bibnamefont {Singh}, \bibfnamefont
  {S.}}, \bibinfo {author} {\bibnamefont {Battiato}, \bibfnamefont {M.}}, and\
  \bibinfo {author} {\bibnamefont {Myong}, \bibfnamefont {R.~S.}},\ }\bibfield
  {title} {\enquote {\bibinfo {title} {{Impact of the bulk viscosity on flow
  morphology of shock-bubble interaction in diatomic and polyatomic gases}},}\
  }\href@noop {} {\bibfield  {journal} {\bibinfo  {journal} {Physics of
  Fluids}\ }\textbf {\bibinfo {volume} {33}},\ \bibinfo {pages} {066103}
  (\bibinfo {year} {2021})}\BibitemShut {NoStop}%
\bibitem [{\citenamefont {Singh}\ and\ \citenamefont
  {Jaleli}(2024)}]{Singh2024ScienceChina}%
  \BibitemOpen
  \bibfield  {author} {\bibinfo {author} {\bibnamefont {Singh}, \bibfnamefont
  {S.}}and\ \bibinfo {author} {\bibnamefont {Jaleli}, \bibfnamefont {D.~T.}},\
  }\bibfield  {title} {\enquote {\bibinfo {title} {{Investigation of coupling
  effect on the evolution of Richtmyer-Meshkov instability at double heavy
  square bubbles}},}\ }\href@noop {} {\bibfield  {journal} {\bibinfo  {journal}
  {Science China Physics, Mechanics \& Astronomy}\ }\textbf {\bibinfo {volume}
  {67}},\ \bibinfo {pages} {214711} (\bibinfo {year} {2024})}\BibitemShut
  {NoStop}%
\bibitem [{\citenamefont {Singh}\ \emph {et~al.}(2025)\citenamefont {Singh},
  \citenamefont {Msmali}, \citenamefont {Tamsir},\ and\ \citenamefont
  {Ahmadini}}]{Singh2025ccc}%
  \BibitemOpen
  \bibfield  {author} {\bibinfo {author} {\bibnamefont {Singh}, \bibfnamefont
  {S.}}, \bibinfo {author} {\bibnamefont {Msmali}, \bibfnamefont {A.~H.}},
  \bibinfo {author} {\bibnamefont {Tamsir}, \bibfnamefont {M.}}, and\ \bibinfo
  {author} {\bibnamefont {Ahmadini}, \bibfnamefont {A.~A.~H.}},\ }\bibfield
  {title} {\enquote {\bibinfo {title} {Mechanisms of coupling-induced
  instabilities in shock-accelerated tandem light square bubbles},}\
  }\href@noop {} {\bibfield  {journal} {\bibinfo  {journal} {Physics of
  Fluids}\ }\textbf {\bibinfo {volume} {37}},\ \bibinfo {pages} {082101}
  (\bibinfo {year} {2025})}\BibitemShut {NoStop}%
\bibitem [{\citenamefont {Singh}\ and\ \citenamefont
  {Torrilhon}(2023)}]{singh2023shock}%
  \BibitemOpen
  \bibfield  {author} {\bibinfo {author} {\bibnamefont {Singh}, \bibfnamefont
  {S.}}and\ \bibinfo {author} {\bibnamefont {Torrilhon}, \bibfnamefont {M.}},\
  }\bibfield  {title} {\enquote {\bibinfo {title} {On the shock-driven
  hydrodynamic instability in square and rectangular light gas bubbles: A
  comparative study from numerical simulations},}\ }\href@noop {} {\bibfield
  {journal} {\bibinfo  {journal} {Physics of Fluids}\ }\textbf {\bibinfo
  {volume} {35}},\ \bibinfo {pages} {012117} (\bibinfo {year}
  {2023})}\BibitemShut {NoStop}%
\bibitem [{\citenamefont {Singh}\ and\ \citenamefont
  {Torrilhon}(2025)}]{singh2025shock}%
  \BibitemOpen
  \bibfield  {author} {\bibinfo {author} {\bibnamefont {Singh}, \bibfnamefont
  {S.}}and\ \bibinfo {author} {\bibnamefont {Torrilhon}, \bibfnamefont {M.}},\
  }\bibfield  {title} {\enquote {\bibinfo {title} {{Shock-accelerated dynamics
  of heavy rectangular bubbles: Influence of aspect ratio on Richtmyer-Meshkov
  instability}},}\ }\href@noop {} {\bibfield  {journal} {\bibinfo  {journal}
  {Physics of Fluids}\ }\textbf {\bibinfo {volume} {37}},\ \bibinfo {pages}
  {122129} (\bibinfo {year} {2025})}\BibitemShut {NoStop}%
\bibitem [{\citenamefont {Sohn}(2003)}]{Sohn2003}%
  \BibitemOpen
  \bibfield  {author} {\bibinfo {author} {\bibnamefont {Sohn}, \bibfnamefont
  {S.~I.}},\ }\bibfield  {title} {\enquote {\bibinfo {title} {{Simple
  potential-flow model of Rayleigh-Taylor and Richtmyer-Meshkov instability for
  all density ratios}},}\ }\href@noop {} {\bibfield  {journal} {\bibinfo
  {journal} {Phys. Rev. E}\ }\textbf {\bibinfo {volume} {67}},\ \bibinfo
  {pages} {026301} (\bibinfo {year} {2003})}\BibitemShut {NoStop}%
\bibitem [{\citenamefont {Taylor}(1950)}]{Taylor1950}%
  \BibitemOpen
  \bibfield  {author} {\bibinfo {author} {\bibnamefont {Taylor}, \bibfnamefont
  {G.~I.}},\ }\bibfield  {title} {\enquote {\bibinfo {title} {The instability
  of liquid surfaces when accelerated in a direction perpendicular to their
  plane},}\ }\href@noop {} {\bibfield  {journal} {\bibinfo  {journal}
  {Proceedings of the Royal Society of London. Series A. Mathematical and
  Physical Sciences}\ }\textbf {\bibinfo {volume} {201}},\ \bibinfo {pages}
  {192--196} (\bibinfo {year} {1950})}\BibitemShut {NoStop}%
\bibitem [{\citenamefont {Thomson}(1871)}]{Thomson1871}%
  \BibitemOpen
  \bibfield  {author} {\bibinfo {author} {\bibnamefont {Thomson}, \bibfnamefont
  {W.~L.~K.}},\ }\bibfield  {title} {\enquote {\bibinfo {title} {Hydrokinetic
  solutions and observations},}\ }\href@noop {} {\bibfield  {journal} {\bibinfo
   {journal} {Philosophical Magazine}\ }\bibinfo {series} {4},\ \textbf
  {\bibinfo {volume} {42}},\ \bibinfo {pages} {362--377} (\bibinfo {year}
  {1871})}\BibitemShut {NoStop}%
\bibitem [{\citenamefont {Vandenboomgaerde}, \citenamefont {M\"{u}gler},\ and\
  \citenamefont {Gauthier}(1998)}]{Vandenboomgaerde1998}%
  \BibitemOpen
  \bibfield  {author} {\bibinfo {author} {\bibnamefont {Vandenboomgaerde},
  \bibfnamefont {M.}}, \bibinfo {author} {\bibnamefont {M\"{u}gler},
  \bibfnamefont {C.}}, and\ \bibinfo {author} {\bibnamefont {Gauthier},
  \bibfnamefont {S.}},\ }\bibfield  {title} {\enquote {\bibinfo {title}
  {{Impulsive model for the Richtmyer-Meshkov instability}},}\ }\href@noop {}
  {\bibfield  {journal} {\bibinfo  {journal} {Physical Review E}\ }\textbf
  {\bibinfo {volume} {58}},\ \bibinfo {pages} {1874} (\bibinfo {year}
  {1998})}\BibitemShut {NoStop}%
\bibitem [{\citenamefont {Vandenboomgaerde}\ \emph {et~al.}()\citenamefont
  {Vandenboomgaerde}, \citenamefont {Souffland}, \citenamefont {Mariani},
  \citenamefont {Biamino}, \citenamefont {Jourdan},\ and\ \citenamefont
  {Houas}}]{Vandenboomgaerde2014}%
  \BibitemOpen
  \bibfield  {author} {\bibinfo {author} {\bibnamefont {Vandenboomgaerde},
  \bibfnamefont {M.}}, \bibinfo {author} {\bibnamefont {Souffland},
  \bibfnamefont {D.}}, \bibinfo {author} {\bibnamefont {Mariani}, \bibfnamefont
  {C.}}, \bibinfo {author} {\bibnamefont {Biamino}, \bibfnamefont {L.}},
  \bibinfo {author} {\bibnamefont {Jourdan}, \bibfnamefont {G.}}, and\ \bibinfo
  {author} {\bibnamefont {Houas}, \bibfnamefont {L.}},\ }\bibfield  {title}
  {\enquote {\bibinfo {title} {{An experimental and numerical investigation of
  the dependency on the initial conditions of the Richtmyer-Meshkov
  instability}},}\ }\href@noop {} {\ }\BibitemShut {NoStop}%
\bibitem [{\citenamefont {Wheeler}\ and\ \citenamefont
  {Harkness}(1990)}]{wheeler1990type}%
  \BibitemOpen
  \bibfield  {author} {\bibinfo {author} {\bibnamefont {Wheeler}, \bibfnamefont
  {J.~C.}}and\ \bibinfo {author} {\bibnamefont {Harkness}, \bibfnamefont
  {R.~P.}},\ }\bibfield  {title} {\enquote {\bibinfo {title} {{Type I
  supernovae}},}\ }\href@noop {} {\bibfield  {journal} {\bibinfo  {journal}
  {Reports on Progress in Physics}\ }\textbf {\bibinfo {volume} {53}},\
  \bibinfo {pages} {1467--1557} (\bibinfo {year} {1990})}\BibitemShut {NoStop}%
\bibitem [{\citenamefont {Wouchuk}\ and\ \citenamefont
  {Nishihara}(1997)}]{wouchuk1997asymptotic}%
  \BibitemOpen
  \bibfield  {author} {\bibinfo {author} {\bibnamefont {Wouchuk}, \bibfnamefont
  {J.~G.}}and\ \bibinfo {author} {\bibnamefont {Nishihara}, \bibfnamefont
  {K.}},\ }\bibfield  {title} {\enquote {\bibinfo {title} {{Asymptotic growth
  in the linear Richtmyer-Meshkov instability}},}\ }\href@noop {} {\bibfield
  {journal} {\bibinfo  {journal} {Physics of Fluids}\ }\textbf {\bibinfo
  {volume} {9}},\ \bibinfo {pages} {2074--2081} (\bibinfo {year}
  {1997})}\BibitemShut {NoStop}%
\bibitem [{\citenamefont {Wu}, \citenamefont {Ma},\ and\ \citenamefont
  {Zhou}(2015)}]{Wu2015}%
  \BibitemOpen
  \bibfield  {author} {\bibinfo {author} {\bibnamefont {Wu}, \bibfnamefont
  {J.-Z.}}, \bibinfo {author} {\bibnamefont {Ma}, \bibfnamefont {H.-Y.}}, and\
  \bibinfo {author} {\bibnamefont {Zhou}, \bibfnamefont {M.-D.}},\ }\href@noop
  {} {\emph {\bibinfo {title} {Vortical Flows}}}\ (\bibinfo  {publisher}
  {Springer},\ \bibinfo {address} {Berlin, Heidelberg},\ \bibinfo {year}
  {2015})\BibitemShut {NoStop}%
\bibitem [{\citenamefont {Xu}\ \emph {et~al.}(2019)\citenamefont {Xu},
  \citenamefont {Gao}, \citenamefont {Deng}, \citenamefont {Liu},\ and\
  \citenamefont {Liu}}]{Xu2019}%
  \BibitemOpen
  \bibfield  {author} {\bibinfo {author} {\bibnamefont {Xu}, \bibfnamefont
  {W.}}, \bibinfo {author} {\bibnamefont {Gao}, \bibfnamefont {Y.}}, \bibinfo
  {author} {\bibnamefont {Deng}, \bibfnamefont {Y.}}, \bibinfo {author}
  {\bibnamefont {Liu}, \bibfnamefont {J.}}, and\ \bibinfo {author}
  {\bibnamefont {Liu}, \bibfnamefont {C.}},\ }\bibfield  {title} {\enquote
  {\bibinfo {title} {{An explicit expression for the calculation of the Rortex
  vector}},}\ }\href@noop {} {\bibfield  {journal} {\bibinfo  {journal}
  {Physics of Fluids}\ }\textbf {\bibinfo {volume} {31}},\ \bibinfo {pages}
  {095102} (\bibinfo {year} {2019})}\BibitemShut {NoStop}%
\bibitem [{\citenamefont {Yosef-Hai}\ \emph {et~al.}(2003)\citenamefont
  {Yosef-Hai}, \citenamefont {Sadot}, \citenamefont {Kartoon}, \citenamefont
  {Oron}, \citenamefont {Levin}, \citenamefont {Sarid}, \citenamefont {Elbaz},
  \citenamefont {Ben-Dor},\ and\ \citenamefont {Shvarts}}]{Yosef-Hai2003}%
  \BibitemOpen
  \bibfield  {author} {\bibinfo {author} {\bibnamefont {Yosef-Hai},
  \bibfnamefont {A.}}, \bibinfo {author} {\bibnamefont {Sadot}, \bibfnamefont
  {O.}}, \bibinfo {author} {\bibnamefont {Kartoon}, \bibfnamefont {D.}},
  \bibinfo {author} {\bibnamefont {Oron}, \bibfnamefont {D.}}, \bibinfo
  {author} {\bibnamefont {Levin}, \bibfnamefont {L.~A.}}, \bibinfo {author}
  {\bibnamefont {Sarid}, \bibfnamefont {E.}}, \bibinfo {author} {\bibnamefont
  {Elbaz}, \bibfnamefont {Y.}}, \bibinfo {author} {\bibnamefont {Ben-Dor},
  \bibfnamefont {G.}}, and\ \bibinfo {author} {\bibnamefont {Shvarts},
  \bibfnamefont {D.}},\ }\bibfield  {title} {\enquote {\bibinfo {title}
  {{Late-time growth of the Richtmyer–Meshkov instability for different
  Atwood numbers and different dimensionalities}},}\ }\href@noop {} {\bibfield
  {journal} {\bibinfo  {journal} {Laser and Particle Beams}\ }\textbf {\bibinfo
  {volume} {21}},\ \bibinfo {pages} {363--368} (\bibinfo {year}
  {2003})}\BibitemShut {NoStop}%
\bibitem [{\citenamefont {Zabusky}(1999)}]{Zabusky1999VORTEX}%
  \BibitemOpen
  \bibfield  {author} {\bibinfo {author} {\bibnamefont {Zabusky}, \bibfnamefont
  {N.~J.}},\ }\bibfield  {title} {\enquote {\bibinfo {title} {{VORTEX PARADIGM
  FOR ACCELERATED INHOMOGENEOUS FLOWS: Visiometrics for the Rayleigh-Taylor and
  Richtmyer-Meshkov Environments}},}\ }\href@noop {} {\bibfield  {journal}
  {\bibinfo  {journal} {Annual Review of Fluid Mechanics}\ }\textbf {\bibinfo
  {volume} {31}},\ \bibinfo {pages} {495--536} (\bibinfo {year}
  {1999})}\BibitemShut {NoStop}%
\bibitem [{\citenamefont {Zhai}\ \emph {et~al.}(2011)\citenamefont {Zhai},
  \citenamefont {Si}, \citenamefont {Luo},\ and\ \citenamefont
  {Yang}}]{zhai2011evolution}%
  \BibitemOpen
  \bibfield  {author} {\bibinfo {author} {\bibnamefont {Zhai}, \bibfnamefont
  {Z.}}, \bibinfo {author} {\bibnamefont {Si}, \bibfnamefont {T.}}, \bibinfo
  {author} {\bibnamefont {Luo}, \bibfnamefont {X.}}, and\ \bibinfo {author}
  {\bibnamefont {Yang}, \bibfnamefont {J.}},\ }\bibfield  {title} {\enquote
  {\bibinfo {title} {On the evolution of spherical gas interfaces accelerated
  by a planar shock wave},}\ }\href@noop {} {\bibfield  {journal} {\bibinfo
  {journal} {Physics of Fluids}\ }\textbf {\bibinfo {volume} {23}},\ \bibinfo
  {pages} {084104} (\bibinfo {year} {2011})}\BibitemShut {NoStop}%
\bibitem [{\citenamefont {Zhai}\ \emph {et~al.}(2014)\citenamefont {Zhai},
  \citenamefont {Wang}, \citenamefont {Si},\ and\ \citenamefont
  {Luo}}]{Zhai2014lp}%
  \BibitemOpen
  \bibfield  {author} {\bibinfo {author} {\bibnamefont {Zhai}, \bibfnamefont
  {Z.}}, \bibinfo {author} {\bibnamefont {Wang}, \bibfnamefont {M.}}, \bibinfo
  {author} {\bibnamefont {Si}, \bibfnamefont {T.}}, and\ \bibinfo {author}
  {\bibnamefont {Luo}, \bibfnamefont {X.}},\ }\bibfield  {title} {\enquote
  {\bibinfo {title} {On the interaction of a planar shock with a light
  polygonal interface},}\ }\href@noop {} {\bibfield  {journal} {\bibinfo
  {journal} {Journal of Fluid Mechanics}\ }\textbf {\bibinfo {volume} {757}},\
  \bibinfo {pages} {800–816} (\bibinfo {year} {2014})}\BibitemShut {NoStop}%
\bibitem [{\citenamefont {Zhai}\ \emph {et~al.}(2018)\citenamefont {Zhai},
  \citenamefont {Zou}, \citenamefont {Wu},\ and\ \citenamefont
  {Luo}}]{zhai2017review}%
  \BibitemOpen
  \bibfield  {author} {\bibinfo {author} {\bibnamefont {Zhai}, \bibfnamefont
  {Z.}}, \bibinfo {author} {\bibnamefont {Zou}, \bibfnamefont {L.}}, \bibinfo
  {author} {\bibnamefont {Wu}, \bibfnamefont {Q.}}, and\ \bibinfo {author}
  {\bibnamefont {Luo}, \bibfnamefont {X.}},\ }\bibfield  {title} {\enquote
  {\bibinfo {title} {{Review of experimental Richtmyer-Meshkov instability in
  shock tube: From simple to complex}},}\ }\href@noop {} {\bibfield  {journal}
  {\bibinfo  {journal} {Proceedings of the Institution of Mechanical Engineers,
  Part C: Journal of Mechanical Engineering Science}\ }\textbf {\bibinfo
  {volume} {232}},\ \bibinfo {pages} {2830--2849} (\bibinfo {year}
  {2018})}\BibitemShut {NoStop}%
\bibitem [{\citenamefont {Zhang}\ and\ \citenamefont
  {Guo}(2016)}]{Zhang2016Universality}%
  \BibitemOpen
  \bibfield  {author} {\bibinfo {author} {\bibnamefont {Zhang}, \bibfnamefont
  {Q.}}and\ \bibinfo {author} {\bibnamefont {Guo}, \bibfnamefont {W.}},\
  }\bibfield  {title} {\enquote {\bibinfo {title} {{Universality of finger
  growth in two-dimensional Rayleigh-Taylor and Richtmyer-Meshkov instabilities
  with all density ratios}},}\ }\href@noop {} {\bibfield  {journal} {\bibinfo
  {journal} {Journal of Fluid Mechanics}\ }\textbf {\bibinfo {volume} {786}},\
  \bibinfo {pages} {47--61} (\bibinfo {year} {2016})}\BibitemShut {NoStop}%
\bibitem [{\citenamefont {Zhang}\ and\ \citenamefont
  {Sohn}(1997)}]{zhang1997nonlinear}%
  \BibitemOpen
  \bibfield  {author} {\bibinfo {author} {\bibnamefont {Zhang}, \bibfnamefont
  {Q.}}and\ \bibinfo {author} {\bibnamefont {Sohn}, \bibfnamefont {S.-I.}},\
  }\bibfield  {title} {\enquote {\bibinfo {title} {Nonlinear theory of unstable
  fluid mixing driven by shock wave},}\ }\href@noop {} {\bibfield  {journal}
  {\bibinfo  {journal} {Physics of Fluids}\ }\textbf {\bibinfo {volume} {9}},\
  \bibinfo {pages} {1106--1106} (\bibinfo {year} {1997})}\BibitemShut {NoStop}%
\bibitem [{\citenamefont {Zhou}(2017{\natexlab{a}})}]{zhou2017rayleigh}%
  \BibitemOpen
  \bibfield  {author} {\bibinfo {author} {\bibnamefont {Zhou}, \bibfnamefont
  {Y.}},\ }\bibfield  {title} {\enquote {\bibinfo {title} {{Rayleigh-Taylor and
  Richtmyer-Meshkov instability induced flow, turbulence, and mixing. I}},}\
  }\href@noop {} {\bibfield  {journal} {\bibinfo  {journal} {Physics Reports}\
  }\textbf {\bibinfo {volume} {720--722}},\ \bibinfo {pages} {1--136} (\bibinfo
  {year} {2017}{\natexlab{a}})}\BibitemShut {NoStop}%
\bibitem [{\citenamefont {Zhou}(2017{\natexlab{b}})}]{Zhou2017partII}%
  \BibitemOpen
  \bibfield  {author} {\bibinfo {author} {\bibnamefont {Zhou}, \bibfnamefont
  {Y.}},\ }\bibfield  {title} {\enquote {\bibinfo {title} {{Rayleigh–Taylor
  and Richtmyer–Meshkov instability induced flow, turbulence, and mixing.
  {II}}},}\ }\href@noop {} {\bibfield  {journal} {\bibinfo  {journal} {Physics
  Reports}\ }\textbf {\bibinfo {volume} {723-725}},\ \bibinfo {pages} {1--160}
  (\bibinfo {year} {2017}{\natexlab{b}})}\BibitemShut {NoStop}%
\bibitem [{\citenamefont {Zhou}(2024)}]{Zhou2024}%
  \BibitemOpen
  \bibfield  {author} {\bibinfo {author} {\bibnamefont {Zhou}, \bibfnamefont
  {Y.}},\ }\href@noop {} {\emph {\bibinfo {title} {{Hydrodynamic Instabilities
  and Turbulence: Rayleigh–Taylor, Richtmyer–Meshkov, and
  Kelvin–Helmholtz Mixing}}}},\ \bibinfo {edition} {1st}\ ed.\ (\bibinfo
  {publisher} {Cambridge University Press},\ \bibinfo {address} {Cambridge,
  United Kingdom},\ \bibinfo {year} {2024})\BibitemShut {NoStop}%
\bibitem [{\citenamefont {Zhou}, \citenamefont {Sadler},\ and\ \citenamefont
  {Hurricane}(2025)}]{zhou2025instabilities}%
  \BibitemOpen
  \bibfield  {author} {\bibinfo {author} {\bibnamefont {Zhou}, \bibfnamefont
  {Y.}}, \bibinfo {author} {\bibnamefont {Sadler}, \bibfnamefont {J.~D.}}, and\
  \bibinfo {author} {\bibnamefont {Hurricane}, \bibfnamefont {O.~A.}},\
  }\bibfield  {title} {\enquote {\bibinfo {title} {{Instabilities and Mixing in
  Inertial Confinement Fusion}},}\ }\href@noop {} {\bibfield  {journal}
  {\bibinfo  {journal} {Annual Review of Fluid Mechanics}\ }\textbf {\bibinfo
  {volume} {57}},\ \bibinfo {pages} {197--225} (\bibinfo {year}
  {2025})}\BibitemShut {NoStop}%
\bibitem [{\citenamefont {Zhou}\ \emph {et~al.}(2021)\citenamefont {Zhou},
  \citenamefont {Williams}, \citenamefont {Ramaprabhu}, \citenamefont {Groom},
  \citenamefont {Thornber}, \citenamefont {Hillier}, \citenamefont {Mostert},
  \citenamefont {Rollin}, \citenamefont {Balachandar}, \citenamefont {Powell},
  \citenamefont {Mahalov},\ and\ \citenamefont {Attal}}]{zhou2021rayleigh}%
  \BibitemOpen
  \bibfield  {author} {\bibinfo {author} {\bibnamefont {Zhou}, \bibfnamefont
  {Y.}}, \bibinfo {author} {\bibnamefont {Williams}, \bibfnamefont {R.~J.}},
  \bibinfo {author} {\bibnamefont {Ramaprabhu}, \bibfnamefont {P.}}, \bibinfo
  {author} {\bibnamefont {Groom}, \bibfnamefont {M.}}, \bibinfo {author}
  {\bibnamefont {Thornber}, \bibfnamefont {B.}}, \bibinfo {author}
  {\bibnamefont {Hillier}, \bibfnamefont {A.}}, \bibinfo {author} {\bibnamefont
  {Mostert}, \bibfnamefont {W.}}, \bibinfo {author} {\bibnamefont {Rollin},
  \bibfnamefont {B.}}, \bibinfo {author} {\bibnamefont {Balachandar},
  \bibfnamefont {S.}}, \bibinfo {author} {\bibnamefont {Powell}, \bibfnamefont
  {P.~D.}}, \bibinfo {author} {\bibnamefont {Mahalov}, \bibfnamefont {A.}},
  and\ \bibinfo {author} {\bibnamefont {Attal}, \bibfnamefont {N.}},\
  }\bibfield  {title} {\enquote {\bibinfo {title} {{Rayleigh--Taylor and
  Richtmyer--Meshkov instabilities: A journey through scales}},}\ }\href@noop
  {} {\bibfield  {journal} {\bibinfo  {journal} {Physica D: Nonlinear
  Phenomena}\ }\textbf {\bibinfo {volume} {423}},\ \bibinfo {pages} {132838}
  (\bibinfo {year} {2021})}\BibitemShut {NoStop}%
\end{thebibliography}%

\end{document}